%% file: Thesis.tex
\documentclass[a4paper,12pt]{book}%
\usepackage[utf8]{inputenc}
\usepackage{amsmath,amssymb,mathrsfs,latexsym}
\usepackage{graphicx}
\usepackage{slashed}
\usepackage{bbm}
\usepackage{fancyhdr}
\usepackage{a4wide}
\usepackage[usenames]{color}
\usepackage{lscape}
\usepackage{cite}

\fancypagestyle{diploma}{
\fancyhf{}
\fancyhead[RO,LE]{\thepage}
\fancyhead[LO]{\rightmark}
\fancyhead[RE]{\leftmark}
}
\input{Commands.tex}

\begin{document}

\begin{titlepage}
\begin{center}
\newcommand{\Rule}{\rule{\textwidth}{1mm}} 
\Rule
\vspace{5mm}

\Huge{Systematic Study \\ of Hadronic Molecules \\ in the Heavy Quark Sector}

\vspace{1mm}
\Rule
\vfill
\vspace{0.5cm}
\LARGE
Dissertation\\
\vspace{0.5cm}
\Large
zur\\\vspace{0.5cm}
\LARGE
Erlangung des Doktorgrades (Dr. rer. nat.)\\\vspace{0.5cm}
\Large
der\\\vspace{0.5cm}
\LARGE
Mathematisch-Naturwissenschaftlichen Fakultät\\\vspace{0.5cm}
\Large
der\\\vspace{0.5cm}
\LARGE
Rheinischen Friedrich-Wilhelms-Universität Bonn\\\vspace{0.5cm}
\vspace{1cm}
\Large
vorgelegt von\\
\LARGE
Martin Cleven\\
\vspace{0.5cm}
\Large
aus\\
Geldern\\
\vspace{1cm}
Bonn Oktober 2013
\end{center}
\end{titlepage}

\newpage

\normalsize
\thispagestyle{empty}
\pagenumbering{roman}
\phantom{xxx}
\newpage
\thispagestyle{empty}
\pagenumbering{roman}

\input{Abstract.tex}

\tableofcontents
           
\pagestyle{diploma}
\clearpage
\pagenumbering{arabic}

\newpage

\input{Introduction.tex}

\input{Theory.tex}

\input{Ds0andDs1.tex}

\input{Zb.tex}
\input{Summary.tex}
\input{Appendix.tex}

\bibliography{Thesis}
\bibliographystyle{Thesis}

\end{document}

%% file: Commands.tex
\newcommand{\non}{\nonumber}
\newcommand{\dszero}{D_{s0}^*}
\newcommand{\dsone}{D_{s1}}
\newcommand{\bszero}{B_{s0}}
\newcommand{\bsone}{B_{s1}}

\newcommand{\lag}{\mathscr L}
\newcommand{\Lag}{\mathcal{L}}
\newcommand{\Amp}{\mathcal{A}}
\newcommand{\Eng}{\mathcal{E}}
\newcommand{\lang}{\left\langle}
\newcommand{\rang}{\right\rangle}
\newcommand{\lr}{ \overleftrightarrow{\partial}}
\newcommand{\zbs}{Z_b^{(\prime)}}
\newcommand{\zb}{Z_b(10610)}
\newcommand{\zbp}{Z_b(10650)}
\newcommand{\lqcd}{\Lambda_{QCD}}

\newcommand{\intxxx}{\int^1_0\!\mathrm d^3x\;\delta\!\left(1-\sum x_i\right)}
\newcommand{\intx}{\int^1_0\!\mathrm dx }
\newcommand{\intl}{\int\!\frac{\mathrm d^d l}{(2\pi)^d}}
\newcommand{\intll}{\int\!\frac{\mathrm d^{d-1} l}{(2\pi)^{d-1}}}

\newcommand{\gmunu}{g^{\mu\nu}}
\newcommand{\mn}{{\mu\nu}}

\newcommand{\projv}{\frac{\mathbbm{1}+\slashed v}{2}}

\newcommand{\mev}{~\mathrm{MeV}}
\newcommand{\gev}{~\mathrm{GeV}}
\newcommand{\kev}{~\mathrm{keV}}
\newcommand{\prop}{[(p-l)^2-M_1^2][(q-l)^2-M_2^2][l^2-M_3^2]}

\newcommand{\propq}{[(q-l)^2-M_1^2][l^2-M_2^2]}

\newcommand{\Jzero}{J^{(0)}}
\newcommand{\Joneone}{J^{(1)}_1}
\newcommand{\Jonetwo}{J^{(1)}_2}
\newcommand{\Jtwozero}{J^{(2)}_0}
\newcommand{\Jtwoone}{J^{(2)}_1}
\newcommand{\Jtwotwo}{J^{(2)}_2}
\newcommand{\Jtwothree}{J^{(2)}_3}

\newcommand{\Ionep}{I^{(1)}\left(p^2,M_1^2,M_3^2\right)}

\newcommand{\Ionek}{I^{(1)}\left((p-q)^2,M_1^2,M_2^2\right)}
\newcommand{\Izerop}{I^{(0)}\left(p^2,M_1^2,M_3^2\right)}
\newcommand{\Izeroq}{I^{(0)}\left(q^2,M_2^2,M_3^2\right)}
\newcommand{\Izerok}{I^{(0)}\left((p-q)^2,M_1^2,M_2^2\right)}

\newcommand{\cdk}{\mathcal C_{DK}}
\newcommand{\cdstark}{\mathcal C_{D^*K}}
\newcommand{\cbk}{\mathcal C_{BK}}
\newcommand{\cbstark}{\mathcal C_{B^*K}}
\newcommand{\dk}{D^{(*)}K}

%% file: Abstract.tex
\section*{Abstract}
In this work we have studied properties of hadronic molecules in the heavy quark sector. These have become increasingly important since from the beginning of this century a large number of states have been measured that for different reasons do not fit the predictions of simple quark models. Theorists have proposed different explanations for these states including tetraquarks, hybrids, hadro-quarkonia and, subject of this work, hadronic molecules. The study of these new states promises to provide insights in an important field of modern physics, the formation of matter by the strong force. 
Hadronic molecules are bound systems of hadrons in the same way two nucleons form the deuteron. For this the molecular states need to be located close to $S$-wave thresholds of their constituents. The dynamics of their constituents will have a significant impact on the molecules which allows us to make predictions that are unique features of the molecular assignement. 

Here we focus on two candidates in the open charm sector, $\dszero$ and $\dsone$, and two candidates in the bottomonium sector, $\zb$ and $\zbp$. 
The $D_{sJ}$ are located similarly far below the open charm thresholds $DK$ and $D^*K$. Since the spin dependent interactions of the pseudoscalar and vector charmed mesons are suppressed in the heavy quark limit the interpretation of these two states as mainly $DK$ and $D^*K$ bound states naturally explains their similarities.
The more recently discovered states $\zb$ and $\zbp$, located very close to the open bottom thresholds $B^*\bar B+c.c.$ and $B^*\bar B^*$, respectively, are manifestly non-conventional. Being electromagnetically charged bottomonia these states necessarily have at least four valence quarks. We can explain that together with the fact that they decay similarly into final states with $S$- and $P$-wave bottomonia if we assume they are $B^*\bar B+c.c.$ and $B^*\bar B^*$ molecules, respectively. 
Since the current experimental situation in both cases does not allow for final conclusions we try to point out quantities that, once measured, can help to pin down the nature of these states. 

For the $D_{sJ}$ we can make use of the fact that the interactions between charmed mesons and Goldstone bosons are dictated by chiral symmetry. This means that we can calculate the coupled channel scattering amplitudes for $DK$ and $D_s\eta$ and their counterparts with charmed vector mesons. $\dszero$ and $\dsone$ can be found as poles in the unitarized scattering amplitudes. We can calculate the dependence of these poles on the the strange quark mass and the averaged mass of up and down quark. This makes the result comparable to lattice calculations. Solving QCD exactly on the lattice can help us to understand the nature of the $D_{sJ}$ states while in the meantime it possibly takes one more decade until the PANDA experiment at FAIR will be able to judge if the molecular assignement is correct.
Furthermore we calculate the radiative and hadronic two-body decays. Here we find that in the molecular picture the isospin symmetry violating decays $\dszero\to D_s\pi$ and $\dsone\to D_s^*\pi$ are about one order of magnitude larger than the radiative decays. This is a unique feature of the molecular interpretation --- compact $c\bar s$ states have extremely suppressed hadronic decay rates. At the same time the radiative decays have comparable rates regardless of the interpretation. In conclusion the hadronic decay widths are the most promising quantities to experimentally determine the nature of $\dszero$ and $\dsone$.

The methods we used in the open charm sector cannot be applied to the bottomonia one-to-one. Since we do not know the interaction strength between open bottom mesons we cannot obtain the state as a pole in a unitarized scattering matrix. We therefore need different quantities to explore the possible molecular nature. In a first attempt we show that the invariant mass spectra provided by the Belle group can be reproduced by assuming the $\zbs$ are bound states located below the $B^*\bar B+c.c.$ and $B^*\bar B^*$ thresholds, respectively. Furthermore we present the dependence of the lineshape on the exact pole position. An important conclusion here is that for near threshold states like the $\zbs$ a simple Breit Wigner parametrization as it is commonly used by experimental analyses is not the appropriate choice. Instead we suggest to use a Flatt\'{e} parametrization in the proximity of open thresholds. 
The second part of the discussion of the $Z_b$ states includes calculations of two-body decays. In particular we present the final states $\Upsilon\pi$ and  $h_b\pi$ which have already been seen by experiment and make predictions for a new final state  $\chi_{b}\gamma$. The rates into this new final state are large enough to be seen at the next-generation $B$-factories.

%% file: Introduction.tex
\chapter{Introduction}

One of the big challenges in modern physics is to understand how matter is formed. It is known that the biggest part of the observable matter --- we will not deal with phenomena like dark matter in this work --- is made of strongly interacting quarks and gluons. Quantum Chromodynamics (QCD) describes the interaction between quarks that carry the charge of the strong interaction, called $color$, by force mediating gluons.
This is similar to the theory of electromagnetic interactions, Quantum Electrodynamics (QED), where the interactions between charged particles are described by the exchange of photons. 
However, QCD is field theoretically a $SU(3)$ gauge theory, instead of the $U(1)$ QED, which leads to nonlinear equations of motion.
As a result gluons can not only couple to quarks but also to themselves and therefore need to carry color charge. The additional self energy correction terms for the gluon arising from this self-interaction make the coupling constant $\alpha_S$ to a color charge behave in  a peculiar way:  it becomes weaker for high energies --- this phenomenon is known as the {\it asymptotic freedom} of QCD. It allows for perturbation theory in terms of $\alpha_S$ for high energies.

However, at smaller energies, i.e. the regime of $1\gev$, the behavior changes: the coupling constant increases until it is of order one and the perturbation series breaks down. Perturbative QCD with quarks and gluons as degrees of freedom is therefore not able to describe interactions in this energy regime. At the same time we find that, while QCD describes the interactions of particles that carry color charge, until today experiments were not able to detect a colored object directly. Instead, only colorless $hadrons$ can be observed --- this phenomenon is called {\it confinement}. Understanding confinement from first principles and the formation of hadrons from quarks and gluons remain to be understood. 

There are many ways to form a colorless object. The simplest ones, called $mesons$, have the structure $\bar qq$: a quark that carries color and an antiquark that carries the corresponding anticolor. The most prominent mesons are formed from the lightest quarks, called pions. 
The second possibility are $baryons$ that contain three quarks with three different colors: $qqq$. The sum of all three colors is is also colorless. The most important baryons are, for obvious reasons, proton and neutron from which the nuclei are built. Baryons will not be part of this work at all. 

Since there are a priori no limits on hadrons besides colorlessness theorists have made predictions of so-called exotic states. 
The exotic mesons include tetraquarks, glueballs, hybrids, hadro-quarkonia and, subject of this work, hadronic molecules. The last ones are bound states of two conventional mesons in the same way two nucleons form the deuteron. 

The discussion about exotic states became more lively when at the beginning of this century the so-called $B$-factories BaBar and Belle started working. Originally designed to study $CP$-violation in $B$-mesons and the weak CKM matrix elements $\left|V_{ub}\right|$ and $\left|V_{db}\right|$, the $B$-factories also became famous for measuring a number of states that challenge the quark models based on simple $\bar qq$ mesons. These models were successful in describing the ground states and some low lying excited states for charmonia and bottomonia, respectively, as well as for  mesons with open charm or bottom. This picture was changed when two narrow resonances with open charm, now referred to as $D_{s0}^*(2317)$ and $D_{s1}(2460)$, and large number of charmonium-like states close to or above the open-charm threshold amongst which the $X(3872)$ is the most famous one were discovered. All these states do not fit in the scheme given by quark-model predictions. The first two are possible candidates for $DK$ and 
$D^{*}K$ bound states, respectively, the latter for an isospin singlet $DD^*$ bound state. 
However, the current data base is insufficient for a definite conclusion on their structure. The $D_{sJ}$ states can be molecules, tetraquarks or conventional mesons while the $X(3872)$ can be a molecular state and a virtual state, or a dynamical state with a significant admixture of a $\bar cc$ state. Most recently high statistics measurements at BESIII from 2013 suggest that $Y(4260)$, previously a prominent candidate for a hybrid, might be a $D_1D$ molecule. Moreover, BESIII also measured a charged charmonium, $Z_c(3900)$, that is a good candidate for a $\bar DD^*$ molecule. 

Due to the large similarities between charmonia and bottomonia that emerges since the QCD Lagrangian becomes flavor independent for $m_Q\to\infty$ one expects similar exotic states here. Indeed, in 2011 the Belle group reported the bottomonium states $Z_b(10610)$ and $Z_b(106510)$. Their exotic nature is manifest since they, being charged bottomonia, have to contain at least four quarks. The later measured $Z_c(3900)$ can therefore be seen as the charmonium partner of $Z_b(10610)$ in accordance with Heavy Quark Flavor Symmetry. 
It is to be expected that once the next-generation $B$-factories like Belle II will start working the number of exotic bottomonia will rise. 
It is our belief that the study of these exotic candidates will help to deepen the understanding of the formation of matter.

%%%%%%%%%%%%%%%%%%%%%%%%%%%%
%
%   This Work
%
%%%%%%%%%%%%%%%%%%%%%%%%%%%
In this work we will focus on the  states $D_{s0}^*(2317)$ and $D_{s1}(2460)$ in the open charm sector and $Z_b(10610)$ and $Z_b(10650)$ in the bottomonium sector as examples for hadronic molecules. For the sake of convenience we will in the following refer to them as $\dszero$, $\dsone$ and $\zbs$. We will present these as examples how states can be formed from meson-meson interactions and lay out ways to test their nature experimentally.

This work is structured as follows. In Ch.~\ref{ch:theory} we will present the theoretical framework. That includes a general discussion of hadronic molecules, in particular in comparison to competing models like tetraquarks, a discussion of the effective field theories that were used, Heavy Meson Chiral Perturbation Theory and Nonrelativistic Effective Theory, and a brief section about Unitarization and the dynamical generation of resonances. 

The main part of this work is divided into a chapter about the open charm states and one about the $Z_b$ states. In Ch.~\ref{ch:charmed} we will obtain both $\dszero$ and $\dsone$ as spin partners with one set of parameters from heavy meson Goldstone boson interactions. Both will be found as dynamically generated poles in unitarized scattering amplitudes. 
Since of late more and more effort is being put into lattice calculations on these matters, we will present light quark mass dependent calculations of relevant quantities like the pole position and the binding energy. These can when calculated on the lattice provide a way to distinguish between hadronic molecules and other explanations like e.g. tetraquarks and help us to pin down the nature of $\dszero$ and $\dsone$. In a second attempt we present calculations of the radiative and hadronic decays of $\dszero$ and $\dsone$. We assume that these are driven mainly by $D^{(*)}K$ loops. The rates are so far predictions, experimental data only exists for ratios 
of hadronic and radiative decays. Finally we will make predictions for similar open bottom states. Since Heavy Quark Effective Theory tells us that the interactions of charm and bottom quarks with light degrees of freedom are the same up to small flavor symmetry breaking effects the experimental discovery of these is a crucial test of our theoretical model.

In Ch.~\ref{ch:zb} we will investigate the properties of the $Z_b$ states assuming they are hadronic molecules formed from $B^*\bar B+B\bar B^*$ and $B^*\bar B^*$ interactions, respectively. First we will show that the measured spectra are also compatible with bound states, that is pole positions below the corresponding thresholds. In the experimental analysis simple Breit-Wigner shapes were used for $\zbs$. We will show that due to the very close proximity of the bottom meson thresholds this is not the proper choice and propose a Flatt\'{e} parametrization instead. Further we will calculate various two-body decays of $\zbs$ with conventional bottomonia as final states. These include the so far not measured final states $\chi_{b}\gamma$. We are also able to make parameter free predictions with nontrivial statements that can be tested by experiment.

In Ch.~\ref{ch:summary} we will summarize the main results of this work and present future tasks.

%% file: Theory.tex
\chapter{Theory}\label{ch:theory}
%%%%%%%%%%%%%%%%%%%%%%%%%%%%%%%%%%%%%%%%%%%%%%%%
%
\section{Hadronic Molecules}\label{sec:molecules}
%
%%%%%%%%%%%%%%%%%%%%%%%%%%%%%%%%%%%%%%%%%%%%%%%%

\begin{figure}[t]
 \centering
 \includegraphics[width=\textwidth]{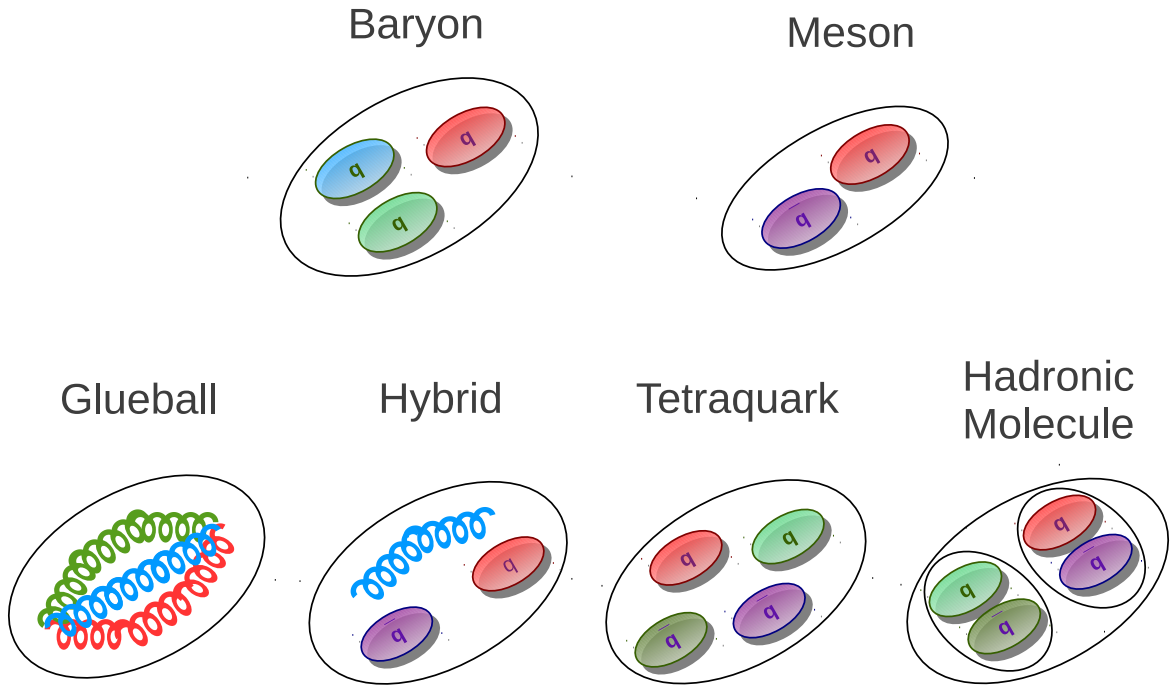}
 \caption{Conventional and exotic colorfree states}
 \label{fig:exotics}
\end{figure}

A peculiar feature of QCD is that only color neutral states, hadrons, can be observed, an effect one refers to as confinement. In the most naive picture we can form colorless states as two quark states $(mesons)$ and three quark states $(baryons)$, both shown in the first line of Fig.~\ref{fig:exotics}. For the first one  the antiquark has the corresponding anticolor to the color of the quark, for the latter the three quarks all have different colors. In both cases this results in a color neutral object. As mentioned this is only the most naive picture. In a more elaborate description of hadrons one also needs to consider sea quarks, quark-antiquark pairs that are created and annihilated, and gluons that couple to the quarks. However, for our discussion these are not necessary and so we will only refer to the valence quarks --- $\bar qq$ in the case of mesons and $qqq$ for baryons. This work focuses on mesons only, so there will be no further discussion of baryons. But all issues of this discussion can 
also be applied to the baryonic sector. 

There are a priori no restrictions on how to form a colorless hadron. We will therefore generalize the concept and call all objects with an integer spin mesons. States with valence quarks $\bar qq$ are then {\it conventional mesons}. However, one can imagine various additional states, shown in Fig.~\ref{fig:exotics}, labeled as {\it exotics} to be distinguished from these conventional mesons. Hybrids are made up from two quarks and gluonic excitations. This can lead to two quark states with quantum numbers that are not possible within conventional quark models. Even more exotic are the so-called glueballs: it is possible to form a state exclusively from the massless gluons, see~\cite{Close:2002zu,Close:2005vf}. Morningstar found a whole spectrum in pure Yang-Mills lattice QCD~\cite{Morningstar:1999rf}. For an experimental review on hybrids and glueballs see~\cite{Klempt:2007cp}.
Due to the binding energy these states can then obtain a mass. Also the number of quarks is not constricted by any known principle. Candidates for mesons with four quarks are tetraquarks and hadronic molecules which need to be distinct. Tetraquarks are compact states with two quarks and two antiquarks. A comprehensive review of tetraquarks in the heavy quarkonium sector is given by Faccini et al.~\cite{Faccini:2012pj}.
Hadronic molecules, on the other hand, consist of two $\bar qq$ states that are bound by the strong force. In the course of this work we will present possible candidates for hadronic molecules and suggest methods to test the molecular nature experimentally or by lattice calculations. 

Since hadronic molecules are the main subject of this work we need to go into more detail here. In the sense we are using the name a hadronic molecule can be any object formed of two hadrons. The name of this state differs depending on the position of the pole in the $S-$matrix. A $bound$ $state$ pole of two hadrons $h_1$ and $h_2$ is located on the physical sheet below the open threshold $m_{h_1}+m_{h_2}$ by a binding energy $\epsilon$. If the state is located above the open threshold on the second sheet it is called a $resonance$ in the $h_1$-$h_2$ system. A $virtual$ $state$ is also located above threshold but on the second Riemann sheet. All three states, $bound$ $states$, $resonances$ and $virtual$ $states$ are called hadronic molecules. The position of the pole is visible in the lineshape of the state. In Ch.~\ref{ch:zb} we will discuss this on the example of the $Z_b$ states. 

The central tool to model-independently identify hadronic molecules is Weinberg's time-honored approach~\cite{Weinberg:1962hj,Weinberg:1963zza}. It was originally introduced to unambiguously determine the most dominant component of the deuteron wave function.  The approach was generalized in Ref.~\cite{Baru:2003qq} to the case  of inelastic channels, as well as to the case of an above-threshold resonance. When being applied to the $f_0(980)$, strong evidence was found for a dominant $\bar KK$ molecule inside the $f_0(980)$, while the situation was not that clear for the $a_0(980)$~\cite{Hanhart:2007wa,Kalashnikova:2004ta,Branz:2008ha}. 

%%%%%%%%%%%%%%%%%%%%%%%%%%%%%%%%%%%%%%%%%%%%%%%%
%
%   Simple QM
%
%%%%%%%%%%%%%%%%%%%%%%%%%%%%%%%%%%%%%%%%%%%%%%%%
We can derive the central formula, a relation between the coupling of the molecule to its constituents and the probability that the physical state is the molecular state, from very basic quantum mechanical principles. Suppose the physical state $\big|\psi\big>$ contains a two-hadron state $\big|h_1h_2\big>=\big|\vec q\big>$ and a state we call $\big|\psi_0\big>$ which contains everything else.  This state fulfills the time independent Schrödinger equation:
\begin{equation}
 \left(\hat H_0+\hat V\right)\big|\psi\big>=-E_B\big|\psi\big>
\end{equation}
with the binding energy $E_B>0$ and the free Hamiltonian $\hat H_0$ that fulfills 
\begin{equation}
 \hat H_0\big |\vec q\big>=\frac{\vec q\;^2}{2\mu}\big|\vec q\big >.
\end{equation}
If we multiply this with $\big<\vec q\big |$ we find 
\begin{equation}
 \left<\vec q\big|\psi\right>=-\frac{\left<\vec q\big|\hat V\big|\psi\right>}{E_B+\vec q\,^2/2\mu}. 
\end{equation}
To obtain the probability $\lambda$ to find the physical state in the continuum state we simply need to integrate the absolute squared of this amplitude over $\vec q$:
\begin{equation}
 \lambda^2=\int \!\mathrm d^3q\left|\left<\vec q\big|\psi\right>\right|^2=\int \!\mathrm d^3q\frac{\left|\left<\vec q\big|\hat V\big|\psi\right>\right|^2}{\left[E_B+\vec q\,^2/2\mu\right]^2}.
\end{equation}
We can rewrite this equation introducing $g_{NR}(\vec q):=\left|\left<\vec q\big|\hat V\big|\psi\right>\right|^2$ which will later be related to the coupling of the molecular state to its constituents. We find
\begin{equation}\label{eq:lambdaintegral}
 \lambda^2=4\mu^2\int\!\mathrm d\Omega\int\!\mathrm dqq^2\frac{g_{NR}^2(\vec q)}{(q^2+2\mu E_B)^2}. 
\end{equation}
We define the binding momentum as $\gamma:=\sqrt{2\mu E_b}$. If the binding momentum is small compared to the range of forces $R$, $\gamma\ll 1/R$ we can expand $g_{NR}(\vec q)$ as 
\begin{equation}
 g^2_{NR}(\vec q)=q^{2L}g^2_{NR}(0)+\mathcal O(R\gamma).
\end{equation}
The range of forces $R$ in the case of deuteron would be the inverse mass of the exchange particle, so $\mathcal O(M_\pi^{-1})$ which is indeed small compared to the binding momentum. 
The integral in Eq.~(\ref{eq:lambdaintegral}) is convergent only for $S-$wave interactions, i.e. $L=0$. Then a model independent statement is possible.
If we carry out the integration for this case we find
\begin{equation}\label{eq:lambdaQM}
 \lambda^2=\frac{4\pi\mu^2}{\gamma}g_{NR}^2(0)+\mathcal O(R\gamma) 
\end{equation}
So far everything was deduced using only basic quantum mechanic principles. In a relativistic quantum field theory the interaction Lagrangian reads
\begin{equation}
 \lag=g\psi^\dagger h_1h_2+{\mathrm h.c.}
\end{equation}
where $\psi^\dagger$ creates the physical field and $h_i$ annihilate the constituent fields. If we want to translate the coupling $g_{NR}$ in  Eq.~(\ref{eq:lambdaQM}) to the relativistic $g$ we need to consider different normalizations. We find
\begin{equation}
 g=(2\pi)^{3/2}\sqrt{2m_1}\sqrt{2m_2}\sqrt{2(m_1+m_2)}g_{NR}(0)
\end{equation}
which means that the coupling $g$ contains the probability $\lambda^2$ earlier introduced. Since this probability is between 0 for a completely compact state and 1 for a purely molecular state we find
\begin{equation}\label{eq:weinberg}
 g^2\simeq16\pi\lambda^2\frac{(m_1+m_2)^2}{\mu}\gamma\leq16\pi\frac{(m_1+m_2)^2}{\mu}\gamma
\end{equation}
which finally allows us to judge the molecular portion in an arbitrary state. 
If we examine the two cases $\lambda=1$, where the physical state is a purely bound state and $\lambda=0$ for no bound state component we observe that the coupling to the bound states components becomes maximal ($\lambda=1$) or vanishes completely $(\lambda=0)$. 

Instead of the coupling constant $g$ we may as calculate the scattering length $a$ for the constituents \cite{Weinberg:1965zz,Hanhart:2006nr}:
\begin{equation}
 -\frac1a=\frac{2E_b}{g}+\sqrt{\mu E_b}
\end{equation}
or
\begin{equation}\label{eq:scatteringlength}
 a=-2\left(\frac{1-Z}{2-Z}\right)\frac1{\sqrt{\mu E_b}}.
\end{equation}

%%%%%%%%%%%%%%%%%%%%%%%%%%%%%%%%%%%%%%%%%%%%%%%%%%%%%%%%%%
%
\section{Effective Theories}\label{sec:effective_theories}
%
%%%%%%%%%%%%%%%%%%%%%%%%%%%%%%%%%%%%%%%%%%%%%%%%%%%%%%%%%%
Certain physics theories that are successful in describing the behavior at e.g. very high energies, while they turn out to be unnecessarily complicated at lower energies and can be replaced by simpler, so--called \textit{effective field theories}. They are able to describe all effects of the underlying theory at this scale and are a well defined limit of it.

A famous example for an effective field theory is the Fermi contact interaction for weak interactions. For energies well below the $W$-boson mass one does not need a dynamical $W$-exchange for weak interactions a contact interaction is sufficient.

The situation is quite different, when it comes to the theory of strong interaction. In the standard model $QCD$ is a $SU(3)$ gauge theory described by the Lagrangian

\begin{equation}
 \lag_{\rm QCD}=\sum_i \bar q_i(\slashed D-m_i )q_i-\frac14 G_{\mu\nu}^aG^{\mu\nu}_a.
\end{equation}
where $D_\mu=\partial_\mu+ igA_\mu^a\lambda_a/2$ is the covariant derivative with the gluon fields $A_\mu^a$ and the $SU(3)$ group generators $\vec \lambda/2$. $q_i$  represent the six quarks and $G_{\mu\nu}^a$ is the gluon field strength tensor. Without going into too much detail,  we state that due to the non-abelian nature of the gauge field and the resulting gluon self-energy corrections the strong coupling obeys
\begin{eqnarray}
 \alpha_s=\frac{12\pi}{(33-2N_q)\log(\mu^2/\lqcd^2)}.
\end{eqnarray} 
where $N_q$ is the number of quark flavors and $\mu$ the scale we are looking at. One can see, that this coupling is small for energies much larger than the scale--independent constant $\lqcd$, which is experimentally $\sim 220$~MeV, and therefore allows for perturbation theory. This is known as asymptotic freedom.  As a result the high--energy behavior of the strong interaction can be described by \textit{perturbative Quantum Chromodynamics} with quarks and gluons as degrees of freedom. 

When we examine the energy region, where $\mu\rightarrow \lqcd$, we find that the coupling constant diverges. This is called the non--perturbative region of QCD. Because of the non--Abelian nature of the QCD, there is the well--known empirical phenomenon confinement: No objects carrying color can be observed in nature, instead only bound states without color charge.

Therefore a different approach is required to perform calculations in this region. The only known model--independent solution besides lattice QCD is to use effective field theories which obey the symmetries of the fundamental theory but allow for a perturbative expansion. 
The known approaches make use of the fact that for the first three quarks -- up, down and strange -- the masses are significantly smaller than $\lqcd$, while the masses of charm and bottom are much larger:
\begin{equation}
 m_u,m_d,m_s\ll \lqcd\qquad {\rm and} \qquad m_c,m_b\gg \lqcd.
\end{equation}
The sixth quark's mass is of course even larger than $m_c$ and $m_b$ and therefore even better suited for an expansion. However, the top quark decays so fast that it cannot form strongly bound states of any kind. The resulting expansion parameters are $m_q/\lqcd$ for the light quarks and $\lqcd/m_Q$ for the heavy quarks. 

These properties are the basis of the perturbative approaches we will be using in this work. 
In Sec.~(\ref{sec:chpt}) we will make use of the additional chiral symmetry that one obtains in the limit of massless quarks. This was first proposed by Weinberg in \cite{Weinberg:1978kz} and worked out in detail by Gasser and Leutwyler~\cite{Gasser:1983yg}. The Goldstone bosons that arise from spontaneous breaking of this additional $SU(3)_L\times SU(3)_R$ symmetry will be identified with the light pseudoscalar mesons $\pi$, $K$ and $\eta$. 
In Sec.~(\ref{sec:hqs}) we will investigate the limit of infinitely heavy quark masses and find that two important consequences are {\it heavy quark spin symmetry} and {\it heavy quark flavor symmetry}: For systems with one heavy quark the exact  mass, i.e. flavor, does not change the interactions as long as the mass is large enough and all spin-dependent interactions are suppressed. 

Sec.~(\ref{sec:hmchpt}) presents the framework most of our calculations are based on: {\it Heavy Meson Chiral Perturbation Theory} (HMChPT). The findings of both previous sections are including there. The heavy meson fields obey {\it heavy quark spin  and flavor symmetry} and the Goldstone bosons are used as gauge fields so that the theory obeys chiral symmetry up to the leading symmetry breaking effects. {\it Nonrelativistic Effective Theory (NREFT)}, presented in Sec.~(\ref{sec:nreft}), can be used for heavy-heavy systems. Thus it allows to describe the dynamics of quarkonia with heavy meson loops. 
%%%%%%%%%%%%%%%%%%%%%%%%%%%%%%%%%%%%%%%%%%%
%
%   Chiral Symmetry
%
%%%%%%%%%%%%%%%%%%%%%%%%%%%%%%%%%%%%%%%%%%%
\subsection{Chiral Symmetry}\label{sec:chpt}
\begin{figure}
 \centering
 \includegraphics[width=0.6\textwidth]{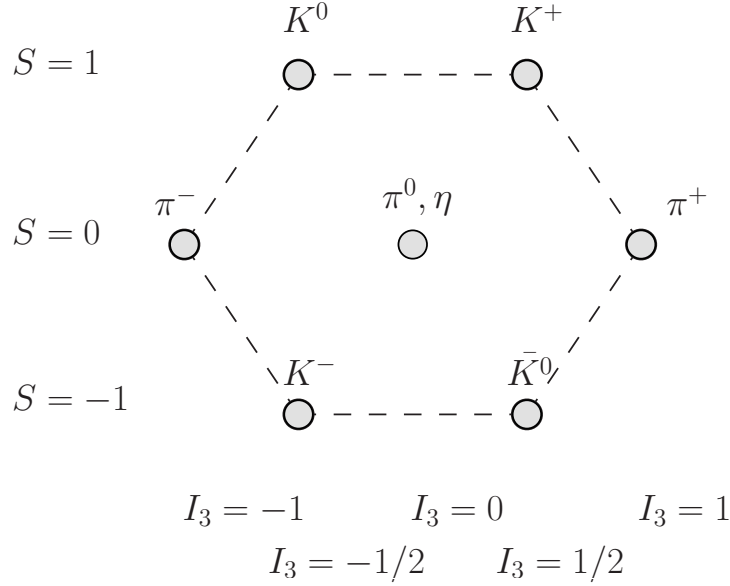}
 \caption{The SU(3) Goldstone boson octet. The x-axis shows the third component of the isospin, the y-axis shows the strangeness}
 \label{fig:octet}
\end{figure}

We want to study the properties of QCD at low energies for light quarks. The approach here follows the introductory Refs.~\cite{Scherer:2002tk,Kubis:2007iy}
The first important symmetry is the additional {\it chiral symmetry} that arises in the limit of massless quarks. For the three lightest quarks, up, down and strange, we find masses that are small compared to the scale of nonperturbative QCD, $\lqcd$. This justifies treating them as massless. Here one may treat the quark masses as perturbations. The light quark part of the QCD Lagrangian reads
\begin{eqnarray}
 \mathscr{L}=\bar q i\slashed D q -  \bar qM_q q...~,
\end{eqnarray}
where ... denote the heavy quark and gluon fields. $q=(u,d,s)$ are the light quark fields, $M_q={\rm Diag}(m_u,m_d,m_s)$ is the light quark mass matrix. Chiral projectors allow us to separate the quark fields into left-handed ($L$) and right-handed ($R$) fields $q_{L/R}=\frac{1}{2}(1\pm\gamma_5)q$. With this we may write
\begin{eqnarray}\label{eq:lag_light_quark}
 \mathscr{L}_{light}=\bar q i\slashed Dq=\bar q_L i\slashed D q_L + \bar q_R i\slashed D q_R-\bar q_R M_q q_L-\bar q_LM_q q_R,
\end{eqnarray}
In the limit of vanishing quark masses the last two terms disappear, left- and right-handed fields decouple and we find the additional chiral symmetry under transformations
\begin{eqnarray}
 q_L\rightarrow Lq_L \qquad q_R\rightarrow Rq_R,
\end{eqnarray}
where $L\in SU(3)_L$ and $R\in SU(3)_R$. This symmetry is spontaneously broken by the vacuum expectation value of quark bilinears
\begin{eqnarray}
 \left<0|\bar q_R^aq_L^b|0\right>=  \delta^{ab} B F^2.
\end{eqnarray}
for $F\neq0$. One can identify $F$ at leading order with the pion decay constant $F_\pi$.
The indices $a,b$ are light flavor indices. An arbitrary $SU(3)_L\times SU(3)_R$ $q\rightarrow q'$ transformation gives us
\begin{eqnarray}
 \left<0\left| \bar{q}_R^{\prime a}{q}_L^{\prime b}\right|0\right>= (LR^\dagger)^{ab} B F^2.
\end{eqnarray}
Thus the vacuum expectation value is unchanged if $L=R$ and the $SU(3)_L\times SU(3)_R$ symmetry is spontaneously broken to its diagonal subgroup $SU(3)_V$. We now consider the composite field $\bar q^a_R q^b_L$. The broken generators of $SU(3)_L\times SU(3)_R$ leave the potential energy unchanged. Since there are eight of these broken generators we obtain eight massless Goldstone bosons. These Goldstone bosons are represented by one $3\times3$ special unitary matrix $U$. Since $U_{kj}\propto \bar q_R^aq_L^b$ it transforms under $SU(3)_L\times SU(3)_R$ as
\begin{eqnarray}
 U\rightarrow LUR^\dagger.
\end{eqnarray}
Using $U$ we want to create an effective Lagrangian that is symmetric under chiral transformations ${SU(3)_L\times SU(3)_R}$. The only possible term without derivatives reads $UU^\dagger$ and is constant and therefore irrelevant; to get non-trivial terms derivatives need to be included. A derivative corresponds to a momentum. The typical momenta at low energies are small compared to the scale $\lqcd$. Since only even numbers of derivatives are possible for purely pionic systems due to the Lorentz-structure, the Lagrangian is expanded in terms momenta as $p^2/\lqcd^2+p^4/\lqcd^4+...$ or
\begin{equation}
 \lag=\lag^{(2)}+\lag^{(4)}+...
\end{equation}
One can immediately see that there is only one possible term at leading order and the effective Lagrangian can be written as
\begin{eqnarray}\label{eq:lag_kinU}
 \mathscr{L}_{U}= \frac{F^2_\pi}{4}Tr[(\partial^\mu U)( \partial_\mu U)]+\text{higher derivative terms},
\end{eqnarray}
where $F$ denotes the pion decay constant in the chiral limit. We can parametrize $U$ as an exponential using the hermitian matrix $\phi$:
\begin{eqnarray}
 U=\exp\left(\frac{\sqrt{2}i\phi}{F_\pi}\right).
\end{eqnarray}
This matrix $\phi$ is a linear combination of the generators of SU(3) that can be identified as the Goldstone boson fields:
\begin{eqnarray} \nonumber
&\phi= \left( \begin{array}{ccc}
\frac{1}{\sqrt{2}}\pi^0+\frac{1}{\sqrt{6}}\eta & \pi^+ & K^+ \\
\pi^- & -\frac{1}{\sqrt{2}}\pi^0+\frac{1}{\sqrt{6}}\eta &  K^0 \\
K^- & \bar{K^0} & -\frac{2}{\sqrt{6}}\eta \end{array} \right),\\
\end{eqnarray}
where $\eta=\eta_8$ the color octet field has been taken. The $SU(3)$ weight diagram of the Goldstone bosons is shown in Fig.~\ref{fig:octet}. Expanding Eq.~(\ref{eq:lag_kinU}) to leading non-vanishing order we find the correct spin-0 kinematic terms for the Goldstone bosons. 

Parallel to the expansion in terms of momenta we can also incorporate explicit $SU(3)$ violation by introducing the light quark mass matrix in Eq.~(\ref{eq:lag_light_quark}) as a perturbation in the chiral Lagrangian. In order to obtain a Lagrangian that is fully consistent with chiral symmetry we pretend that the light quark mass matrix transforms as $M_q\rightarrow LM_qR^\dagger$. Then $U^\dagger M_q$ and $M_q^\dagger U$ are terms allowed in the effective Lagrangian.
We include them as $\chi=2BM_q$, where the factor $2B$ from the chiral quark condensate is included for later convenience. 
At leading order we find
\begin{eqnarray}\label{eq:lag_U}
 \mathscr{L}_{eff}= \frac{F^2_\pi}{4}Tr[\partial^\mu U \partial_\mu U]+\frac{F^2_\pi}{4}Tr[\chi^\dagger U+ U^\dagger \chi].
\end{eqnarray}
From this Lagrangian we can see that  $\chi$ scales as $M_\pi^2$ and the pion mass fits into our earlier expansion as $M_\pi\sim p$.
At subleading order the number of possible terms increases significantly so we stick to the leading term in the chiral expansion which is sufficient for our purposes. 
Expanding the fields $U$ in terms of the Goldstone boson fields collected in $\phi$ up to leading non-trivial order and evaluating the traces gives us the Klein-Gordon kinematic terms for the light fields. Matching the  Goldstone boson masses in terms of the quark masses yields
\begin{eqnarray}%
&M_{\pi^\pm}^2=B(m_u+m_d) \quad M_{K^0}^2=B(m_d+m_s) \quad M_{K^\pm}^2=B(m_u+m_s)\\ \nonumber
\end{eqnarray}
and the $\pi^0-\eta$ mass matrix:
\begin{eqnarray}%\nonumber
M_{\pi^0\eta}=\left(\begin{array}{cc}
B(m_u+m_d) & B\frac{m_u-m_d}{\sqrt{3}}\\
B\frac{m_u-m_d}{\sqrt{3}} & B\left(\frac{4}{3}m_s+\frac{1}{3}m_u+\frac{1}{3}m_d\right)\\
\end{array}\right).
\end{eqnarray}
Obviously the $\pi^0$ and the $\eta$ mix for $m_u\neq m_d$. We can use a rotation by $\epsilon_{\pi^0\eta}$, the $\pi^0-\eta$ mixing angle, to get the diagonal mass matrix and the mass eigenstates $\tilde \pi^0$ and $\tilde\eta$
\begin{eqnarray}\nonumber\label{mixing} \nonumber
&&M_{\pi^0\eta}=\left(\begin{array}{cc}
\cos \epsilon_{\pi^0\eta} & -\sin \epsilon_{\pi^0\eta}\\
\sin \epsilon_{\pi^0\eta} & \;\;\:\cos \epsilon_{\pi^0\eta}\\
\end{array}\right)
 \left(\begin{array}{cc}
M_{\pi^0}^2 & 0\\
0 & M_\eta^2\\
\end{array}\right)
 \left(\begin{array}{cc}
\;\;\:\cos \epsilon_{\pi^0\eta} & \sin \epsilon_{\pi^0\eta}\\
-\sin \epsilon_{\pi^0\eta} & \cos \epsilon_{\pi^0\eta}\\
\end{array}\right)\\ \nonumber
\\ 
&&\left(\begin{array}{c}
      \tilde \pi^0 \\ \tilde\eta\\
     \end{array}\right)
=\left(\begin{array}{cc}
        \;\;\:\cos \epsilon_{\pi^0\eta} & \sin \epsilon_{\pi^0\eta}\\
	-\sin\epsilon_{\pi^0\eta} & \cos\epsilon_{\pi^0\eta}\\
       \end{array}\right)
\left(\begin{array}{c}
       \pi^0 \\ \eta
     \end{array}\right).\\ \nonumber
\end{eqnarray} 
From this we can deduce the $\pi^0-\eta$ mixing angle and the masses of $\pi^0$ and $\eta$ in terms of quark masses:
\begin{eqnarray}\nonumber
&\epsilon_{\pi^0\eta}=\frac{1}{2}\arctan\left(\frac{\sqrt{3}}{2}\frac{m_d-m_u}{m_s-\hat m}\right),\quad \hat m=(m_u+m_d)/2\\ \nonumber
&m_{\pi^0}^2=B(m_u+m_d)-\mathscr{O}((m_u-m_d)^2) \\ & m_\eta^2=\frac{B}{3}(4m_s+m_u+m_d)+\mathscr{O}((m_u-m_d)^2).\\ \nonumber
\end{eqnarray} 
This angle $\epsilon_{\pi^0\eta}$ is connected to  the isospin breaking ratio of light quark masses. Lattice averages \cite{Golterman:2009kw} give
\begin{eqnarray}\label{eq:epspieta}
  \frac{m_d-m_u}{m_s-\hat m} = 0.0299 \pm 0.0018, \quad \epsilon_{\pi^0\eta}=(0.01275\pm0.00075).
\end{eqnarray}
So $\epsilon_{\pi^0\eta}$ is sufficiently small and can be considered zero in all isospin conserving calculations. When we calculate the hadronic width of $\dszero$ and $\dsone$ which emerges from isospin violation Eq.~(\ref{eq:epspieta})  will be used. 

%%%%%%%%%%%%%%%%%%%%%%%%%%%%%%%%%%%%%%%%%%%
%
%   u and (axial-)vector current
%
%%%%%%%%%%%%%%%%%%%%%%%%%%%%%%%%%%%%%%%%%%%
We can further define an additional quantity that will turn out useful in constructing a chiral invariant Lagrangian for heavy mesons. The field $u$ with $u^2=U$ has the nontrivial chiral transformation properties
\begin{eqnarray}
 u\rightarrow LuK^\dagger=KuR^\dagger.
\end{eqnarray}  
Here $K$ is not only a function of $L$ and $R$ but also of $\phi$ which makes it space-time dependent. One can show that the heavy meson multiplet we will introduce in Sec.~(\ref{sec:hmchpt}) transform exclusively via $K$ or $K^\dagger$, $H\to HK$ and $\bar H\to K^\dagger \bar H$, respectively. To get chiral invariant objects we need operators that transform as $\hat O\to K\hat O K^\dagger$ for then $H\hat O\bar H$ is invariant.
Two possible linear combinations with one derivative are
\begin{eqnarray}
  \Gamma_\mu=\frac{1}{2}\left(u^\dagger\partial_\mu u+u\partial_\mu u^\dagger\right)\qquad u_\mu=i\left(u^\dagger\partial_\mu u-u\partial_\mu u^\dagger\right)
\end{eqnarray}
which transform as
\begin{eqnarray}\label{eq:trafo_(axial)vector}
 u^\mu\rightarrow Ku^\mu K^\dagger \qquad \Gamma^\mu\rightarrow K\Gamma^\mu K^\dagger+K\partial^\mu K^\dagger.
\end{eqnarray}
While $u_\mu$ has the desired transformation properties we can use $\Gamma_\mu$ to to construct a chiral covariant derivative acting on a heavy field
\begin{equation} \label{eq:CovDeriv}
 D_\mu=\partial_\mu+\Gamma_\mu, \qquad D_\mu H\to KD_\mu H.
\end{equation}
Because of their transformation quantities we refer to them as the axialvector current $u^\mu$ and the vector current $\Gamma^\mu$. Expanding in terms of the light fields we find that the axialvector current contains an odd number of light fields while the vector current contains an even number.

We can further use the fields $u$ to construct explicit chiral symmetry breaking terms due to degenerate light quark masses and electromagnetic charge:
\begin{eqnarray}
 &&\chi_\pm=u^\dagger\chi u^\dagger\pm u\chi^\dagger u \qquad \chi_\pm\rightarrow K\chi_\pm K^\dagger \non \\
 &&Q_\pm=\frac 12 \left(u^\dagger Q u \pm u Q u^\dagger\right) \qquad Q_\pm\rightarrow KQ_\pm K^\dagger
\end{eqnarray}
with $\chi=2BM_q$ as introduced in Eq.~\ref{eq:lag_U} and the light quark charge matrix $Q=\mathrm{Diag}(2/3,-1/3,-1/3)$. It can be useful to use to define a traceless quantity:
\begin{equation}\label{eq:chitilde}
 \tilde\chi=\chi-{\rm Tr}\chi
\end{equation}

If we use $D_\mu$, $u_\mu$, $ \chi_\pm$ and $Q_\pm$ as the light building blocks of the effective Lagrangian  chiral constraints are implemented automatically.
%%%%%%%%%%%%%%%%%%%%%%%%%%%%%%%%%%%%%%%%%%%
%
%   Heavy Quark Symmetry
%
%%%%%%%%%%%%%%%%%%%%%%%%%%%%%%%%%%%%%%%%%%%
\subsection{Heavy Quark Symmetry}\label{sec:hqs}
In this section we will explore the consequences of the charm and bottom quark masses being larger than the scale of {\it nonperturbative QCD}, $\lqcd$. This discussion follows the textbook by Manohar and Wise~\cite{Manohar:2000dt}.
Experimentally we find 
\begin{equation}
 \lqcd/m_c\sim 0.2 \text{ and }\lqcd/m_b\sim 0.05
\end{equation}
 which justifies an expansion in terms of $\lqcd/m_Q\ll1$. Once again, the top quark, while being even better suited for an expansion like that, decays before it can form bound states and is therefore not suited for a effective low-energy theory. The heavy quark part of the $QCD-$Lagrangian needs to be rewritten in terms of the expansion parameter:
\begin{eqnarray}
\mathscr L_{HQET}=\mathscr L_0+\frac{\lqcd}{m_Q}\mathscr L_1+ \frac{\lqcd^2}{m_Q^2}\mathscr L_2+...
\end{eqnarray}
A useful quantity when talking about heavy systems is the velocity $v$ of the heavy quark, defined as $v=p/m_Q$. The typical momentum transfer between the two quarks is of order $\lqcd$. As a result, the velocity $v$ of the heavy quark is almost unchanged by interactions with the light quark:
\begin{eqnarray}\label{heavyvelocity}
 \Delta v=\frac{\Delta p}{m_Q}\rightarrow 0 \text{ for }m_Q\rightarrow \infty.
\end{eqnarray}
Therefore the velocity is a constant of motion in the limit of infinite quark mass. 

We start with the relevant part of the QCD Lagrangian for the heavy quarks
\begin{eqnarray}
 \mathscr L_{c,b}=\sum_{c,b}\bar Q(i\slashed D)Q-\bar Q m_Q Q.
\end{eqnarray} 
The first step is to rewrite the quark field
\begin{eqnarray}
 Q(x)\rightarrow e^{-im_Qv\cdot x}[P_+Q(x)+P_-Q(x)]=e^{-im_Qv\cdot x}[Q_{+,v}(x)+Q_{-,v}(x)].\\ \nonumber
\end{eqnarray} 
with the projection operators $P_\pm=(1\pm \slashed v)/2$ which project on the (anti)particle part of $Q(x)$. Now we can insert this in the original Lagrangian, integrate out the antiparticle field $Q_{-,v}$ and expand up--to the order $\mathscr O(1/m_Q^2)$
\begin{figure}[t]
 \centering
 \includegraphics{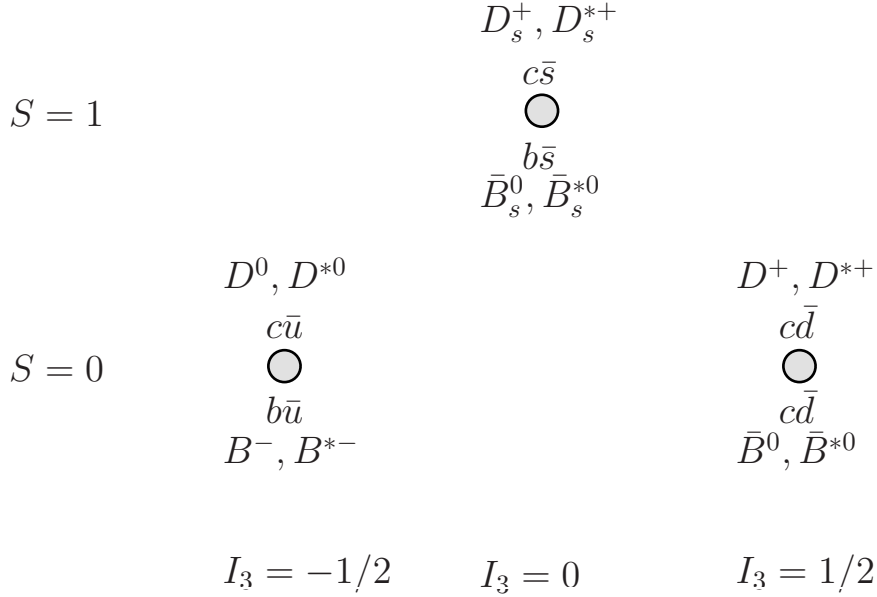}
 \caption{Flavor SU(3) weight diagram for charmed and bottomed pseudoscalar and vector mesons.  The x-axis shows the third component of the isospin, the y-axis shows the strangeness. The quark content is given below the members of the triplet. In the limit $m_Q\to\infty$ they all form a degenerate $SU(2)_S\times SU(2)_F\times SU(3)_V$ multiplet.}
 \label{fig:triplet}
\end{figure}

\begin{eqnarray}\nonumber
 \mathscr L&=&\bar Q_{+,v}(iv\cdot D)Q_{+,v}+\frac{1}{2m_Q}\bar Q_{+,v}(i(D_\mu-v_\mu v\cdot D)^2)Q_{+,v}\\
&&-\frac{g}{4m_Q}\bar Q_{+,v}\sigma_{\alpha\beta}G^{\alpha\beta}Q_{+,v}+\mathscr O(1/m_Q^2).\\ \nonumber
\end{eqnarray} 

The only strong interactions of the heavy quark, that are allowed by the QCD-Lagrangian, are the ones with gluons. These are separated into interactions via the chromoelectric charge and via the chromomagnetic charge. The interactions via the chromoelectric charge are spin-- and flavor--independent and also present in the limit $m_Q\rightarrow \infty$. The interactions via the chromomagnetic charge, which are spin-dependent, are on the other hand proportional to the chromomagnetic moment of the quark and so of order 1/$m_Q$ and vanish in the heavy quark limit. This gives us \textit{heavy quark spin} and \textit{flavor symmetry}. 

The main consequence of spin symmetry is that the spin of the light quarks in a hadron with a heavy quark can be neglected. In the case of charmed or bottomed mesons this means that the pseudoscalar $D(B)$-mesons and the vector $D^*(B^*)$-mesons each form a degenerate $SU(2)_S$ spin multiplet. The corrections for heavy quark spin symmetry breaking are proportional to $1/m_Q$ and therefore different for each heavy quark flavor. 

For the case of charm and bottom as heavy quarks we also find an $SU(2)_F$ symmetry: the interactions of charmed and bottom mesons are identical up to heavy quark flavor symmetry breaking effects of the order $1/m_c-1/m_b$. 

On top of the $SU(2)_S\times SU(2)_F$ symmetry obtained from the limit $m_Q\to\infty$ the heavy mesons also obey the $SU(3)_V$ symmetry for the light quarks. In Fig.~\ref{fig:triplet} we show all mesons that are degenerate in the limit of infinitely heavy $c$- and $b$-quarks and identical light quark masses. 

%%%%%%%%%%%%%%%%%%%%%%%%%%%%%%%%%%%%%%%%%%%
%
%   Heavy Meson ChPT
%
%%%%%%%%%%%%%%%%%%%%%%%%%%%%%%%%%%%%%%%%%%%
\subsection{Heavy Meson Chiral Perturbation Theory}\label{sec:hmchpt}

In this section we will introduce an effective theory that includes all four additional symmetries of QCD at low energies that we have identified before: $SU(3)$ flavor symmetry, {\it chiral symmetry}, {\it heavy quark spin symmetry} and {\it heavy quark flavor symmetry}. For the charmed mesons this means that the pseudoscalar chiral triplet $P=(D^0,D^+,D_s)$ forms a spin multiplet together with its vector spin partner $V=(D^{*0},D^{*+},D^*_s)$. Note that in the literature one often finds the notation $P^*$ instead of $V$. We will stick to this notation to make it consistent with Nonrelativistic Effective Theory, introduced in the next section, and avoid unnecessary confusion. 
The corresponding weight diagram with the relevant quantum numbers is given in Fig.~(\ref{fig:triplet}). The standard way to combine vector and pseudoscalar fields into bispinors was introduced in ~\cite{Falk:1990yz}. The spin multiplets read
\begin{eqnarray}\nonumber
 H_v=\frac{1+\slashed{v}}{2}\left[V_v^\mu\gamma_\mu+\gamma_5 P_v\right]\qquad
 \bar{H_v}=\gamma^0H_v^\dagger\gamma^0=\left[ V_{v,\mu}^{\dagger}\gamma^\mu-\gamma_5P^{\dagger}_v\right] \frac{1+\slashed{v}}{2}.
\end{eqnarray}
As a vector field, $V^{(Q)}_{v,\mu}(x)$ has a polarization vector $\epsilon$ with the properties $\epsilon\cdot \epsilon=-1$ and $\epsilon\cdot v=0$. From here on the index $v$ that denotes the velocity dependence of the fields will be dropped for simplicity. 
The normalization of the heavy fields needs some attention. For relativistic fields  the convention 
\begin{eqnarray}
 \left<H(p')|H(p)\right>=2E_p(2\pi)^3\delta^3(\vec p-\vec p')
\end{eqnarray} 
is commonly used. Instead of this one we use the convention
\begin{eqnarray}
 \left<H(v',k')|H(v,k)\right>=2v^0\delta_{vv'}(2\pi)^3\delta^3(\vec k-\vec k'),
\end{eqnarray} 
where $k=p-m_Qv$ with p being the momentum of the heavy quark is the so--called residual momentum. These two normalizations differ by a factor of $\sqrt{M_H}$,
\begin{eqnarray}
 |H(p)\big>=\sqrt{M_H}[|H(v,k)\big>+\mathscr O(1/m_Q)].
\end{eqnarray} 
Once this trivial dependence on the heavy mass is removed, the newly introduced fields depend on the heavy mass only very weakly. As stressed before, in order to construct the effective Lagrangian, we use building blocks whose transformation laws contain only $K$ and $K^\dagger$. The field $H$ does not have uniquely defined transformation properties under ${SU(3)_L\times SU(3)_R}$, see e.g. \cite{Georgi:1985kw}. We can redefine the field by multiplying a complex phase without changing any observable. In \cite{Manohar:2000dt} it is shown that one can choose this phase such that the field $H$ obeys the desired transformation law to construct chiral invariant terms:
\begin{eqnarray}
 H_a\rightarrow H_b K^\dagger_{ba}.
\end{eqnarray}
Here $a,b$ denote flavor indices. 
%%%%%%%%%%%%%%%%%%%%%%%%%%%%%%%%%%%%%%%%%%%
%
%   Kinetic Energy
%
%%%%%%%%%%%%%%%%%%%%%%%%%%%%%%%%%%%%%%%%%%%

The simplest term we can construct that is chirally invariant is the one without derivatives:
\begin{equation}
 M_H\mathrm{Tr}[\bar H_aH_a].
\end{equation}
This serves as a mass term. However, we can  scale the field $H$ with a complex phase $e^{-iM_H v{\cdot}x}$. Then the kinetic energy 
\begin{equation}
 \lag_\mathrm{Kin}=-i\mathrm{Tr}[\bar H_a v{\cdot}D_{ba}H_b]=-i\mathrm{Tr}[\bar H_a v{\cdot}(\partial+\Gamma)_{ba}H_b]
\end{equation}
removes the mass term. The chiral covariant derivative in the kinetic energy ensures chiral symmetry. 

%%%%%%%%%%%%%%%%%%%%%%%%%%%%%%%%%%%%%%%%%%%
%
%   spin symmetry breaking
%
%%%%%%%%%%%%%%%%%%%%%%%%%%%%%%%%%%%%%%%%%%%
So far  pseudoscalar and vector mesons are degenerate. To incorporate the fact that this is not realized perfectly in nature we introduce the leading $SU(3)$ and spin symmetry breaking terms. The first spin symmetry breaking term appears at order $1/m_Q$:
\begin{eqnarray}
 \delta\mathscr L^{(1)}=\frac{\lambda}{m_Q}Tr[\bar H_a\sigma_{\mu\nu}H_a\sigma^{\mu\nu}].
\end{eqnarray} 
It leads to the mass difference
\begin{eqnarray}
 \Delta=m_{V}-m_P=-8\frac{\lambda}{m_Q}.
\end{eqnarray}
In the calculations this will be taken into account by using the physical masses for the $D_{(s)}$ and $D_{(s)}^*$, respectively.
%%%%%%%%%%%%%%%%%%%%%%%%%%%%%%%%%%%%%%%%%%%
%
%   SU(3) breaking
%
%%%%%%%%%%%%%%%%%%%%%%%%%%%%%%%%%%%%%%%%%%%
The leading $SU(3)$ flavor symmetry breaking effects for the heavy mesons enter at NLO in the chiral expansion:
\begin{eqnarray}\label{eq:charmmasses} \nonumber
 \delta\lag^{(2)}=h_1Tr[\bar H_a(\chi_+)_{ba}H_a]=-2\frac{h_1}{m_Q}P_a^\dagger \chi_{ba} P_b+...
\end{eqnarray}
where the dots denote the vector meson terms not relevant for this discussion. In the isospin symmetric case $\hat m=m_u=m_d$ we get  
\begin{eqnarray} 
 M^2_{D_s}-M^2_{D}=4Bh_1(m_s-\hat m).
\end{eqnarray}
We can determine the LEC $h_1$ using the relation $B(m_s-\hat m)=(M^2_{K^0}+M^2_{K^+})/2-M^2_{\pi^0}$ as done in \cite{Hofmann:2003je}. This has been measured with high precision, we use the value $h_1=0.42$. 

%%%%%%%%%%%%%%%%%%%%%%%%%%%%%%%%%%%%%%%%%%%
%
%   EM interactions
%
%%%%%%%%%%%%%%%%%%%%%%%%%%%%%%%%%%%%%%%%%%%

There are more possibilities.
To calculate electro-magnetic transitions we also need the operators
\begin{eqnarray}
 F^{\mu\nu}_\pm=-2F^{\mu\nu}Q_\pm,\quad Q_\pm=\frac 12 \left(u^\dagger Q u \pm u Q u^\dagger\right)
\end{eqnarray}
with the electromagnetic field strength tensor $F^{\mu\nu}=\partial^\mu A^\nu-\partial^\nu A^\mu$ and the light quark charge matrix $Q=\mathrm{Diag}(2/3,-1/3,-1/3)$.

%%%%%%%%%%%%%%%%%%%%%%%%%%%%%%%%%%%%%%%%%%%
%
%   NREFT
%
%%%%%%%%%%%%%%%%%%%%%%%%%%%%%%%%%%%%%%%%%%%
\subsection{Nonrelativistic Effective Theory}\label{sec:nreft}
For the second half of this work --- the study of the $Z_b$ resonances --- we will rely on the well-established framework of Nonrelativistic Effective Theory (NREFT), for references see e.g. \cite{Fleming:2008yn,Guo:2010ak}. We can simply obtain the two-component Lagrangians in NREFT from the four-component Lagrangians in HMChPT. So far we have used the heavy meson fields
\begin{equation}
 H_4^{(Q)}=\projv \left[\slashed V-\gamma_5 P\right],\qquad \bar H_4^{(Q)}=\gamma_0 H_4^{(Q)\dagger}\gamma_0=\left[\slashed V^{\dagger}+\gamma_5 P^\dagger\right]\projv.
\end{equation}
In the rest frame of the heavy particle the velocity is given by $v=(1,0,0,0)$. Using the Dirac-Representation of the gamma matrices,
\begin{eqnarray}
 \gamma^0=\left(\begin{array}{cc}\mathbbm{1} & 0 \\ 0 & \mathbbm{1}  \end{array}\right),\quad  \gamma^i=\left(\begin{array}{cc} 0 & \sigma_i \\ -\sigma_i & 0 \end{array}\right),\quad \gamma^5=\left(\begin{array}{cc} 0 & \mathbbm1 \\ \mathbbm1 & 0 \end{array}\right)
\end{eqnarray}
where $\sigma_i$ are the Pauli matrices, we find
\begin{equation}
 H_4^{(Q)}=\left(\begin{array}{cc} V^{0} & -\vec\sigma\cdot\vec V -P \\ 0 & 0 \end{array}\right)\simeq\left(\begin{array}{cc}0 & -\vec\sigma\cdot\vec V-P \\ 0 & 0 \end{array}\right)=\left(\begin{array}{cc}0 & -H_2 \\ 0 & 0 \end{array}\right).
\end{equation}
In the last step we have used the fact that the time-like component of the heavy mesons is suppressed by a factor of $\vec p/m_Q$. Similarly the antifield becomes
\begin{equation}
 \bar H_4^{(Q)}=\left(\begin{array}{cc} 0 & 0 \\ - \vec \sigma\cdot\vec V^{\dagger} -P^\dagger & 0 \end{array}\right).
\end{equation}
As a result we find the NREFT heavy meson fields
\begin{equation}
 H_2=\vec{V}\cdot\vec{\sigma}+P, \qquad H^\dagger_2=\vec{V}^\dagger\cdot\vec{\sigma}+P^\dagger.
\end{equation}
Throughout this work we will use the four-component notation for the $D_{sJ}$ mesons in Ch.~\ref{ch:charmed} and NREFT in Ch.~\ref{ch:zb} for the study of the $Z_b$ mesons. We therefore drop the indices $2,4$ in both cases. 

Contrary to the $D_{sJ}$ chapter where only heavy mesons of the kind $Q\bar q$ are considered the coupling to bottomonia requires mesons with heavy antiquarks $q\bar Q$. The convention for the charge conjugation $\zeta$ is
\begin{equation}
 P^{*(\bar Q)}_i=-\zeta V^{(Q)}_i\zeta^{-1},\qquad P^{(\bar Q)}=\zeta P^{(Q)}\zeta^{-1}.
\end{equation}
With this we find
\begin{equation}
 H_2^{(\bar Q)}=-\vec\sigma\cdot \vec V^{(\bar Q)}+P^{(\bar Q)},\qquad H_2^{(\bar Q)\dagger}=-\vec\sigma\cdot \vec V^{(Q)\dagger}+P^{(\bar Q)\dagger}
\end{equation}
where $P^{(\bar Q)}=\left(B^{+},B^{0}\right)$ and $V^{(\bar Q)}=\left(B^{*+},B^{*0}\right)$ in the case of the $B$-mesons. In Ch.~\ref{ch:zb} where no confusion is possible we will simply refer to them as $\bar P(\vec {\bar V})$.

%%%%%%%%%%%%%%%%%%%%%%%%%%%%%%%%%%%%%%%%%%%%%%%%%%%%%%%%%%%%%%%%%%
\begin{table}
\centering
\small
% \scriptsize
\renewcommand{\arraystretch}{1.5}
\begin{tabular}{|c c c c c|}
\hline\hline
Rotations & Heavy Quark Spin & Parity & Charge Conjugation & $SU_L(3)\times SU_R(3)$\\ \hline
$H_a\to UH_a U^\dagger$ & $H_a\to SH_a$ & $H_a\to -H_a$ & $H_a\to \sigma_2 \bar H_a^T\sigma_2$ & $H_a\to H_b V_{ba}^\dagger$ \\
$\bar H_a\to U \bar H_a U^\dagger$ & $\bar H_a\to \bar H_a\bar S\dagger$ & $\bar H_a\to -\bar H_a$ & $\bar H_a\to \sigma_2 H_a^T\sigma_2$ & $\bar H_a\to V_{ab} \bar H_b$ \\
$\Upsilon^i\to R^{ij}U^\dagger \Upsilon^j U$ & $\Upsilon^i\to S\Upsilon^i\bar S^\dagger$ & $\Upsilon^i\to \Upsilon^i$ & $\Upsilon^i\to -\sigma_2(\Upsilon^i)^T\sigma_2$ & $\Upsilon^i\to\Upsilon^i$ \\
$\chi^i\to R^{ij}U^\dagger \chi^j U$ & $\chi^i\to S\chi^i\bar S^\dagger$ & $\chi^i\to \chi^i$ & $\chi^i\to -\sigma_2(\chi^i)^T\sigma_2$ & $\chi^i\to\chi^i$ \\
$Z_b^i\to R^{ij}U^\dagger Z_b^j U$ & $Z_b^i\to SZ_b^i\bar S^\dagger$ & $Z_b^i\to Z_b^i$ & $Z_b^i\to -\sigma_2(Z_b^i)^T\sigma_2$ & $Z_b^i\to Z_b^i$ \\
\hline\hline
\end{tabular}
\caption{\label{tab:fields}Transformation properties of the various multiplets. $U(R^{ij})$ are $2\times2(3\times3)$ are rotation matrices related by $U^\dagger\sigma^i U=R^{ij}\sigma^i$, $S(\bar S)$ a rotation matrix acting on the heavy quark (heavy antiquark) spin and $V_{ab}$ is an $SU(3)$ matrix.}
\end{table}
We can introduce the bottomonium fields in this framework. The $P-$wave bottomonia are given by
\begin{equation}
 \chi^i=h_b^i+\sigma^j\left(\frac{1}{\sqrt{3}}\delta^{ij}\chi_{b0}-\frac{1}{\sqrt{2}}\varepsilon^{ijk}\chi_{b1}^k-\chi_{b2}^{ij}\right),
\end{equation}
the $S-$wave bottomonia by 
\begin{equation}
 \Upsilon=\eta_b+\vec \sigma\cdot \vec \Upsilon
\end{equation}
In accordance with the experimental data the $J^P=1^+$ fields $Z_b$ will be introduced as 
\begin{equation}
 Z^i_{ba} = \left(
         \begin{array}{cc}
           \frac1{\sqrt{2}} Z^{0i} & Z^{+i} \\
           Z^{-i} & - \frac1{\sqrt{2}} Z^{0i} \\
         \end{array}
       \right)_{ba}.
\end{equation}
In Tab.~(\ref{tab:fields}) we show the transformation properties relevant for the construction of the interaction Lagrangians. 

%%%%%%%%%%%%%%%%%%%%%%%%%%%%%%%%%%%%%%%%%%%
%
\section{Unitarization}\label{sec:unitarization}
%
%%%%%%%%%%%%%%%%%%%%%%%%%%%%%%%%%%%%%%%%%%%
\begin{figure}[h]
 \centering
 \includegraphics[width=0.9\textwidth]{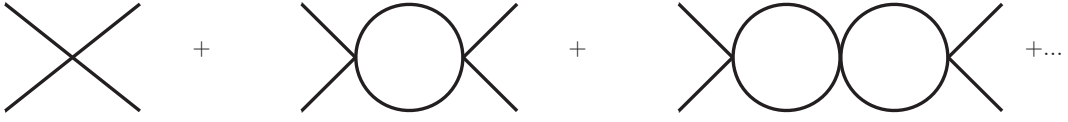}
 \caption{Lippmann-Schwinger resummation for two-body scattering}
 \label{fig:lippmannschwinger}
\end{figure}
In this section we will introduce a theoretical method called unitarization. This is necessary to obtain resonances like e.g. the $\dszero$ in heavy meson-Goldstone boson scattering. Analogously to the treatment of kaon-nucleon scattering by Meißner and Oller~\cite{Oller:2000fj} the scattering potential $V(s)$ is expanded and the result is used in scattering equations. 
Since such an amplitude $V(s)$ always is a polynomial of the center-of-mass energy $s$ one can never find poles without introducing the resonances as explicit fields. 
We can resum the scattering potential in such a way that the resummed amplitude  is unitary and thus called the unitarized amplitude $T(s)$.

We will have a closer look at this in the most general terms. Generally one can define the $S-$Matrix as
\begin{eqnarray}%\nonumber
 S_{fi}=\left <f|S|i\right >.
\end{eqnarray} 
This means that $S_{fi}$ is the amplitude for the scattering process from the initial state $|i\big>$ to the final state $|f\big>$. Since the probability for $|i\big>$ to become any of the possible final states is one, the S-Matrix has to be unitary. 
We define the transition operator $T$ as
\begin{eqnarray}\label{eq:smatrix}
 S=1-iT\rho,
\end{eqnarray} 
where $\rho(s)=|\vec p|/(8\pi \sqrt{s})$ is the phase space factor. From unitarity it follows
\begin{eqnarray}
 S^\dagger S=1 \Rightarrow {\rm Im} T=-2|T|^2\rho.
\end{eqnarray} 
We notice that the transition operator occurs linearly on the left-handed side and quadratically on the right-handed side. Thus any finite expansion of $T$ would result in an inequality here and consequently unitarity can only be realized perturbatively.

However, a method exists that leads to unitary amplitudes. At the same time the possibility of having dynamically generated poles in the amplitudes arises. The approach for resumming the amplitude we will use throughout this work is presented in \cite{Oller:2000fj}. In Fig.~\ref{fig:lippmannschwinger} the resummation process is shown schematically. If one denotes the tree-level $S-$wave amplitude as $V(s)$ and the two-particle loop as $G(s)$ the infinite sum $T(s)$ of all scattering and rescattering amplitudes reads
\begin{equation}
 T(s)=V(s)+V(s)G(s)V(s)+V(s)G(s)V(s)G(s)V(s)+...
\end{equation}
The geometrical series for this infinite sum is given by the Lippmann-Schwinger equation
\begin{equation}\label{eq:Tmatrix}
  T(s)=V(s)[1-G(s)\cdot V(s)]^{-1}
\end{equation}
The fact, that this amplitude is unitary is easily proven. From Eq.~\eqref{eq:smatrix} we can derive
\begin{eqnarray}%\nonumber
{\rm Im}\left[T(s)^{-1}\right]=\rho(s)
\end{eqnarray} 
with the phase space factor $\rho(s)$ introduced earlier. We can therefore write
\begin{eqnarray}\label{eq:Tinv}
 T^{-1}(s)=K(s)+i\rho(s),
\end{eqnarray} 
where $K(s)$ is a real function of $s$. On the other hand Cutkosky's rules \cite{Cutkosky:1960sp} relate the imaginary part of the loop function to the phase space factor
\begin{eqnarray}
 {\rm Im}\left[G(s)\right]=-\rho(s).
\end{eqnarray} 
Using Eq.~(\ref{eq:Tmatrix}) gives
\begin{eqnarray} \nonumber
 T(s){=}[V^{-1}(s)-G(s)]^{-1}=[V^{-1}(s)-{\rm Re}(G(s))-i{\rm Im}(G(s))]^{-1}.
\end{eqnarray} 
Comparison with Eq.~(\ref{eq:Tinv}) leads to $K(s)=V^{-1}(s)-{\rm Re}(G(s))$. That means, as long as $V(s)$ is real unitarity is guaranteed. The formalism can further be generalized to complex scattering amplitudes that appear at one loop order. See, e.g., Oller and Meißner \cite{Oller:2000fj}. Notice that in the case of coupled channel amplitudes, for $\dszero$ we have $DK$ and $D_s\eta$, the Lippmann-Schwinger Equation in Eq.(\ref{eq:Tmatrix}) becomes a matrix equation.

As mentioned before the Lippmann-Schwinger Equation does not only provide a way to make an amplitude from a finite expansion in ChPT unitary, it also allows the dynamical generation of poles. While  $V(s)$  does not contain any poles, the resummed amplitude $T(s)$ can have poles of the form $s=(M+i\Gamma/2)^2$ with mass $M$ and width $\Gamma$. We will make use of this when we extract the masses and hadronic widths of $\dszero$ and $\dsone$ as poles in unitarized scattering amplitudes for the coupled channels $D^{(*)}K$ and $D^{(*)}_s\eta$, respectively.

%%%%%%%%%%%%%%%%%%%%%%%%%%%%%%%%%%%%%%%%%%%
%
\section{Power Counting}\label{sec:power_counting}
%
%%%%%%%%%%%%%%%%%%%%%%%%%%%%%%%%%%%%%%%%%%%
One of the most essential ingredients of an effective theory is a proper {\it power counting} scheme. This serves several purposes. First we notice effective Lagrangian are fundamentally different from the interaction Lagrangians in renormalizable field theories. In $QED$ for example all interactions consist of the coupling of a photon to a charged particle with the dimensionless coupling constant $\alpha$ that fulfills $\alpha\ll1$. So all vertices are of the same order and the more vertices a diagram it has the smaller its contribution becomes. Since in an effective Lagrangian as the ones constructed before many different interactions and various numbers of derivatives are possible. They all come with different coupling constants that even have different dimensions. One needs to find a way to sort the Lagrangian and ultimately the transition amplitudes in such a way that the amplitudes at one order contribute similarly. Such a scheme is referred to as a  {\it power counting} scheme. Further the issue of 
renormalizability needs to be addressed. While it can be shown that any gauge field theory can be renormalized this can not be said for effective theories. However, it can be shown that an effective theory is renormalizable at any given order in its power counting individually. Another important benefit of a power counting scheme is that it provides a way to estimate the theoretical uncertainty from higher order diagrams.

%%%%%%%%%%%%%%%%%%%%%%%%%%%%%%%%%%%%%%%%%%%
%
\subsection{Chiral Perturbation Theory}
%
%%%%%%%%%%%%%%%%%%%%%%%%%%%%%%%%%%%%%%%%%%%
\begin{figure}[ht]
 \centering
 \includegraphics[width=0.6\textwidth]{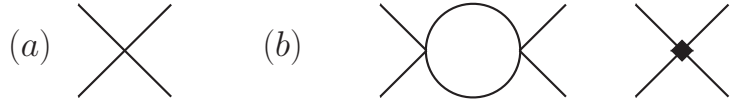}
 \caption{Leading (a) and subleading (b) contributions to light meson-scattering in ChPT}
 \label{fig:pipiscattering}
\end{figure}

For the power counting in ChPT we have in principle to distinguish between two ingredients: the dimension of vertices and loops. The full effective Lagrangian in Eq.~(\ref{eq:lag_U}) is expanded in the number of derivatives and light meson masses, both scale as $p$. Therefore vertices in ChPT can scale as $p^2$, $p^4$, etc. This means we have to count the number $V_d$ of vertices of dimension $p^d$.  Propagators in the loop scale as $\left(p^{-2}\right)^I$ with $I$ the number of internal lines in a diagram and the integral measures for a diagram with $L$ loops as $(p^4)^L$. Now we have all the ingredients necessary to find the so-called chiral dimension $\nu$ for any amplitude $\mathcal A\propto p^\nu$:
\begin{equation}
 \mathcal{A}\propto\int (d^4p)^L\frac{1}{(p^2)^I}\prod_d(p^d)^{V_d}
\end{equation}
One can easily find $\nu=4L-2I+\sum_d dV_d$ which can further be simplified using the fact that the number of loops is connected to the number of internal lines and vertices via $L=I-\sum_dV_d+1$. The final formula for the chiral dimension $\nu$ reads
\begin{equation}\label{eq:pc_chpt}
 \nu=\sum_d V_d(d-2)+2L+2.
\end{equation}
To get a converging expansion the momenta $p$ have to be small compared to the chiral symmetry breaking scale $\Lambda_\chi=4\pi F_\pi\simeq 1.2\gev$. This means ChPT is a good approximation for light meson momenta $p\ll 1.2\gev$ and cannot safely be applied to higher momenta.

As an example we look at the leading and subleading contributions for light meson scattering. The diagrams are shown in Fig.~(\ref{fig:pipiscattering}). We notice that at leading order we can only find a single contribution from tree-level scattering. The effective Lagrangian has only a single vertex with two derivatives and one with the light quark masses in $\chi$. This makes up the full leading order $\sim p^2\sim m_\pi^2$.  At subleading order one can find two contributions: the leading loop with vertices from $\lag^{(2)}$ and next-to-leading order tree-level scattering, both scale as $p^4$. To ensure that the theory is renormalizable at all orders individually one of the subleading tree-level diagrams needs to serve as a counter term for each divergent leading loop. This one together with the other low energy constants needs to be fit to experimental data. 

We further need to include heavy mesons in the power counting scheme. This adds heavy mesons-Goldstone boson vertices of the type $V^{\pi H}_{d'}$ to the power counting. Note that the first heavy meson-Goldstone boson vertex is of dimension 1 and thus $d'\geq1$ compared to Eq.~(\ref{eq:pc_chpt}). The chiral dimension becomes
\begin{equation}
 \nu=2L+1+\sum_dV^{\pi\pi}_d(d-2)+\sum_{d'}V^{\pi H}_{d'}(d'-1).
\end{equation}
This is similar to the inclusion of Baryons into ChPT, see e.g. the review by Bernard et al. \cite{Bernard:1995dp}.
%%%%%%%%%%%%%%%%%%%%%%%%%%%%%%%%%%%%%%%%%%%
%
\subsection{Nonrelativistic Effective Field Theory}\label{sec:powercountingnreft}
%
%%%%%%%%%%%%%%%%%%%%%%%%%%%%%%%%%%%%%%%%%%%
\begin{figure}[t]
 \centering
 \includegraphics[width=0.6\textwidth]{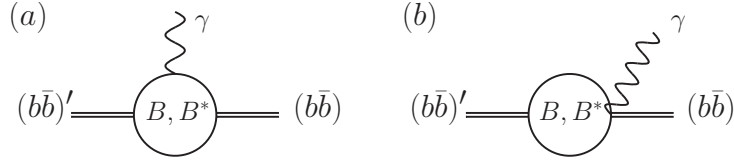}
 \caption{(a) Three-point and (b) two-point loop contributions for $(b\bar b)'\to(b\bar b)\gamma$}
 \label{fig:bbarbgamma}
\end{figure}
Here we follow Refs.~\cite{Guo:2010ak,Guo:2009wr,Guo:2010zk} where the authors lay out in detail a power counting scheme that can be applied to the transitions of heavy quarkonia via heavy meson loops in NREFT. The power counting is performed in terms of the typical velocity $v$ of a heavy meson inside the loop. This velocity  for a $B-$meson in the decay of a bottomonium with mass $M_{b\bar b}$ is given by
\begin{equation}
 v_B\sim \sqrt{|M_{b\bar b}-2M_B|/M_B}.
\end{equation}
In nonrelativistic expansion the propagator  gets replaced by 
\begin{equation}
 \left[l^2-M_B^2\right]^{-1}\to\left[l^0-\frac{\vec l^2}{2M_B}-M_B\right]^{-1}.
\end{equation}
Since $\vec l^2/(2M_B)\sim M_B v_B^2$ propagators count as $v_B^{-2}$. Having this said we may look at the integral measure. In the nonrelativistic expansion it is useful to perform the $l^0-$integration separately. Since this removes one propagator it is clear the full measure counts in four dimensions as
\begin{equation}
 \int\frac{\mathrm d^4l}{(2\pi)^4} \sim\frac{ v_B^5}{(4\pi)^2}.
\end{equation}
Of course the integration of other degrees of freedom like photons or pions introduce additional scales like the photon energy $E_\gamma$.
As a basic example we have a look at the emission of a photon by a bottomonium in the process $(b\bar b)'\to(b\bar b)\gamma$ via heavy meson loops. Two possible diagrams are shown in Fig.~(\ref{fig:bbarbgamma}): in $(a)$ the photon couples to the magnetic moment of the heavy meson, this coupling scales as $E_\gamma$. The other couplings are assumed to be in $S-$wave and therefore count as $1$. We find
\begin{equation}
 v_B^5\left(\frac{1}{v_B^2}\right)^3E_\gamma\sim \frac{E_\gamma}{v_B}.
\end{equation}
The naturalness of couplings says that the $BB(b\bar b)\gamma$ coupling should be of the same order as $BB(b\bar b)$ and need to give an additional photon momentum. This yields
\begin{equation}
 v_B^5\left(\frac{1}{v_B^2}\right)^2E_\gamma\sim E_\gamma v_B.
\end{equation}
The ratio $(b)/(a)$ gives $v_B^2$ which for $v_B\ll 1$ is heavily suppressed. If these two were the most important contributions here we could simply calculate Diag.~$(a)$ and  estimate the uncertainty from Diag.~$(b)$ as $v_B^2$. However, as we will see in the discussion of $\zbs$ decays, higher order loops can also have significant impact here.

%% file: Ds0andDs1.tex
%%%%%%%%%%%%%%%%%%%%%%%%%%%%%%%%%%%%%%%%%%%%%%%%%%%%%%%%%%%%%%%%%%%%%%%%%%%%%%%%%%%%%%
%
\chapter{$\dszero$ and $\dsone$ as $D^{(*)}K$ molecules}\label{ch:charmed}
%
%%%%%%%%%%%%%%%%%%%%%%%%%%%%%%%%%%%%%%%%%%%%%%%%%%%%%%%%%%%%%%%%%%%%%%%%%%%%%%%%%%%%%%

\begin{figure}[t]
 \centering
 \includegraphics[width=0.7\textwidth]{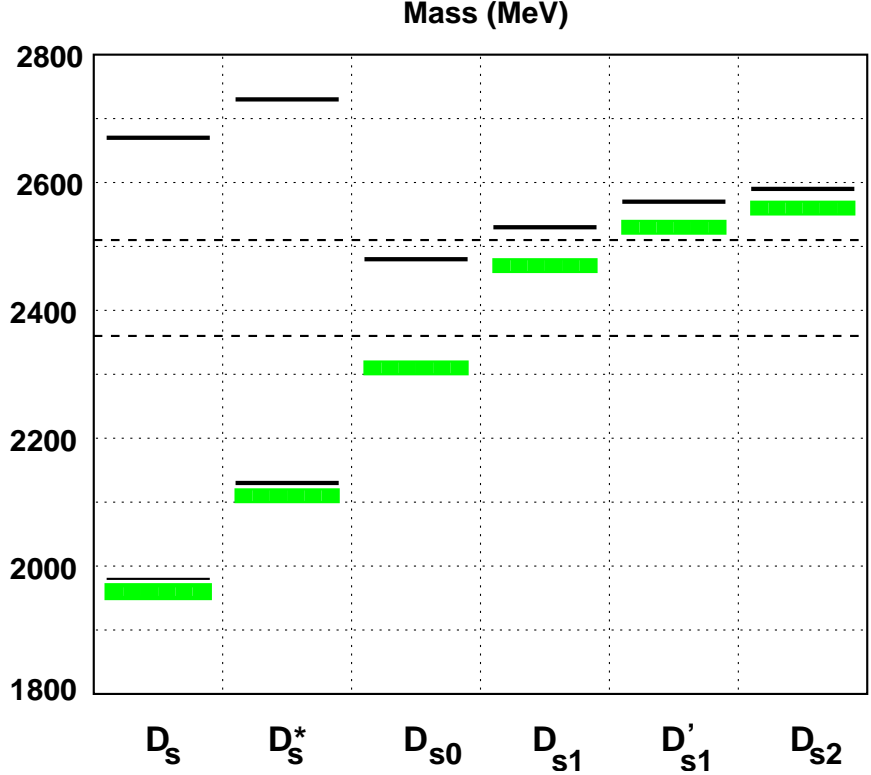}
  \caption{The $D_s$ spectrum: green boxes are the experimental results, black lines are theoretical predictions from \cite{Godfrey:1985xj}, dashed lines are the $D^{(*)}K$ thresholds. The figure was taken from \cite{Swanson:2006st}}
 \label{fig:DsSpect}
\end{figure}

The discovery of the two narrow states with open charm and strangeness, $\dszero$ and $\dsone$, by BaBar and Cleo-c~\cite{Aubert:2003fg,Besson:2003cp} marked, together with the discovery of the $X(3872)$, the starting point of a new era in open and hidden charm spectroscopy. For the first time quark models that were so far successful in describing the ground states and low excited states failed to predict these newly measured states. In Fig.~\ref{fig:DsSpect} we show the theoretical prediction by the Godfrey-Isgur model~\cite{Godfrey:1985xj} together with experimental results. As one can clearly see especially the mass of $\dszero$ is much lower than the prediction, approximately 160~MeV. But also $\dsone$, being 70~MeV below the prediction, seems not compatible with a naive quark model. 

This lead to a large number of theoretical explanations. Barnes and Close~\cite{Barnes:2003dj} and van Beveren and Rupp~\cite{vanBeveren:2003kd} were the first ones to point out the possible molecular nature of these resonances, i.e. that they are $D^{(*)}K$ bound states instead of conventional $c\bar s$ mesons. This also gives a natural explanation to another interesting implication:
\begin{eqnarray}
M_{D_{s1}}{-}M_{D_{s0}^*}\simeq M_{D^*}{-}M_D  .
\end{eqnarray}
In the molecular approach the spin--dependent interactions for a heavy quark are suppressed by $\Lambda_{QCD}/m_Q$, with $m_Q$ for the mass of a heavy quark. Thus $D_{s0}^*$ and $D_{s1}$ form a multiplet in the same fashion $D$ and $D^*$ do. 
A molecular assignment was used by many authors, however, using different approaches.
The works include hadrogenesis conjecture in Ref.~\cite{Lutz:2007sk}, model calculations in Refs.~\cite{Faessler:2007gv,Faessler:2007us}, dynamical generation from $\dk$ scattering in Refs.~\cite{Guo:2006fu,Guo:2006rp} and calculations based on $SU(4)$ flavor symmetry in Refs.~\cite{Gamermann:2006nm,Gamermann:2007fi}.

However, as appealing as the idea of hadronic molecules in this light might be, the mere difference between measured and predicted masses is not proof enough to rule out a conventional quark model. In addition, it was shown in Refs.~\cite{Bardeen:2003kt,Nowak:2003ra} that a parity doubling model can also obtain the correct masses. Later Mehen and Springer extended this calculation to the one-loop level in Ref.~\cite{Mehen:2005hc}. Naturally one  has to look at the decay patterns to find observables where where the molecular nature of $\dszero$ and $\dsone$ manifests itself. Unfortunately, at this moment none of the possible branching fractions has been measured precisely. Only upper limits exist, e.g.
\begin{equation}
 \frac{\Gamma(\dszero \rightarrow D_s^*\gamma)}{\Gamma(\dszero\rightarrow D_s\pi^0)}<0.059 .
\end{equation}
It is remarkable by itself that both states have been seen in hadronic decays, $\Gamma(\dszero\to D_s\pi)$ and $\Gamma(\dsone\to D_s^*\pi)$. Since both are isospin $I=0$ states these decays violate  isospin symmetry and thus should be heavily suppressed. 

Calculations of the relevant electromagnetic decays have been performed in Refs.~\cite{Faessler:2007gv,Faessler:2007us,Gamermann:2007bm,Lutz:2007sk}, the hadronic decays in~\cite{Faessler:2007gv,Lutz:2007sk,Bardeen:2003kt}.

Besides hadronic molecules and conventional mesons also other approaches exist. Most prominently, a number of publications assume the states to be tetraquarks, i.e. compact states with four valence quarks~\cite{Cheng:2003kg,Terasaki:2003qa,Terasaki:2003dj,Terasaki:2004yx,Vijande:2003wk}. Since some of their features are similar to the molecular picture it remains to point out features that make a distinction possible. We will present some in the following chapter. 

%%%%%%%%%%%%%%%%%%%%%%%%%%%%%%%%%%%%%%%%%%%%%%%%%%%%%%%%%%%%%%%%%%%%%%%%%%%%%%%%%%%%
%%%%%%%%%%%%%%%%%%%%%%%%%%%%%%%%%%%%%%%%%%%%%%%%%%%%%%%%%%%%%%%%%%%%%%%%%%%%%%%%%%%%

We will incorporate the most important experimental facts, the similarity of the pole position to the open threshold and the significant isospin violating width, by assuming that the states $\dszero$ and $\dsone$ are $DK$ and $D^*K$ bound states, respectively, instead of compact $c\bar s$. 
This explains the similarity of their features naturally. Especially the equidistance to the open thresholds simply reflects the fact that their binding energies are the same. Further we have a natural explanation for the possibly enhanced isospin violating decays: the decays are driven by $D^{(*)}K$ loops and the difference between charged and neutral meson masses is large enough to give a significant contribution here. On the other hand, a compact $c\bar s$ state will, due to the complete absence of light valence quarks, only produce vanishing isospin violating decay widths. 

The calculations and discussions in this chapter, based on the molecular assumption for $\dszero$ and $\dsone$, are taken from the publications in Refs.~\cite{Cleven:2010aw} and \cite{Cleven:2013rd}. The first one focuses on the dynamical generation of the states and the light quark mass dependence. The object of the second one are the decays, both radiative and hadronic. We will organize this chapter as follows: in Sec.~\ref{sec:poles} we discuss the dynamical generation of $\dszero$ and $\dsone$ in unitarized heavy meson chiral perturbation theory together with the prediction of open bottom partners. Sec.~\ref{sec:hadronic_decays} contains the hadronic decays, Sec.~\ref{sec:radiative_decays} the radiative decays and a discussion of both together with the relative importance and their ratios. In Sec.~\ref{sec:quarkmassdependence} we present a calculation of the masses and binding energies in dependence of the light quark masses expressed as pion and kaon mass dependence. These provide predictions 
that can be tested in lattice calculations. 

%%%%%%%%%%%%%%%%%%%%%%%%%%%%%%%%%%%%%%%%%%%
%
\section{Dynamical Generation of the states}\label{sec:poles}
%
%%%%%%%%%%%%%%%%%%%%%%%%%%%%%%%%%%%%%%%%%%%
In this section the molecular states $\dszero$ and $\dsone$ will be generated dynamically as poles in the $S-$matrix. We have done this in both publications on the subject. The coupled channel potentials for the channels with isospin $I=0$ and strangeness $S=1$ $\big(D^{(*)}K$ and $D_s^{(*)}\eta\big)$ are calculated. Then these amplitudes are unitarized as described in Sec.~\ref{sec:unitarization}. After the remaining parameters are fixed to observables we are able to obtain the masses of the molecular states as poles in the $S-$matrix.

In the first approach we use a nonrelativistic and manifestly spin symmetric approach to obtain both poles simultaneously and from the same set of parameters. Notice that due to the still rather poor experimental data we had to drop large-$N_c$ suppressed terms for the subleading contributions to reduce the number of low energy constants. We will see that this together with a relativistic loop, which was used in previous calculations,  violates spin symmetry: it is not possible to obtain both poles simultaneously with the same set of parameters.

However, in the second approach we were able to use new lattice calculations for charmed meson-Goldstone boson scattering. Note that the scattering amplitudes are calculated in channels different from the ones used here, i.e. strangeness $S=1$ and isospin $I=0$. In this way all subleading low energy constants can be fit at the same time. When this is done, it is possible to obtain both masses in agreement with experiment.
%%%%%%%%%%%%%%%%%%%%%%%%%%%%%%%%%%%%%%%%%%%
%
\subsection{Relativistic and Nonrelativistic Lagrangian}
%
%%%%%%%%%%%%%%%%%%%%%%%%%%%%%%%%%%%%%%%%%%%
The leading order chiral covariant Lagrangian \cite{Wise:1992hn,Burdman:1992gh} for pseudoscalar charmed mesons is given by 
\begin{equation}\label{eq:LagrLORel}
 \lag_\mathrm{LO}^\mathrm{Rel}=\left({\cal D}_\mu P\right)\left({\cal D}^\mu P^\dagger\right)-M^2PP^\dagger
\end{equation}
with $P=(D^0,D^+,D_s^+)$ and the chiral covariant derivative introduced in the previous chapter ${\cal D}_\mu=\partial_\mu+\Gamma_\mu$. Since the vector current $\Gamma_\mu$ provides an even number of Goldstone boson fields we can find the leading contribution to the scattering of Goldstone bosons off heavy mesons from it:
\begin{equation}
 \frac{i}{2F^2}\left[(\partial^\mu P)[\phi,\partial_\mu \phi]P^\dagger + P[\phi,\partial_\mu \phi](\partial^\mu P^\dagger) \right].
\end{equation}
This part provides the so-called Weinberg-Tomozawa term in the amplitude. 
The next-to-leading order contribution was introduced by Guo et al. in Ref.~\cite{Guo:2008gp}:
\begin{eqnarray}\nonumber\label{eq:LagrNLORel}
  \mathscr{L}^\mathrm{Rel}_{\rm NLO}&=&P\left(-h_0\langle\chi_+\rangle - h_1\chi_+ + h_2\left\langle u_{\mu}u^{\mu} \right\rangle - h_3u_{\mu}u^{\mu}\right) P^\dagger \non\\
& & +  {\cal D}_{\mu}P \bigl( h_4\langle u^{\mu}u^{\nu}\rangle - h_5 \{u^{\mu},u^{\nu}\} - h_6 [u^{\mu},u^{\nu}] \bigr) {\cal D}_{\nu}{P^\dagger}.
\end{eqnarray}
where the light quark masses enter explicitly via $\chi$. $\left<...\right>$ denotes the trace of light flavor indices, $\{\dots\}$ and $[...]$ the anticommutator and commutator, respectively. 
At this point we have six low energy constants that need to be fit to experimental data or lattice calculations. It is possible to reduce the number of terms as was pointed out by Lutz and Soyeur in Ref.~\cite{Lutz:2007sk}. The terms with $h_0$, $h_2$ and $h_4$ contain more than one flavor trace and can therefore be dropped in the large $N_c$ limit of QCD. We can drop the term $h_6$ because it contributes at the next order due to the commutator structure. 
The low energy constant $h_1$ is here the only one we can fit to data right away. In Sec.~\ref{sec:hmchpt} we found $h_1=0.42$. 

The extension of these Lagrangians to the heavy vector mesons is in principle straightforward: one simply needs to replace $P$ by $V^\nu$ in Eq.~(\ref{eq:LagrLORel}) and Eq.~(\ref{eq:LagrNLORel}). However, this is not manifestly spin symmetric and may contain heavy quark spin symmetry breaking terms. The safe way to enforce spin symmetry is to use the Heavy Meson Chiral Perturbation Theory fields introduced in Sec.~\ref{sec:hmchpt}. Here the Lagrangian for the kinetic energy of the heavy mesons reads
\begin{equation}
 \mathscr L^\mathrm{NR}_\mathrm{LO}=-i\mathrm{Tr}[\bar H_a v_\mu {\cal D}^\mu_{ba} H_b]
\end{equation}
where $a,b$ denote the light flavor indices. This way the pseudoscalar and vector mesons are treated equally. The difference to the relativistic approach becomes visible when we carry out the trace over the Lorentz matrices and look at the Weinberg-Tomozawa-like term for the pseudoscalars:
\begin{equation}
\frac{i}{2F^2}v^\mu P_a[\phi,\partial_\mu \phi]P_a^\dagger-\frac{i}{2F^2}v^\mu V_{a,\nu}[\phi,\partial_\mu \phi]V_a^{\dag\nu}.
\end{equation}
The derivative acting on the heavy field has been replaced by the heavy meson velocity. This corresponds to an expansion of the heavy momenta in the fashion $q^\mu=M_D v^\mu+\mathcal O[(\vec q/M_D)^2]$. The nonrelativistic NLO Lagrangian reads
\begin{eqnarray}\nonumber\label{Eq:LagrNLO}
  \mathscr{L}^\mathrm{NR}_\mathrm{NLO}&=&-h_0Tr[\bar H_a(\chi_+)_{bb}H_a]+h_1Tr[\bar H_a(\tilde\chi_+)_{ba}H_a]+h_2Tr[\bar H_a (u_\mu u^\mu)_{bb}H_a]\\ \nonumber
  &&+h_3Tr[\bar H_a (u_\mu u^\mu)_{ba}H_b]+h_4Tr[\bar H_a (-iv_\mu)(u^\mu u^\nu)_{bb} (iv_\nu)H_a]\\
  &&+h_5Tr[\bar H_a (-iv_\mu)\{u^\mu, u^\nu\}_{ba} (iv_\nu)H_b].
\end{eqnarray}
Notice that for this calculation $\tilde\chi$ defined in Eq.~(\ref{eq:chitilde}) was used. 

The masses used in this chapter are taken from the PDG~\cite{Behringer:2012ab}. The light mesons are
\begin{eqnarray}\label{eq:light-meson-masses}
&& M_{\pi^+} = (139.57018 \pm 0.00035)\mev, \qquad M_{ \pi^0}= (134.9766 \pm 0.0006) \mev\non\\
&& M_{K^+ }= (493.677 \pm 0.016)\mev,       \qquad M_{K^0 }=(497.614 \pm 0.024)\mev \non\\
&& M_{\eta }= (547.853 \pm 0.024)\mev,
\end{eqnarray}
the heavy mesons are
\begin{eqnarray}\label{eq:heavy-meson-masses}
 && M_{D^0}=(1864.86 \pm 0.13)\mev,\qquad   M_{D^+ }=(1869.62 \pm 0.15)\mev \non\\
 && M_{D_s^+ }= (1968.49 \pm 0.32)\mev ,\qquad M_{D^{*0}}= (2006.98 \pm 0.15)\mev \non\\
 && M_{D^{*+}}=(2010.28 \pm 0.13)\mev,\qquad  M_{D_s^{*+} }= (2112.3 \pm 0.5)\mev \non\\
 && M_{B^+}=(5279.25 \pm 0.17)\mev,\qquad   M_{B^0 }= (5279.58 \pm 0.17)\mev \non\\
 && M_{B^0 }=( 5366.77 \pm 0.24)\mev ,\qquad  M_{B^{*+} }=(5325.2 \pm 0.4)\mev \non\\
 && M_{B^{*0} }=( 5325.2 \pm 0.4)\mev,\qquad   M_{B_s^{*0} }= \left(5415.4^{+2.4}_{-2.1}\right)\mev .
\end{eqnarray}
Except for the hadronic decay widths of $\dszero$ and $\dsone$ the calculations are not sensitive to the isospin violation and we will use $M_D=(M_{D^+}+M_{D^0})/2$, etc. for simplicity. 
%%%%%%%%%%%%%%%%%%%%%%%%%%%%%%%%%%%%%%%%%%%
%
\subsection{Nonrelativistic Calculation}\label{sec:polesnonrel}
%
%%%%%%%%%%%%%%%%%%%%%%%%%%%%%%%%%%%%%%%%%%%
In this first attempt we will use Heavy Meson Chiral Perturbation Theory to describe the scattering of Goldstone bosons off heavy mesons. This was originally done as part of my Diploma Thesis~\cite{diploma}. The task there was to extend the work  by Guo et al. in Ref.~\cite{Guo:2009ct} from $\dszero$ to $\dsone$ in a consistent way. The choice is to use HMChPT so that charmed pseudoscalar and vector mesons are  treated on the same footage. These calculations were extended for the publication \cite{Cleven:2010aw}.

For contact interactions we use the leading and next-to-leading order contributions from the Lagrangians discussed in the previous section. Further we will consider exchange contributions that stem from the axialvector coupling introduced in Sec.~\ref{sec:hmchpt}:
\begin{eqnarray}
 \lag_\pi&=&\frac{ g_\pi}{2}\mathrm{Tr}\left[\bar H_a H_b i\gamma_\nu\gamma_5 u_{ba}^\nu\right]
\end{eqnarray}

\begin{figure}%[htbp]
\centering
\includegraphics[width=\linewidth]{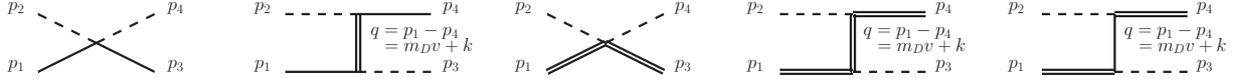}
\caption{Diagrams contributing at LO and NLO to Goldstone boson scattering off 
$D$- and $D^*$-mesons and the pertinent kinematics. Dashed lines denote Goldstone
bosons, solid lines charmed pseudoscalar mesons and solid double lines charmed
vector mesons, in order.} \label{fig:diagrams_scattering}
\end{figure}

In order to calculate the scattering amplitudes for the diagrams shown in Fig.~\ref{fig:diagrams_scattering}. In this work we use physical values for the meson masses and thus the mentioned term is considered automatically. Note, since $M_{D_s*}-M_{D_s}\neq M_{D*}-M_{D}$ (c.f. Eqs.~(\ref{eq:charmmasses})), in this way we also include an effect of simultaneous spin symmetry and SU(3) violation, which is formally of  next-to-next-to leading order (N$^2$LO). We come back to the quantitative role of this subleading effect below.
Since we are interested in the masses of the resonances, isospin breaking can be
neglected. Hence, we take averaged values for charged and neutral particles.

We will calculate the scattering of the Goldstone bosons off the pseudoscalar
$D$-mesons ($0^+$ channel) as well as off vector $D^*$-mesons ($1^+$ channel).  The
diagrams at the  order we are working are shown in Fig.~\ref{fig:diagrams_scattering}.
The diagrams are evaluated in the isospin basis, the relation between the isospin
basis and the  particle basis was derived in  Ref.~\cite{Guo:2009ct}:
\begin{eqnarray}
 V_{DK\to DK}(s,t,u) \! &= & \! 2V_{D^0K^+\to D^0K^+}(s,t,u) -  \! V_{D^+K^+\to D^+K^+}(s,t,u), \non\\
 V_{D_s\eta\to D_s\eta}(s,t,u) \! &= & \! V_{D_s^+\eta\to D_s^+\eta}(s,t,u) , \non\\
 V_{D_s\eta\to DK}(s,t,u) \! &= & \! -\sqrt{2}V_{D_s^+\eta\to D^0K^+}(s,t,u).
\end{eqnarray}

%%%%%%%%%%%%%%%%%%%%%%%%%%%%%%%%%%%%%%%%%%%%%%%%%%%%%%%%%%%%%%%%%%%%%%%%%%%%%%%%%%%%
%%%%%%%%%%%%%%%%%%%%%%%%%%%%%%%%%%%%%%%%%%%%%%%%%%%%%%%%%%%%%%%%%%%%%%%%%%%%%%%%%%%%

Evaluating the diagrams for the leading order contact interaction we find that
the contributions for the $0^+$ channel and the $1^+$ channel are the same up to
the different masses for the pseudoscalar and vector charmed mesons, as expected
from heavy quark spin symmetry:
\begin{eqnarray}\label{TLO}%\nonumber
V^{0^+}_{\rm LO}=V^{1^+}_{\rm LO}
=C_{LO}\sqrt{M_1M_3}\frac{1}{2F^2}(E_2+E_4).
\end{eqnarray}
The constants $C_0$ for the various channels are listed in
Tab.~\ref{tab:constants}. Further, $M_1$ and $M_3$ are the masses of the
in-coming and out-going charmed meson, respectively, cf.
Fig.~\ref{fig:diagrams_scattering}, and $E_2$ and $E_4$ are the energies of the in-coming and out-going light
mesons.

For the NLO contact interactions  we again find the same contributions except
for the different pseudoscalar and vector charmed meson masses:
\begin{eqnarray}\nonumber
  V_{\rm NLO}^{0^+}&=&V_{\rm NLO}^{1^+}=\sqrt{M_1M_3}\frac{1}{2F^2}\bigg[C_1\frac{4}{3}h_1 +C_{35}(h_3 p_2 {\cdot} p_4+ 2h_5 E_2 E_4)\bigg] \ ,
\end{eqnarray}
which also preserves spin symmetry. Its S-wave projection is
\begin{eqnarray}\nonumber\label{eq:LO}
 V^{0^+}_{s}&=&V^{1^+}_{s} =\sqrt{M_1M_3}\frac{2}{F^2}\bigg[C_{LO} (E_2+E_4)+C_1\frac{4}{3}h_1 +C_{35}\frac{1}{M_D}\tilde h_{35}  E_2 E_4\bigg]
\end{eqnarray}
with $h_{35}=h_3+2h_5\equiv {{\tilde h}_{35}/M_D}$. This is the only free
parameter at this order. We fit it to the mass of the $D_{s0}^*(2317)$. We
get $\tilde h_{35}=0.35$ for the dimensionless parameter, which is of natural
size.

The $s$--channel exchange needs a P-wave interaction and thus does not need to
be considered. The only contribution from exchange diagrams that does not have
an vanishing S-wave projection is the $u$--channel exchange of a charmed
pseudoscalar meson in the $1^+$ channel:
\begin{eqnarray}\label{eq:exch}
V_{\rm PS-Ex}^{1^+}=-C_u \frac{4g^2_\pi}{3F^2}\bigg[\frac{E_4}{M_1}|\vec p|^2+  \frac{E_2}{M_3}|\vec p\,'|^2\bigg] \frac{1}{(v{\cdot} k)}\sqrt{M_1M_3},
\end{eqnarray}
where the constant $C_u$ is also given in Tab.~\ref{tab:constants}, and
$\vec{p}$ and $\vec{p}\,'$ are the three-momenta of the in-coming and out-going
mesons in the center-of-mass frame and the energy transfer, $(v{\cdot} k)$, is defined in Fig.~\ref{fig:diagrams_scattering}.
Observe, since $k$, $\vec p$ and $\vec p\,'$ are counted of order of the $E_i$,
this potential appears formally at NLO.
Since it acts in the $1^+$ channel, while there is no counter--part 
in the $0^+$ channel, it provides formally the leading source of spin symmetry violation.
However, in practice its contribution turns out to be very small, due to three
reasons: first of all $g_\pi^2=0.1$, second, $|\vec p|\sim \sqrt{M_K\epsilon}$,
 with $\epsilon\equiv
M_{D^*}+M_K-M_{D_{s1}(2460)}\approx M_{D}+M_K-M_{D_{s0}^*(2317)}$ being the
binding energy, which is significantly smaller than $M_K$ and, most
importantly, it does not operate in the $KD^*\to KD^*$ channel (in that
channel $C_u=0$ --- c.f. Tab.~\ref{tab:constants}). In total it gives
a negligible contribution.

For SU(3) calculations, the uncertainty from the chiral expansion up to NLO is $\mathscr O(M_K^2/\Lambda_\chi^2)$, where $\Lambda_\chi\approx4\pi F_\pi$ sets the hard
scale in the chiral expansion, compared with the LO contributions. So we use
\begin{equation}\label{eq:error}
 V^\pm=\left(1\pm \frac{M_K^2}{\Lambda_\chi^2}\right)V_s,
\end{equation}
where $V_{s}$ is the S-wave projection of the full  NLO amplitude calculated
here for the $0^+$ and $1^+$ channel, respectively. 

\begin{table*}%[htbp]
\centering \caption{Coefficients for the amplitudes for all possible channels
  with total strangeness $S=1$ and total isospin $I=0$. While $C_{LO}$, $C_1$,
  and
$C_{35}$ act in both channels $0^+$ and $1^+$, while $C_u$ acts solely in $1^+$.}
\renewcommand{\arraystretch}{1.3}
\begin{tabular}{|l|l|l|l|l|}
\hline
Channel & $C_{LO}$ & $C_1$ & $C_{35}$ & $C_u$ \\
\hline
 $DK\rightarrow DK$ & $-2$ & $-4M_K^2$ & $\;\;\:2$ & $\;\;\:0$ \\ %\hline
   $D_s\eta\rightarrow D_s\eta$ & $\;\;\:0$ & $-2(2M_\eta^2-M_\pi^2)$ & $\;\;\:\frac 43$  & $\;\;\:\frac 23$ \\ %\hline
   $D_s\eta\rightarrow DK$ & $-\sqrt{3}$ &  $-\frac{\sqrt{3}}{2}(5M_K^2-3M_\pi^2)$ & $\;\;\:\frac{ \sqrt{3}}{ 3}$ &  $-\sqrt{\frac 13}$
 \\\hline
\end{tabular}
\label{tab:constants}
\end{table*}

Dynamical generation of bound states from a theory without bound states as
fundamental fields is a non-perturbative phenomenon. It cannot be provided by
any finite perturbative expansion in the momentum. Thus we have to unitarize, {\it
i.e.} resum, the amplitude we had obtained so far. 
As laid out in Sec.~\ref{sec:unitarization} we can write the $T-$matrix as an infinite sum of loops:
\begin{equation}%\label{Tmatrix}
  T(s)=V^\pm(s)[1-G(s)\cdot V^\pm(s)]^{-1},
\end{equation}
where $G(s)$ is the diagonal matrix with non-vanishing elements being loop
integrals of the relevant channels --- see
Eq.~(\ref{firstloop}) below. The bound state mass is found as a pole of the
analytically continued unitarized scattering matrix $T$. To be specific,
the pole of a bound state, as is the case
here, is located on the real axis below threshold on the first Riemann sheet of the
complex energy plane (note that because we neglect isospin-breaking, these
bound states do not acquire a width). In our approach, both the $D_{s0}^*(2317)$
and the $D_{s1}(2460)$
appear as poles in the $S=1$  isoscalar $I=0$ channel.
Since they couple predominantly to $DK$ and $D^*K$, 
we interpret the $D_{s0}^*(2317)$ and $D_{s1}(2460)$ as  $DK$ and $D^*K$
bound states, respectively.

At this point we have to make some comments on the loop matrix. In previous
works on this subject
\cite{Gamermann:2006nm,Gamermann:2007fi,Guo:2009ct}, %Kolomeitsev:2003ac,
the relativistic two-particle loop function
\begin{eqnarray}\label{firstloop}
 I_\mathrm{Rel}^{(0)}\left(s,m_D^2,M_\phi^2\right){=}i\int \!  \frac{d^4l}{(2\pi)^4}\frac{1}{[l^2{-}M_\phi^2{+}i\epsilon][(P-l)^2{-}M_D^2{+}i\epsilon]}
\end{eqnarray}
with $P^2=s$ was used in dimensional regularization with a subtraction constant $a(\mu)$
which absorbs the scale dependence of the integral and where $M_\phi$ and $M_D$ are the masses of
the light meson and charmed meson involved in the loop~\cite{Oller:2000fj}. However, this relativistic loop function violates spin
symmetry. Particularly, using the parameters fitted to the mass
of the $D_{s0}^*(2317)$, we find the mass of the $D_{s1}(2460)$ at
2477~MeV. From a field theoretical point of view, using a relativistic
propagator for a heavy meson and using the Lagrangian from heavy quark
expansion simultaneously is inconsistent. There is a well-known problem in
relativistic baryon chiral perturbation theory for pion--nucleon scattering,
which is closely related to the case here. The relativistic nucleon propagator
in a loop explicitly violates power counting, see \cite{Gasser:1987rb}. In our
case this effect appears only logarithmically as $\log\left({M_D^2}/{\mu^2}
\right)$.
However, it induces a violation of spin symmetry in the interaction already at
NLO and thus violates the power counting.

There are different ways to deal with this. In \cite{Kolomeitsev:2003ac} a
subtraction scheme is used that identifies the scale $\mu$ with the mass of the
$D$- and $D^*$-mesons, respectively, and thus the spin symmetry violating terms
disappear. In Refs.~\cite{Gamermann:2006nm,Gamermann:2007fi} simply different
subtraction constants are used for the two channels. In this work we use a
static propagator for the heavy meson, as this is consistent with the
Lagrangian in the
heavy quark expansion. This also allows us to use the
same subtraction constant and scale of regularization for both channels. The
heavy meson propagator for the vector $D^*$-meson then takes the form
\begin{eqnarray}\label{eq:prop}
 \frac{\{P^{\mu\nu},1\}}{q^2-M_D^2+i\epsilon}\rightarrow  \frac{\{P^{\mu\nu},1\}}{2M_D(v{\cdot} k
   -\Delta+i\epsilon)} \ ,
\end{eqnarray}
where $P^{\mu\nu}$ is a projector for spin-1. Dropping $\Delta$ in the
denominator and using the one in the numerator of Eq.~({\ref{eq:prop}) amounts
to the propagator  for the pseudoscalar $D$-meson.

With this the integral $I_\mathrm{Rel}^{(0)}$ defined in Eq.~(\ref{firstloop}) turns into the heavy-light (HL) scalar loop
\begin{eqnarray}\nonumber\label{nonrelloop}
 I^{(0)}_\mathrm{HL}(\sqrt s,M_D,M_\phi^2)=i\intl \frac{1}{2M_D}\frac{1}{[l^2-M_\phi^2+i\epsilon][v{\cdot} l+\sqrt s+i\epsilon]} \phantom{xxxxxxxxxxxx} \\
 =\frac{1}{16\pi^2  M_D}\bigg\{(\sqrt s{-}M_D)\left[a(\mu){+}\log\left(\frac{M_\phi^2}{\mu^2}\right)\right]\\  \nonumber
{+}2\sqrt{(\sqrt s{-}M_D)^2{-}M_\phi^2}\cosh^{{-}1}\left(\frac{\sqrt s{-}M_D}{M_\phi}\right){-}2\pi i|\vec p_{\rm cms}| \bigg\}+ \mathcal{O}\left(\frac{M_\phi}{M_D}\right),
\end{eqnarray}
with $a(\mu)$ a subtraction constant, $\mu$ the scale of dimensional
regularization, which is fixed to the averaged mass of $D$ and $D^*$, and $|\vec
p_{\rm cms}|=\sqrt{(\sqrt s-M_D)^2-M_\phi^2}$. Note, the use 
of the modified loop function of Eq.~(\ref{nonrelloop}) leads only
to tiny changes from relativistic effects in the observables discussed in
Refs.~\cite{Guo:2008gp,Guo:2009ct}. The subtraction constant $a(\mu)$ is fitted to
the mass of the $D_{s0}^*(2317)$. We find $a(\mu\!= \! 1936~{\rm
MeV})\!=\!-3.034$. Using the same parameters for the axial-vector channel we
find the pole at
\begin{equation}\label{Ds1result}
M^\mathrm{NR}_{D_{s1}(2460)}=(2459.6\pm0.6\pm1.8)~{\rm MeV} \ ,
\end{equation}
 where the first uncertainty stems from the experimental
uncertainty for the mass of the $D_{s0}^*(2317)$, which was used for fitting
the parameters, and the second 4\% uncertainty estimates the higher orders
from the heavy quark expansion --- since
there is no significant spin symmetry breaking term in the scattering amplitudes at NLO,
spin symmetry breaking effect should appear at $\mathscr
O([\Lambda_{\rm QCD}/m_c]^2)$. Hence the second uncertainty can be estimated by
$(\Lambda_{\rm QCD}/m_c)^2 \epsilon$ with the binding energy $\epsilon$. In this context it is interesting to note
that the spin--symmetry violation induced by $M_{D_s*}-M_{D_s}\neq
M_{D*}-M_{D}$, which is of N$^2$LO, indeed contributes to 
$M_{D_{s1}(2460)}$ less than 1 MeV, consistent with the uncertainty estimate.

 The result in Eq.~(\ref{Ds1result}) is in
perfect agreement with the experimental result $$M^{\rm
  exp}_{D_{s1}(2460)}=(2459.5\pm0.6)~{\rm MeV} \ .$$  To the order we are working
instead of $F_\pi$ we could as well have used $F_K = 1.18 F_\pi$. However,
this change does not have any impact on the mass calculated for the
$D_{s1}(2460)$ once the subtraction constant is re-adjusted to reproduce the
mass for the scalar state.
\subsection{Relativistic}\label{sec:polesrel}
%%%%%%%%%%%%%%%%%%%%%%%%%%%%%%%%%%%
%
%     Molecular Fields
%
%%%%%%%%%%%%%%%%%%%%%%%%%%%%%%%%%%%
For the second approach we can follow Liu et al. in Ref.~\cite{Liu:2012zya}. They performed a lattice calculation of charmed meson--Goldstone boson scattering. With these scattering lengths they were able to fit the low energy constants in Eq.~(\ref{eq:LagrNLORel}). 
This means we have now the full NLO Lagrangian for $DK$ and $D^*K$ scattering available. As we will see using the full Lagrangian without dropping the large $N_c$ suppressed terms allows us to obtain the correct pole position for $\dsone$ in a completely relativistic framework. We will later calculate the residues of the unitarized amplitude to introduce $\dszero$ and $\dsone$ as explicit fields with couplings to their constituents. These can be used for the radiative decays. Having calculated the amplitudes relativistically ensures consistency here. 

The amplitudes for Goldstone boson--light meson scattering are given by
\begin{equation}
 \label{eq:v}
V(s,t,u) = \frac1{F^2} \bigg[\frac{C_{\rm LO}}{4}(s-u) - 4 C_0 h_0 +
2 C_1 h_1 - 2C_{24} H_{24}(s,t,u) + 2C_{35} H_{35}(s,t,u) \bigg],
\end{equation}
where
\begin{equation}
 H_{24}(s,t,u) = 2 h_2 p_2{{\cdot}} p_4 + h_4 (p_1{\cdot} p_2\, p_3{\cdot} p_4 +p_1{\cdot} p_4 \, p_2{\cdot} p_3), 
\end{equation}
and
\begin{equation}
H_{35}(s,t,u) = h_3 p_2{\cdot} p_4 + h_5 (p_1{\cdot} p_2 \, p_3{\cdot} p_4 + p_1{\cdot} p_4 \, p_2{\cdot} p_3). 
\end{equation}
The coefficients $C_i$  are given in Tab.~\ref{tab:ci}. As we have shown before the exchange of heavy mesons, which is formally also a NLO contribution, is heavily suppressed. So we have the full NLO amplitude here. 

%%%%%%%%%%%%%%%%%%%%%%%%%%%%%%%%%%%%%%%%%%%%%%%%%%%%%%%%%%%%%%%%%%
\begin{table}
\centering
\begin{tabular}{| c | c c c c c |}
\hline\hline
Channels             & $C_{\rm LO}$ & $C_0$      & $C_1$                           & $C_{24}$ & $C_{35}$ \\ \hline
$D K\to D K$         & $-2$         & $M_K^2$    & $-2M_K^2$                       & 1        & 2  \\
$D_s\eta\to D_s\eta$ & 0            & $M_\eta^2$ & $-2M_\eta^2+2M_\pi^2/3$         & 1        & $\frac43$ \\
$D K\to D_s\eta$     & $-\sqrt{3}$  & 0          &  $-\sqrt{3}(5M_K^2-3M_\pi^2)/6$ & 0        & $\frac1{\sqrt{3}}$ \\
\hline\hline
\end{tabular}
\caption{\label{tab:ci}The coefficients in the scattering amplitudes $V(s,t,u)$ for all channels with strangeness ($S=1$) and isospin ($I=0$). }
\end{table}
%%%%%%%%%%%%%%%%%%%%%%%%%%%%%%%%%%%%%%%%%%%%%%%%%%%%%%%%%%%%%%%%%%

The unitarization process is exactly the same as in the previous section. The unitarized amplitude reads
\begin{equation}
 \label{eq:t}
T(s) = V(s) [1-G(s) V(s)]^{-1},
\end{equation}
where $V(s)$ is the $S-$wave projection of Eq.~(\ref{eq:v}). For the diagonal loop matrix $G(s)$  we can now use the relativistic two-point scalar integral. The explicit expression can be given as
\begin{eqnarray}
 I^{(0)}_{\rm Rel}(s,m_1^2,m_2^2) \!=\! \frac{1}{16\pi^2}\bigg\{\tilde{a}(\mu)+\ln{\frac{m_2^2}{\mu^2}} +\frac{m_1^2-m_2^2+s}{2s}\ln{\frac{m_1^2}{m_2^2}}
+\frac{\sigma}{2s}\left[\ln({s-m_1^2+m_2^2+\sigma}) \right. \non\\ 
 \left. -\ln({-s+m_1^2-m_2^2+\sigma}) + \ln({s+m_1^2-m_2^2+\sigma})-\ln({-s-m_1^2+m_2^2+\sigma}) \right] \bigg\},\phantom{xxx}
\end{eqnarray}
with $\sigma=\left\{[s-(m_1+m_2)^2][s-(m_1-m_2)^2]\right\}^{1/2}$. The loop is regularized using dimensional regularization. $\tilde a(\mu)$ is the subtraction constant at the scale of regularization $\mu$ and fixed to produce the mass of $\dszero$. So via construction the mass $M_{D_{s0}^*}=2317.8\mev$ is a pole of the $T-$matrix for $DK$ scattering in this coupled channel approach. The transition to the axialvector state can simply be made by replacing $M_D$ and $M_{D_s}$ by $M_{D^*}$ and $M_{D^*_s}$. This yields to
\begin{equation}
 M_{D_{s1}}^\mathrm{Rel}=(2457.8\pm 6.7\pm1.8)\mev
\end{equation}
The first error is obtained using a $1\sigma$ environment for the best fit of the $h_i$, the second error stems from higher order contributions in the same fashion as in the last section. The experimental uncertainty of $0.6\mev$ used there can be dropped here. The higher uncertainty in the second approach reflects the fact that we take the uncertainties of the LECs into account here. This value matches perfectly both the experimental value as well as the one calculated in the previous section. 

%%%%%%%%%%%%%%%%%%%%%%%%%%%%%%%%%%%%%%%%%%%%%%%%%%%%%%%%%%%%%%%%%%%%%%%%
%
\subsection{Extension to Hypothetical Bottom Partners}
%
%%%%%%%%%%%%%%%%%%%%%%%%%%%%%%%%%%%%%%%%%%%%%%%%%%%%%%%%%%%%%%%%%%%%%%%%
%
%      Nonrelativistic
%
%%%%%%%%%%%%%%%%%%%%%%%%%%%%%%%%%%%%%%%%%%%%%%%%%%%%%%%%%%%%%%%%%%%%%%%%
The formalism used here can easily be applied to the meson sector with open bottom as well. 
Heavy quark flavor symmetry implies that all the other parameters stay the
same up to higher order corrections. 
In order to make the transition to the bottom sector, we just have to replace the $D$-mesons with the corresponding ${\bar B}$-mesons. 
This allows us to predict the masses of the ${\bar B}K$ and ${\bar B}^*K$ bound states in the ($S=1,I=0$) channel. The predicted
masses of both approaches are given in Table~\ref{tab:comparison}, together with a comparison with previous
calculations~\cite{Kolomeitsev:2003ac,Guo:2006fu,Guo:2006rp}.~\footnote{The
$B^{(*)}$ mesons in the calculations of Ref.~\cite{Kolomeitsev:2003ac,Mehen:2005hc} should be
understood as ${\bar B}^{(*)}$ mesons which contain a $b$ quark rather than
$\bar b$.} Our results are in reasonable agreement with the previous ones.
\begin{table}[t]
\centering \caption{Comparison of our predictions of the masses of the ${\bar
B}K$ and ${\bar B}^*K$ bound states with those in
Refs.~\cite{Kolomeitsev:2003ac,Guo:2006fu,Guo:2006rp,Mehen:2005hc}. All masses are given in
MeV. The uncertainties given in the first  column originate from the residual
scale dependence and an estimate of higher order effects, respectively.}
\renewcommand{\arraystretch}{1.3}
\begin{tabular}{|l|l l l l l|}
\hline
 & NR & Rel & Ref.~\cite{Kolomeitsev:2003ac} & Refs.~\cite{Guo:2006fu,Guo:2006rp} & \cite{Mehen:2005hc} \\
\hline
 $M_{B^*_{s0}}$ & $5696\pm 20 \pm 30$ & $5663\pm20\pm30$ & $5643$ & $5725\pm39$ & $5667$ \\ %\hline
 $M_{B_{s1}}$   & $5742\pm 20 \pm 30$ & $5712\pm20\pm30$ & $5690$ & $5778\pm7$  & $5714$
 \\ \hline
\end{tabular}
\label{tab:comparison}
\end{table}
 In our calculation we have two sources of uncertainties, both shown
 explicitly in the table. One stems from higher order effects, which may be
 obtained from multiplying the binding energy by 20\%, estimated as ${\cal
   O}(\Lambda_{\rm QCD}/m_c)$ since heavy flavor symmetry was used to relate
 the LECs in charm and bottom sectors. The other one originates from the
 intrinsic scale dependence of the result: To come to a fully renormalization
 group invariant amplitude, a complete one--loop calculation is
 necessary for the transition amplitude.  This is, however, beyond the scope
 of this work. Thus, when connecting the charm to the bottom sector, a
 residual scale dependence remains --- to quantify it we varied the scale
 parameter $\mu$ (see Eq. (\ref{nonrelloop})) from the averaged mass of $D$
 and $D^*$ to that of $\bar B$ and $\bar B^*$, while keeping $a(\mu)$ fixed.
 Combining the two uncertainties in quadrature gives a total uncertainty 40
 MeV which is about 1 \% for the masses.  When again switching from $F_\pi$ to
 $F_K$  the predicted masses change by 8 MeV only, well consistent
 with our uncertainty estimates. 

We can see  in Tab.~\ref{tab:comparison} that our values agree within the, admittedly large, uncertainties with previous calculations as well as with each other in both approaches. 
 
It should be stressed that the uncertainties quoted for $M_{B^*_{s0}}$ and
$M_{B_{s1}}$ are highly correlated. As already explained for the charmed
system, within the molecular scenario the relation
\begin{equation}
M_{B_{s1}}{-}M_{B_{s0}^*}\simeq M_{B^*}{-}M_B \ 
\end{equation}
should hold up to corrections of $\mathscr O([\Lambda_{QCD}/m_b]^2)$ --- c.f.
discussion below Eqs.~(\ref{eq:exch}) and (\ref{Ds1result}). Thus we predict
\begin{equation}\label{split}
M_{B_{s1}}{-}M_{B_{s0}^*}=46\pm 0.4 \pm 1 \ \mbox{MeV} \ , 
\end{equation}
where the first uncertainty comes from the current experimental uncertainty in
$M_{B^*}{-}M_B$ and the second from the estimated spin breaking effects in the
formation of the molecule. Clearly, all mass differences deduced from
Tab.~\ref{tab:comparison} are consistent with this value.

$M_{B_{s1}}$ and $M_{B_{s0}^*}$ have not been measured so far. Note that
their existence in the deduced mass range and, especially, with the mass
splitting of Eq.~(\ref{split}), is a crucial and highly non--trivial test for
the theory presented and especially for the molecular nature of both states.

\subsection{Lagrangian and Explicit Fields}
%%%%%%%%%%%%%%%%%%%%%%%%%%%%%%%%%%%
%
%     Molecular Fields
%
%%%%%%%%%%%%%%%%%%%%%%%%%%%%%%%%%%%
Finally we can use the unitarized amplitudes to introduce the molecular states as explicit fields. These will be used in the  calculation of the radiative decays. 
Gauge invariance can be built in unitarized Chiral Perturbation Theory, see \cite{Borasoy:2005zg}. However, here we will adapt a different approach. We will introduce the molecular fields explicitly and calculate the coupling constants as the residues of the unitarized amplitude. This is equivalent to setting up an effective theory which is expanded around the pole of the amplitude. Since we will use this only to calculate two-body decays this is completely justified. 
The Lagrangian for the generated $\dszero$, $\dsone$ states and their coupling to $D^{(*)}K$ is given by
\begin{eqnarray}\label{eq:LagrMol}
 \lag_\mathrm{Mol}&=&\frac{g_{DK}}{\sqrt2}D_{s0}^*(D^{+\dagger}K^{0\dagger}+D^{0\dagger}K^{+\dagger})+g_{D_s \eta}D_{s0}^*D_s^\dagger\eta^\dagger \non \\
                   &&+\frac{g_{D^*K}}{\sqrt2}D^\mu_{s1}(D^{*+\dagger}_\mu K^{0\dagger}+D^{*0\dagger}_\mu K^{+\dagger} )+g_{D^*_s \eta}D_s^{*\dagger}\eta^\dagger+\mathrm{h.c.}
\end{eqnarray}
The extension to the open bottom sector is natural:
\begin{eqnarray}\label{eq:LagrMolB}
 \lag_{\mathrm{Mol},B}&{=}&\frac{g_{BK}}{\sqrt2}B_{s0}^*(B^{-\dagger}K^{+\dagger}+\bar B^{0\dagger}K^{0\dagger} )+g_{B_s \eta}B_{s0}^*B_s^\dagger\eta^\dagger \non\\
 &&+\frac{g_{B^*K}}{\sqrt2}B^\mu_{s1}(B^{*-\dagger}_\mu K^{+\dagger}+\bar B^{*0\dagger}_\mu K^{0\dagger} )+g_{B^*_s \eta}B_{s1}B_s^{*\dagger}\eta^\dagger+\mathrm{h.c.}
\end{eqnarray}
Here we have adapted the notation for the charmed strange mesons. 
We need to determine the coupling constants in Eq.~(\ref{eq:LagrMol}). The unitarized amplitude for $DK\to DK$ must be equal to the $S-$channel exchange of $\dszero$:
\begin{equation}
 T^{DK\to DK}=g_{DK}\frac{1}{s-M_{D_{s0}}^2}g_{DK}
\end{equation}
So we can determine the coupling constant by calculating
\begin{equation}
 g^2_{DK}=\lim_{s\to m_{D_{s0}}^2}(s-m_{D_{s0}}^2)T^{DK\to DK}
\end{equation}
The results are
\begin{eqnarray}\label{eq:couplings}
 &&g_{DK}=\phantom{0}(9.0\pm0.5)\gev,\qquad g_{D^*K}=(10.0\pm0.3)\gev,\non\\ &&g_{BK}=(29.1\pm0.7)\gev,\qquad g_{B^*K}=(28.3\pm1.0)\gev
\end{eqnarray}
\subsection{Conclusion}
%%%%%%%%%%%%%%%%%%%%%%%%%%%%%%%%%%%
%
%     Summary and Comparison
%
%%%%%%%%%%%%%%%%%%%%%%%%%%%%%%%%%%%
We have calculated the masses of $\dszero$ and $\dsone$ as dynamically generated poles in the $S-$matrix. This was done in one calculation with nonrelativistic heavy mesons and a fully relativistic one.
The approaches are slightly different in both calculations, also due to new, elaborate lattice data. While in both cases the mass of $\dszero$ is fixed to the experimental value $2317.8~\mev$ we obtain the mass of $\dsone$ as
\begin{equation}
 M_{\dsone}^{\rm NR}=(2459.6\pm0.6\pm1.8)~{\rm MeV}\qquad {\rm and} \qquad M_{\dsone}^{\rm Rel}=(2457.8\pm 6.7)\mev
\end{equation}
Both are in perfect agreement with the experimental value quoted by PDG \cite{Behringer:2012ab}
$$M^{\rm exp}_{D_{s1}(2460)}=(2459.5\pm0.6)~{\rm MeV} \ .$$
as well as with each other. The fact that it is possible to obtain both as spin partners without any further assumptions or any other spin symmetry breaking effects besides the use of explicit masses for the vector and pseudoscalar heavy mesons and the perfect agreement with experiment support our interpretation.

We can make predictions for so far hypothetical open bottom partners with masses
\begin{eqnarray}
 M_{B^*_{s0}}^\mathrm{NR}=(5696\pm36)\mev\qquad M_{B_{s1}}^\mathrm{NR}=(5742\pm36)\mev \\
 M_{B^*_{s0}}^\mathrm{Rel}=(5663\pm48)\mev\qquad M_{B_{s1}}^\mathrm{Rel}=(5712\pm48)\mev.
\end{eqnarray}
Experimental confirmation of the existence of these states is a crucial test of our theory. 

%%%%%%%%%%%%%%%%%%%%%%%%%%%%%%%%%%%%%%%%%%%%%%%%%%%%%%%%%%%%%%%%%%%%%%%%%%%%%%%%%%%%%%%%%%
%
\section{Hadronic Decays}\label{sec:hadronic_decays}
%
%%%%%%%%%%%%%%%%%%%%%%%%%%%%%%%%%%%%%%%%%%%%%%%%%%%%%%%%%%%%%%%%%%%%%%%%%%%%%%%%%%%%%%%%%%
\begin{figure}
\centering
\includegraphics[width=.7\linewidth]{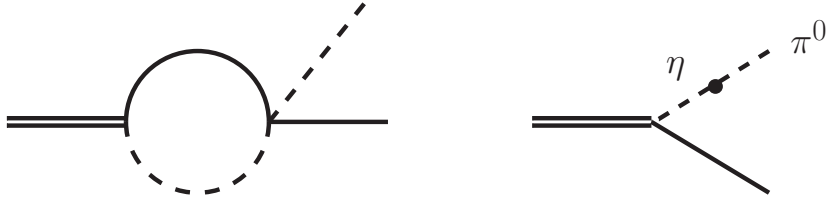}
\caption{\label{fig:diagrams_hadronic} The possible decay modes for $\dszero\to D_s\pi^0$ are shown: different loops for charged and neutral mesons and $\pi^0-\eta$-mixing}
\label{fig:hadronic_decays}
\end{figure}
In this section we calculate the hadronic decay widths for the open charm and bottom molecules, i.e. $\dszero\to D_s\pi^0$ and $\dsone\to D_s^*\pi^0$ and their corresponding bottom partners. Since experimental analyses  favors isospin $I=0$ for both $\dszero$ and $\dsone$ these decays are isospin violating and thus should be suppressed. If, however, one assumes that the states are $D^{(*)}K$ bound states, all decays have to go via $D^{(*)}K$ loops. The close proximity of the $DK$ and $D^*K$ thresholds to $\dszero$ and $\dsone$, respectively, leads to an enhancement of these loops non-analytical in the quark masses.
This will result in larger decay widths as one gets for compact $c\bar s$ states without any up or down quarks, e.g. Colangelo et al.~\cite{Colangelo:2012xi}. Thus a measurement of the hadronic widths is one of best channels to experimentally differentiate between a compact $c\bar s$ state and a molecule. 

The current experimental data does not give numbers for the individual decay rates. Only ratios between radiative and hadronic decay widths are provided. Thus a comparison to available data is only 
possible by calculating both on the same footage, the radiative decays and the discussion of the ratios will be subject of the next section. 

The isospin violating decays here are driven by two mechanisms: 
\begin{enumerate}
 \item The mass differences between charged and neutral mesons are small, but non\--va\-ni\-shing:
\begin{eqnarray}
 && M_{D^+}-M_{D^{0}}=4.8\mev,\qquad M_{D^{*+}}-M_{D^{*0}}=3.3\mev,\\
 && M_{K^{0}}-M_{K^{+}}=3.9\mev.
\end{eqnarray}
In isospin conserving calculations this small mass difference does not have any impact. For isospin violating decays this contributes because e.g. loops with charged particles do not cancel the contribution from neutral particles. 
\item As discussed in Sec.~\ref{sec:chpt} the isospin eigenstates of $\pi^0$ and $\eta$ mix with a small mixing angle. Once again, this contribution is suppressed for isospin conserving processes.
\end{enumerate}

We will calculate the decay width using two slightly different methods. The first one, proposed by Liu et al. in Ref.~\cite{Liu:2012zya} is based on the calculation of unitarized amplitudes as done in the previous section. The authors calculated the unitarized amplitude for the coupled channels $DK$ and $D_s\eta$ with isospin symmetric masses and no $\pi^0-\eta$ mixing and fix the remaining parameter, the subtraction constant of the two-body loop,  to the mass of $\dszero$ on the first Riemann sheet. After this the isospin violating mechanisms are included and the pole moves into the complex plane to the second Riemann sheet. They found $\Gamma(\dszero\to D_s\pi^0)=(133\pm22)\kev$. This approach is equivalent to the second approach, the calculation of the diagrams in Fig.~\ref{fig:hadronic_decays} with the four-body vertex put on shell.

We will start with the method of Liu et al.. The amplitude given in Eq.~(\ref{eq:v}) was calculated with isospin symmetric masses $M_D=(M_{D^+}-M_{D^0})/2$ and without $\pi^0-\eta$ mixing. All low energy constants can be fit to heavy meson-Goldstone boson scattering which makes the subtraction constant from the two-body scalar loop the only remaining unknown. This can be fit to the mass of $\dszero$. Here $\dszero$ is stable and the pole lies on the first Riemann sheet. We can now include the isospin breaking effects. First we use physical masses for the charged and neutral mesons. As a result
\begin{equation}
 I^{(0)}\left(M^2_{\dszero},M^2_{D^+},M^2_{K^0}\right)-I^{(0)}\left(M^2_{\dszero},M^2_{D^0},M^2_{K^+}\right)\neq0
\end{equation}
which gives a non-vanishing contribution for the sum of two diagrams in Fig.~\ref{fig:hadronic_decays}. 
As discussed in detail in Sec.~\ref{sec:chpt} the isospin eigenstates of $\pi^0$ and $\eta$ mix as
\begin{equation}
 \left(\begin{array}{c}
      \tilde \pi^0 \\ \tilde\eta\\
     \end{array}\right)
=\left(\begin{array}{cc}
        \;\;\:\cos \epsilon_{\pi^0\eta} & \sin \epsilon_{\pi^0\eta}\\
	-\sin\epsilon_{\pi^0\eta} & \cos\epsilon_{\pi^0\eta}\\
       \end{array}\right)
\left(\begin{array}{c}
       \pi^0 \\ \eta
     \end{array}\right)
\end{equation}
with the mixing angle $\epsilon_{\pi^0\eta}=\frac{1}{2}\arctan\left(\frac{\sqrt{3}}{2}\frac{m_d-m_u}{m_s-\hat m}\right)\simeq 0.013$. Including this corresponds to the calculation of the diagram in Fig.~\ref{fig:hadronic_decays}(b). 
One has to look for poles in the $S-$matrix of the form $m+i\Gamma/2$ on the second Riemann sheet. The second Riemann sheet is defined as the one where the sign of the imaginary part of the lightest of the coupled channels changes its sign. In this case this means 
\begin{equation}
 I^{(0)}\left( s,M^2_{D_s},M^2_{\pi^0}\right)\to I^{(0)}\left( s,M^2_{D_s},M^2_{\pi^0}\right)-2i \mathrm{Im}\left[I^{(0)}\left( s,M^2_{D_s},M^2_{\pi^0}\right)\right].
\end{equation}
The results are
\begin{equation}\label{eq:hadronicwidth}
 \Gamma(\dszero\to D_s\pi^0)=(133\pm 22)\kev ,\qquad \Gamma(\dsone\to D_s^*\pi^0)=(69\pm26)\kev.
\end{equation}
Note that, while the first one,  $\Gamma(\dszero\to D_s\pi^0)$, was published by Liu et al.~\cite{Liu:2012zya} and only the latter was new, we discuss both simultaneously as they are obtained with the same parameters. 

We can now make the same transition to the open bottom sector we made in Sec.~\ref{sec:polesrel}: we keep all parameters the same, i.e. the subtraction constant $a$ and the low energy constants $h_i$, and replace the scale of regularization $\mu_D$ with $\mu_B=\left(M_{B^+}+M_{B^0}+M_{B_s}\right)/3$.
The decay widths for the open bottom states are of the order $1\kev$ in this approach. 

The detailed results are listed in the Tab.~(\ref{tab:hadronic_widths}). There we distinguish between the contribution that arises from the two isospin symmetry violating processes explicitly.
We will discuss the results in detail later together with the other approaches.

%%%%%%%%%%%%%%%%%%%%%%%%%%%%%%%%%%%%%%%%%%%%%%%%%%%%%%%%%%%%%%%%%%%%%%%%%%%%%%%%%%%%%%%%%%%%%%%%%%%%
\begin{table}[t]
\centering
\renewcommand{\arraystretch}{1.3}
\begin{tabular}{| c | c c |}
\hline\hline
$\Gamma(\dszero\to D_s\pi^0)$	&       		&   \\ \hline
Masses 				& $(36\pm2)\kev$ 	& $(26\pm3)\kev$ \\ 
$\pi^0-\eta$ mixing		& $(31\pm5)\kev$	& $(23\pm3)\kev$ \\
Both				& $(133\pm 22)\kev$ 	& $(97\pm9)\kev$ \\ \hline
$\Gamma(\dsone\to D^*_s\pi^0)$ &       		& \\ \hline
Masses 				& $(12\pm4)\kev$ 	& $(20\pm4)\kev$ \\
$\pi^0-\eta$ mixing		& $(23\pm4)\kev$	& $(19\pm3)\kev$ \\
Both				& $(69\pm 26)\kev$ 	& $(78\pm10)\kev$ \\ \hline
$\Gamma(\bszero\to B_s\pi^0)$  &  			&    \\ \hline
Masses 				& $(0.2\pm0.1)\kev$ 	& $(28\pm15)\kev$ \\
$\pi^0-\eta$ mixing		& $(2.4\pm0.8)\kev$	& $(16\pm14)\kev$ \\
Both				& $(1.3\pm0.5)\kev$ 	& $(85\pm28)\kev$ \\ \hline
$\Gamma(\bsone\to B^*_s\pi^0)$ &			&    \\ \hline
Masses 				& $(0.2\pm0.1)\kev$ 	& $(24\pm14)\kev$ \\
$\pi^0-\eta$ mixing		& $(2.6\pm0.8)\kev$	& $(23\pm13)\kev$ \\
Both				& $(1.4\pm0.6)\kev$ 	& $(91\pm38)\kev$ \\
\hline\hline
\end{tabular}
\caption{\label{tab:hadronic_widths}Hadronic Decay width. The left column gives the values extracted as poles in the $S$-matrix, the right column the values calculated from diagrams.}
\end{table}
%%%%%%%%%%%%%%%%%%%%%%%%%%%%%%%%%%%%%%%%%%%%%%%%%%%%%%%%%%%%%%%%%%%%%%%%%%%%%%%%%%%%%%%%%%%%%%%%%%%%

Alternatively, with the Lagrangians set up, c.f.~(\ref{eq:LagrMol}), and the coupling of the molecular states to a heavy mesons--kaon pair we have the necessary tools to calculate the hadronic decays explicitly as shown in Fig.~\ref{fig:diagrams_hadronic}. 
Since the low energy constants in the $NLO$ Lagrangian were obtained neglecting the off-shell contributions we will set the momenta for this vertex in Diag.~(a) in Fig.~\ref{fig:diagrams_hadronic} on-shell. This should reproduce the numbers from the approach with the unitarized amplitude. The sum of all diagrams gives
\begin{eqnarray}
 &&\Gamma(\dszero\to D_s\pi^0)=(97\pm 9)\kev\qquad \Gamma(\dsone\to D^*_s\pi^0)=(78\pm10)\kev\\
 &&\Gamma(\bszero\to B_s\pi^0)=(85\pm28)\kev\qquad \Gamma(\bsone\to B^*_s\pi^0)=(91\pm38)\kev.
\end{eqnarray}
In Tab.~\ref{tab:hadronic_widths} we give the full results. 

%%%%%%%%%%%%%%%%%%%%%%%%%%%%%%%%%%%
%
%   Comparison and Discussion
%
%%%%%%%%%%%%%%%%%%%%%%%%%%%%%%%%%%%

When we look at the individual contributions for the decays of the charmed mesons we find that the contributions for both mechanisms are of the same order for $\dszero$. For $\dsone$ $\pi$--$\eta$ mixing is the larger contribution. The reason for that is the smaller mass difference:
\begin{equation}
 M_{D^{*+}}-M_{D^{*0}}<M_{D^{+}}-M_{D^{0}}.
\end{equation}
The results in both calculations end up in the order of $100\kev$. We will use the results in Eq.~(\ref{eq:hadronicwidth}) in the next section for comparison to the radiative decays. 

The situation is different for the open bottom mesons. In the first method we find numbers of about $1\kev$ while at the same time we find $(85\pm28)\kev$ and $(91\pm38)\kev$ which is about two orders of magnitude larger. The conclusion here is that we are not able to make any statements on these decay widths. The low energy constants obtained in charmed meson--Goldstone boson scattering obviously cannot be applied to the bottom sector. An additional lattice calculation for this will be necessary. 
We are therefore not able to make reliable predictions for the hadronic width of the open bottom molecules.

There have been a number of calculations before, using both a molecular and a compact approach on the resonances. We will start with the molecular interpretations. 
Faessler et al.  calculated the strong width of the $\dszero$ to be $(79.6\pm 33)~\kev$ in Ref.~\cite{Faessler:2007gv} and the one of $\dsone$ to be $50.1-79.2\kev$. They did not consider $\pi^0-\eta$ mixing which explains the difference to our results. They also find the latter one to be smaller. 
In \cite{Lutz:2007sk} Lutz and Soyeur found a width of $140~\kev$ for both the $D_{s0}^*(2317)$ and the $D_{s1}(2460)$. The authors included both mechanisms, $\pi^0-\eta$ mixing and charged and neutral mass differences. 

A calculation in a parity doubling model by Bardeen et al. \cite{Bardeen:2003kt} gives significantly smaller numbers, their widths are $21.5\kev$ for both $c\bar s$ decays. Similarly Colangelo and De Fazio find $\Gamma(\dszero\to D_s\pi)=7\kev$ \cite{Colangelo:2003vg}.
This is easily understood because a compact $c\bar s$ state does not decay via loops in the way we described it for molecular states. This makes a measurement of these decays a promising channel to judge if $\dszero$ and $\dsone$ are $c\bar s$ states or more complicated objects like molecules or tetraquarks. 
%%%%%%%%%%%%%%%%%%%%%%%%%%%%%%%%%%%
%
\section{Radiative Decays of $\dszero$ and $\dsone$}\label{sec:radiative_decays}
\subsection{Lagrangians}
%
%%%%%%%%%%%%%%%%%%%%%%%%%%%%%%%%%%%
In this section we discuss the Lagrangians used for the radiative decays of $\dszero$ and $\dsone$ in some detail. First of all we use relativistic Lagrangians for the calculations. This is the recommended approach because in a relativistic framework gauge invariance is manifest and can simply be tested with the current conservation
\begin{equation}
 k^\mu \mathcal M_\mu=0.
\end{equation}

%%%%%%%%%%%%%%%%%%%%%%%%%%%%%%%%%%%%%%%%%%%%%%%%%%%%%%%%%%%%%%%%%%%%%%
%
%  Building Blocks
%
%%%%%%%%%%%%%%%%%%%%%%%%%%%%%%%%%%%%%%%%%%%%%%%%%%%%%%%%%%%%%%%%%%%%%%

There are several building block we can construct the Lagrangians from. We start with the heavy fields in a spin symmetric formulation. In Sec.~\ref{sec:hmchpt} we introduced the pseudoscalar and vector heavy mesons as
\begin{equation}
 H=\projv \left[\slashed V+\gamma_5 P\right], \quad \bar H=\gamma^0H^\dagger \gamma^0=\left[{\slashed V}^{\dagger}-\gamma_5 P^\dagger\right]\projv
\end{equation}
where the pseudoscalar fields $P$ and the vector fields $V$ are collected in 
\begin{equation}
 P(V)=\left(D^{(*)0},D^{(*)+},D^{(*)+}_s\right).
\end{equation}

We can now construct a similar multiplet with the scalar $\dszero$ and the axialvector $\dsone$. The $SU(3)$ multiplets only have the molecules as fields:
\begin{equation}
 \mathcal S=\left(0,0,D^*_{s0}\right), \quad \mathcal A=\left(0,0,D_{s1}\right).
\end{equation}
The spin symmetric multiplet $R$ reads
\begin{equation}
 R=\projv\left[\mathcal S+\gamma_5\slashed{\mathcal A}\right], \quad \bar R=\left[\mathcal S^\dagger-\gamma_5\slashed{\mathcal A}^\dagger\right]\projv.
\end{equation}
The chiral and gauge covariant derivative on the heavy pseudoscalar field is (for vector fields simply replace $P^\dagger$ by $V^{\dagger}_{\nu}$)
\begin{eqnarray}
 D_\mu P^\dagger&=&\partial_\mu P^\dagger+\Gamma_\mu P^\dagger-ie A_\mu(P^\dagger Q_h-Q_lP^\dagger)\non\\
 &&\partial_\mu P^\dagger-ieA_\mu Q_HP^\dagger+...
 \end{eqnarray}
where $A_\mu$ is the electromagnetic vector potential, $Q_h$ is the heavy quark's charge, $Q_c=2/3$ or $Q_b=-1/3$, respectively, and $Q_l=\mathrm{Diag}(2/3,-1/3,-1/3)$ is the light quark charge matrix. $Q_H$ is the heavy meson charge matrix with $Q_D={\rm Diag} (0,1,1)$ and $Q_B={\rm Diag} (-1,0,0)$. Further we need the field strength tensor defined as  $V_{\mu\nu}=D_\mu V_\nu-D_\nu V_\mu$. 
The kinetic energy term reads
\begin{equation}
 \lag_\mathrm{Kin}=(D_\mu P^\dagger_a)(D^\mu P_a)-m_P^2P^\dagger_aP_a -\frac12 V_a^{\mu\nu} V_{a,\mu\nu}^{\dagger}+\frac12m_V^{2} V_a^{\mu} V_{a,\mu}^{\dagger}-(D_\mu V_a^{\mu})(D^\nu V_{a,\mu}^{\dagger}).
\end{equation}
In App.~\ref{app:stueckelberg} we explain the last term and its consequences. The kinetic terms for $\mathcal{S}$ and $\mathcal{A}$ are obtained equally.

We also need to consider the light fields as discussed in Sec.~\ref{sec:hmchpt}.
The leading order Lagrangian for the light fields reads, cf. Eq.~(\ref{eq:lag_U}),  
\begin{equation}
 \mathscr L_\phi^{(2)}=\frac{4}{F^2_\pi}\left<(D_\mu U)(D^\mu U)\right>+\frac{F_\pi^2}{4}\left<\chi_+\right>,
\end{equation}
where the gauge-covariant derivative is given by
\begin{equation}
 D_\mu u=\partial_\mu u+ieA_\mu[Q_l,u].
\end{equation}

%%%%%%%%%%%%%%%%%%%%%%%%%%%%
%
% Axialvector Coupling
%
%%%%%%%%%%%%%%%%%%%%%%%%%%%%
The interaction Lagrangians are derived from the spin symmetric multiplets. We need the axialvector coupling, i.e. the emission of a light field from a heavy field: 
\begin{eqnarray}\label{eq:lag_AV}
\lag_{\mathrm{AV}}&{=}& \sqrt{M_DM_{D^*}}\frac {g_\pi}2\mathrm{Tr}\left[ \bar H_b H_a \gamma_5\gamma^\mu u_{ba\mu} \right]  \non\\
	  &{=}& ig_\pi \sqrt{M_DM_{D^*}}\left(P^*_{\mu}u^\mu P^\dagger-P u^\mu V^{\dagger}_{\mu}+\varepsilon_{\mu\nu\alpha\beta}\;v^\mu V^{\alpha}u^\nu V^{\dagger\beta}\right)
\end{eqnarray}
Since the decay $D^{*+}\to D^0\pi$ has been measured we can deduce $g_\pi$ from experiment:
\begin{equation}
 g_\pi=0.61\pm0.07.
\end{equation}
Notice that one could in principle use a fully relativistic Lagrangian as done in ~\cite{Cheng:1992xi}:
\begin{equation}
  \mathscr L=g_\pi\left[i\sqrt{m_D m_{D^*}}\left(V_\mu u^\mu P^\dagger-P u^\mu V^{\dagger}_\mu\right)+\frac14\varepsilon^{\mu\nu\alpha\beta}\left(V_{\mu\nu} u_\alpha V^{\dagger}_\beta+V_\beta u_\alpha V^{\dagger}_{\mn}\right)\right].
\end{equation}
However, we found that this violates spin symmetry too strongly. While after mass renormalization all other decays are finite, for the decay $\dsone\to D_s^*\gamma$ a divergence remains only for the intermediate $D^{*0}K^+$. Since we do not have a counter term at hand that only works in that channel, we will here and in the following use the Lagrangians with the heavy meson velocity $v$ as in Eq.~(\ref{eq:lag_AV}) to get results consistent with spin symmetry.

%%%%%%%%%%%%%%%%%%%%%%%%%%%%%%%%%%%
%
%     Magnetic Moments
%
%%%%%%%%%%%%%%%%%%%%%%%%%%%%%%%%%%%
Photons can not only couple to the electric charge, but also to the magnetic moments of the heavy mesons. This needs to be included as well. We follow references \cite{Amundson:1992yp,Hu:2005gf}. in our notation the Lagrangians read
\begin{eqnarray}\label{LagrMagnMom}
 \lag_{\rm Magn.Mom.}&{=}&\sqrt{M_DM_{D^*}}F^{\mu\nu}\left\{\frac{e\beta}{2}\mathrm{Tr}\left[\bar H_aH_b\sigma_{\mu\nu}Q_{ab}\right]+\frac{eQ'}{2m_Q}\mathrm{Tr}\left[ \bar H_a\sigma_{\mu\nu} H_b\right] \right\} \non\\
 &{=}&\frac i2 e F_{\mu\nu}\sqrt{m_Dm_{D^*}}\left[\varepsilon^{\mu\nu\alpha\beta}v_\alpha\left(\beta Q+\frac{Q'}{m_Q}\right)_{ab}\left(P_aV^{\dagger\beta}_b-V^{\beta}_aP^\dagger_b \right) \right.\\
 &&\left.+V_a^{\mu} V^{\dagger\nu}_b\left(\beta Q-\frac{Q'}{m_Q}\right)_{ab} \right]\non
\end{eqnarray}
The first term is the magnetic moment of the light degrees of freedom, the second one the magnetic moment coupling of the heavy quark. $\beta$ and $m_Q$ can be fixed from experimental data for  $\Gamma(D^{*0}\to D^0\gamma)$ and $\Gamma(D^{*+}\to D^+\gamma)$. We will use the values quoted by Hu and Mehen in \cite{Hu:2005gf}:
\begin{equation}
 1/\beta=379\mev\qquad m_c=1863\mev
\end{equation}
and $m_b=4180\mev$ for the bottom mesons. 

%%%%%%%%%%%%%%%%%%%%%%%%%%%%%%%%%%%
%
%     Molecular Fields
%
%%%%%%%%%%%%%%%%%%%%%%%%%%%%%%%%%%%
In the molecular picture the $\dszero$ and $\dsone$ couple to the $DK$ and $D^*K$ most strongly, see Refs.~\cite{Guo:2006fu,Guo:2006rp}, and the other components including the $D_s\eta$ and $D^*_s\eta$ only provide a correction. 
Thus, here we will only consider the direct coupling of the $D_{sJ}$ to $\dk$. Since they are not far from the thresholds, both $D^{(*)}$ and $K$ are nonrelativistic. In this case, we can use the Lagrangians in Eqs~\ref{eq:LagrMol} and \ref{eq:LagrMolB} and drop the $D_s^{(*)}\eta$ and $B_s^{(*)}\eta$ part, respectively.
\begin{equation}
 \lag_\mathrm{Mol}=\frac{g_{DK}}{\sqrt2}D_{s0}^*\left(D^{+\dagger}K^{0\dagger}+D^{0\dagger}K^{+\dagger}\right)+\frac{g_{D^*K}}{\sqrt2}D^\mu_{s1}\left(D^{*+\dagger}_\mu K^{0\dagger}+D^{*0\dagger}_\mu K^{+\dagger}\right)+\mathrm{h.c.}.
\end{equation}
The extension to the open bottom sector is natural:
\begin{equation}
 \lag_{\mathrm{Mol},B}=\frac{g_{BK}}{\sqrt2}B_{s0}^*\left(B^{-\dagger}K^{+\dagger}+\bar B^{0\dagger}K^{0\dagger}\right)+\frac{g_{B^*K}}{\sqrt2}B^\mu_{s1}\left(B^{*-\dagger}_\mu K^{+\dagger}+\bar B^{*0\dagger}_\mu K^{0\dagger}\right)+\mathrm{h.c.}
\end{equation}
%%%%%%%%%%%%%%%%%%%%%%%%%%%%%%%%%%%%%%%%%%%%%%%%%%%%
%
%  Mass Renormalization
%
%%%%%%%%%%%%%%%%%%%%%%%%%%%%%%%%%%%%%%%%%%%%%%%%%%%%

\begin{figure}%[htbp]
\centering
\includegraphics[width=.75\linewidth]{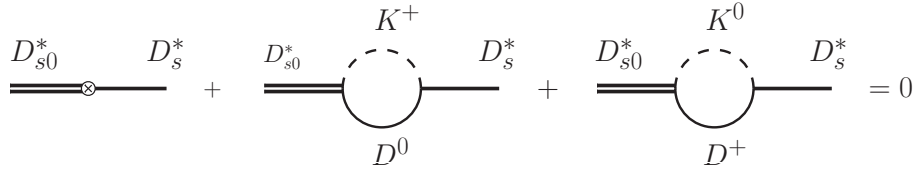}
\caption{\label{fig:mixing} The mass renormalization mechanism that ensures that $\dszero$ and $D_s^*$ do not mix for the physical particle.}
\end{figure}
So far we have simply included the molecular fields with their physical masses on Lagrangian level. This is somehow problematic because the longitudinal component of the vector $D_s^*$ can mix with the scalar $\dszero$; $D_s$ and $\dsone$ behave similarly. For physical processes like $\dszero\to D_s^*\gamma$ we need to make sure that the mass matrix for $\dszero$ and $D_s^*$ is diagonal. 
The standard  mass renormalization procedure would be to start with a Lagrangian with bare masses and  construct a mass renormalization matrix including the mixing via $DK$ loops as off-diagonal elements. This matrix would then be diagonalized to find the proper states for this problem. However, we propose a different method that will turn out to be simpler. We introduce the Lagrangian
\begin{eqnarray}\label{eq:LagrMR}
 \mathscr L_{\rm MR}&{=}& \frac{\kappa}{2}\mathrm{Tr}\left[\bar S \slashed\partial  R  \right] \non\\
 &{=}&\kappa_{DK} D_{s0}^*\left(\partial_\mu D_s^{*\dagger\mu}\right)-\kappa_{D^*K}\left[\left(\partial_\mu D_{s1}^{*\mu}\right)D_s^\dagger-i\varepsilon_{\mu\nu\alpha\beta}v^\alpha D_{s1}^\mu D_s^{*\dagger\beta} \right].
\end{eqnarray}
The coupling constants $\kappa_{DK}$ and $\kappa_{D^*K}$ can be fixed by demanding the equality shown in Fig.~\ref{fig:mixing}. 
This is justified since we use the physical masses in the Lagrangian. Note that we allow for different constants $\kappa_{DK}$ and $\kappa_{D^*K}$. This is necessary because we allow for spin symmetry violation by using physical masses for $D$ and $D^*$, respectively.
Having fixed the couplings we obtain three additional diagrams to the transition via gauging the interaction in Eq.~(\ref{eq:LagrMR}), also shown in Fig.~\ref{fig:diagrams}. 

%%%%%%%%%%%%%%%%%%%%%%%%%%%%%%%%%%%%%%%%%%%%%%%%%%%%
%
%  Contact Interaction
%
%%%%%%%%%%%%%%%%%%%%%%%%%%%%%%%%%%%%%%%%%%%%%%%%%%%
In accordance with spin symmetry, one finds a contact interaction:
\begin{eqnarray}
 \lag_{\rm Contact}&{=}& \frac{\lambda}{2}F_{\mu\nu}\mathrm{Tr}\left[ \bar H\sigma^{\mu\nu}R \right] \non\\
 &{=}&\lambda F_{\mu\nu}\left[v^\mu \dszero D_s^{*\dagger\nu}+\dsone^\mu v^\nu D_s^\dagger+\varepsilon^{\mu\nu\alpha\beta}{\dsone}_\alpha D^{*\dagger}_{s\beta}\right].
\end{eqnarray}
This means that up to higher order in spin symmetry breaking we can use this contact interaction in the channels $\dszero\to D_s^*\gamma$, $\dsone\to D_s\gamma$ and $\dsone\to D_s^*\gamma$ with the same interaction strength. This means after fixing it to an observable for one channel we still have predictive power in the remaining ones. 

This is different for the process $\dsone\to\dszero\gamma$. This contact interaction cannot be related to the others which means we lack predictive power for this decay. The contact term is
\begin{eqnarray}
 \lag_{\rm Contact,2}=\tilde\lambda \varepsilon^{\mu\nu\alpha\beta}F_{\mu\nu}v_\beta D_{s1,\alpha} D_{s0}^{*\dagger}. 
\end{eqnarray}
In the following section we will introduce a power counting scheme to determine at what order the contact term and the loops with coupling to the electric charge and magnetic moments enter. 
\subsection{Amplitudes}\label{sec:amplitudes}
%
%%%%%%%%%%%%%%%%%%%%%%%%%%%%%%%%%%%%%%%%%%%%%%%%%%%%
\begin{figure*}%[htbp]
% \captionsetup{width=.88\textwidth}
\centering
\includegraphics[width=.95\linewidth]{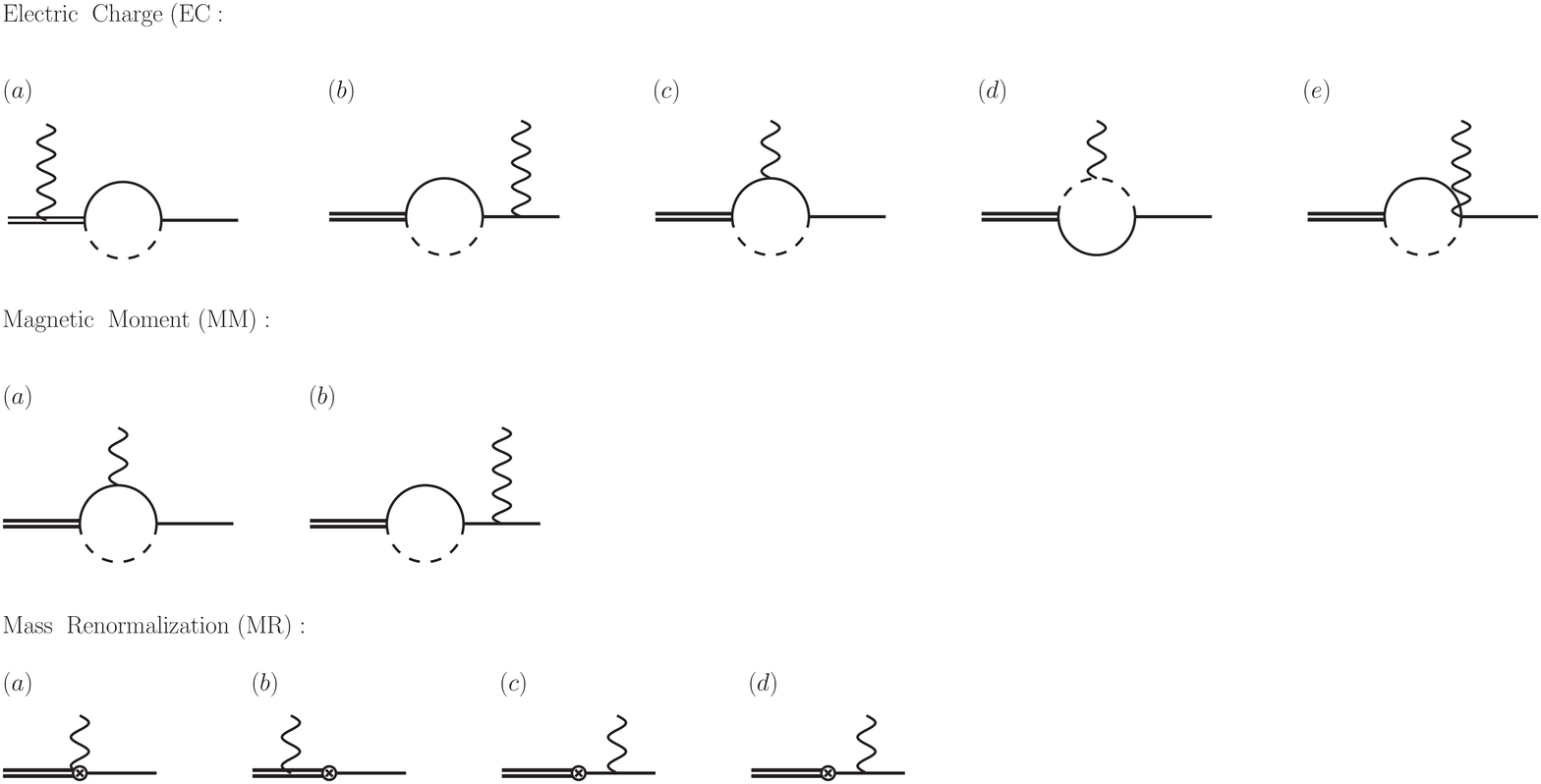}
\caption{\label{fig:diagrams}  All possible diagrams. Double lines denote the molecules, single lines the charmed mesons and dashed lines the kaons. In the last line (c) and (d) denote the diagrams where the photon couples to the electric charge and magnetic moment of the outgoing heavy meson, respectively.}\vspace{6mm}
\end{figure*}
%%%%%%%%%%%%%%%%%%%%%%%%%%%%%%%%%%%
%
%     Power Counting
%
%%%%%%%%%%%%%%%%%%%%%%%%%%%%%%%%%%%
We will use the standard power counting scheme of Chiral Perturbation Theory, see e.g. Ref~\cite{Scherer:2002tk} and the discussion in Sec.~\ref{sec:power_counting}. The relevant scale is $l\sim\sqrt{2\mu E}$  with the binding energy of the molecule $E$ and the reduced mass of its constituents $\mu$. The integration measure counts as $l^4$, light meson propagators as $l^{-2}$ and heavy ones as $l^{-1}$. Similarly the coupling of a photon to the electric charge gives $l^2$ for light and $l$ for heavy mesons. The field strength tensor of the photon, relevant for the coupling to the magnetic moments and the contact interaction, enters as $l^2$. 

First we look at the one loop diagrams where the photon couples to the electric charge of the involved mesons. For a photon emission inside the loop from a light meson, Fig~\ref{fig:diagrams} EC(c), we find two light and one heavy propagators $l^{-5}$, the axial vector coupling with $l$ and the coupling of a photon to the electric charge of a light meson $l^2$ and so
\begin{equation}
 l^4 \frac{1}{l^5}l^3\sim l^2.
\end{equation}
The same process with a charged intermediate heavy meson, Fig~\ref{fig:diagrams} EC(d), gives
\begin{equation}
 l^5\frac{1}{l^4}l^2\sim l^2.
\end{equation}
All other diagrams that contribute to the same set give the same contribution as it is required by gauge invariance. The diagrams where the photon couples to the magnetic moment enter one order higher. Fig.~\ref{fig:diagrams}~MM(a) gives
\begin{equation}
 l^4\frac{1}{l^4}l^3\sim l^3.
\end{equation}
Finally we need to worry about the contact interactions. Being proportional to the photon field strength tensor they also contribute as $l^2$. This means that we do not have an enhancement of the loop diagrams compared to the contact term. The contact interaction may be seen as a term that incorporates physics that we have not considered with the loop diagrams so far. In particular we have not considered the $D_s\eta$ in the loops depicted in Fig.~\ref{fig:diagrams}. The contact terms also include contributions from other compact components, such as $c\bar s$, tetraquark. In other calculations the $D_s\eta$ channel gives a significant contribution. We will use the available experimental data to fix the contact interaction once we have calculated the one loop diagrams. 

The most relevant higher order diagrams for us include the exchange of an additional pion in the loop. These diagrams give additional factors of $(l/\Lambda_\chi)^2$ which means we can safely ignore higher order loops. The largest subleading contribution stems from the NLO term for the axial vector coupling. This is suppressed by one order $l/\Lambda_\chi$ and can be used as a theoretical uncertainty for the amplitudes. 

In summary we can conclude that we have contact interactions and loop diagrams at the same order. Ultimately this lowers our predictive power since only the loop diagrams are fully controlled. The diagrams where the photon couples to the magnetic moment of the heavy mesons are necessary for the decay $\dsone\to\dszero\gamma$ since they give the leading loop contribution here. We will also calculate them for the other processes to test the convergence of the chiral expansion.

%%%%%%%%%%%%%%%%%%%%%%%%%%%%%%%%%%%
%
%     Electric Charge
%
%%%%%%%%%%%%%%%%%%%%%%%%%%%%%%%%%%%

Having set up the theoretical framework we are now able to  calculate the matrix elements for the radiative transitions.
The diagrams in the first line of Fig.~\ref{fig:diagrams} show the full gauge invariant set of diagrams where the photon couples to the electric charge. These are obtained by gauging the kinetic terms and the axial vector coupling. 
In App.~\ref{app:amplitudes_radiative} we give the intermediate meson pairs for all possible transitions. Explicit calculation confirms that the two subsets ($D^0,K^+$ and $D^+,K^0$ for $D_{s0}^*$) are gauge invariant separately. However, these diagrams alone still have a remaining divergence. This is due to the mass renormalization mechanism we introduced previously. If we also include the diagrams shown in the last line of Fig.~\ref{fig:diagrams} all divergences vanish and we have a finite amplitude. 

%%%%%%%%%%%%%%%%%%%%%%%%%%%%%%%%%%%
%
%     Magnetic Moment
%
%%%%%%%%%%%%%%%%%%%%%%%%%%%%%%%%%%%

Finally we need to consider diagrams that arise from the coupling of a photon to the magnetic moment of the heavy mesons. We notice that these are the only ones that contribute to the transition $\dsone\to\dszero\gamma$. The Lagrangian for the   coupling  $D^*\to D\gamma$ is given in Eq.~(\ref{LagrMagnMom}). The resulting diagrams are shown in the second line of Fig.~\ref{fig:diagrams}. One has to note here that due to the mass renormalization mechanism described before the diagrams where the photon couples to the magnetic moment of the outgoing $D_s^{(*)}$ are canceled exactly by the corresponding mass renormalization term. 

All amplitudes are given explicitly in App.~\ref{app:amplitudes_radiative}.
%%%%%%%%%%%%%%%%%%%%%%%%%%%%%%%%%%
%
\subsection{Results}
%
%%%%%%%%%%%%%%%%%%%%%%%%%%%%%%%%%%%
%
%     Ratios
%
%%%%%%%%%%%%%%%%%%%%%%%%%%%%%%%%%%%
The current experimental data is rather limited. For some decay widths only upper limits exist, some are not yet measured at all, see Tab.~\ref{tab:ratios}. This makes comparison to measurements somewhat complicated at the moment. The only available quantities are the following ratios:
\begin{eqnarray} 
 &&R_1:=\frac{\Gamma(\dszero \rightarrow D_s^*\gamma)}{\Gamma(\dszero\rightarrow D_s\pi^0)} \qquad
   R_2:=\frac{\Gamma(\dsone\rightarrow D_s\gamma)}{\Gamma(\dsone\rightarrow D_s^*\pi^0)}\non\\ \non
 &&R_3:=\frac{\Gamma(\dsone\rightarrow D^*_s\gamma)}{\Gamma(\dsone\rightarrow D_s^*\pi^0)}\qquad
   R_4:=\frac{\Gamma(\dsone\rightarrow \dszero\gamma)}{\Gamma(\dsone\rightarrow D_s^*\pi^0)}\\ \non
 &&R_5:=\frac{\Gamma(\dsone\rightarrow D_s^*\pi^0)}{\Gamma(\dsone\rightarrow D_s^*\pi^0)+\Gamma(\dsone\rightarrow \dszero\gamma)}\\ \non
 &&R_6:=\frac{\Gamma(\dsone\rightarrow D_s\gamma)}{\Gamma(\dsone\rightarrow D_s^*\pi^0)+\Gamma(\dsone\rightarrow \dszero\gamma)}\\ \non
 &&R_7:=\frac{\Gamma(\dsone\rightarrow D^*_s\gamma)}{\Gamma(\dsone\rightarrow D_s^*\pi^0)+\Gamma(\dsone\rightarrow \dszero\gamma)}\\ 
 &&R_8:=\frac{\Gamma(\dsone\rightarrow \dszero\gamma)}{\Gamma(\dsone\rightarrow D_s^*\pi^0)+\Gamma(\dsone\rightarrow \dszero\gamma)}.
\end{eqnarray}

\begin{table}[t]%[htbp]
\centering
\caption{Results for all relevant decay channels from PDG \cite{Behringer:2012ab}}
\renewcommand{\arraystretch}{1.3}
\begin{tabular}{|c|c|c|c|c|c|}
\hline\hline
      & Our Result     & Exp            &       & Our Result    &     Exp          \\ \hline\hline
$R_1$ & $0.08\pm0.04$ 	&  $<0.059$      & $R_5$ & $0.98\pm0.01$ 	&  $0.93 \pm 0.09$ \\ \hline
$R_2$ & $0.38\pm0.23$ 	&  $0.38\pm0.05$ & $R_6$ & $0.37\pm0.22$	&  $0.35 \pm 0.04$ \\ \hline
$R_3$ & $0.38\pm0.22$	&  $<0.16$       & $R_7$ & $0.38\pm0.21$ 	&  $<0.24$         \\ \hline
$R_4$ & $0.019\pm0.010$	&  $<0.22$       & $R_8$ & $0.018\pm0.010$ 	&  $<0.25$         \\ \hline\hline
\end{tabular}
\label{tab:ratios}
\end{table}
In Tab.~\ref{tab:ratios} we show both our results and the experimental values. We have chosen the ratio $R_2$ to fix the remaining free parameter, the strength of the contact interaction. Note that we can relate the contact interactions for the three decays $\dszero\to D_s^*\gamma$, $\dsone\to D_s\gamma$ and $\dsone\to D_s^*\gamma$ but not $\dsone\to\dszero\gamma$. So we cannot make final statements on the ratios $R_4$ and $R_8$. However, although we do not have a contact term here our results are compatible with the upper limits. Within the theoretical uncertainty our results are compatible with all measured ratios or upper limits. The ratios $R_6$, $R_7$ and $R_8$ have almost the same values as $R_2$, $R_3$ and $R_4$ since in our calculation $\Gamma(\dsone\rightarrow D_s^*\pi^0)\gg\Gamma(\dsone\rightarrow \dszero\gamma)$. However, this seems to agree with the experimental values if we look at $R_2$ and $R_6$. 
%%%%%%%%%%%%%%%%%%%%%%%%%%%%%%%%%%%
%
%     Partial Widths
%
%%%%%%%%%%%%%%%%%%%%%%%%%%%%%%%%%%%
\begin{table}[htbp]
\centering
\caption{\label{tab:results_comparison}Results for the radiative decay widths: the first column gives our result with all uncertainties added in quadrature. The numbers in the second column are from a parity doubling model by Bardeen et al., the ones in the third column from light-cone sum rules by Colangelo et al.. The fourth column gives results by Lutz and Soyeur. They give two values where they have made reasonable estimations for their remaining free parameter.}
\renewcommand{\arraystretch}{1.3}
\begin{tabular}{|l|c|c|c|c|}
\hline\hline
Decay Channel               & Our Results      		& \cite{Bardeen:2003kt} &  \cite{Colangelo:2005hv}	& \cite{Lutz:2007sk}	\\ \hline\hline
$\dszero\to D_s^*\gamma$    & $(10.1\pm4.5)\kev$		& $1.74\kev$            & $4-6~\kev$			& $1.94(6.47) \kev$	\\ \hline 
$\dsone\to D_s\gamma$       & $(26.2\pm12.4)$ keV		& $5.08\kev$            & $19-29 \kev$			& $44.50 (45.14) \kev$ 	\\ \hline
$\dsone\to D^*_s\gamma$     & $(26.5\pm11.0)$ keV		& $4.66\kev$            & $0.6-1.1\kev $	 	& $21.8 ( 12.47)~\kev$	\\ \hline
$\dsone\to \dszero \gamma$  & $(1.3\pm0.5)$ keV		& $2.74\kev$            & $0.5-0.8\kev$		& $0.13 (0.59)\kev$	\\ \hline
$B_{s0}\to B_s^*\gamma$     & $(40.0\pm11.0) $ keV		& $58.3\kev$            & --				& --			\\ \hline
$B_{s1}\to B_s\gamma$       & $(3.0\pm0.9)$ keV		& $39.1\kev$            & --				& -- 			\\ \hline
$B_{s1}\to B_s^*\gamma$     & $(77.0\pm22.0)$ keV		& $56.9\kev$            & --				& --			\\ \hline
$B_{s1}\to B_{s0}\gamma$    & $(0.03\pm0.01)$ keV 		& $0.0061\kev$          & -- 				& --			\\ \hline\hline
\end{tabular}
\end{table}

In Tab.~\ref{tab:results_comparison} we give the explicit numbers for the individual radiative decays.
We cannot compare these values to experimental data since no numbers for these widths have been published. However, comparison to previous calculations is possible, we will compare ours to some of them. Bardeen et al. used a parity doubling model in Ref.~\cite{Bardeen:2003kt}, Colangelo et al. \cite{Colangelo:2005hv} used light-cone sum rules. Both have in common that they assume the $D_{sJ}$ to be conventional $c\bar s$ states. 
The other  approach we list  works in a  molecular picture similar to ours. In  Ref.~\cite{Lutz:2007sk} Lutz and Soyeur calculate all radiative decays for the charmed states using a Hadrogenesis Conjecture.
All their results are listed together with ours in Tab.~\ref{tab:results_comparison}.  

Another calculation in the molecular picture was done by Gamermann  et al. Ref.~\cite{Gamermann:2007bm}. They use an $SU(4)$ Lagrangian and give $0.475^{+0.831}_{-0.290}$ keV for $\dszero\to D_s^*\gamma$. Model calculations by Faessler et al.~\cite{Faessler:2007gv,Faessler:2007us} give $0.55-1.41\kev$ for $\Gamma(\dszero\to D_s\gamma)$  and $2.37-3.73\kev$ for $\Gamma(\dsone\to D_s^*\gamma)$.

The results by Lutz and Soyeur are the closest to our results, the other calculations give generally smaller numbers. However, the differences are not huge. The range is from $10\kev$ for our result for $\dszero\to D^*_s\gamma$ to the smaller values of about $1-2\kev$. In other words, the calculations, performed in different frameworks, all end up in the same ballpark in the single $\kev$ area. 
In fact, all calculations are compatible with naive estimations for this decay to be of the order $\mathcal O(\alpha)$. This is different of course for the isospin violating decays. Here naively the contact interaction is proportional to $m_u-m_d$ and thus heavily suppressed while our calculations give results driven by the loops that are of the order $100\kev$. Calculations performed for compact $c\bar s$ states predict significantly smaller values here.

In our calculation we find that the decays in the bottom flavor sector have significantly broader widths. This can be understood by looking at the coupling constants in Eq.~(\ref{eq:couplings}). In the $B-$sector the couplings are larger by roughly a factor 3 which results in one order of magnitude for the width. In a model calculation by Faessler et al. in Ref.~\cite{Faessler:2008vc} the author finds results for the radiative decays of the $BK$ molecules in the order of $2\kev$. Bardeen et al. predict them in a parity doubling model to be $\Gamma(\bszero\to B_s^*\gamma)=58.3\kev$, $\Gamma(\bsone\to B_s\gamma)=39.1\kev$, $\Gamma(\bsone\to B_s^*\gamma)=56.9\kev$ and $\Gamma(\bsone\to \bszero\gamma)=0.0061\kev$. We see that our results agree better with the parity doubling model. 

% Colangelo:2005hv

In summary we can state that radiative decays are not the proper channel to pin down the nature of $\dszero$ and $\dsone$ since the results of the different models are too similar. The isospin violating decays $\dszero\to D_s\pi^0$ and $\dsone\to D_s^*\pi^0$ are more promising channels. 

\begin{table}[htbp]
\centering
\caption{The results for all relevant channels are given. The uncertainties stem from the chiral expansion, the experimental uncertainty of the couplings $g_\pi$, the uncertainty of the coupling of the molecules to their constituents and the strength of the contact interaction (used only in the first three channels)}
\renewcommand{\arraystretch}{1.3}
\begin{tabular}{|l|c|}
\hline\hline
Decay Channel               & Partial Width                         \\ \hline\hline
$\dszero\to D_s^*\gamma$    & $(10.1\pm3.3\pm0.9\pm0.4\pm3.0)\kev$ \\ \hline 
$\dsone\to D_s\gamma$       & $(26.2\pm8.5\pm1.8\pm1.0\pm8.8)$ keV \\ \hline
$\dsone\to D^*_s\gamma$     & $(26.5\pm8.6\pm3.4\pm1.9\pm5.6)$ keV  \\ \hline
$\dsone\to \dszero \gamma$  & $(1.3\pm0.4\pm0\pm0.2)$ keV      \\ \hline 
$B_{s0}\to B_s^*\gamma$     & $(40.4\pm22.2\pm9.3\pm1.8) $ keV     \\ \hline
$B_{s1}\to B_s\gamma$       & $(3.0\pm1.6\pm0.7\pm0.2)$ keV     \\ \hline
$B_{s1}\to B_s^*\gamma$     & $(77.1\pm41.8\pm17.7\pm5.4)$ keV      \\ \hline
$B_{s1}\to B_{s0}\gamma$    & $(0.03\pm0.01\pm0.01\pm0.0)$ keV  \\ \hline\hline
\end{tabular}
\label{tab:results}
\end{table}
When we analyze the theoretical uncertainties we see that for the first three channels the uncertainty for the contact interaction is the largest. This one again is driven mainly by the large uncertainty of the hadronic width for $\dsone$.
The second largest impact has the axialvector coupling $g_\pi$ which is determined from the pionic decay of $D^*$. Here improved experimental data on the width of $D^*$ would be helpful.

%%%%%%%%%%%%%%%%%%%%%%%%%%%%%%%%%%%
%
%     Single Contributions
%
%%%%%%%%%%%%%%%%%%%%%%%%%%%%%%%%%%%

\begin{table}[htbp]
\centering
% \captionsetup{width=0.9\textwidth}
\caption{In this table we show the widths calculated only from the coupling to the electric charge (EC), the magnetic moments (MM) and the contact term (CT) compared to the sum of all three.}
\renewcommand{\arraystretch}{1.3}
\begin{tabular}{|l|c|c|c|c|}
\hline\hline
                   Decay Channel &   EC   	&    MM		&    CT		&   Sum     \\ \hline\hline
$\dszero\rightarrow D_s^*\gamma$ & $2.0 $ keV	& $0.03$ keV	& $3.76$ keV	& $10.1$ keV \\ \hline
$\dsone\to D_s\gamma$            & $4.2$ keV	& $0.23$ keV	& $12.7$ keV	& $26.2$ keV  \\ \hline
$\dsone\to D^*_s\gamma$          & $9.4$ keV	& $0.45$ keV	& $11.6$ keV	& $26.5$ keV  \\ \hline
$\dsone\to \dszero \gamma$       & --		& $1.3$ keV	&     -- 	& $1.3$ keV   \\ \hline\hline
\end{tabular}
\label{tab:results_2}
\end{table}
Our amplitudes consist of three different contributions. For the decays via $D^{(*)}K$ loops we considered the coupling of the photon to the electric charges as well as  to the magnetic moment of the heavy mesons. In addition we need to consider the contact interaction. In Tab.~\ref{tab:results_2} we show the central values for the width  calculated using one of the three contributions exclusively. We can see that with having the contact interaction fixed so that the ratio $R_2$ matches the experimental value, the largest contribution comes from the loop diagrams where the photon couples to the electric charges. The coupling of photons to magnetic moments gives only tiny contributions. Therefore we have a good convergence in the chiral expansion. 

The fact that the transition $\dsone\to\dszero\gamma$, despite being only from this coupling, has a similar width is easy to understand. While the other diagrams are proportional to the axial vector coupling $g_\pi$ this decay is proportional to the product of the resonance couplings $g_{DK}g_{D^*K}$.

%%%%%%%%%%%%%%%%%%%%%%%%%%%%%%%%%%%%%%%%%%%%%%%%%%%%%%%%%%%%%%%%%%%%%%
%
\section{Light Quark Mass dependence of $\dszero$ and $\dsone$}\label{sec:quarkmassdependence}
%
%%%%%%%%%%%%%%%%%%%%%%%%%%%%%%%%%%%%%%%%%%%%%%%%%%%%%%%%%%%%%%%%%%%%%%
To test the nature of the resonances,  we can
also compare our results to lattice calculations (for a corresponding study
in the light meson sector see Ref.~\cite{Hanhart:2008mx}). To do so, we extend
the calculations from the physical world to unphysical quark masses which are
frequently used in lattice calculations.

Varying the light quark masses is equivalent to varying the pion mass.
Although the physical strange quark mass is nowadays routinely used in lattice
calculations, we emphasize that by varying the strange quark mass, or
equivalently varying the kaon mass, one can learn a lot about the nature of
some hadrons, as will be discussed below. Therefore we have to express all results in
terms of pion and kaon masses, respectively.
In the following chapter only the charmed mesons are discussed explicitly,
however, analogous arguments apply to their bottom partners as well.

In order to proceed we need to assume that the subtraction constant $a(\mu)$
does not depend on the light quark masses. We stress, however, that even
allowing for a quark mass dependence of $a(\mu)$ would not change the general
features of the results, but might slightly enhance the uncertainties.

%%%%%%%%%%%%%%%%%%%%%%%%%%%%%%%%%%%
%
\subsection{Pion mass dependence}
%
%%%%%%%%%%%%%%%%%%%%%%%%%%%%%%%%%%%

\begin{figure}[hbtp]
\centering
\begin{minipage}{0.4\linewidth}
% \centering
\includegraphics[width=1\linewidth]{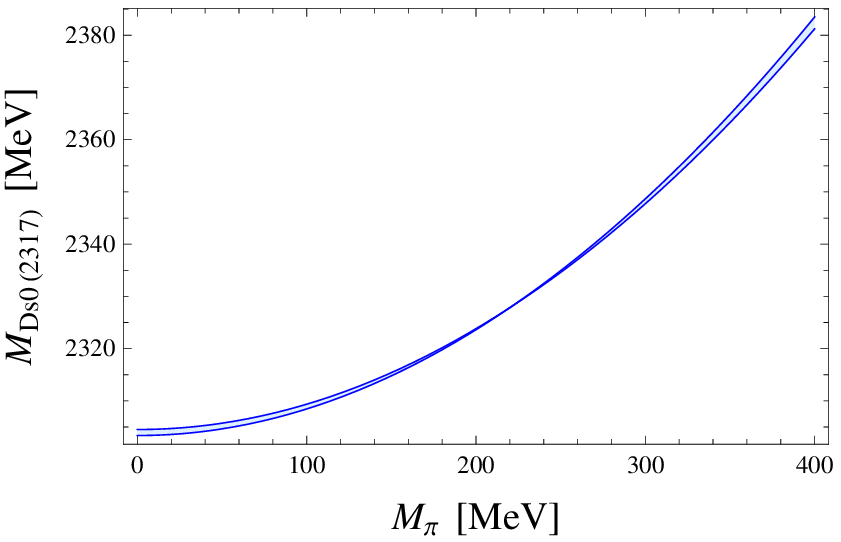}
\end{minipage}
\hspace{1cm}
\begin{minipage}{0.4\linewidth}
% \centering
\includegraphics[width=1\linewidth]{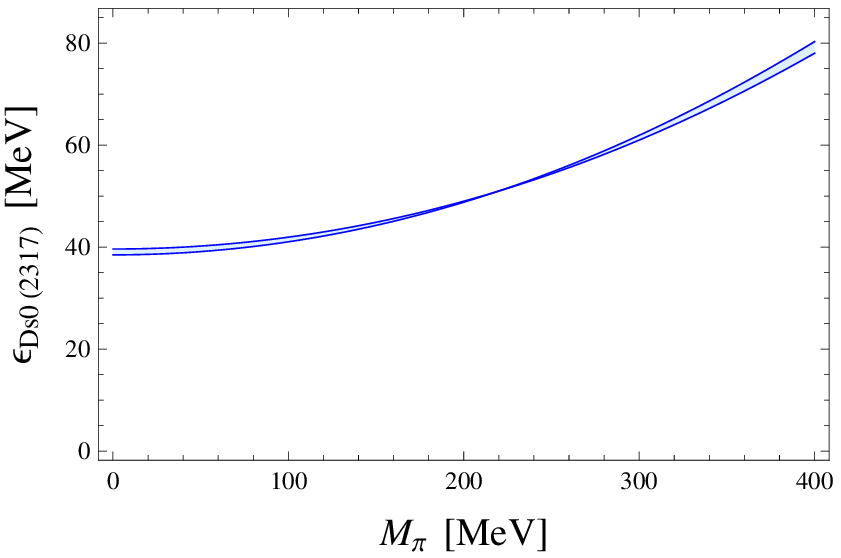}
\end{minipage}

\vspace{1cm}

\begin{minipage}{0.4\linewidth}
% \centering
\includegraphics[width=1\linewidth]{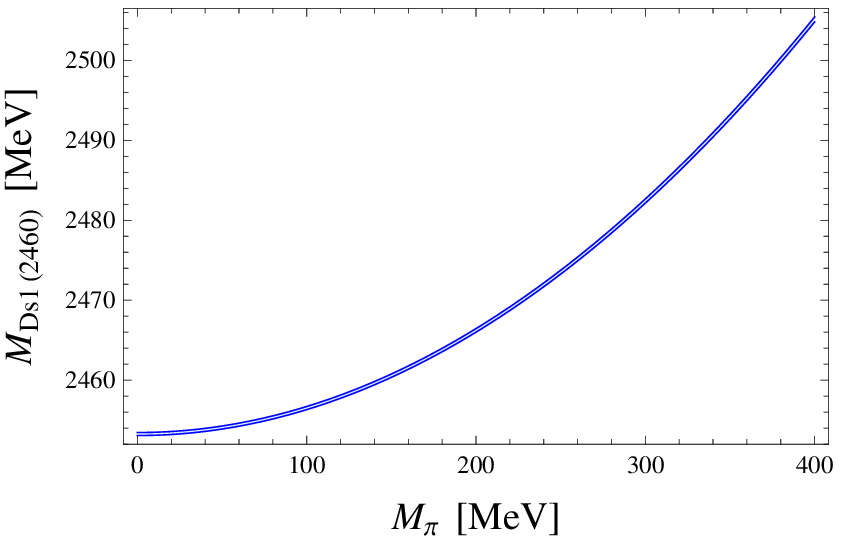}
\end{minipage}
\hspace{1cm}
\begin{minipage}{0.4\linewidth}
% \centering
\includegraphics[width=1\linewidth]{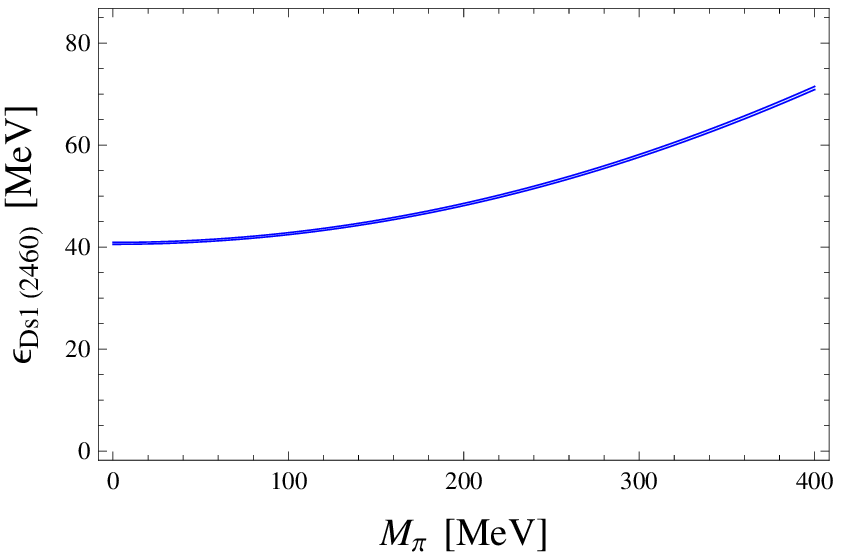}
\end{minipage}
\caption{\label{fig:PlotsPion} The masses and binding energies of the $D^*_{s0}(2317)$ and the $D_{s1}(2460)$
as a function of the pion mass.} 
\end{figure}

Lattice QCD calculations are often performed at larger quark masses than
realized in nature --- for the $KD$ system of interest here exploratory
lattice
studies are presented in Ref.~\cite{Liu:2008rz}.
In addition, as we will argue here, the quark mass dependence of a state
contains important information on its nature. Varying $u$ and $d$
quark masses can be expressed by varying the pion mass. In \cite{Jenkins:1992hx},
masses of the charmed mesons  were expanded up to one loop order in the chiral expansion.
Nevertheless for our purposes the expansion up to $\mathscr O(M_\pi^2)$ is
sufficient.

Using the Lagrangian given in Eq.~(\ref{eq:charmmasses}) we find the NLO correction
to the charmed meson masses to be
\begin{eqnarray}
 \delta M_{D^{(*)}}^2=4 h_1 B \hat m \ , \qquad \delta M_{D^{(*)}_s}^2=4 h_1 B
 m_s \ ,
\end{eqnarray}
with $\hat m = (m_u+m_d)/2$ the average light quark mass.
When studying the pion mass dependence, we consider the physical value for the
strange quark mass here, thus we can use the physical mass for the
$D_s^{(*)}$. Using $M_\pi^2=2B\hat m$ yields, up to $\mathscr
O(M_\pi^2)$,
\begin{eqnarray}
M_{D^{(*)}} = \overset{_\circ}{M}_{D^{(*)}} + h_1\frac{M_\pi^2}{\overset{_\circ}{M}_{D^{(*)}}},
\end{eqnarray}
where $\overset{_\circ}{M}_{D^{(*)}}$ is the charmed meson mass in the SU(2)
chiral limit $m_u=m_d=0$ with $m_s$ fixed at its physical value. For the kaon
and the eta mass we can find similar expressions by using ${M_K^2=B(\hat
m+m_s)}$, ${\overset{_\circ}{M}_K^2=Bm_s}$ and ${M_\eta^2=B\frac{2}{3}(\hat
m+2m_s)}$, ${\overset{_\circ}{M}_\eta^2=B\frac{4}{3}m_s}$ from the LO chiral
expansion:
\begin{eqnarray}
M_K = \overset{_\circ}{M}_K + \frac{M_\pi^2}{4\overset{_\circ}{M}_K}, \qquad
M_\eta = \overset{_\circ}{M}_\eta + \frac{M_\pi^2}{6\overset{_\circ}{M}_\eta}.
\end{eqnarray}

In Fig.~\ref{fig:PlotsPion} we show the mass of both resonances,
$D_{s0}^*(2317)$ and $D_{s1}(2460)$, as well as their binding energies as a
function of the pion mass. Note that the observed rather strong pion mass
dependence is specific for a molecular state. The corresponding dependence for a
quark state should be a lot weaker. To understand this, one notices that a pure
$c\bar s$ state does not contain any $u,~\bar u,~d,~\bar d$ quarks. 
The leading
term containing these light quarks is $1/N_c$ suppressed.  Thus, for the
quark state, the pion mass dependence can only enter via $D^{(*)}K$ loops ---
as in case of the molecular state. However, their contribution should be a lot
smaller for the quark state than for the molecular state. To see this, observe
that the loop contributions from a particular meson pair  is
proportional to the squared coupling of that meson pair to the re\-so\-nance. As
shown in Refs.~\cite{Weinberg:1965zz,Baru:2003qq} and Sec.~\ref{sec:molecules}, this coupling is
proportional to the probability to find the molecular state in the physical
state.  Thus, the pion mass dependence for $D_{s0}^*(2317)$ and $D_{s1}(2460)$
should be maximal if both are pure molecules. This case is depicted
in Fig.~\ref{fig:PlotsPion}. On the other hand it should be very weak for the
admittedly unrealistic scenario of a pure $c\bar s$ state.  Note that the
mentioned relation between the coupling and the structure of the state holds
only in leading order in an expansion in $\sqrt{M_K\epsilon}/\beta$ (see
Ref.~\cite{Hanhart:2007wa} for a detailed discussion), with the inverse range of
forces $\beta\sim m_\rho$. For the scalar and axial
vector open charm mesons this gives an uncertainty of the order of 20\%.
As a result of the larger binding, for the bottom analogs we even find 30\%.
In addition in the analysis we neglected terms of ${\cal{O}}(1/N_c)$. 
Thus, from this kind of analysis at most statements like `the state is
predominantly molecular/compact' are possible.

Furthermore we notice that the plots show an almost identical behavior in the
scalar and the axial-vector channel. So the spin symmetry breaking effects are
only very weak here.  We see that the binding energy in both cases varies from
about 40 MeV to about 80 MeV.

After our publication on this subject Mohler et al. calculated the $S$-wave $DK$ scattering  length in dependence of the pion mass on the lattice and found a linear relation~\cite{Mohler:2013rwa}. Using Eq.~(\ref{eq:scatteringlength}) we find that this linear dependence for the scattering length translates into a quadratic dependence of the binding energy just as we have predicted.

%%%%%%%%%%%%%%%%%%%%%%%%%%%%%%%%%%%
%
\subsection{Kaon mass dependence}
%
%%%%%%%%%%%%%%%%%%%%%%%%%%%%%%%%%%%

Before going into details of the calculations, let us make some general
statements about  the $M_K$--dependence of the mass of a bound state
of a kaon and some other hadron. The mass of such a kaonic bound state
is given by
\begin{equation}
M = M_K+M_h-\epsilon,
\end{equation}
where $M_h$ is the mass of the other hadron, and $\epsilon$ denotes the binding
energy. Although both $M_h$
and $\epsilon$ have some kaon mass dependence, it is expected to be a lot
weaker
than that of the kaon itself.
Thus, the important implication of this simple formula is that the leading
kaon mass dependence of a kaon--hadron bound state is {\em linear, and the slope is unity}. The only
exception to this argument is if the other hadron
is also a kaon or anti-kaon.~\footnote{The $f_0(980)$ was proposed to be such a
$K\bar K$ bound state~\cite{Weinstein:1990gu,Baru:2003qq}.} In this case, the
leading kaon mass dependence is still linear but with the slope being changed to two.
Hence, as for the $DK$ and $D^*K$ bound states discussed here, we expect that
their masses are linear in the kaon mass, and the slope is approximately one. As we
will see, our explicit calculations confirm this expectation.

\begin{figure}
\centering
\begin{minipage}{0.39\linewidth}
% \centering
\includegraphics[width=1\linewidth]{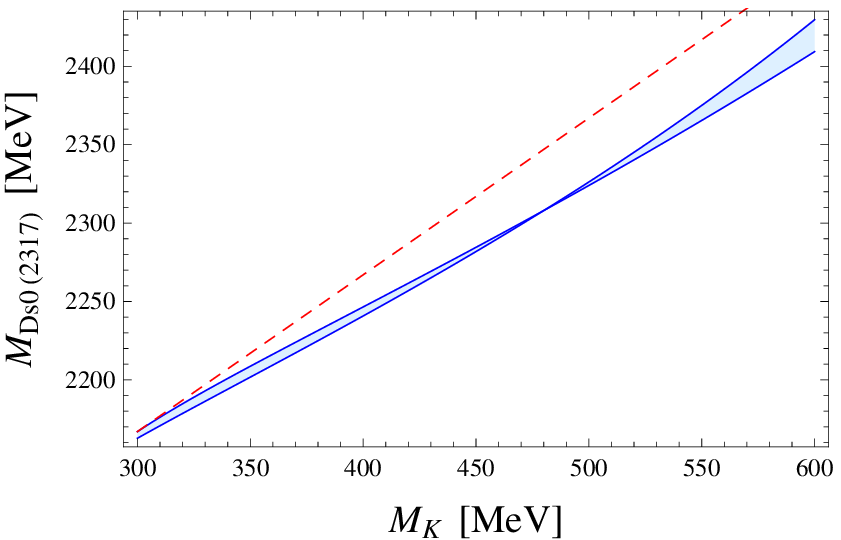}
\end{minipage}
\hspace{1cm}
\begin{minipage}{0.39\linewidth}
% \centering
\includegraphics[width=1\linewidth]{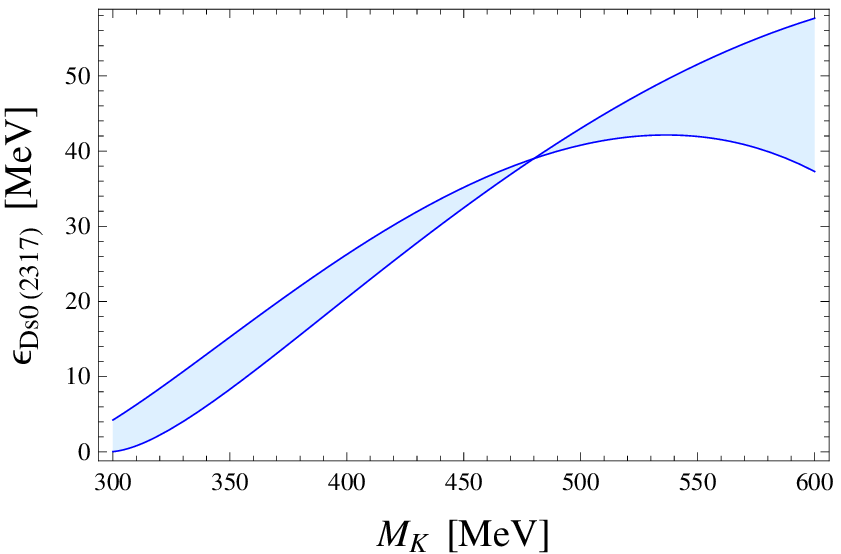}
\end{minipage}

\vspace{1cm}

\begin{minipage}{0.39\linewidth}
% \centering
\includegraphics[width=1\linewidth]{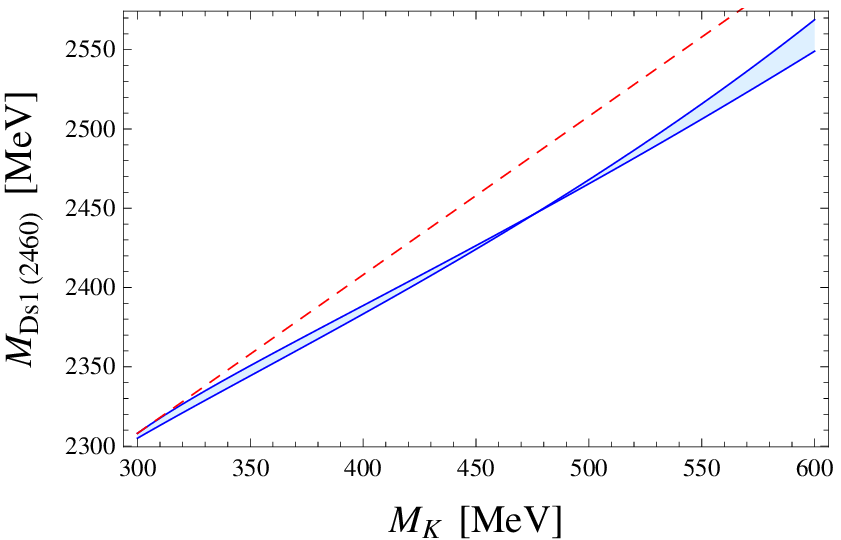}
\end{minipage}
\hspace{1cm}
\begin{minipage}{0.39\linewidth}
% \centering
\includegraphics[width=1\linewidth]{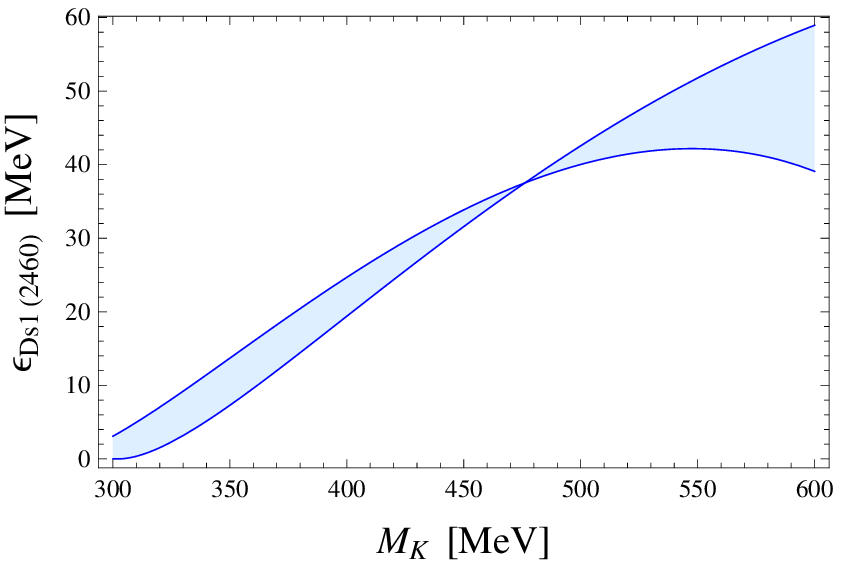}
\end{minipage}
\caption{The blue band shows the masses and binding energies of
  $D^*_{s0}(2317)$ and $D_{s1}(2460)$ in dependence of the kaon mass, the red
  dashed line shows the threshold.}
\label{fig:PlotsKaon}
\end{figure}

For calculating the kaon mass dependence, we use physical masses for the pion
and the charmed mesons without strangeness.~\footnote{Certainly, lattice
simulations use unphysical values for up and down quarks. However, the
conclusion of this subsection will not be affected by the value of pion mass.
For definiteness, we choose to use physical masses for the pion and non-strange
charmed mesons.} To express the strange quark mass dependence in terms
of kaon masses, we write
\begin{eqnarray}
 M_K^2=B\left(m_s+\hat m\right)~, \quad
\overset{_\circ}{M}_K^2=B\hat m= \frac{1}{2}M_\pi^2~,
\end{eqnarray}
where $\overset{_\circ}{M}_K^2$ is the mass of the kaon in the limit $m_s=0$.
The charmed strange meson mass is
\begin{eqnarray}
M_{D^{(*)}_s} = \overset{_\circ}{M}_{D_s^{(*)}} + 2 h_1\frac{M_K^2}{\overset{_\circ}{M}_{D_s^{(*)}}}-h_1\frac{M_\pi^2}{\overset{_\circ}{M}_{D_s^{(*)}}},
\label{mqdepofD}
\end{eqnarray}
where $\overset{_\circ}{M}_{D_s^{(*)}}$ is the charmed strange meson mass in the
limit $m_s=0$. Finally the eta mass is given to this order by the
Gell-Mann--Okubo relation
\begin{eqnarray}
 M_\eta^2=B\left(\frac 43 m_s+\hat m\right)=\frac 43 M_K^2-\frac 13 M_\pi^2.
\end{eqnarray}

For analyzing the kaon mass dependence we calculate the masses of the $D_{s0}^*(2317)$ and
$D_{s1}(2460)$ and their binding energies as before. The corresponding results
are shown in Fig.~\ref{fig:PlotsKaon}. Furthermore we also show the $D^{(*)}K$
threshold in dependence on the kaon mass.

For the $D_{s0}^*(2317)$, we find that the kaon mass dependence of the mass is
almost perfectly linear, especially in the region $M_K=300 - 500~{\rm MeV}$. For
higher strange quark masses the chiral expansion  is no longer valid and the
results become less reliable.

The kaon mass dependence of the mass and the $DK$ threshold have a similar
slope. The slope of the threshold is one, since the $D$-meson mass is
independent of $M_K$ to the order we are working, and the slope of 
$M_{D^*_{s0}(2317)}$ is about 0.85. This finding is perfectly consistent with
the expectation formulated in the beginning of this subsection. The deviation
of the slope from one can be understood from the kaon mass dependence of the
binding energy (see right column of Fig.~\ref{fig:PlotsKaon}) and from the
effects of the coupling to the $D_s\eta$ channel. The situation for the
$D_{s1}(2460)$ is similar.

On the contrary, if $D_{s0}^*(2317)$ and $D_{s1}(2460)$ are assumed to be quark
states, their masses necessarily depend quadratically on the kaon mass ---
analogous to Eq.~(\ref{mqdepofD}). Thus, an extraction of the kaon mass
dependence of the masses of $D_{s0}^*(2317)$ and $D_{s1}(2460)$ from lattice
data is of high interest to pin down the nature of these states.

\section{Summary}
In this chapter we have studied various properties of the open charm states $\dszero$ and $\dsone$. The fact that both states are  located equally far below the corresponding thresholds suggests that they are $DK$ and $D^*K$ bound states rather than compact $c\bar s$ states. We have presented consequences of this hypothesis as well as ways to test it. When it comes to the decays we state that the isospin violating decays with a $\pi^0$ in the final state is the best channel to pin down the nature of $\dszero$ and $\dsone$: we predict the decay widths to be $\Gamma(\dszero\to D_s\pi)=(133\pm22)\kev$ and $\Gamma(\dsone\to D_s^*\pi)=(69\pm22)\kev$. These are about two orders of magnitude larger than values obtained for compact $c\bar s$ states. The decay widths for the radiative decays are predicted to be one order of magnitude smaller. However, the numbers we give are compatible with the ones calculated in different frameworks. As a result, these will not 
be the quantities that can decide whether $\dszero$ and $\dsone$ are hadronic molecules rather than compact states. The PANDA experiment at the FAIR collider is expected to reach the necessary statistic and detector resolution down to $100\kev$. This will hopefully be sufficient to make a final judgment. 

Besides experiments also lattice data can provide important information about the true nature of $\dszero$ and $\dsone$. We presented calculations of the light quark mass dependence of the dynamically generated molecular states. These can be calculated in lattice calculations. In our calculation we investigate the pion and kaon mass dependence of mass and binding energy of $\dszero$ and $\dsone$. Especially the kaon mass dependence is interesting: while a compact $c\bar s$ state has a quadratic dependence on the kaon (or strange quark) mass we find that the molecular state has a linear dependence with a slope about unity. A calculation of this dependence by the lattice community is strongly recommended. 

Finally we can use heavy quark flavor symmetry to make predictions for similar molecular states with open bottom. In the same way heavy quark spin symmetry predicts $\dsone$ to be the spin partner of $\dszero$ we predict their flavor partners to be found at the masses $M_{\bszero}=(5663\pm48)\mev$ and $M_{B_{s1}}=(5712\pm48)\mev$. The discovery of these flavor partners is a crucial test for our theory.

%% file: Zb.tex
%%%%%%%%%%%%%%%%%%%%%%%%%%%%%%%%%%%%%%%%%%%%
%
\chapter{$Z_b(10610)$ and $Z_b(10650)$}\label{ch:zb}
%
%%%%%%%%%%%%%%%%%%%%%%%%%%%%%%%%%%%%%%%%%%%%

The second half of this work is dedicated to the new bottomonium states, $Z_b(10610)$ and $Z_b(106510)$, discovered by the  Belle collaboration in 2011~\cite{Belle:2011aa}. 
Both states were discovered in the invariant mass spectra for the decays of $\Upsilon(5S)$ into $\Upsilon(mS)\pi^+\pi^-$ and $h_b(nP)\pi^+\pi^-$ with $m=1,2,3$ and $n=1,2$. 
Further experimental analysis suggests that part of the pionic decays occur in cascades and thus the intermediate state needs to be charged: 
\begin{equation}
 \Upsilon(5S)\to Z_b^+\pi^-\to \Upsilon(mS)\pi^+\pi^- \quad {\rm and} \quad \Upsilon(5S)\to Z_b^+\pi^-\to h_b(nP)\pi^+\pi^-.
\end{equation}
Being charged bottomonium states the $Z_b$ states are necessarily formed of at least four valence quarks. This makes them manifestly exotic  compared to the $D_{sJ}$ states which despite some experimental facts could still be conventional $c\bar s$ states. 

\begin{figure}[t]
\begin{center}
  \includegraphics[width=0.31\textwidth]{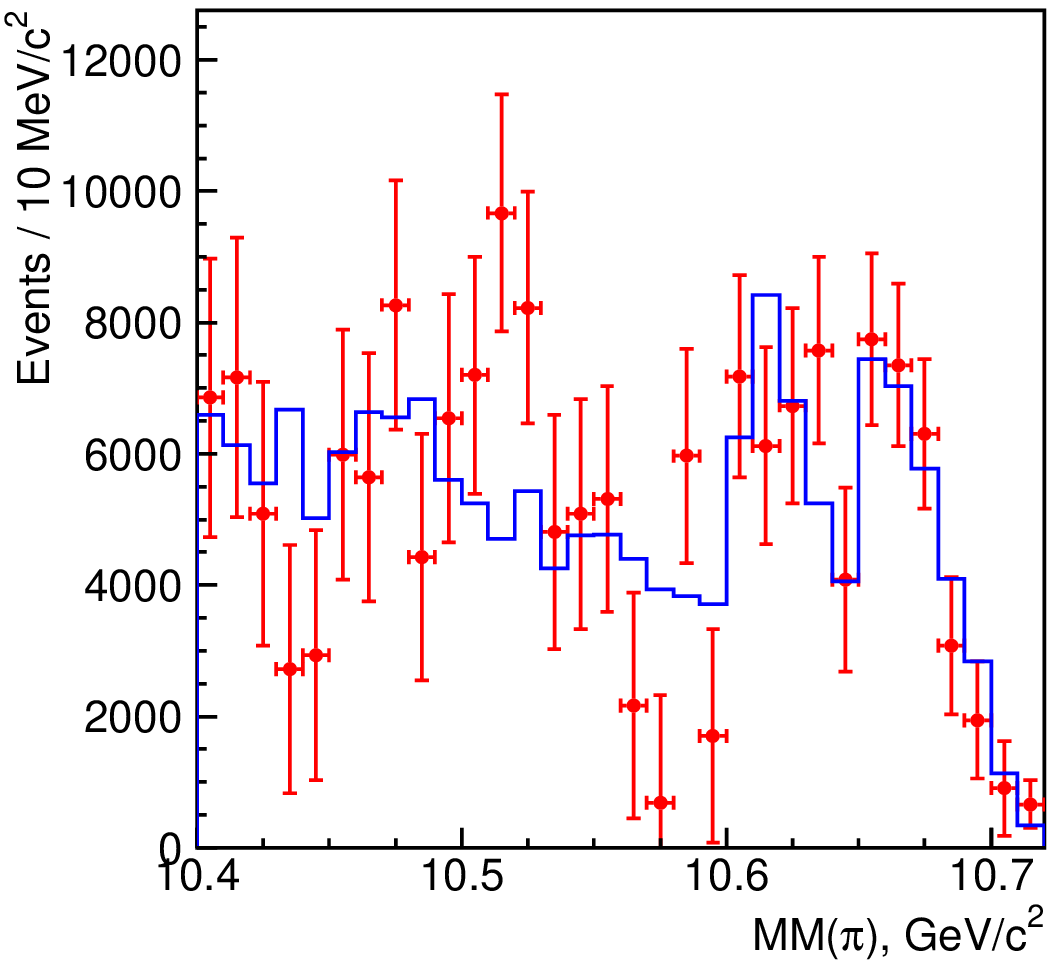}\hspace{.3cm}
  \includegraphics[width=0.31\textwidth]{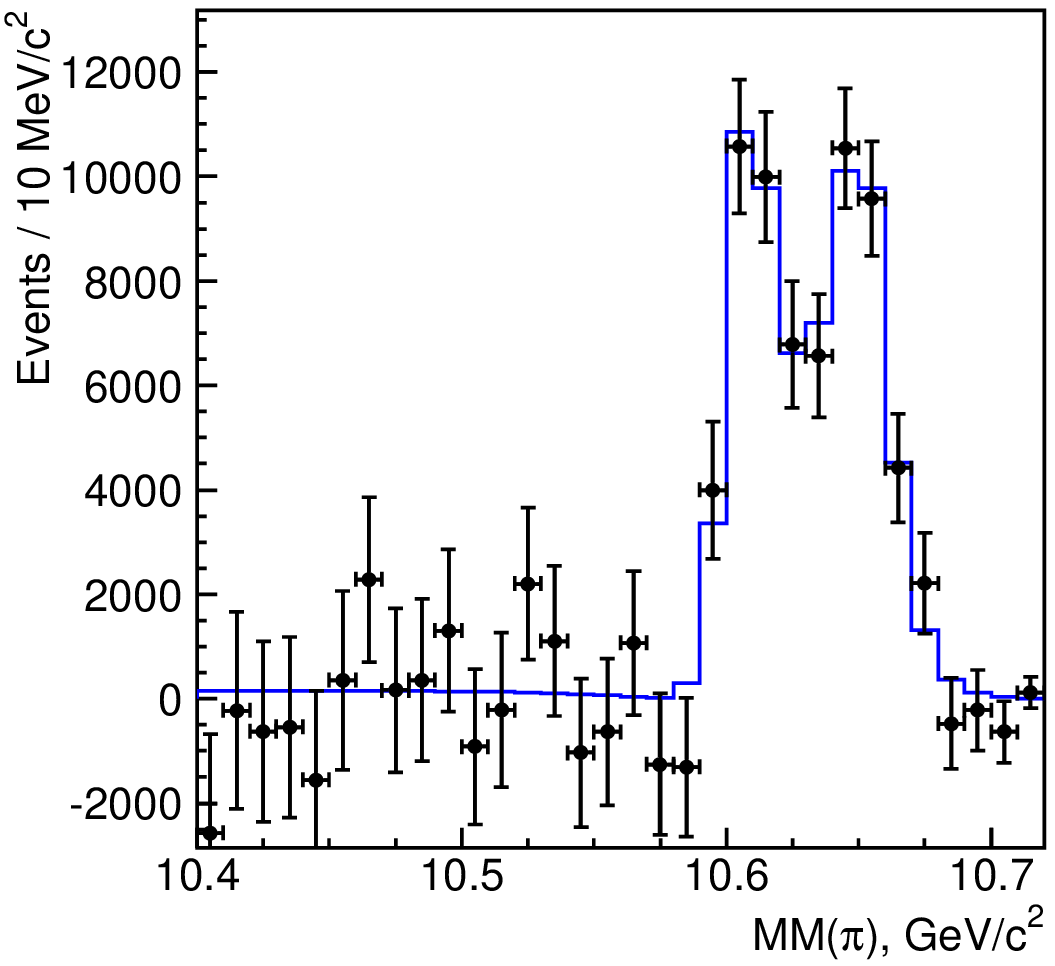}\hspace{.3cm}
  \includegraphics[width=0.31\textwidth]{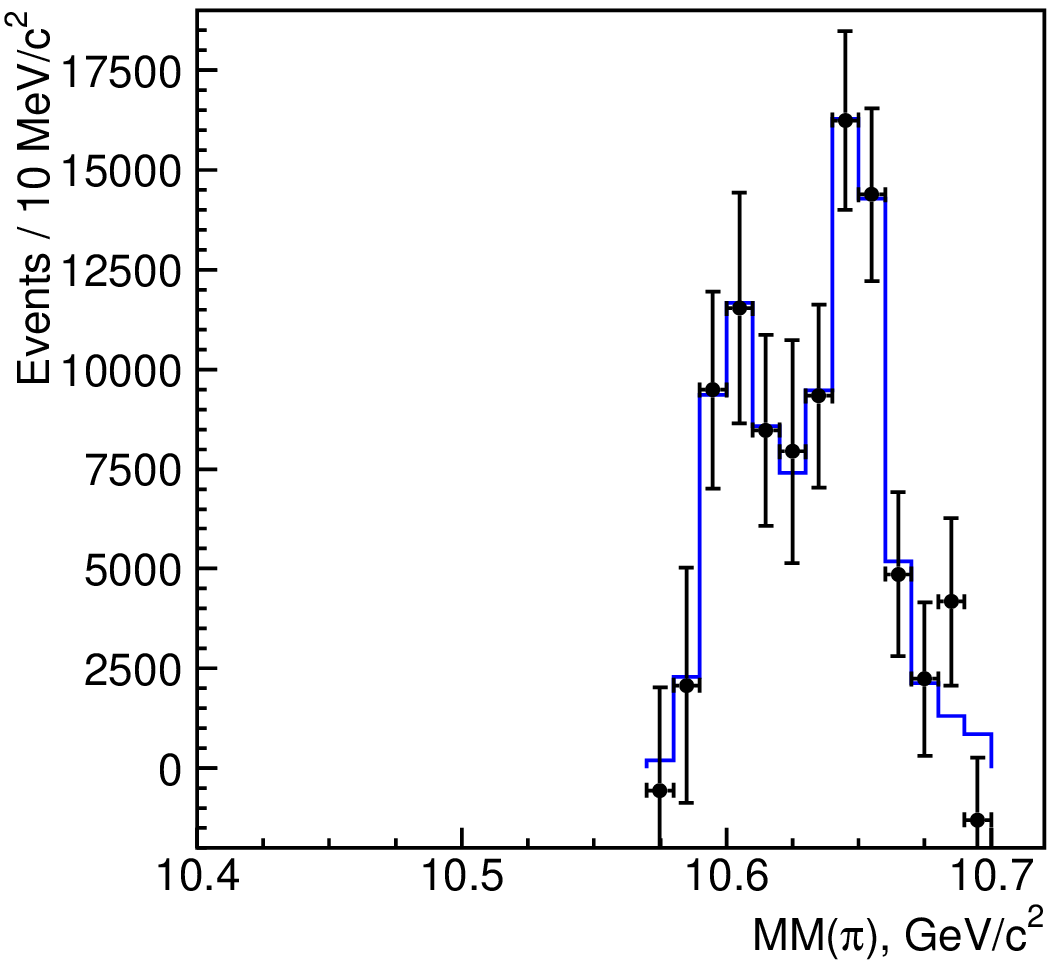}
  \caption{Invariant mass plots for the final states $\Upsilon(2S)\pi$, $h_b(1P)\pi$ and $h_b(1P)\pi$. The figure is taken from  \cite{Belle:2011aa}. \label{fig:zbexp}}
\end{center}
\end{figure}

The second peculiar fact about the $Z_b$ states concerns the discovery channels. If we look the the invariant mass plots for the final states $\Upsilon(2S)\pi^+\pi^-$, $h_b(1P)\pi^+\pi^-$ and $h_b(2P)\pi^+\pi^-$ in Fig.~\ref{fig:zbexp} we see that the number of events are of the same order. Even if we consider different efficiency, background, etc. it seems that the rates in all final states are comparable.
This is at first glance surprising: The $\Upsilon$ states are spin triplets since $S_{b\bar b}=1$ while the $h_b$ has $S_{b\bar b}=0$.  Therefore a heavy quark spin flip is necessary which should be suppressed  by $\Lambda_{\rm QCD}/m_b$ according to  Heavy Quark Effective Theory. A possible solution was first pointed out by Bondar et al. in Ref.~\cite{Bondar:2011ev}. Hadronic molecules of $B^*\bar B+B\bar B^*$ (we will use $B\bar B^*$ here from now on for brevity) and $B^*\bar B^*$, respectively, contain both possible  spin configurations for the heavy quarks at the same time:
\begin{equation}
 |Z_b\rangle  = \frac1{\sqrt{2}}\left( 0^-_{b\bar b}\otimes1^-_{\bar qq} + 1^-_{b\bar b}\otimes0^-_{\bar qq}\right), \quad
 |Z_b'\rangle = \frac1{\sqrt{2}}\left( 0^-_{b\bar b}\otimes1^-_{\bar qq} - 1^-_{b\bar b}\otimes0^-_{\bar qq}\right).
\end{equation}
Therefore an intermediate $Z_b$ state formed as a hadronic molecule can drive the decay into final states with $P$-wave bottomonia. This assumption is in agreement with the fact that the experimental analysis favors $I^G(J^P)=1^+(1^+)$ for both states, compatible with a molecular interpretation. 

Similar to the situation for the open charm states discussed in the previous chapter the mass difference between the states reflects the mass difference between pseudoscalar and vector $B$-mesons:
\begin{equation}
 M_{Z'_b}-M_{Z_b}\simeq M_{B^*}-M_B.
\end{equation}
In a subsequent publication the Belle collaboration  provides  additional data on the $Z_b$ states~\cite{Adachi:2011ji}.
Since the so far discussed charged $Z_b$ states are isospin triplet states an electromagnetically neutral state with $I_3=0$ is to be expected. In their second paper on the subject the Belle collaboration reported a candidate for such a state at $M_{Z_b^0}=10609^{+8}_{-6} \pm6\mev$.

Furthermore they find that  $Z_b$ and $Z'_b$ predominantly decay into $B^*\bar B$ and $B^*\bar B^*$, respectively. Especially the fact that the heavier $Z'_b$ decays almost exclusively into what we believe to be its constituents, $B^*\bar B^*$, and is not seen in $B^*\bar B$ despite the larger phase space strongly supports the molecular picture. 

In summary, the evidence that the newly discovered $Z_b$ states are hadronic molecules formed from $B$-meson interactions is quite compelling. Naturally this lead to a large number of publications on this subject, see Refs.~\cite{Nieves:2011zz,Zhang:2011jja,Yang:2011rp,Sun:2011uh,Ohkoda:2011vj,Ke:2012gm}. However, it seems possible to describe the most important features with the $Z_b$ states being $\bar b\bar q bq$ tetraquarks, see Refs.~\cite{Li:2012wf,Guo:2011gu,Ali:2011ug}. Thus one of the most important tasks of this work is to make model independent predictions that can prove or falsify the molecular picture. 

In this chapter we will investigate the properties of the $Z_b$ states under the assumption that they are hadronic molecules formed of $B$-mesons and how the dynamics of the $B$-mesons affect their behavior. Within this framework we focus mainly on two subjects: the location of the poles in the $S$-matrix and two-body decays, published in Refs.~\cite{Cleven:2011gp} and \cite{Cleven:2013sq}, respectively. 

We will start by introducing the Nonrelativistic Effective Theory (NREFT) Lagrangians used in this chapter in Sec.~\ref{sec:lagrangians_nreft}. In Sec.~\ref{sec:loc_sing} the location of the poles in the complex plane and their impact on the lineshapes is investigated. We will show that it is possible to obtain agreement with experimental results assuming that the states are located below the corresponding thresholds and therefore could be bound states rather than resonances. By now the updated experimental data by Belle \cite{Adachi:2012cx} indicates that both $Z_b$ and $Z'_b$ predominantly decay into their constituents and are therefore to be seen as resonances. By the time our work was published in 2011 \cite{Cleven:2011gp} such data was not available. However, a position below threshold is still not ruled out.

Sec.~\ref{sec:zbdecays} contains the results of \cite{Cleven:2013sq}. Incorporating the latest experimental results, including the decay rates of the charged $Z_b$ and the discovery of the neutral isospin partner of $Z_b$, we calculate radiative and hadronic transitions to conventional bottomonia. These include $\zbs\to h_b\pi$, $\zbs\to\Upsilon\pi$ and  $\zbs\to\chi_b\gamma$. The latter transition is a prediction and remains to be discovered. We will give a range for the expected branching fractions. 
%%%%%%%%%%%%%%%%%%%%%%%%%%%%%%%%%%%%%%%%%%%%
%
\section{Lagrangians in NREFT}\label{sec:lagrangians_nreft}
%
%%%%%%%%%%%%%%%%%%%%%%%%%%%%%%%%%%%%%%%%%%%%
%
%
%%%%%%%%%%%%%%%%%%%%%%%%%%%%%%%%%%%%%%%%%%%%
%
%           Zb
%
%%%%%%%%%%%%%%%%%%%%%%%%%%%%%%%%%%%%%%%%%%%%
The NREFT framework was introduced in Sec.~\ref{sec:effective_theories}. The general idea is to use it to describe the dynamics of bottomia via loops with the open bottom flavor mesons. Those are collected in one heavy field $H$:
\begin{equation}
 H=P+\sigma^iV^i
\end{equation}
with the Pauli matrices $\sigma^i$ and vector indices $i$. The pseudoscalar and vector mesons are collected in the fields
\begin{equation}
 P=(B^-,\bar B^0),\qquad V^i=(B^{*-},\bar B^{*0}).
\end{equation}
As a next step we include the new states, $Z_b$ and $Z'_b$. 
Since they were discovered in the final states $\Upsilon\pi$ and $h_b\pi$ the $C$ parity of the neutral
$Z_b$ states should be negative. This means under parity and charge conjugation,
the fields annihilating the $Z_b$ states should transform as
\begin{equation}
Z^i \overset{\mathcal P}{\rightarrow} Z^i, \qquad Z^i \overset{\mathcal
C}{\rightarrow} -Z^{iT},
\end{equation}
and the heavy quark spin symmetry transformation is given by $Z^i
\overset{\mathcal S}{\rightarrow} S Z^i{\bar S}^\dag$ with $S$ and $\bar S$
acting on the bottom and anti-bottom quark fields, respectively. With these
transformation properties, one can construct the Lagrangian for an $S$-wave
coupling of the $Z_b$ states to the bottom and anti-bottom mesons,
\begin{eqnarray}\label{eq:Lz1}
\Lag_Z &=& i \frac{z}{2} \lang Z^{\dag i}_{ba} H_a \sigma^i{\bar H}_b\rang + {\rm H.c.}\\
&=&z\varepsilon^{ijk}\bar V^{\dagger i} Z^j V^{\dagger k}+z\left[\bar V^{\dagger i}Z^iP^\dagger-\bar P^\dagger Z^iV^{\dagger i}\right]+ {\rm H.c.}
\end{eqnarray}
where the three different charged states are collected in a $2\times2$ matrix as
$$ Z^i_{ba} = \left(
         \begin{array}{cc}
           \frac1{\sqrt{2}} Z^{0i} & Z^{+i} \\
           Z^{-i} & - \frac1{\sqrt{2}} Z^{0i} \\
         \end{array}
       \right)_{ba}.
$$
In Sec.~\ref{sec:loc_sing} we use for both $Z_b$ and $Z'_b$ the Lagrangian
\begin{equation}
 \Lag_Z =z^\mathrm{bare}\varepsilon^{ijk}\bar V^{\dagger i} Z^j V^{\dagger k}+z^\mathrm{bare}\left[\bar V^{\dagger i}Z^iP^\dagger-\bar P^\dagger Z^iV^{\dagger i}\right]+ {\rm H.c.}
\end{equation}
where $z^{\rm bare}$ is the bare coupling constant, which will be renormalized to
the physical one via $z=z^{\rm bare}\sqrt{Z}$. Here $Z$ is the wave function
renormalization constant. The physical couplings then remain free parameters.

%%%%%%%%%%%%%%%%%%%%%%%%%%%%%%%%%%%%%%%%%%%%
%
%          Axialvector Coupling
%
%%%%%%%%%%%%%%%%%%%%%%%%%%%%%%%%%%%%%%%%%%%%
The nonrelativistic version of the axialvector coupling Lagrangian that we will use to describe the scattering of a pion off a heavy meson reads
\begin{eqnarray}\label{eq:NREFTgpi}
 \lag^{NREFT}_\pi&=&g_\pi\frac{1}{\sqrt2F_\pi}\left<H_a^\dagger H_b\vec \sigma\cdot \nabla\phi_{ba}\right>+c.c.\non\\
&=&\frac{\sqrt2g_\pi}{F_\pi}\left[i\varepsilon^{ijk}V_b\partial^jV_a^{\dagger k}+V^i_b\partial^i\phi_{ba} P_a^\dagger+P_b\partial^i\phi_{ba}V_a^{\dagger i}\right]+c.c..
\end{eqnarray}
For the coupling of the bottom mesons we cannot compare to experiment, since the available phase space is not sufficient for a decay $B^*\to B\pi$. A number of lattice calculations is summarized in \cite{Li:2010rh}. Almost all the determinations fall in the range between 0.3 and 0.7, we will use $g_\pi=0.5$ here. Considering flavor symmetry violation this is consistent with the one determined for the charmed mesons, $g_\pi=0.61\pm 0.07$. 

%%%%%%%%%%%%%%%%%%%%%%%%%%%%%%%%%%%%%%%%%%%%
%
%       Magnetic Moments
%
%%%%%%%%%%%%%%%%%%%%%%%%%%%%%%%%%%%%%%%%%%%%
As we will see from power counting arguments later the radiative decays will be dominated by processes where the photon couples to the magnetic moment of the heavy mesons. The corresponding Lagrangian is taken from Hu and Mehen \cite{Hu:2005gf}:
\begin{eqnarray}\label{eq:LagrMagnMom}
 \lag&=&\frac{e\beta}{2}\left<H_a^\dagger H_b\vec\sigma\cdot \vec {\cal B}Q_{ab}\right>+\frac{e}{2m_Q}Q'\left<H_a^\dagger\vec \sigma\cdot \vec {\cal B}H_a\right>+c.c.\non\\
 &=&\left[e\beta Q+\frac{eQ'}{m_Q}\right]_{ab}\!\!\!{\cal B}^i\left(V_a^{\dagger i} P_b+P_a^\dagger V_b^i\right)+\left[e\beta Q-\frac{eQ'}{m_Q}\right]_{ab}i\varepsilon^{ijk}{\cal B}^iV_a^{\dagger j}V_b^k+c.c.
\end{eqnarray}
where $\vec {\cal B}$ is the magnetic field, $Q=\mathrm{diag}(2/3,-1/3,-1/3)$ is the light quark charge matrix and $Q'$ is the charge of the heavy quark, i.e. $-1/3$ for the bottom quark. So the first term is the magnetic moment coupling to the light degrees of freedom, the second one to the heavy degrees of freedom. We already fixed $\beta$ and $m_c$ from experimental data for $D-$meson decays and will now use the same value for $\beta$ and change $m_c$ to the $b-$quark mass: 
\begin{equation}
 1/\beta=0.3759\gev \qquad \text{and} \qquad m_b=4.18\gev
\end{equation}

%%%%%%%%%%%%%%%%%%%%%%%%%%%%%%%%%%%%%%%%%%%%
%
%        Bottomonia
%
%%%%%%%%%%%%%%%%%%%%%%%%%%%%%%%%%%%%%%%%%%%%
Finally we need to cover the couplings of bottomonia to heavy mesons, see Ref.~\cite{Guo:2010ak}. We collect the $P-$wave bottomonia in the field $\chi^i$
\begin{equation}
 \chi^i=h_b^i+\sigma^j\left(\frac{1}{\sqrt{3}}\delta^{ij}\chi_{b0}-\frac{1}{\sqrt{2}}\varepsilon^{ijk}\chi_{b1}^k-\chi_{b2}^{ij}\right).
\end{equation}
The Lagrangian for the coupling of  bottom mesons to $P$-wave bottomonia reads
\begin{eqnarray}
{\cal L}_{\chi_b,h_b} &=& i\frac{g_1}{2}\left<\chi^{\dag i}H_a\sigma^i\bar H_a\right> + {\rm H.c.}\non\\
&=&ig_1h_b^{\dag i}\left[i\varepsilon^{ijk}V_a^j\bar V_a^k-P_a\bar V_a^i+V_a^i\bar P_a\right]+\frac{i}{\sqrt3}g_1\chi_{b0}^\dag\left[\vec V_a\cdot \vec {\bar V}_a+3 P_a\bar P_a\right]\\
&&+\sqrt2g_1\chi_{b1}^{\dag i}\left(V_a^i\bar P_a+P_a\bar V_a^i\right)+ig_1\chi_{b2}^{\dag ij}\left[V_a^j\bar V_a^i+V_a^i\bar V_a^j\right] + {\rm H.c.}\non.
\end{eqnarray}
For readers who are not familiar with tensor mesons we list the most important properties in App.~\ref{app:tensor}.
We can similarly introduce the  $S-$wave bottomonia as 
\begin{equation}
 \Upsilon=\eta_b+\vec \sigma\cdot \vec\Upsilon
\end{equation}
and the coupling to a heavy meson anti-meson pair by
\begin{eqnarray}\label{eq:LUps}
\Lag_{\Upsilon}& =&  i \frac{g_2}{2} \lang \Upsilon^\dag H_a \vec{\sigma}\cdot \!\overleftrightarrow{\partial}\!{\bar H}_a\rang + {\rm H.c.}\non\\
&=&ig_2\eta_b^\dag \left[i\varepsilon^{ijk}V_a^i\lr^j \bar V_a^k +V_a^i\lr^i \bar P_a-P_a\lr ^i\bar V_a^i \right]\non\\
&&+ig_2\Upsilon^{\dag i} \left[\left(\delta^{ik}\delta^{jl}-\delta^{ij}\delta^{kl}+\delta^{il}\delta^{jk}\right)V_a^j\lr^k\bar V_a^l+P_a\lr^i\bar P_a\right.\non\\
&&\left. +i\varepsilon^{ijk} V_a^i\lr^j\bar V_a^k+V_a^i\lr^i\bar P_a-P_a\lr^i\bar V_a^i\right] + {\rm H.c.}
\end{eqnarray}
where $A\overleftrightarrow{\partial}\!B\equiv A(\vec{\partial}B)-(\vec{\partial}A)B$. 
Unfortunately we can determine neither $g_1$ nor $g_2$ in a model independent way. So when it comes to calculating observables we have to refrain to ratios. 

\begin{table}[t]
\centering
\scriptsize
% \small
\caption{Left: Preliminary measurements of the branching ratios for $Z_b^{(\prime)}$ from the Belle Collaboration~\cite{Adachi:2012cx}. Right:
Masses of the various particles used here~\cite{Adachi:2011ji,Behringer:2012ab}.}
\medskip
\renewcommand{\arraystretch}{1.3}
\begin{tabular}{|l|c|c|}
\hline
Branching ratio ($\%$) & $Z_b(10610)$ & $Z'_b(10650)$ \\ \hline\hline
$\Upsilon(1S)\pi^+$ & $ 0.32\pm0.09$ & $ 0.24\pm0.07$ \\ \hline
$\Upsilon(2S)\pi^+$ & $ 4.38\pm1.21$ & $ 2.40\pm0.63$ \\ \hline
$\Upsilon(3S)\pi^+$ & $ 2.15\pm0.56$ & $ 1.64\pm0.40$ \\ \hline
$h_b(1P)\pi^+$ & $ 2.81\pm1.10$ & $ 7.43\pm2.70$ \\ \hline
$h_b(2P)\pi^+$ & $ 4.34\pm2.07$ & $ 14.82\pm6.22$ \\ \hline
$B^+\bar B^{*0}+\bar B^0 B^{*+}$ & $ 86.0\pm3.6$ & $-$ \\ \hline
$B^{*+}\bar B^{*0}$ & $-$ & $ 73.4\pm7.0$ \\ \hline\hline
\end{tabular}
\hspace{0.4cm}
\begin{tabular}{|l|c|l|c|}
\hline
  & Mass [GeV] & & Mass [GeV] \\ \hline\hline
$\Upsilon(1S)$ & $\phantom{0}9.460$ & $\chi_{b0}(1P)$ & \phantom{0}9.859 \\ \hline
$\Upsilon(2S)$ & $ 10.023$ & $\chi_{b1}(1P)$ & \phantom{0}9.893 \\ \hline
$\Upsilon(3S)$ & $ 10.355$ & $\chi_{b2}(1P)$ & \phantom{0}9.912 \\ \hline
$h_b(1P)$ & $ \phantom{0}9.899$ & $\chi_{b0}(2P)$ & 10.233 \\ \hline
$h_b(2P)$ & $10.260$ & $\chi_{b1}(2P)$ & 10.255 \\ \hline
$B$ & $\phantom{0}5.279$ & $\chi_{b2}(2P)$ & 10.269 \\ \hline
$B^{*}$ & $\phantom{0}5.325$ & $\pi$ & \phantom{0}0.138 \\ \hline\hline
\end{tabular}
\label{table-Belle-branching-ratios}
\end{table}
%%%%%%%%%%%%%%%%%%%%%%%%%%%%%%%%%%%%%%%%%%%%
%
\section{Location of the Singularities}\label{sec:loc_sing}
\subsection{Propagator of the $Z_b$ states}\label{sec:prop}
%
%%%%%%%%%%%%%%%%%%%%%%%%%%%%%%%%%%%%%%%%%%%%
\begin{figure}[ht]
\begin{center}
\includegraphics[width=0.68\textwidth]{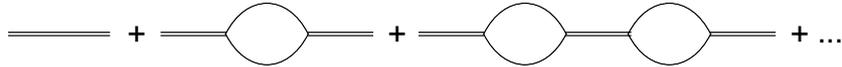}
\caption{Expansion of the two-point Green's function. Double lines and bubbles represent the bare propagators and self-energies, respectively. \label{fig:prop}}
\end{center}
\end{figure}
The states $\zbs$ were originally discovered as narrow resonances in the decays of $\Upsilon(5S)$ into $\Upsilon(mS)\pi^+\pi^-$ and $ h_b(nP)\pi^+\pi^-$.
In this section we will explore the consequences of the decays being driven by bottom meson loops with an intermediate $\zbs$ propagator.
The propagator of the $Z_b$ states is given by the two-point Green's function
\begin{equation}
\delta^{ij} \delta^{ab} G_Z(E) \equiv \int d^4x e^{-iEt}\lang 0\left| T\{Z^{i}_a(x) Z^{\dag j}_b(0) \} \right|0\rang,
\end{equation}
where $i,j$ and $a,b$ are the indices for spin and isospin, respectively, and $T$ denotes time ordering operator. Figure~\ref{fig:prop} illustrates the renormalization of
the Green's function to one loop order.

The bare propagator $i/[2(E-\Eng_0)]$ is dressed by the self-energy $-i\Sigma$. Therefore, the full propagator can be written as
\begin{align}
G_Z(E) = \frac12 \frac{i}{E-\Eng_0-\Sigma(E)}.
\end{align}
It is instructive to start with the relativistic self-energy loop. 
The scalar two-point loop integral is given by
\begin{equation}\label{eq:Sigma}
 \Sigma^\mathrm {Rel}(P^2)=2i \intl \frac1{[(P-l)^2-m_1^2+i\epsilon][l^2-m_2^2+i\epsilon]}
\end{equation}
with the $Z_b$ momentum $P=(P^0,\vec 0)$ in its rest-frame.
Notice that a factor of 2 has been multiplied in this definition in order to take into account both $B\bar B^*$ and its charge conjugated channel. The same
factor appears in the $B^*\bar B^*$ self-energy due to a different reason: we obtain $\epsilon_{ijk}\epsilon_{i'jk}=2\delta^{ii'}$ from the vertices.
To stay consistent with the nonrelativistic Lagrangians, we want to use the nonrelativistic version. 
Our convention is such that the nonrelativistic normalization differs from the relativistic one by a factor of $1/\sqrt{M}$ for the fields. Further the nonrelativistic and relativistic self-energies differ by a factor of $2M_Z$, $\Sigma^\mathrm{Rel}=2M_Z\Sigma$, which cancels the factor $M_Z$ from the vertices. The additional factor $m_1m_2$ gets canceled by the factors $1/(m_i)$ from the nonrelativistic expansion of the propagators. Ultimately the self-energy integral in $d$-dimensional space-time reads
\begin{align}
\Sigma(E) {=} i\frac{(z^{\rm bare})^2}{4} \int\!\! \frac{d^dl}{(2\pi)^d}\,\frac1{[l^0-\vec{l}^2/(2m_1)+i\epsilon][E-l^0-\vec{l}^2/(2m_2)+i\epsilon]}.\label{eq:sigma0}
\end{align}
where $E=P^0-m_1-m_2$. We can carry out the $l^0$ integration and rewrite the integral as 
\begin{equation}
 \Sigma(E) {=} -\frac{(z^{\rm bare})^2}{2}\int\!\! \frac{d^{d-1}l}{(2\pi)^{d-1}}\, \frac {\mu_{12}}{[\vec l^2-2\mu_{12}E-i\epsilon]}
\end{equation}

In dimensional regularization with the $\overline{\rm MS}$ subtraction scheme, this integration is finite for $d=4$. One has
\begin{eqnarray}\label{eq:sigma}
\Sigma^{\overline{\rm MS}}(E)&=&(z^{\rm bare})^2 \frac{\mu}{8\pi} \sqrt{-2\mu E-i\epsilon}\non\\
&=&(z^{\rm bare})^2 \frac{\mu}{8\pi} \left[\sqrt{{-}2\mu E}\theta({-}E) {-} i \sqrt{2\mu E}\theta(E) \right], 
\end{eqnarray}
where $\theta(E)=1$ for positive $E$ and 0 for negative $E$ is the Heaviside step function. The superscript $\overline{\rm MS}$ denotes the subtraction scheme (modified minimal subtraction).

The bare energy $\Eng_0$ is renormalized to the physical energy, $\Eng$, at the mass of the $Z_b$. $\Eng$ is connected to the mass of the $Z_b$ as
$\Eng=M_Z-m_1-m_2$ with $m_{1,2}$ being the masses of mesons in the loop. The renormalization condition is such that $\Eng$ is a zero of the real part of the
denominator of the propagator. Thus, $\Eng=\Eng_0+{\rm Re}\Sigma(\Eng)$, and expanding the real part of
the self-energy $\Sigma(E)$ around $E=\Eng$ gives
\begin{align}
G_Z(E) =\frac12\frac{i}{(E-\Eng)[1-{\rm Re}(\Sigma'(\Eng))]-\widetilde{\Sigma}(E)} \equiv \frac12 \frac{iZ}{E-\Eng-Z\widetilde{\Sigma}(E)} , \label{eq:gz0}
\end{align}
where the wave function renormalization constant is $Z\equiv\left[1-{\rm Re}(\Sigma'(\Eng))\right]^{-1}$ with $\Sigma'(\Eng)$ representing the
derivative of $\Sigma(E)$ with respect to $E$ at $E=\Eng$, and
$$\widetilde{\Sigma}(E)=\Sigma(E)-{\rm Re}(\Sigma(\Eng))-(E-\Eng){\rm Re}(\Sigma'(\Eng)) \ .$$
By construction, ${\rm Re}(\widetilde{\Sigma}(\Eng))={\rm Re}(\widetilde{\Sigma}'(\Eng))=0$ holds.
In the standard scenario (absence of nearby thresholds; stable states)
$\widetilde{\Sigma}(E)$ is dropped, however, here, due to the very close
branch point singularity at $E=0$, this function not only
acquires an imaginary part for $E>0$ but also  varies rapidly. It therefore needs to be kept in the
propagator.
 Eq.~(\ref{eq:gz0}) is valid for both $\Eng>0$
as well as $\Eng<0$ (but ill defined for $\Eng=0$). The expression for $Z$
follows from Eq.~(\ref{eq:sigma}),
\begin{equation}
Z = \left[ 1+\frac{\mu^2(z^{\rm bare})^2}{8\pi\gamma} \right]^{-1}
\theta(-\Eng) + 1\times \theta(\Eng).
\end{equation}

If the $Z_b$ ($Z_b'$) is a pure $B\bar B^*$ ($B^*\bar B^*$) bound state, the
wave function renormalization constant should be 0 since $1-Z$ measures the
probability of finding a bound state in the physical state, see Sec.~\ref{sec:molecules}. This means the bare coupling $z^{\rm bare}$ goes to
infinity. However, the physical effective coupling is finite,
\begin{equation}
(z^{\rm eff})^2 = \lim_{|z|\to\infty} Z (z^{\rm bare})^2 =
\frac{8\pi}{\mu^2}\gamma \ , \label{eq:zeff}
\end{equation}
with the binding momentum $\gamma = \sqrt{-2\mu \Eng}$. Eq.~(\ref{eq:zeff})
coincides with the one for an $S$-wave loosely bound state derived in Sec.~\ref{eq:weinberg} taking into account the factor $2$ as discussed below Eq.~(\ref{eq:Sigma}).

Furthermore, the $Z_b$ states can in principle also decay into channels other than the bottom
and anti-bottom mesons, such as $\Upsilon(nS)\pi~(n=1,2,3)$, $h_b(1P,2P)\pi$,
$\eta_b\rho$ and so on. For the complete propagator we therefore need to write
\begin{align}
G_Z(E) = \frac12\frac{iZ}{E-\Eng-Z\widetilde{\Sigma}(E) + i\Gamma^{\rm phys}(E)/2}.
\end{align}

\begin{figure}[t]
\begin{center}
\includegraphics[width=0.8\textwidth]{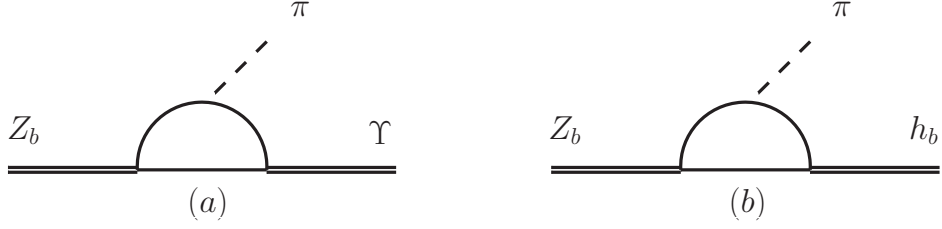}
\caption{One loop diagrams for the subprocesses $\Upsilon(5S)\to Z_b\pi$ (a) and $Z_b\to h_b\pi$ (b). Solid lines in the loops
represent bottom and anti-bottom mesons. \label{fig:Zboneloop}}
\end{center}
\end{figure}

Following the formalism set up in Sec.~\ref{sec:power_counting}, we can analyze the power counting of higher order loops. Two examples are  shown in Fig.~\ref{fig:Zboneloop}. For processes with intermediate heavy meson loops, if the virtuality of these intermediate heavy mesons is not large, their three-momenta are small compared with their masses. Hence these heavy mesons can be dealt with non relativistically, and one can set up a power counting in terms of the velocity of the heavy mesons, $v_B$. In this power counting, the momentum and energy of the intermediate mesons scale  as $v_B$ and $v_B^2$, respectively, and hence the measure of one-loop integration scales as $\int d^4l\sim v_B^5$.
The transition amplitude between
the $Z_b$ states, which couple to the bottom mesons in an $S$-wave, and $S$-wave
bottomonia as shown in Fig.~\ref{fig:Zboneloop}(a) should scale as 
\begin{equation}\label{eq:pconeloop1}
 v_B^5 \frac1{(v_B^2)^3}q^2\frac1{M_B^2}\sim \frac{q^2}{(M_B^2v_B)},
\end{equation}
with $q$ being the external momentum, and $v_B$ the velocity of the intermediate bottom mesons. The factor $1/M_B^2$ has been introduced to get a dimensionless quantity.  Since
$q\ll M_Bv_B^{1/2}$, this is a suppression factor. On the other hand, the
transition amplitudes to $P$-wave bottomonia $h_b(1P,2P)$ scales as 
\begin{equation}\label{eq:pconeloop2}
 v_B^5 \frac1{(v_B^2)^3}q\frac1{M_B}\sim \frac{q}{(M_Bv_B)}.
\end{equation}
Since $q\ll M_B$, one would expect the decays of $Z_b$ to $h_b(1P,2P)\pi$ to be more frequent than those to $\Upsilon(nS)\pi$. Therefore we assume that $\Gamma^{\rm phys}$ is saturated by the former channels.

%%%%%%%%%%%%%%%%%%%%%%%%%%%%%%%%%%%%%%%%%%%%%%%%
%
\subsection{Power counting of two-loop diagrams}\label{sec:twoloop}
%
%%%%%%%%%%%%%%%%%%%%%%%%%%%%%%%%%%%%%%%%%%%%%%%%

So far we have only considered one-loop diagrams. There can be more loops by adding four--$B$-meson contact terms or exchanging pions between (anti-) bottom mesons. 

\begin{figure}[t]
\begin{center}
\includegraphics[width=0.9\textwidth]{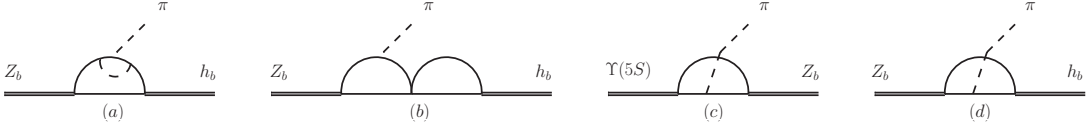}
\caption{Two loop diagrams for the subprocesses $\Upsilon(5S)\to Z_b\pi$ (a) and $Z_b\to h_b\pi$ (b). Solid lines in the loops
represent bottom and anti-bottom mesons. \label{fig:Zbtwoloop}}
\end{center}
\end{figure}

There are two different topologies to be distinguished. The first one arises from the insertion of a contact operator followed by a loop with two heavy mesons  as shown in Fig.~\ref{fig:Zbtwoloop}(a). The short-range operator scales as $v_B^\lambda$ with $\lambda\geq 0$. So the contribution for $Z_b\to h_b\pi$ is
\begin{equation}
 \left( v_B^5 \right)^2\left(\frac1{v_B^2}\right)^5 \frac q{M_B}v_B^\lambda\sim\frac{q}{M_B}v_B^\lambda
\end{equation}
which yields a suppression of $v_B^{\lambda+1}$. The second topology are vertex corrections, i.e. terms where an additional pion is exchanged between the heavy mesons inside the loop. One example is shown in Fig.~\ref{fig:Zbtwoloop}(b). The contribution is
\begin{equation}
 \left(v_B^5\right)\left(\frac{1}{v_B^2}\right)^5 v_B^2\frac{q}{M_B}\sim\frac{q}{M_B} v_B.
\end{equation}
So far both higher loop diagrams are suppressed by at least one order of $v_B$. This can be generalized to higher orders of the same kind and is also applicable for the decay $Z_b\to \Upsilon\pi$. However, one needs to pay special attention to the two-loop diagrams of the kind shown in Fig.~\ref{fig:Zbtwoloop}(c) and Fig.~\ref{fig:Zbtwoloop}(d), where, e.g., a pion gets produced on one heavy meson and rescatters off the
other one before going on-shell. These diagrams need to be analyzed case by case.

Let us first analyze the power counting of the diagram Fig.~\ref{fig:Zbtwoloop}(c)
for $\Upsilon(5S)\to Z_b\pi$.  The leading order amplitude for the bottom
meson--pion scattering formally scales as ${(E_{\pi1}+E_{\pi2})/F_\pi^2}$, see
e.g.~\cite{Guo:2009ct}, with $E_{\pi1,2}$ the energies of the two pions.  Due to
some subtle cancellation mechanisms also the energy of the exchanged pion gets
put on-shell in this vertex --- in analogy to what happens in the reaction
$NN\to NN\pi$~\cite{Lensky:2005jc}.  For the numerical estimates below we use
$E_\pi=M_{\Upsilon (5S)}-M_{Z_b^{(\prime)}}\simeq 250$ MeV.  There are two
$P$-wave couplings in the two-loop diagrams: the coupling of the $\Upsilon(5S)$
to the bottom and anti-bottom mesons and the vertex emitting a pion inside the
loop.  The $Z_b$ bottom meson vertex is in an $S$-wave. The momenta from the two
$P$-wave vertices can contract with each other, and hence scale as $v_B^2$
(recall in the one-loop case, the $\Upsilon(5S)B\bar B$ $P$-wave vertex must
contract with the external momentum, and hence scales as $q$~\cite{Guo:2010ak}).
There are five propagators, and each one of them has a contribution of order
$v_B^{-2}$. Therefore, the power counting of the two-loop diagram of
Fig.~\ref{fig:Zbtwoloop}(c) reads 
\begin{equation} \frac{(v_B^5)^2 v_B^2}{(v_B^2)^5}
\frac{E_\pi}{16\pi^2F_\pi^2} M_B = \frac{v_B^2E_\pi M_B}{\Lambda_\chi^2}, \label{eq:pctwoloop1} 
\end{equation}
where the factor $1/16\pi^2$ appears because there is one more loop compared to the
one-loop case, and the chiral symmetry breaking scale is denoted as
$\Lambda_\chi=4\pi F_\pi$. One may estimate $v_B\sim \sqrt{(\hat{M}-2\hat{M}_B)/\hat{M}_B}\simeq 0.15$ with
$\hat{M}=(M_{\Upsilon(5S)}+M_Z)/2$ and $\hat{M}_B$ the average mass of the bottom
mesons $B$ and $B^*$. This is to be compared to the one-loop diagram, which scale
as $q^2/(M_B^2v_B)$ --- see the discussion in Eqs.~(\ref{eq:pconeloop1},\ref{eq:pconeloop2}) and Refs.~\cite{Guo:2010zk,Guo:2010ak}. Numerically, it is of similar size as or even smaller than the two-loop
diagram given in Eq.~(\ref{eq:pctwoloop1}). Hence, the two-loop diagram could be
more important than the one-loop diagrams. However, the same mechanisms that
suppress vertex corrections should also suppress three-- or more--loop diagrams.

The situation for the transition $Z_b\to h_b\pi$ is different. The two-loop
diagram is shown in Fig.~\ref{fig:Zbtwoloop}(d). Now there is only one $P$-wave
vertex, which is the bottom meson--pion vertex. All the other three vertices are
in an $S$-wave. Due to the $P$-wave nature of the decay $Z_b\to h_b\pi$, the
amplitude must be proportional to the external momentum. Therefore, the only
$P$-wave vertex should scale as $q$, and the product of the other three vertices
scales, again, as $E_{\pi}$. Taking into account further the  integral
measure of the loop and the propagators, the power counting for this diagram reads
\begin{equation}
\frac{(v_B^5)^2}{(v_B^2)^5} \frac{E_\pi}{16\pi^2F_\pi^2} q M_B =
q\frac{M_B E_\pi}{\Lambda_\chi^2}, \label{eq:pctwoloop2}
\end{equation}
where $M_B$ is again introduced to render the scaling dimensionless. The
one-loop diagrams scale like the transitions between two $P$-wave heavy
quarkonia. According to Ref.~\cite{Guo:2010ak}, the power counting is given by
$q/v_B$, which is numerically much larger than the scaling for the two-loop
diagram given in Eq.~(\ref{eq:pctwoloop2}). Hence, the two-loop diagrams can be
safely neglected for the $Z_b\to h_b\pi$.

A more proper power counting considering different velocities in the two loops will be discussed in the next section. As will be shown there, the transition $Z_b\to h_b\pi$ is at the edge of the applicable range of NREFT. However, we will still analyze the data at this channel due to limited experimental information.

In addition, contact-terms of the kind $\Upsilon Z_b\pi$ need to be considered.
On the one hand  they are needed to absorb the divergences of the loop diagrams just
discussed and are expected to be of the same importance as the loops. 
On the other hand they are necessary to parametrize the physics not considered explicitly. The importance of these counter terms is controlled by dimensional analysis. 
This is similar to what we found in Sec.~\ref{sec:hadronic_decays}. 
The corresponding Lagrangian to leading order of the chiral expansion reads
\begin{equation}
\Lag_{\Upsilon Z_b\pi} = c \Upsilon^i Z_{ba}^{\dag i} \partial^0\phi_{ab} + {\rm
H.c.} \label{eq:Lconstant}
\end{equation}
 The analogous counter-terms for $Z_b h_b \pi$ can
be dropped here for they are suppressed compared to the one loop diagram.
\begin{figure}[t]
\begin{center}
\includegraphics[height=3cm]{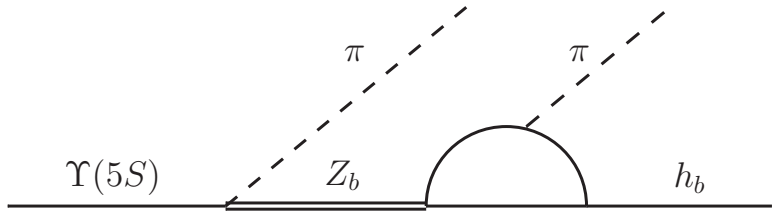}
\caption{Decay mechanism of the process $\Upsilon(5S)\to Z_b\pi\to h_b\pi\pi$. Solid lines in the loop
represent bottom and anti-bottom mesons. \label{fig:oneloop}}
\end{center}
\end{figure}
Assuming the $Z_b$ and $Z_b'$ are spin partners of each
other~\cite{Bondar:2011ev}, one can use the same coupling constant for them.

A full analysis would not only require the evaluation of all the diagrams
mentioned above but also those where there are no $Z_b^{(\prime)}$ present in
the processes. Here we take a more pragmatic point of view and simply represent the whole $\Upsilon
Z_b\pi$ transition by the single contact term of Eq.~(\ref{eq:Lconstant}). The
full transition is illustrated in Fig.~\ref{fig:oneloop}.

One comment is in order: the Lagrangian of Eq.~(\ref{eq:Lconstant}) could as
well mimic a compact component of the $Z_b^{(\prime)}$. This is important
especially because both the two-loop and one-loop scaling given by
Eq.~(\ref{eq:pctwoloop1}) and $q^2/(M_B^2 v_B)$ are much smaller than 1, and
hence a small compact component could be more important than the bottom meson
loops in the $\Upsilon Z_b\pi$ vertex. Thus, since we expect this contact term
to appear at leading order, it seems as if we would not be able to disentangle a
compact, say, tetraquark component from a molecular one. However, since in the
molecular scenario the transition $Z_b\to \pi h_b$ is dominated by the loop, the
structure can indeed be tested, since for molecules the dynamics appears to be
quite restricted. Thus, in the present approach the $\Upsilon (5S)\pi$ vertex
provides a source term for the $Z_b^{(\prime)}$, while their decays are the
subject of this study.

\subsection{Results}
\label{sec:constaint}

The relevant vertices follow from these Lagrangians. The decay amplitudes for
the two-body transitions can be found in Appendix~\ref{app:AmplitudesZb_1}. On one hand,
the $\Upsilon(5S)$ is only about 120 and 70~MeV above the $Z_b\pi$ and $Z_b'\pi$
thresholds, respectively. On the other hand, it has a large width
$110\pm13$~MeV, and the experimental data were taken in an energy range around
the $\Upsilon(5S)$ mass. Therefore, when calculating its decay widths, one has
to take into account the mass distribution of the $\Upsilon(5S)$. Its three-body
decay width can then be calculated using
\begin{align}
\Gamma(\Upsilon(5S))_{\rm 3-body}= \frac1{W}\int_{(M_\Upsilon-2\Gamma_\Upsilon)^2}^{(M_\Upsilon+2\Gamma_\Upsilon)^2}
\!\!ds\, \frac{(2\pi)^4}{2\sqrt{s}} \int d\Phi_3 |\Amp|^2  \frac1{\pi}{\rm Im}\left(\frac{-1}{s-M_\Upsilon^2+iM_\Upsilon\Gamma_\Upsilon}\right) ,
\end{align}
where $M_\Upsilon=10.865$~GeV and $\Gamma_\Upsilon=0.11$~GeV are the mass and
width of the $\Upsilon(5S)$, respectively, and $\int d\Phi_3$ denotes the
three-body phase space, see e.g.~\cite{Behringer:2012ab}. The function $\Amp$ contains
all the physics and can be easily obtained using the loop amplitudes given
explicitly in the appendix. Both positively and negatively charged $Z_b$ and
$Z_b'$ states should be considered. The factor $1/W$ with
$$W = \int_{(M_\Upsilon-2\Gamma_\Upsilon)^2}^{(M_\Upsilon+2\Gamma_\Upsilon)^2}
\!\!ds\, \frac1{\pi}{\rm
Im}\left(\frac{-1}{s-M_\Upsilon^2+iM_\Upsilon\Gamma_\Upsilon}\right)$$ is
considered in order to normalize the spectral function of the $\Upsilon(5S)$.

We consider the case that the $Z_b$ and $Z_b'$ couple only to the $B\bar
B^*+{\rm c.c.}$ and $B^*\bar B^*$ channels, respectively. Coupled channel
effects should be suppressed because $|\Eng| \ll M_{B^*}-M_B$ for both $Z_b$
states.

The parameters of the model are the normalization factors $N$, chosen
individually for the two final states, the physical couplings, which are
products of $\sqrt{Z}$ and the bare couplings, for the $Z_b$ states to the
relevant open bottom channels, $z_1$ and $z_2$ (or equivalently the binding
energies of the $Z_b^{(\prime)}$ in Eq.~(\ref{eq:zeff})), and those couplings for the $h_b$ and
$h_b(2P)$, denoted by $g_1$ and $g_1'$, respectively. The parameter $c$ of
Eq.~(\ref{eq:Lconstant}) is absorbed into the overall normalization factor. Both
$g_1$ and $g_1'$ only appear in a product with the $z_i$. In order to reduce the
number of free parameters, we assume $g_1'=g_1$ in the following. Note that
neither of them can be measured directly, since the masses of $h_b$ and
$h_b(2P)$ are below the $\bar B B^*$ threshold. In the actual fit we will adjust
$z_1$, $r_z\equiv z_2/z_1$, $g_1z_1$ and the two normalization constants.

Using the amplitudes of Eqs.~(\ref{Aeq:ahb1},\ref{Aeq:ahb2}), we fit
 the parameters to the  invariant mass spectra of both $h_b\pi^+$ and
$h_b(2P)\pi^+$  from 10.56 GeV to 10.70~GeV in the
missing mass spectrum $MM(\pi)$. In the chosen region, there are 14 data points
for the $\Upsilon(5S)\to h_b\pi^+\pi^-$ and 13 for the $\Upsilon(5S)\to
h_b(2P)\pi^+\pi^-$.

The decay widths of the $Z_b$ and $Z_b'$ into $h_b\pi$ are obtained to be
\begin{eqnarray}\label{eq:widhb}
\Gamma(Z_b\to h_b\pi) &=& 4.8 \left(\frac{gg_1z_1}{F_\pi} {\rm GeV}^2 \right)^2~{\rm MeV} = 140 (g_1z_1 {\rm GeV})^2~{\rm MeV}, \non\\
\Gamma(Z_b^{\prime}\to h_b\pi) &=& 5.8 \left(\frac{gg_1z_2}{F_\pi} {\rm GeV}^2 \right)^2~{\rm MeV}=169  (g_1z_2 {\rm GeV})^2~{\rm MeV}. 
\end{eqnarray}
Due to smaller phase space, the widths for the decays $Z_b^{(\prime)}\to
h_b(2P)\pi$ get smaller numerical factors. We find
\begin{align}
\Gamma(Z_b\to h_b(2P)\pi) = 30 (g_1'z_1 {\rm GeV})^2~{\rm MeV}, \non\\
\Gamma(Z_b^{\prime}\to h_b(2P)\pi) = 46  (g_1'z_2 {\rm GeV})^2~{\rm MeV}.
\label{eq:widhbp}
\end{align}

If the $Z_b$ and $Z_b'$ states are $S$-wave bound states of the $B\bar B^*$ and
$B^*\bar B^*$, respectively, their coupling strengths to the bottom and
anti-bottom mesons are related to the binding energies. The relation has been
derived in Eq.~(\ref{eq:zeff}). The coupling constants $z_i$ appear in
Eqs.~(\ref{eq:widhb}), which enter the $Z_b$ propagators, as well as in the
transition amplitudes --- c.f. Eqs.~(\ref{Aeq:ahb1},\ref{Aeq:ahb2}).

\begin{figure}[t]
\begin{center}
\begin{minipage}{0.48\textwidth}
 \includegraphics[width=\textwidth]{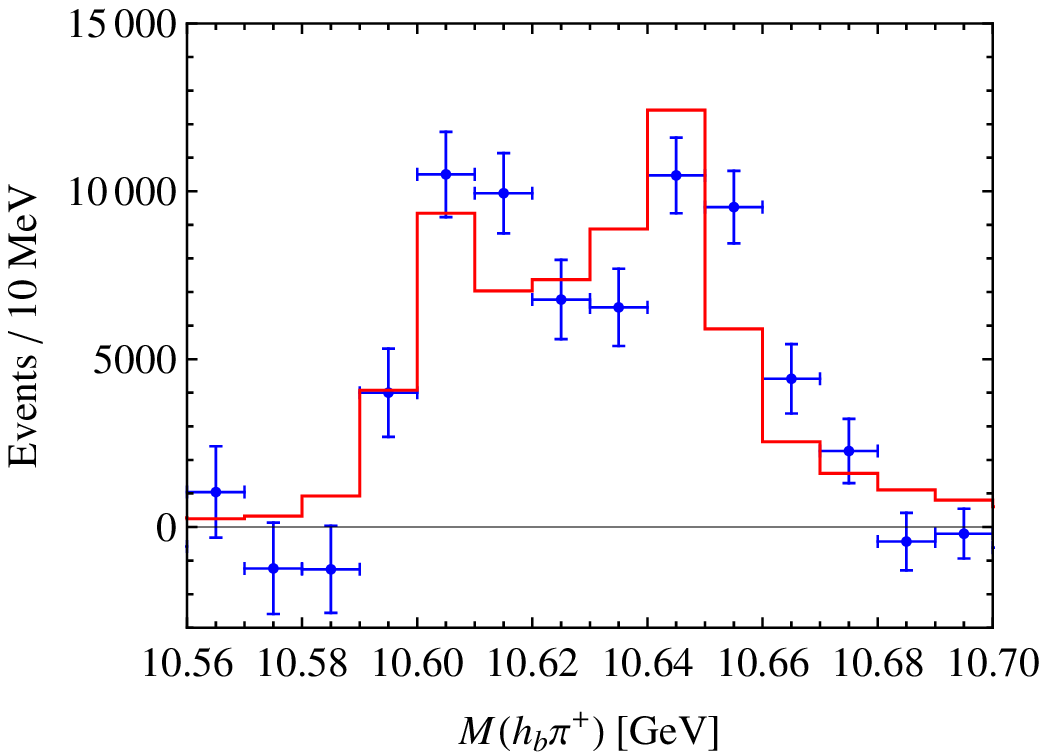}
\end{minipage}
% \\[5mm]
\begin{minipage}{0.48\textwidth}
 \includegraphics[width=\textwidth]{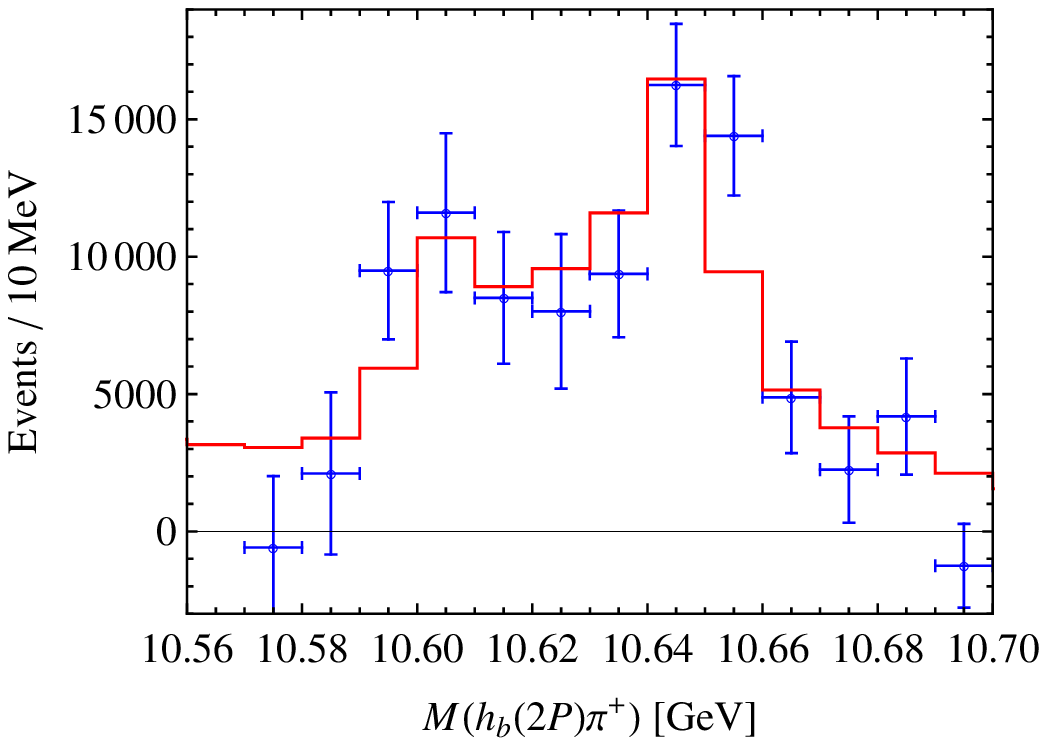}
\end{minipage}
\caption{Comparison of the calculated invariant mass spectra of $h_b\pi^+$ and $h_b(2P)\pi^+$ with
the measured missing mass spectra $MM(\pi)$. \label{fig:resbs}}
\end{center}
\end{figure}

In order to fit to the data which are events collected per 10~MeV, we integrate
the invariant mass spectra for each bin corresponding to the measurements. This
is important for narrow structures. The best fit results in $\chi^2/{\rm
d.o.f.}=54.1/22=2.45$.
\begin{eqnarray}
z_1 = 0.75^{+0.08}_{-0.11}~{\rm GeV}^{-1/2}, \qquad r_z = -0.39^{+0.06}_{-0.07},\qquad g_1 z_1 = (0.40\pm0.06)~{\rm GeV}^{-1}.
\end{eqnarray}
The results from the best fit are plotted in Fig.~\ref{fig:resbs} together with
the experimental data. Using Eq.~(\ref{eq:zeff}) the
     couplings can be converted to binding energies. Especially we find
     \begin{equation}
     \Eng_{Z_b}= -4.7^{+2.2}_{-2.3}~{\rm MeV}, \qquad \Eng_{Z_b'}=
     -0.11^{+0.06}_{-0.14}~{\rm MeV} \ .
     \end{equation}

Although the $Z_b$ and $Z_b'$ are supposedly spin partners, a value of
$r_z=-0.4$ is not completely unreasonable: the fine tuning necessary to put a
bound state as close as 0.1~MeV to a threshold is extremely sensitive to even a
small variation in the scattering potential, driven by spin symmetry violations.
In effect this can give significant differences in the binding energies and, via
Eq.~(\ref{eq:zeff}), also in the coupling constants. However, a microscopic
calculation, which goes beyond the scope of this study,
 would be necessary to check this hypothesis.

One can obtain a better fit to the data if one would either release the bound
state condition given as Eq.~(\ref{eq:zeff}), and allowing the masses of the
$Z_b$ states to float freely, or allow for non-resonant terms. But this is not
the purpose of our work --- we here only want to demonstrate that the data are
consistent with the bound state picture, which implies the masses of the
$Z_b^{(\prime)}$ states to be located below the corresponding thresholds.

It is interesting to look at the $Z_b$ line shape or the absolute value of
$G_Z(E)$ for different locations of the $Z_b$ pole. This is similar to the discussion of states close to the $K\bar K$ threshold by Flatt\'{e} in \cite{Flatte:1976xu}.
\begin{figure}[t]
\begin{center}
\includegraphics[width=0.6\textwidth]{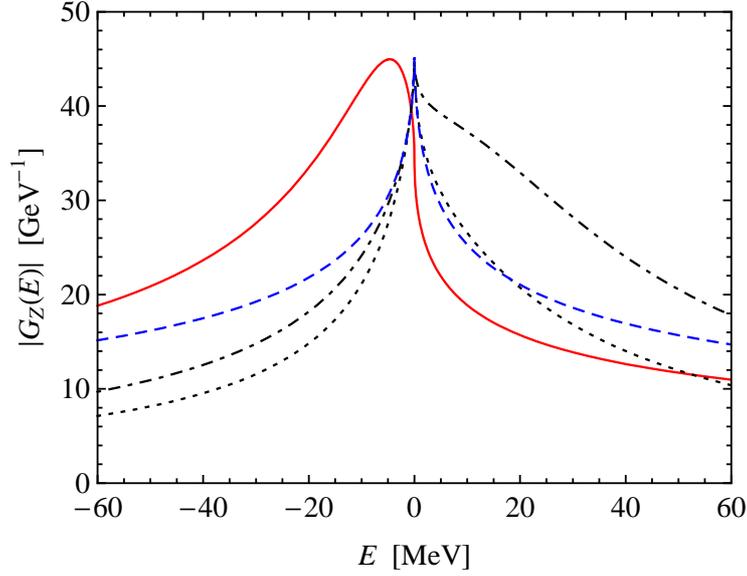}
\caption{The absolute value of $G_Z(E)$.
The solid (red) and dashed (blue) are for a bound state and virtual state, respectively,
with the same mass, $\Eng=M_{Z_b}-M_B-M_{B^*}=-4.7$~MeV. The
dotted and dot-dashed (black) lines are for resonances with $\Eng=8$ and 20~MeV, respectively.
The maxima of the resonance and virtual state curves have been normalized to the bound state one. \label{fig:gz}}
\end{center}
\end{figure}
The function $|G_Z(E)|$ using the parameters from the best fit are plotted as
the solid line in Fig.~\ref{fig:gz}. In this case, the $Z_b$ is a $B\bar B^*$
bound state with a binding energy of $-4.7$~MeV. Keeping $z_1=0.75~{\rm
  GeV}^{-1/2}$, and $g_1 z_1=0.4$~GeV$^{-1}$ fixed we also plot as the dashed
line the line shape for the virtual state with the same value of
$\Eng$.\footnote{The curve for virtual state is obtained using
  Eq.~(\ref{eq:gz0}) but with $\widetilde{\Sigma}(E)=\Sigma(E)-{\rm
    Re}(\Sigma_{\rm II}(\Eng))-(E-\Eng){\rm Re}(\Sigma'(\Eng)) \ ,$ where, in
  the $\overline{\rm MS}$ scheme, ${\rm Re}(\Sigma_{\rm II}(\Eng)) = -{\rm
    Re}(\Sigma(\Eng))$ is the self-energy in the second Riemann sheet.}
 The dotted and dot-dashed lines are for a
resonance with a mass above the $B\bar B^*$ threshold by 8 and 20~MeV,
respectively. From the figure, one sees that the bound state produces a bump
below the threshold, and a small cusp at the threshold, while the virtual state
produces a prominent cusp at the threshold and no structure below. Above the
threshold, the energy dependence of the virtual and bound state curves are
exactly the same. The bump in the bound state case reflects the pole position.
Hence, if we reduce the value of binding energy to, say about 0.1~MeV, which is
the case for the $Z_b'$ and the $X(3872)$, then the bump below threshold would
be invisible for typical experiments with a resolution larger than $1\mev$, and the cusp dominates the structure. In this case, it is hard to
distinguish between the bound state and virtual state scenarios. For more
discussions on the shape of a virtual state, see e.g.~\cite{Baru:2004xg}.

One important feature of the line shapes of dynamically generated states
 is shown in Fig.~\ref{fig:gz}: for poles slightly
above the threshold, since the coupling to the opening channel is strong,  the
position of the peak is \emph{locked to the threshold}, as can be seen from the
dotted line. Increasing the resonance mass, the effect of the cusp is smeared
out, and shape is approaching a normal Breit-Wigner resonance --- in the
dot-dashed line one starts to see a bump above threshold developing for a mass
as large as 20~MeV above the $B\bar B^*$ threshold. However, even then the peak
is still located at the threshold. We  therefore  conclude that a
Breit-Wigner parametrization, as was used in the experimental analysis, should
not be used when analyzing structures that emerge from dynamically generated
states.

\section{Hadronic and Radiative Decays of $\zbs$}\label{sec:zbdecays}
In this section we will explore the consequences of assuming that $\zbs$ are $B^*\bar B(B^*\bar B^*)$ resonances, i.e. hadronic molecules with the pole locations above the threshold for their constituents. This distinguishes it slightly from the approach of the last section where both were assumed to be located below threshold. Still, being very close to or almost compatible with the thresholds this does not change the conclusions of last section. However, now we can assume that the total width is saturated by the decays into the constituents and  deduce the corresponding coupling constant directly from data. This leads to the interesting consequence that two-body decays calculated via bottom meson loops have only one free parameter left. Therefore we will be able to make parameter free predictions for ratios of said two-body decays. These can be measured by future experiments and thus allow to test the molecular nature of the $Z_b$ states. Further we will incorporate the later discovered neutral 
partner for $Z_b$ into our analysis. 

We will use the improved experimental data and incorporate the experimental observation that the $Z_b^{(\prime)}$
couples only to $B^{(*)}\bar B^*$ via the Lagrangian,
\begin{equation}\nonumber
 \mathcal{L}_{Z,Z'}=z'\varepsilon^{ijk}\bar V^{\dagger i} Z^{\prime j} V^{\dagger k}+z\left[\bar V^{\dagger i}Z^iP^\dagger-\bar P^\dagger Z^iV^{\dagger i}\right]+ {\rm H.c.}.
\end{equation}
From this we can deduce the tree-level decay width
\begin{eqnarray}
 \Gamma[Z_b\to B^{*+} \bar B^0+\bar B^{*0}B^+]&{=}&2\frac{1}{8\pi}\frac{|\vec q|}{M_{Z_b}^2}\frac13\sum_{Pol}\left|\varepsilon^a(Z_b)\varepsilon^b(B^*) g^{ab}z\right|^2M_BM_{B^*}M_{Z_b}\\
 &{=}&\frac{1}{4\pi}\frac{|\vec q|}{M_{Z_b}}\left|z\right|^2M_BM_{B^*}
\end{eqnarray}
where we averaged the incoming and summed over the outgoing polarizations. The masses are due to the normalization of the heavy fields and the factor 2 was multiplied to consider the two final states $B^{*+} \bar B^0$ and $\bar B^{*0}B^+$. For the decay $Z'_b\to B^{*+}\bar B^{*0}$ we find the same factor 2 due to the sum over the Levi-Civita symbols. 
Fitting to the experimental data we find
\begin{equation}
 z=(0.79\pm0.05)~{\rm GeV}^{-1/2}, \qquad z'=(0.62\pm0.07)~{\rm GeV}^{-1/2} \label{eq:couplings-NREFT}.
\end{equation}
% Applying the Weinberg criterion as discussed in Sec.~\ref{sec:molecules} we find the molecular component of the $Z_b$ states to be about $0.85\%$ and $0.73\%$, respectively. 
Moreover the ratio of the two couplings is
\begin{equation}
 \frac{z}{z'}= 1.27 \pm0.16 \  \label{eq:coupling-ratio}
\end{equation}
which deviates from unity by 2$\sigma$. This deviation indicates a significant amount of spin symmetry violation, which, however, is not unnatural for very-near-threshold states where small differences in masses  imply huge differences in binding energies resulting in significantly different effective couplings, as we will also discuss in the following section.

\subsection{Power Counting}\label{sec:nonrel}
The theoretical framework including Lagrangians is already set up and the calculation of triangle diagrams within the framework of NREFT has been shown to be a straightforward matter. We need to have a closer look at the  power counting since the very close proximity of the open thresholds to the $\zbs$ adds an additional difficulty here to what has been discussed in general terms in Sec.~\ref{sec:power_counting}.
We will use the same formalism here  together with an improved discussion for the higher loop diagrams.

In general, the heavy meson velocities relevant for the decay of
some particle $X$ may be estimated as
$v_X\sim\sqrt{|{M_X}-2M_B|/M_B}$, where the absolute value indicates
that the formula can be used for both bound systems as well as
resonances. The analogous formula holds when the two heavy mesons
merge to a quarkonium in the final state. According to the rules of
a nonrelativistic effective field theory~\cite{Kaplan:2005es} (for a review,
see e.g.~\cite{Brambilla:2004jw}), the momentum and nonrelativistic
energy count as $v_X$ and $v_X^2$, respectively. For the integral
measure one finds $v_X^5/(4\pi)^2$. The heavy meson propagator
counts as $1/v_X^2$. The leading order $S$-wave vertices do not have
any velocity dependence. The case for the $P$-wave vertices is
more complicated: it scales either as $v_X$ when the momentum due to
$P$-wave coupling contracts with another internal momentum, or as
the external momentum $q$ when $q$ is contracted.

\begin{figure}
\centering
\includegraphics[width=\linewidth]{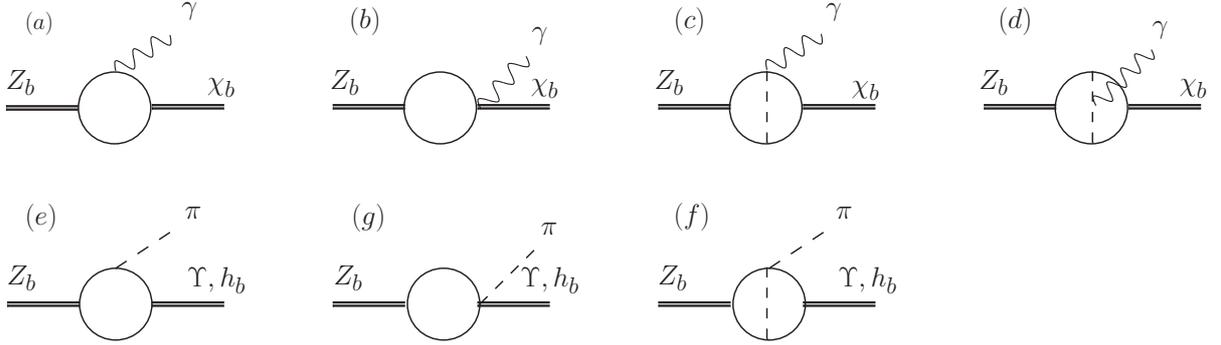}
\caption{Schematic one- and two-loop diagrams of the transitions $Z_b\to \Upsilon\pi,~h_b\pi$ and $\chi_{bJ}\gamma$.}
\label{fig:PowerCounting1}
\end{figure}
We start with the radiative transitions as shown in the upper row of
Fig.~\ref{fig:PowerCounting1}. If the $Z_b$ states are molecular
states, their spin wave functions contain both  $s_{b\bar b}=0$ and
1 components~\cite{Bondar:2011ev}, where $s_{b\bar b}$ is the total
spin of the $b\bar b$ component. Thus, the radiative decays of the
$Z_b$ states into the spin-triplet $\chi_{bJ}$ can occur without
heavy quark spin flip and survive in the heavy quark limit. This is
different from the M1 transitions between two $P$-wave heavy
quarkonia which have been analyzed in Ref.~\cite{Guo:2011dv}. Both
the couplings of $Z_b$ and $\chi_{bJ}$ to a pair of heavy mesons are
in an $S-$wave, and the photon coupling to the bottom mesons is
proportional to the photon energy $E_\gamma$. For the diagram of
Fig.~\ref{fig:PowerCounting1}~(a) the amplitude therefore scales as
\begin{equation}
 \frac{\bar v^5}{(4\pi)^2}\frac{1}{(\bar v^2)^3}E_\gamma\sim
 \frac{E_\gamma}{(4\pi)^2 \bar v} \ ,
 \label{eq:Apower}
\end{equation}
where  the velocity that appears is $\bar v=(v_Z+v_\chi)/2\simeq
v_\chi/2$~\cite{Guo:2012tg}, since $v_Z\sim v_{Z'}\simeq 0.02$, if the central
values of the measured $Z_b^{(\prime)}$ masses, 10607.2~MeV and 10652.2~MeV, are
used, while $v_\chi$ ranges from 0.12 for the $\chi_{bJ}(3P)$ to 0.26 for the
$\chi_{bJ}(2P)$ to 0.37 for the $\chi_{bJ}(1P)$. Here we used the mass of the
$\chi_{bJ}(3P)$, 10.53~GeV, as reported by the ATLAS
Collaboration~\cite{Aad:2011ih}. In the following, we will count $\bar v$ as
$\mathcal{O}(v_\chi)$. The scaling ensures that the amplitude gets larger when the bottomonium in the
final state is closer to the open-bottom threshold. Thus, we expect for the
absolute value of the decay amplitude from this diagram
\begin{eqnarray}
\left|\frac{\mathcal{A}_{\chi_{bJ}(1P)\gamma}}{E_\gamma}\right|:
\left|\frac{\mathcal{A}_{\chi_{bJ}(2P)\gamma}}{E_\gamma}\right|:\left|\frac{\mathcal{A}_{\chi_{bJ}(3P)\gamma}}{E_\gamma}\right|
\sim \frac1{v_{1P}} : \frac1{v_{2P}} : \frac1{v_{3P}}
=  1:1.4:3.1,
\label{eq:Aratio}
\end{eqnarray}
if the $\chi_{bJ}(nP)B\bar B$ coupling constants take the same value. Diagram
(a) can be controlled easily in theory. Thus, clear predictions can be made
whenever diagram (a) dominates. In the following we will identify such dominant
decays based on the power counting for the NREFT.

As for Fig.~\ref{fig:PowerCounting1}~(b), the coupling $\chi_b B
\bar B \gamma$ cannot be deduced by gauging the coupling of
$\chi_{bJ}$ to a $B\bar B$-meson pair. Thus, it has to be gauge
invariant by itself and proportional to the electromagnetic field
strength tensor $F^{\mu\nu}$. This gives a factor of photon energy
$E_\gamma$ and the amplitude of Fig.~\ref{fig:PowerCounting1}~(b)
scales as
\begin{equation}
\label{eq:leadrad}
 \frac{v_Z^5}{(4\pi)^2}\frac{1}{(v_Z^2)^2} E_\gamma\sim
 \frac{E_\gamma v_Z}{(4\pi)^2} \ ,
\end{equation}
where we have assumed that the corresponding coupling is of natural size. Thus,
diagram (b) is suppressed compared to diagram (a) at least by a factor of $v_Z
v_\chi<0.01$ for the decay to $\chi_{bJ}(1P)$ and even smaller for the excited
states.

The situation is more complicated for the graph displayed
in Fig.~\ref{fig:PowerCounting1}~(c). Here we
have a two-loop diagram, so that the velocities running in different loops are
significantly different --- the one in the loop connected to the $Z_b$ is $v_Z$,
and the other is $v_\chi$. It is important to count them separately since
$v_Z\ll v_\chi$ due to the very close proximity of the $Z_b$ to the
threshold\footnote{The concept applied here is analogous to the scheme by now
well established for the effective field theory for reactions of the type $NN\to
NN\pi$ --- see Ref.~\cite{Hanhart:2003pg,Baru:2013zpa} for a review.}. The internal pion momentum
scales as the larger loop momentum, and thus the pion propagator should be
$\sim1/(M_B^2 v_\chi^2)$. This leaves us with
\begin{equation}
 \frac{v_Z^5}{(4\pi)^2}\frac{1}{(v_Z^2)^2}\frac{v_\chi^5}{(4\pi)^2}  \frac{1}{(v_\chi^2)^2}\frac{1}{M_B^2v_\chi^2}\frac{E_\gamma g}{F_\pi}  \frac{g}{F_\pi}M_B^4
 \sim \frac{v_Z}{v_\chi} \frac{E_\gamma g^2 M_B^2}{(4\pi)^2\Lambda_\chi^2}~,
\end{equation}
where $F_\pi$ is the pion decay constant in the chiral limit, the
factor $M_B^4$ has been introduced to give the same dimension as the
estimate for the first two diagrams, and the hadronic scale was
introduced via the identification $\Lambda_\chi=4\pi F_\pi$. Thus
the two-loop diagram is suppressed compared to the leading one,
Eq.~(\ref{eq:leadrad}), by a factor $v_Z g^2
M_B^2/\Lambda_\chi^2\sim 0.1$, where we used for the coupling
$B^*\to B\pi$ the value $g=0.5$ and $\Lambda_\chi\sim1$~GeV.
It can easily be seen that Fig.~\ref{fig:PowerCounting1}~(d) gives the same
contribution which also reflects the fact that they are both
required at the same order to ensure gauge invariance. Thus, from
our power counting it follows that the loop diagrams of
Fig.~\ref{fig:PowerCounting1} (b)-(d) provide a correction of at
most 10\%. We will therefore only calculate diagram (a) explicitly
and introduce a 10\%  uncertainty for the amplitudes which
corresponds to 20\% for the branching ratios. Higher loop
contributions are to be discussed later.

Next we consider the hadronic transition $Z_b\to h_b\pi$.  This decay
has already been studied in the previous section in the same
formalism. Again, since the $h_b$ has even parity, its coupling to
the bottom mesons is in an $S$-wave. In addition, the final state
must be in a $P$-wave to conserve parity, such that the amplitude
must be linear in the momentum of the outgoing pion, $q$. We therefore find
for the one-loop contribution of Fig.~\ref{fig:PowerCounting1}~(e)
\begin{equation}
\frac{\bar v^5}{(4\pi)^2}\frac{1}{(\bar  v^2)^3} \frac{gq}{F_\pi}
\sim g\frac{q F_\pi}{\bar v \Lambda_\chi^2} \ ,\label{eq:hbpia}
\end{equation}
while Fig.~\ref{fig:PowerCounting1}~(f)  gives
\begin{equation}
 \frac{v_Z^5}{(4\pi)^2}\frac{1}{(v_Z^2)^2}\frac{q}{F_\pi}\sim \frac{v_Z q
   F_\pi}{\Lambda_\chi^2}.
\end{equation}
Notice that the pion has to be emitted after the loop if the $Z_b$ is a pure
hadronic molecule, so that the velocity in the counting should be $v_Z$ instead
of $v_h$. This is suppressed by $v_hv_Z/g$ which leads to a correction of the
order of  2\% noticing that $v_h\simeq v_\chi$.

The two-loop diagram Fig.~\ref{fig:PowerCounting1}~(g) contributes as
\begin{equation}
\frac{v_Z^5}{(4\pi)^2}\frac{1}{(v_Z^2)^2}\frac{v_\chi^5}{(4\pi)^2}\frac{1}{(v_\chi^2)^2}\frac{1}{M_B^2v_\chi^2}\frac{gq}{F_\pi}
  \frac{E_\pi }{F_\pi^2}M_B^3
=\frac{v_Z}{v_h}\frac{ g F_\pi q^2 M_B}{\Lambda_\chi^4} \ ,
\label{eq:hbpi_g}
\end{equation}
where we have used that the energy from the $\pi B\to \pi B$ vertex can be
identified with the energy of the outgoing pion~\cite{Lensky:2005jc}, $E_\pi\sim
q$.  The $B^*B\pi$ vertex contributes a factor of the external momentum $q$
since the $Z_b\to h_b\pi$ is a $P$-wave decay, and this is the only $P$-wave
vertex. Therefore, this diagram is suppressed compared to the leading loop,
Eq.~\eqref{eq:hbpia}, by a factor $v_Z M_B q/\Lambda_\chi^2$ which is smaller
than 10\%. Thus,  also for the transitions $Z_b^{(\prime)}h_b\pi$ we may only
calculate the leading one-loop diagrams, Fig.~\ref{fig:PowerCounting1} (e), and
assign an uncertainty of 10\% to the rates which gives an uncertainty of 20\%
for the branching ratios. Higher loop contributions are to be discussed later.

Finally, we consider at the decay channel $Z_b\to \Upsilon \pi$. Here the final
state is in an $S$-wave, but the coupling of the $\Upsilon$ to $\bar BB$ is in a
$P$-wave. For diagram (e) the momentum due to this coupling has to scale as the
external pion momentum. Together  with the pionic coupling that is also linear
in the pion momentum, the amplitude is thus proportional to $q^2$. The one-loop
diagram for $Z_b\to \Upsilon\pi$ via Fig.~\ref{fig:PowerCounting1} (e), is
therefore estimated as
\begin{equation}
 \frac{\bar v^5}{(4\pi)^2}\frac{q}{(\bar v^2)^3}\frac{gq}{F_\pi} \sim g\frac{q^2
   F_\pi}{\bar v\Lambda_\chi^2}.
   \label{eq:upsilonpi}
\end{equation}
The diagram Fig.~\ref{fig:PowerCounting1}~(f) on the other hand gives
\begin{equation}
 \frac{v_Z^5}{(4\pi)^2}\frac{1}{(v_Z^2)^2}\frac{E_\pi}{F_\pi} M_B
\sim
\frac{v_Z E_\pi F_\pi M_B}{\Lambda_\chi^2},
\end{equation}
where $M_B$ is introduced in order to get the same dimension as
Eq.~\eqref{eq:upsilonpi}. Compared to the one-loop diagram (e) this
is a relative suppression of order $v_\Upsilon v_Z M_B/q$ which is
less than 10\% for all the $\Upsilon$ states,  where the value of
$v_\Upsilon$ is about 0.46, 0.33 and 0.22 for the $1S$, $2S$ and
$3S$ states, respectively. The two-loop contribution with the
exchange of a pion, Fig.~\ref{fig:PowerCounting1}~(g), is estimated
as
\begin{equation}
  \frac{v_Z^5}{(4\pi)^2}\frac{1}{(v_Z^2)^2}\frac{v_\Upsilon^5}{(4\pi)^2} \frac{1}{(v_\Upsilon^2)^2}\frac{1}{M_B^2v_\Upsilon^2}\frac{v_\Upsilon^2 g}{F_\pi} \frac{E_\pi }{F_\pi^2}M_B^5 
\sim \frac{v_Zv_\Upsilon gE_\pi F_\pi}{\Lambda_\chi^4}M_B^3 \ .
\end{equation}
Thus, the strength of two-loop diagram relative to leading one-loop diagram
 is estimated  for the $\Upsilon\pi$ as
$ v_Z v_\Upsilon^2 m_B^3/(\Lambda_\chi^2 q)$. Numerically, this corresponds to a
factor of around 0.6 for the $\Upsilon(1S,2S)\pi$ and 0.7 for the
$\Upsilon(3S)\pi$ amplitudes. As a consequence, the branching ratios for these
transitions can only be calculated with rather large uncertainties up to 100\%.

\begin{figure}
\centering
\includegraphics[width=0.5\linewidth]{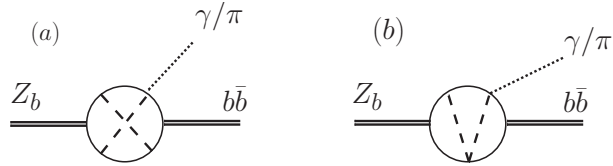}
\caption{Two three-loop diagrams contributing to the decays of the $Z_b$ into a heavy quarkonium and a pion or photon.}
\label{fig:PowerCounting2}
\end{figure}

The heavy meson velocities relevant for the mentioned transitions
range from 0.02 to 0.5 --- in momenta this is a range from 0.1 to
2.5~GeV. While pion contributions are expected to be suppressed
significantly and can be controlled within chiral perturbation
theory for pion momenta of up to 500 MeV and smaller, higher pion
loop contributions might get significant for momenta beyond 1~GeV.
We will now study those higher pion loops within the power counting
scheme outlined above. We will start with the three-loop diagrams,
as shown in Fig.~\ref{fig:PowerCounting2} considering first diagram~(a).
The results can be easily generalized to higher loops as shown
below. Compared to the two-loop diagrams in
Fig.~\ref{fig:PowerCounting1}, there are one more pion propagator
and two more bottom meson propagators in the transition. In addition,
there is no two-bottom-meson unitary cut present. As a consequence,
we have to use a relativistic power counting --- c.f.
Ref.~\cite{Kaplan:2005es}. Then pion momentum and energy are of order $m_B
v_{b\bar b}$, with $v_{b\bar b}$ the velocity of the $B$-meson
connected to the $b\bar b$-meson in the final state. Both the
energies and momenta of the additional bottom mesons are now of the
same order, such that the bottom-meson propagator is counted as
$1/v_{b\bar b}$. The integral measure reads $v_{b\bar
b}^4/(4\pi)^2$. Therefore, the additional factor as compared to the
two-loop diagrams  is
\begin{equation}
\label{higherloop1}
\frac{v_{b\bar b}^4}{(4\pi)^2} \frac1{v_{b\bar b}^4} \frac{\left(g v_{b\bar b}\right)^2}{F_\pi^2}M_B^2
= \left( \frac{g\,M_B v_{b\bar b}}{\Lambda_\chi} \right)^2 \ .
\end{equation}
If $v_{b\bar b}\sim 0.4$, then the three-loop diagram (a) is of
similar size as that of the two-loop diagrams.  This is the case for
the processes with the $h_b(1P), \chi_{bJ}(1P)$, and
$\Upsilon(1S,2S)$. For smaller values of $v_{b\bar b}$, it is
suppressed. In diagram (b) there is only one more bottom meson
propagator and the additional $B\pi B\pi$ vertex is in an $S$-wave.
We obtain
\begin{equation}
\label{higherloop2}
\frac{v_{b\bar b}^4}{(4\pi)^2} \frac1{v_{b\bar b}^3} \frac{v_{b\bar b}}{F_\pi^2}M_B^2
= \left( \frac{M_B v_{b\bar b}}{\Lambda_\chi} \right)^2.
\end{equation}
Four and higher loop diagrams that cannot be absorbed by using
physical parameters may now be estimated by applying a proper number
of factors of the kind of Eqs.~(\ref{higherloop1}) and
(\ref{higherloop2}). It is easy to see that also additional
topologies provide analogous factors. Since $M_B/\Lambda_\chi\sim
5$, higher loops get increasingly important, if $v_{b\bar b}>0.2$.
For $v_{b\bar b}\sim0.2$, the three and more loops are of  the same
order as the two-loop diagrams, which is the case for the $2P$
states, and thus suppressed in comparison with the one-loop
contribution. For the $3P$ bottomonia in the final state, the value
of $v_{b\bar b}$ is even smaller, and the multiple loops are even
suppressed as compared to the two-loop contribution.

To summarize the findings of the power counting analysis, we conclude that the
calculation  of one-loop triangle diagrams as depicted in
Fig.~\ref{fig:PowerCounting1} (a) and (e) is a good approximation, with a
controlled uncertainty, to the transitions of the $Z_b^{(\prime)}$ into the
$\chi_{bJ}(2P,3P)\gamma$ and $h_{b}(2P)\pi$ --- for the $h_b(3P)\pi$ the phase
space is too limited. But similar calculations are not applicable to the decays
into the $1P$ states as well as the $\Upsilon(nS)\pi$. For the decays $Z_b\to
\Upsilon(nS)\pi$, the contribution from the two-loop diagrams is not suppressed
or even enhanced compared to the one-loop contribution. In light of this
discussion, it becomes clear why the pattern of branching fractions for these
channels (see Tab.~\ref{table-Belle-branching-ratios}) $\mathcal
B[\Upsilon(2S)\pi]>\mathcal B[\Upsilon(3S)\pi]\gg\mathcal B[\Upsilon(1S)\pi]$
cannot be reproduced by calculating the three-point diagrams in the NREFT
formalism, which always favors the $1S$ to the $2S$ transition and the $2S$
compared to the $3S$ transition due to the factor $q^2$ in
Eq.~\eqref{eq:upsilonpi}.

\subsection{Branching Fractions and Ratios}\label{sec:results}
In this section, we investigate quantitatively the decays of the $Z_b$ states
through  heavy meson loops. 
 The
nonrelativistic Lagrangian for the $\chi_{bJ}$ coupling to a pair of heavy
mesons can be found in Ref.~\cite{Fleming:2008yn}, and  the one for the magnetic
coupling of heavy mesons in Ref.~\cite{Hu:2005gf}.

With the amplitudes given in App~\ref{app:AmplitudesZb_1} and \ref{app:AmplitudesZb_2}, it is straightforward
to  calculate the decay widths of the $Z_b^{(\prime)}\to h_b(mP)\pi$, which are
proportional to $g_1^2$, where $g_1$ is the $P$-wave bottomonium--bottom meson
coupling constant. In the ratio defined as
\begin{equation}
 \xi_m:=\frac{\Gamma[Z'_b\to h_b(mP)\pi]}{\Gamma[Z_b\to h_b(mP)\pi]},
\label{eq:ratio-hb}
\end{equation}
$g_1$ is canceled out. With the meson masses listed in
Tab.~\ref{table-Belle-branching-ratios}, we obtain
\begin{eqnarray}
 \xi_1=1.21\left|\frac{z'}{z}\right|^2= 0.75,\qquad
 \xi_2=(1.53\pm0.43)\left|\frac{z'}{z}\right|^2= 0.95\pm0.36.
 \label{eq:ratios}
\end{eqnarray}
where the first error in the second term is the theoretical uncertainty due to
neglecting higher order contributions (see the discussion in the previous section),
and the second one also includes the uncertainty of $z'/z$ added in quadrature.
Due to the theoretically uncontrollable higher order contributions for the
decays into the $1P$ states, no uncertainty is given for $\xi_1$.

\begin{table}[t]
\centering
\caption{The ratios $\xi$ of different decay modes in both NREFT and a relativistic framework
  as compared with the experimental data. The NREFT values quoted without uncertainties may be understood
  as order-of-magnitude estimates.}
\medskip
\begin{tabular}{|ccc|}
  \hline\hline
  $\xi$ 		& Our Result 	& Exp. \\\hline
  $\Upsilon(1S)\pi$ 	& $0.7$ 	&  $0.47\pm 0.22$ \\
  $\Upsilon(2S)\pi$ 	& $0.9$ 	&  $0.34\pm 0.15$ \\
  $\Upsilon(3S)\pi$ 	& $2\pm 2$  	& $0.48\pm 0.20$ \\
  $h_b(1P)\pi$ 		& $0.8$ 	& $1.65\pm 0.96$ \\
  $h_b(2P)\pi$ 		& $1.0\pm0.4$ 	& $2.13\pm 1.44$ \\
  \hline\hline
\end{tabular}
\label{tab:xi}
\end{table}
The predictions are consistent with their experimental counterparts
(see Tab.~\ref{table-Belle-branching-ratios})
\begin{eqnarray}
 \xi_1^{\rm Exp}=1.65\pm0.96, \quad
 \xi_2^{\rm Exp}=2.13\pm1.44.
\end{eqnarray}
Here, new measurements with significantly reduced uncertainties
would be very desirable. A collection of ratios for the decays of
the $Z_b$ states to $h_b(mP)\pi$ and $\Upsilon(nS)\pi$,
respectively, is presented in Tab.~\ref{tab:xi}. The uncertainties,
whenever they are under control theoretically, are also included.
The significant deviations for the $\Upsilon(nS)\pi$ results from
the experimental numbers appear natural, given that for those
transitions higher loops were argued to be at least as important as
the one-loop diagram included here, as outlined in detail in the
previous section.  

In Ref.~\cite{Cleven:2013sq} we calculated the same quantities using a Lorentz
covariant formalism with relativistic propagators for all the
intermediate mesons as a cross check of our nonrelativistic
treatment. The results are comparable.  
The difference between relativistic and NREFT calculations here and in the following never exceeds 15\% for
the rates and thus is well below the uncertainty due to higher loops. A more detailed discussion is therefore not necessary here. 

It is important to ask to what extent the above predictions can be
used to probe the nature of the $Z_b$ states. The $Z_b$ and $Z_b'$
are away from the $B\bar B^*$ and $B^*\bar B^*$ thresholds by
similar distances, $M_{Z_b}-M_B-M_{B^*}\simeq M_{Z'_b}-2M_{B^*}$.
Additionally, due to the heavy quark spin symmetry, one may expect
that the loops contribute similarly to the decays of these two
states into the same final state. This is indeed the case. We find
that the ratios aside of $|z'/z|^2$ are basically the phase space
ratios, which are
\begin{equation}
 \frac{|\vec q(Z_b^\prime\to h_b(1P)\pi)|^3}{|\vec q(Z_b\to h_b(1P)\pi)|^3}=1.20,
 \qquad  \frac{|\vec q(Z_b^\prime\to h_b(2P)\pi)|^3}{|\vec q(Z_b\to h_b(2P)\pi)|^3}=1.53.
\end{equation}
This implies that the ratios for the decays into the $h_b(mP)\pi$
are determined by the ratio of the partial decay widths for the
open-bottom decay modes,
\begin{equation}
\xi_m \simeq \frac{\text{PS}'_m}{\text{PS}_m} \frac{\Gamma(Z_b^{\prime+}\to B^{*+}\bar B^{*0})}
{\Gamma(Z_b^+\to B^+\bar B^0+B^0\bar B^{*+})},
\label{xiinterpret}
\end{equation}
where $\text{PS}_m^{(\prime)}$ is the phase space for the decays
$Z_b^{(\prime)}\to h_b(mP)\pi$.  Such a relation will not be obtained if the
$Z_b$ states are of tetraquark structure because the decay of a $\bar b\bar q b q$
tetraquark into the $h_b\pi$ knows nothing about the decay into the open-bottom
channels. Here, if we impose spin symmetry for $z$ and $z'$, one would get for
$\xi_m$ simply the ratio of phase spaces. In reality the second factor which is
the square of the ratio of effective couplings, c.f.
Eq.~(\ref{eq:coupling-ratio}), deviates from unity due to spin symmetry
violations enhanced by the proximity of the $B^{(*)}\bar B^{*}$ thresholds as
discussed above.

\begin{table}
\small \caption{The ratio $\Omega$ and the corresponding branching
fractions for all possible radiative decays. Uncertainties are given,
whenever they can be controlled theoretically (see text). The values
quoted without uncertainties may be understood as order of magnitude
estimates.}
\begin{center}
\begin{tabular}{|ccccc|}
               \hline\hline
                 & \multicolumn{2}{c}{$Z_b$} & \multicolumn{2}{c|}{$Z'_b$} \\
               & $\Omega$ & Branching Fraction & $\Omega$ & Branching Fraction  \\\hline
               $\chi_{b0}(1P)\gamma$ & $5\times10^{-3}$ & $1 \times10^{-4}$ & $4\times 10^{-3}$ & $3\times10^{-4}$ \\
               $\chi_{b1}(1P)\gamma$ & $1\times10^{-2}$ & $3 \times10^{-4}$ & $1\times 10^{-2}$ & $8\times10^{-4}$ \\
               $\chi_{b2}(1P)\gamma$ & $2\times10^{-2}$ & $ 5 \times10^{-4}$ & $2\times 10^{-2}$ & $1\times10^{-3}$ \\
               $\chi_{b0}(2P)\gamma$ & $(6.3\pm1.8)\times10^{-3}$ & $(2.7\pm1.5)\times10^{-4}$ & $(4.2\pm1.2)\times10^{-3}$ & $(6.2\pm 3.2)\times10^{-4}$ \\
               $\chi_{b1}(2P)\gamma$ & $(1.3\pm0.4)\times10^{-2}$ & $(5.6\pm 3.2)\times10^{-4}$ & $(1.3\pm0.4)\times10^{-2}$  & $(1.9\pm 1.0 )\times10^{-3}$ \\
               $\chi_{b2}(2P)\gamma$ & $(1.9\pm0.5)\times10^{-2}$ & $(8.3\pm 4.5)\times10^{-4}$ & $(1.8\pm0.5)\times10^{-2}$  & $(2.7\pm 1.3)\times10^{-3}$ \\
               \hline\hline
             \end{tabular}
\end{center}
\label{tab:radiative-decays}
\end{table}
Heavy quark spin symmetry allows one to gain more insight into the
molecular structure.  Because the $\chi_{bJ}(mP)$ are the
spin-multiplet partners of the $h_b(mP)$, the radiative decays of
the neutral $Z_b^{(\prime)0}$ into $\chi_{bJ}(mP)\gamma$ can be
related to the hadronic decays of the $Z_b^{(\prime)}$, no matter
whether they are
neutral or charged, into the $h_b(mP)\pi$. It is therefore useful to
define the following ratios
\begin{equation}
 \Omega_{\left[Z_b^{(\prime)},\chi_{bJ}(mP)\right]}:=
 \frac{\Gamma(Z_b^{(\prime)0}\to \chi_{bJ}(mP)\gamma)}{\Gamma(Z_b^{(\prime)0}\to h_b(mP)\pi^0)}.
\label{eq:ratio-decay}
\end{equation}
With the coupling constants in the bottom meson--photon Lagrangian
determined from  elsewhere, see for instance Ref.~\cite{Hu:2005gf},
such ratios can be predicted with no free parameters. At the
hadronic level, the $Z_b$ radiative transitions can only be related
to the hadronic ones if the $Z_b$'s are hadronic molecules so that
the two different types of transitions involve the same set of
coupling constants (modulo the bottom meson--pion/photon coupling
which can be determined from other processes or lattice
simulations). Thus, if the branching fractions of the radiative
transitions are large enough to be detected, such a measurement
would provide valuable information on the nature of the $Z_b$
states. The results for the ratios $\Omega$ are collected in
Tab.~\ref{tab:radiative-decays}.

Using the branching ratio $\mathcal{B}(Z_b^{({\prime})}\to h_b(1P,2P)\pi^+)$, we
find that the branching ratios of $Z_b^{(\prime)}\to \chi_{bJ}(1P,2P)\gamma$ are
of order $10^{-4}\sim 10^{-3}$. The largest branching fractions of
the $\chi_{bJ}(mP)$ are those into the $\gamma\Upsilon(nS)$, and the $\Upsilon(nS)$
can be easily measured. Thus, the final states for measuring the
$Z_b^0\to\chi_{bJ}\gamma$ would be the same as those of the
$Z_b^0\to\Upsilon(nS)\pi^0$ because the $\pi^0$ events are selected from photon
pairs. This means that the detection efficiency and background of these two
processes would be similar. 
In the preliminary experimental results~\cite{Adachi:2012im}, the $Z_b(10610)^0$
event number collected in the $\Upsilon(1S,2S)\pi^0$ channels is of order
$\mathcal{O}(100)$.   Given that the luminosity of the future
Super-KEKB could be two orders of magnitude higher than KEKB,
such transitions will hopefully be measured. Furthermore, one may also expect to
measure these radiative transitions at the LHCb. Note that the experimental
confirmation of the ratios given in Tab.~\ref{tab:radiative-decays} would be a
highly nontrivial evidence for the molecular nature of the $Z_b$ states.

Lacking knowledge of the $\chi_{bJ}(3P)B\bar B$ coupling constant,
the transitions into the $3P$ states cannot be predicted
parameter-free. However, they can be used to check the pattern in
Eq.~\eqref{eq:Aratio} predicted by the power counting analysis. The
decay widths of the $Z_b\to\chi_{bJ}(mP)\gamma$ are proportional to
$g_{1,mP}^2$. Taking the same value for $g_{1,mP}^2$, the explicit
evaluation of the triangle loops gives $ 1:1.8:4.4$ for the ratios
defined in Eq.~\eqref{eq:Aratio} with $J=1$. The values are close to
the ones in Eq.~\eqref{eq:Aratio}, and thus confirm the $1/\bar v$
scaling in Eq.~\eqref{eq:Apower} of the amplitudes.

\begin{table}[t]
\caption{The ratios defined in Eq.(\ref{eq:radiative-ratio}) for all
channels. Uncertainties are given, whenever they can be controlled
theoretically.}
\medskip
\centering
\begin{tabular}{|cc|}
               \hline\hline
                 & $(J=0):(J=1):(J=2)$ \\\hline
               $Z_b\to\chi_{bJ}(1P)\gamma$ & $1:2.5: 3.7$ \\
               $Z_b^\prime\to\chi_{bJ}(1P)\gamma$ & $1:2.9: 4.4$ \\
               $Z_b\to\chi_{bJ}(2P)\gamma$ & $1:(2.1\pm0.6):(2.9\pm 0.8)$ \\
               $Z_b^\prime\to\chi_{bJ}(2P)\gamma$ & $1:(3.0\pm 0.9): (4.2\pm 1.2) $ \\
               \hline\hline
             \end{tabular}
\label{tab:radiative-ratio}
\end{table}
As observed in Ref.~\cite{Guo:2010ak}, the NREFT leading loop calculation
preserves  the heavy quark spin structure. Because the $Z_b$ contains both
$s_{b\bar b}=0$ and 1 components, the leading contribution to its transitions
into the normal bottomonia, which are eigenstates of $s_{b\bar b}$, comply with
the spin symmetry. This conclusion should be true no matter what nature the
$Z_b$'s have as long as the spin structure does not change. Thus, one expects
that the branching fraction ratios of the decays of the same $Z_b$ into a
spin-multiplet bottomonia plus a pion or photon, such as
\begin{eqnarray}
\mathcal{B}(Z_b^{(\prime)}\to
\chi_{b0}(mP)\gamma): \mathcal{B}(Z_b^{(\prime)}\to\chi_{b1}(mP)\gamma): \mathcal{B}(Z_b^{(\prime)}\to\chi_{b2}(mP)\gamma),
\label{eq:radiative-ratio}
\end{eqnarray}
are insensitive to the structure of the $Z_b$. This statement may be confirmed
by observing that our results, as shown in Tab.~\ref{tab:radiative-ratio} agree
with the ratios $1:2.6:4.1$ (for the $1P$ states) and $1:2.5:3.8$ (for the $2P$
states) which are obtained solely based on heavy quark spin symmetry in
Ref.~\cite{Ohkoda:2012rj}. One may expect a derivation from the spin symmetry
results, which is due to the mass difference between the $B$ and $B^*$ mesons,
to be of order $\mathcal{O}(2\Lambda_{\rm QCD}/m_B)\sim 10\%$. The central
values given in Tab.~\ref{tab:radiative-ratio} deviate from the spin symmetry
results by at most 20\%, and they are fully consistent considering the
uncertainties.

\subsection{Comparison with other works on {$Z_b$} decays}\label{sec:comparison}

Since their discovery in an impressive number of theoretical works  the
molecular nature of the $Z_b$ states was investigated. In this section we
compare in some detail our approach to the calculations in
Refs.~\cite{Mehen:2011yh,Dong:2012hc,Li:2012uc,Li:2012as} which deal with some
of the decays considered in our paper.
Common to most of these works is that, contrary to our approach, the  second
part of the one loop integral shown in Fig.~\ref{fig:PowerCounting1}~(e) is
either approximated~\cite{Mehen:2011yh,Dong:2012hc} or  calculated differently~\cite{Li:2012uc}.
Especially, by this the analytic structure of the loop gets changed converting
it to a topology of type (f) in that figure, since
 the second $B^{(*)}\bar
B^{(*)}$ cut was removed from the loop. However, our power counting gives
that it is exactly this cut that drives the enhancement of the one loop diagrams
compared to the two-loop diagrams. 

Since the loop of type (f) gives the wave function
at the origin in $r$-space, the formalism applied in Refs.~\cite{Mehen:2011yh,Dong:2012hc,Li:2012uc} is  basically identical to that used in the classic calculations
for the decay of positronium into two photons.  However, as discussed in detail
in Ref.~\cite{Hanhart:2007wa}, it is applicable only if the range of the transition potential
from the constituents to the final state is significantly shorter ranged than the 
potential that formed the molecule --- a scale not to be mixed up with the size
of the molecule which can be very large for a shallow bound state. 
However, the range of the binding momentum of the $Z_b$ states is not
known and might well be of the order of the range of the transition potential (at
least as long as the final bottomonium is not a ground state). In such a situation
in Ref.~\cite{Hanhart:2007wa} it is proposed to calculate the full loop function for the
transitions, as done here in our work, which in effect means to expand around the
limit of a zero range potential that forms the bound state. In that paper it is also shown
that the potentially most important corrections to the transition rate cancel, such
that the uncertainty of the procedure is given by the binding momentum of the
molecule in units of the range of forces and not of the order of the final momenta
in units of the range of forces. This gives an additional justification for the
approach we are using.
We now discuss the formalisms of Refs.~\cite{Mehen:2011yh,Dong:2012hc,Li:2012uc,Li:2012as} 
in some more detail.

In Ref.~\cite{Mehen:2011yh} an effective field theory called X-EFT is
used. It is valid for hadronic
molecules with small binding energies so that the pion mass and the heavy meson
hyperfine splitting are hard scales. The decays of the $Z_b^{(\prime)}$ can be
represented by a bubble with two vertices, one connecting the $B^{(*)}\bar B^*$
to the $Z_b^{(\prime)}$ states and the other is a local operator for the
$B^{(*)}\bar B^*\pi(\bar b b)$ coupling. The coefficient of the local operator
depends on the pion energy, and is obtained by matching to the tree-level diagrams
in heavy hadron chiral perturbation theory.
For more details, we refer to
Refs.~\cite{Fleming:2007rp,Fleming:2008yn}. The ratios of $\Gamma(Z'\to
\Upsilon(3S)\pi)/\Gamma(Z\to \Upsilon(3S)\pi)$ and $\Gamma(Z'\to
h_b(2P)\pi)/\Gamma(Z\to h_b(2P)\pi)$ were calculated in Ref.~\cite{Mehen:2011yh}
assuming $z=z'$. As outlined in the discussion below Eq.~(\ref{xiinterpret}), 
in this case these ratios are just the ratios of phase spaces and thus
our results for them agree for $z=z'$.
The method of Ref.~\cite{Dong:2012hc} is a phenomenological variant of
the approach outlined above.

Also in Ref.~\cite{Li:2012uc} the transition
from the intermediate $B^{(*)}\bar B^{*}$-system to the final $\pi (\bar bb)$
system was assumed to be local, however, here the strength of this local
operator was calculated differently: the authors estimate it via the overlap
integral of the $\bar bb$ component in the $B^{(*)}\bar B^{*}$ wave function
with the outgoing $\bar bb$ pair in the presence of a dipole operator. While
this procedure is certainly justified when there are $1P$ states in the final
state
--- here the relative momenta between the two $B$ mesons are beyond 2.5 GeV and
indeed in this case our effective field theory does not converge anymore (c.f.
Sec.~\ref{sec:nonrel}) --- we regard it as questionable for the $2P$ states.
There is one more difference, namely the fact that in the formalism of
Ref.~\cite{Li:2012uc}, the transitions to the $\gamma \chi_{bJ}(nP)$ final
states are disconnected from those to the $\pi h_b(nP)$ states, while in our
approach they are connected as discussed in detail above. Thus, an experimental
observation of the decay of one of the $Z_b$ states to, say, $\gamma
\chi_{bJ}(2P)$ would allow one to decide on the applicability of our approach.

Similar to our work, in Ref.~\cite{Li:2012as} the full heavy meson loop is
evaluated, but regularized with a form factor. Absolute predictions are given
for the transitions using a model to estimate the $B^{(*)}\bar B^{(*)}(\bar bb)$
coupling --- it is difficult to judge the uncertainty induced by this. In a
first step in that work a cut-off parameter was adjusted to reproduce each
individual transition. It is found that the cut-off parameters needed for the
$h_b\pi$ transitions are typically larger and closer together than those needed
for the $\Upsilon\pi$ transitions. This hints at form factor effects being not
very significant in the former decays. This interpretation is also supported by
the observation that the ratios of decay rates --- the same quantities as
investigated here --- are found to be basically independent of the form factor.
In this sense the phenomenological studies of Ref.~\cite{Li:2012as} provide
 additional support for the effective field theory calculation presented here, 
 although a well-controlled error estimate cannot be expected from such a method.

\section{Conclusion}\label{sec:summary} 
In two approaches we studied properties of the recently discovered $Z_b$ states. 
We were able to show that the data in Ref.~\cite{Belle:2011aa} is consistent with the assumption
that the main components of the lower and higher $Z_b$ states are $S$-wave
$B\bar B^*+{\rm c.c.}$ and $B^*\bar B^*$ bound states, respectively. A small
compact tetraquark component, however, can not be excluded.

It is difficult to distinguish between resonance and bound state scenarios
with  the invariant mass spectra in Ref.~\cite{Belle:2011aa}. However, the fact that in a later publication Belle reported decays of the states into the $B^*\bar B$ and $B^*\bar B^*$, respectively, which indicates  a pole position above the open threshold. 

We have demonstrated that, if the  $Z_b$ states are indeed generated from
non-perturbative  $B\bar B^*+{\rm c.c.}$ and $B^*\bar B^*$ dynamics, the data
should not be analyzed using a Breit-Wigner parametrization. This statement can
also be reversed: if a near threshold state can be described by a Breit-Wigner
form, it is not dynamically generated, as this is possible only, if the coupling
of the resonance to the continuum channel is very weak. At present the data
appears to be consistent with line shapes that result from dynamical states. Therefore a decision about the nature of the $Z_b$ states
will be possible only once data with higher resolution and statistics will be
available.

In the second part we assume that both $Z_b$ and $Z_b^\prime$
are hadronic molecules predominantly coupling to $B\bar B^*$ and $B^*\bar B^*$,
respectively, in line with the data by the Belle Collaboration. As a consequence
of this assumption, the $Z_b$ states can only couple through $B^{(*)}\bar
B^*$--loops. Using NREFT power counting we argue that
\begin{itemize}
\item the decay channels $Z_b^{(\prime)}\to\Upsilon(nS)\pi$ as well as the
    transitions into the ground state $P$-wave bottomonia in the final state
    can not be controlled within the effective field theory, since higher loop
    contributions are expected to dominate the transitions;
\item model-independent predictions can be provided for $Z_b^{(\prime)}\to
    h_b(2P)\pi$ and radiative decays $Z_b^{(\prime)}\to
    \chi_{bJ}(2P)\gamma$.
\end{itemize}
The ratios for $Z_b$ and $Z'_b$ decays into the same final states
$h_b(mP)\pi$ are  consistent with the experimental data. Our results
reflect the fact that those ratios are essentially the ratio of the
corresponding phase space factors times the ratio of the $Z_bB\bar
B^*$ and $Z'_bB^*\bar B^*$ couplings squared. If further
experimental analysis with higher statistics could confirm this
fact, it would be a very strong evidence for the molecular
interpretation since such a relation cannot be obtained from, e.g.,
a tetraquark structure.

Furthermore, we calculate branching fractions for the final states
$\chi_{bJ}(mP)\gamma$. They are predicted to be of order
$10^{-4}\sim10^{-3}$. Although this is clearly a challenge to
experimentalists, a confirmation of these rates would strongly
support the molecular picture. It is noted that the ratios of a
certain $Z_b$ into bottomonia in the same spin multiplet are
insensitive to the structure of the $Z_b$, and may be obtained
solely based on heavy quark spin symmetry.

%% file: Summary.tex
%%%%%%%%%%%%%%%%%%%%%%%%%%%%%%%%%%%%%%%%%%%%%%%%
%
\chapter{Summary and Outlook}\label{ch:summary}
%
%%%%%%%%%%%%%%%%%%%%%%%%%%%%%%%%%%%%%%%%%%%%%%%%
In this work we have studied hadronic molecules in the heavy quark sector. The candidates we have discussed are $\dszero(2317)$ and $\dsone(2460)$ in the open charm sector and $Z_b(10610)$ and $Z_b(10650)$ in the bottomonium sector. Throughout this work we have presented arguments why we believe that the first two are $DK$ and $D^*K$ bound states, respectively, while the latter two are $BB^*$ and $B^*B^*$ resonances, respectively. Naturally, this is not the only possible explanation. The molecular interpretation is challenged by a number of other approaches. Namely, $\dszero$ and $\dsone$ are also claimed to be conventional $c\bar s$ states or tetraquarks. Similarly, features of the $Z_b$ states can also be explained by a tetraquark nature. The aim of this work was to provide ways to identify the most prominent component in the wave function of these states.

In the case of the charm-strange mesons we could make use of the fact that the interaction between charmed mesons and Goldstone bosons is dictated by chiral symmetry. This allowed us to dynamically generate $\dszero$ and $\dsone$ as poles in unitarized scattering amplitudes with coupled channels $D^{(*)}K$ and $D^{(*)}_s\eta$, respectively. The fact that we were able to produce both poles in perfect agreement with the experimental data with the same set of parameters can be seen as a support for our interpretation. Furthermore we were able to make predictions for these states. The most promising channel to distinguish between conventional $c\bar s$ states and bound states are the hadronic decays $\dszero\to D_s\pi$ and $\dsone\to D_s^*\pi$. These isospin violating decays are predicted to be of single $\kev$ order for conventional $c\bar s$ states or tetraquarks. However, we found them to be significantly larger for bound states since the loops there necessarily present lead to enhanced, non-analytic isospin violating terms. The decay widths are
\begin{equation}
 \Gamma(\dszero\to D_s\pi^0)=(133\pm 22)\kev \qquad \Gamma(\dsone\to D^*_s\pi^0)=(69\pm 26)\kev. 
\end{equation}
These values are large enough to be measured by the upcoming PANDA experiment at the FAIR collider which currently is the most promising experiment for these states. The start of measurements there is scheduled to be in the end of 2018.
Additional information may also come from the next generation $B-$factories like Belle II, especially when one considers that $\dsone$ was discovered at a $B-$factory, BaBar. Belle II is currently scheduled to start operating by the end of 2016. 
At the same time we can state that the radiative decays will probably not provide information about the nature of the decaying states. The numbers predicted by different approaches are too close together and also so small that measurements are challenging even for the dedicated experiments like PANDA. 
Besides predictions for future experiments we can also make predictions that can be tested by lattice calculations. The kaon mass dependence of the binding energy or the mass turns out to be the most promising channel here. We predict that bound states will have a linear dependence with a slope of about unity while conventional $c\bar s$ states will have quadratic dependence. We would like to encourage the lattice community to perform these calculations. 

Finally we predict $B^{(*)}K$ bound states. Heavy quark spin and flavor symmetry tells us that the dynamics in all these channels are the same up to higher order effects. We predict their masses to be
\begin{equation}
 M_{\bszero}=(5663\pm48\pm30)\mev, \qquad M_{B_{s1}}=(5712\pm48\pm30)\mev.
\end{equation}
Among the currently operating experiments the LHCb experiment at the LHC is the most promising one to measure these states. 

The second part of this work deals with the newly discovered charged bottomonium states. In direct comparison to the first part we need to state that a dynamical generation of the states is not possible here since chiral symmetry does not dictate the kinematics of $B^{(*)}B^*$ scattering. However, we present other ways to pin down the nature of the $Z_b$ states. We can calculate parameter free, model independent ratios for the currently measured decay channels, i.e. $\Upsilon\pi$ and $h_b\pi$. We find that these ratios are basically the ratio of the decays into the $BB^*$ and $B^*B^*$ channels, respectively. This is something e.g. tetraquark models cannot explain naturally. These ratios can be tested once updated data from Belle is available, the current data is not yet decisive. 

We present a detailed discussion of the lineshape of hadronic molecules for different pole positions. This way bound states, resonances and virtual states can be distinguished. In the case of the $Z_b$ states we find that their lineshapes are consistent with the assumption that they are bound states rather than resonances. However, updated experimental data has shown that they decay primarily into the continuum states and are therefore more likely to be located above threshold. An important remark here is for states very close to open thresholds --- which is the case for the $Z_b$ states --- a simple Breit-Wigner fit is not appropriate. A new fit of the experimental data using a Flatt\'{e} parametrization is desirable here. 

Finally we can predict a new decay channel: $Z^{(\prime)}_b\to\chi_b\gamma$. The branching fractions are of order $10^{-4}\sim 10^{-3}$ which makes them difficult, yet not impossible to measure. Confirmation of these states would be a highly nontrivial confirmation of our interpretation. The luminosity of Super-KEKB could be sufficient to measure it at Belle II. 

%% file: Appendix.tex
\begin{appendix}
%%%%%%%%%%%%%%%%%%%%%%%%%%%
%
\chapter{Kinematics}
%
%%%%%%%%%%%%%%%%%%%%%%%%%%%%
%
\section{Kinematics of Two-Body Decays}
%
%%%%%%%%%%%%%%%%%%%%%%%%%%%%
\begin{figure}[h]
 \centering
 \includegraphics[width=0.6\textwidth]{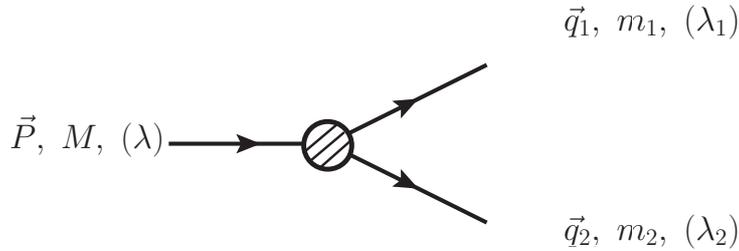}
 \caption{General form of a two-body decay}
 \label{fig:twobodydecay}
\end{figure}
The major part of this work deals with two-body decays. In this section we will discuss some general properties of these processes. Fig.~\ref{fig:twobodydecay} shows the kinematics for the decay of a particle with mass $M$ into two particles with masses $m_1$ and $m_2$. The possible polarizations in the case of (axial-)vector mesons are denoted by $\lambda$, $\lambda_I$ and $\lambda_{II}$. In almost all possible scenarios for two-body decays the rest frame of the decaying particle with $P=(M,\vec 0)$ is the best choice. Throughout this work we use it exclusively. In the rest frame energy conservation gives $M=E_1+E_2$ and momentum conservation gives $\vec q_1=-\vec q_2\equiv \vec q$. Together with $E_i^2=m_i^2+\vec q\;^2$ we find
\begin{equation}
 E_1=\frac{M^2+m_1^2-m_2^2}{2M} \qquad E_2=\frac{M^2+m_2^2-m^2_1}{2M}.
\end{equation}
Taking the square of this and replacing $E_1^2=\vec q\;^2+m_1^2$ we can obtain the absolute of the three momenta of the outgoing particles:
\begin{equation}
 |\vec q|=\frac{\sqrt{[(M^2-(m_1+m_2)^2][(M^2-(m_1-m_2)^2]}}{2M}.
\end{equation}
The decay width is given by
\begin{equation}\label{eq:widthgeneral}
 d\Gamma=\frac1{32\pi^2}\frac{\vec q}{M^2}\left|\mathcal M\right|^2d\Omega
\end{equation}
with the solid angle of particle 1, $d\Omega=d\phi_1d(\cos\theta_1)$. 
In all processes we will calculate at least one of the three particles will be a (axial-)vector particle. In this case $\mathcal M$ is not necessarily independent of the solid angle. But since we are not interested in the polarizations of the (axial-)vector mesons we will average the amplitude over the polarizations of the incoming particle and sum over the polarizations of the outgoing particles. Average means we sum over the three possible polarizations for a spin-1 particle and divide by 3. This is exactly what a detector that cannot measure polarizations see. For the process of a vector meson with polarization $\lambda$ decaying into a vector meson with polarization $\lambda'$ and a scalar meson we can write the matrix element as
\begin{equation}
 \varepsilon^\mu(P,\lambda)\varepsilon^\kappa(q_1,\lambda_1)\mathcal M_{\mu\kappa}.
\end{equation}
One can show that in general the sum over all polarizations of an external particle can be replaced by
\begin{equation}\label{eq:polsum}
 \sum_\lambda \varepsilon^\mu(q,\lambda)\varepsilon^{*\nu}(q,\lambda)\to-g^{\mu\nu}+\frac{q^\mu q^\nu}{q^2}. 
\end{equation}
Note that this is not an equality, but it holds for the matrix element if one rewrites it as $\mathcal M=\varepsilon^{\mu}(P,\lambda)\varepsilon^{\nu}(q,\lambda_1)\mathcal M_{\mu\nu}$ as 
\begin{equation}
 \overline{\left|\mathcal M\right|^2}=\frac13\sum_{\lambda,\lambda_1=1}^3\left|\varepsilon^\mu(P,\lambda)\varepsilon^\nu(q_1,\lambda_1)\mathcal M_{\mu\nu}\right|^2=\mathcal M_{\mu\nu}\mathcal M^*_{\alpha\beta}\left(-g^{\mu\alpha}+\frac{P^\mu P^\alpha}{P^2}\right)
\left(-g^{\nu\beta}+\frac{q^\nu q^\beta}{q^2}\right)
\end{equation}
Note that this replacement explicitly satisfies the relation $\varepsilon_\mu(q)  q^\mu=0$ that one can derive from the Proca Lagrangian. 

The situation is slightly different for the case of photons. Since they are massless particles physical photons are only allowed to have two degrees of freedom. Also, with the photon four momentum $k^2$ a replacement as in Eq.~(\ref{eq:polsum}) would lead to a singular expression. One can however show that any amplitude with a physical photon, $\varepsilon^\mu(\gamma,k)\mathcal M_\mu$ fulfills $k^\mu \mathcal{M}_\mu=0$. This is known as the Ward Identity, see e.g. \cite{Peskin:1995ev}. With this equality one can further show that the sum over the polarizations in the case of the photon results into
\begin{equation}
 \varepsilon_\mu(\gamma,k)\varepsilon^*_\nu(\gamma,k)\mathcal M^{\mu\nu}=(-g^{\mu\nu})M^{\mu\nu}.
\end{equation}

We can now come back to Eq.~(\ref{eq:widthgeneral}). The matrix element where we have averaged over the polarizations is independent of the solid angle $\Omega$ and  we can carry out the angular integration easily. We find
\begin{equation}
 \Gamma=\frac1{8\pi}\overline{\left|\mathcal M\right|^2}\frac{|\vec q|}{M^2}.
\end{equation}

%%%%%%%%%%%%%%%%%%%%%%%%%%%%%%%%%%%%%%%%%%%%%%%%%%%%%%%%%%%%%%%%%%%%%%%%%%%%%%%%%%%%%%%%%%%%%%%%%%%%%%%%%%%%%%%%%%%%
%
\section{Kinetic Energy for (Axial-)Vector Mesons}\label{app:stueckelberg}
%
%%%%%%%%%%%%%%%%%%%%%%%%%%%%%%%%%%%%%%%%%%%%%%%%%%%%%%%%%%%%%%%%%%%%%%%%%%%%%%%%%%%%%%%%%%%%%%%%%%%%%%%%%%%%%%%%%%%%
In this section we present the discussion of massive vector mesons based on Ref.~ \cite{Itzykson:1980rh}.
The standard Lagrangian for vector particles $A$ with mass $M_A$ reads
\begin{equation}
 \lag_A=-\frac12 A^{\mu\nu} A_{\mu\nu}^\dagger+M_A^2 A^\mu A_\mu^\dagger
\end{equation}
with the field strength tensor $A_{\mu\nu}=\partial_\mu A_\nu-\partial_\nu A_\mu$. From this one can derive the Euler-Lagrange equation:
\begin{eqnarray}
 \partial_\mu\frac{\partial\lag}{\partial\left(\partial_\mu A^\dagger_\nu \right)}-\frac{\partial\lag}{\partial A^\dagger_\nu}=0 =\partial_\nu \partial\cdot A-\square A_\nu-M_A^2 A_\nu.
\end{eqnarray}
If we contract this equation with $\partial^\nu$ we find
\begin{equation}
 M_A^2\partial{\cdot}A=0. 
\end{equation}
In other words massive vector bosons have to fulfill $\partial{\cdot}A=0$. This ensures that the physical vector mesons only have three degrees of freedom. The general solution to the equations of motion reads
\begin{equation}
 A_\mu(x)=\int\!\mathrm d^4k\sum_{\lambda=1}^3\left[a^{(\lambda)}(k)\varepsilon_\mu^{(\lambda)}e^{-ik\cdot x}+a^{(\lambda)\dagger}(k)\varepsilon_\mu^{*(\lambda)}e^{ik\cdot x}\right]\bigg|_{k^0=\sqrt{k^2+M_A^2}}
\end{equation}
with the creation and annihilation operators that fulfill the commutator relation
\begin{equation}
 \left[a^{(\lambda)}(k),a^{(\lambda')\dagger}(k')\right]=\delta_{\lambda\lambda'}2k^0(2\pi)^3\delta^{(3)}(\vec k-\vec k').
\end{equation}
The three polarization vectors $\varepsilon_\mu^{(\lambda)}$ are space-like, orthonormal and orthogonal to the time-like $k^\mu$. The resulting relations are
\begin{equation}
 \varepsilon_\mu^{(\lambda)}\cdot \varepsilon^{\mu(\lambda')}=\delta_{\lambda\lambda'}\qquad \sum_\lambda\varepsilon_\mu^{(\lambda)}(k)\varepsilon_\nu^{(\lambda)}(k)=-\left(g_{\mu\nu}-\frac{k_\mu k_\nu}{M_A^2}\right).
\end{equation}
The propagator is given by the two-point Green's function
\begin{equation}
 \left<0|TA_\mu(x)A_\nu(y)|0\right>=-i\int\!\frac{\mathrm d^4k}{(2\pi)^4}e^{-ik\cdot(x-y)}\frac{g_{\mu\nu}-k_\mu k_\nu/M_A^2}{k^2-M_A^2+i\varepsilon}-\frac{2i}{M_A^2}\delta_{\mu 0}\delta_{\nu 0}\delta^{(4)}(x-y). 
\end{equation}
As one can see the last term violates Lorentz covariance. One can solve this problem by introducing an additional term to the Lagrangian:
\begin{equation}
 \lag_A=-\frac12 A^{\mu\nu} A_{\mu\nu}^\dagger+M_A^2 A^\mu A_\mu^\dagger-\lambda \left(\partial_\mu A^\mu\right)\left(\partial_\nu A^{\dagger\nu}\right).
\end{equation}
The propagator resulting from this is
\begin{eqnarray}
 \left<0\left|TA_\mu(x)A_\nu(x)\right|0\right>&=&-i\int\!\frac{\mathrm d^4k}{(2\pi)^4}e^{-ik\cdot(x-y)}\times \\
 &&\times\left(\frac{g_{\mu\nu}-k_\mu k_\nu}{k^2-M_V^2+i\varepsilon}
 +\frac{1-\lambda}{\lambda k^2-M_V^2+i\varepsilon}\frac{k_\mu k_\nu}{k^2-M_V^2+i\varepsilon}\right).
\end{eqnarray}
We are free to choose the parameter $\lambda$. For $\lambda=0$ we recover the propagator from the Proca theory. In momentum space this is
\begin{equation}
 \frac{i}{k^2-M_A^2+i\varepsilon}\left(\frac{k^\mu k^\nu}{M_A^2}-g^{\mu\nu}\right),
\end{equation}
which for $\lambda=1$ simplifies to 
\begin{equation}
 \frac{-ig^{\mu\nu}}{k^2-M_A^2+i\varepsilon}. 
\end{equation}
The additional term does not only effect the propagator but also the coupling of a photon to the electric charge. Since the additional term has derivatives, the emission of a photon with momentum $k$ from an incoming vector meson with momentum $p$ becomes
\begin{equation}
 ie\left(g^{\mu\nu}k^\alpha-g^{\mu\alpha}k^\nu+g^{\alpha\nu}k^\mu\right)\varepsilon_\mu(p)\varepsilon_\nu(q)\varepsilon_\alpha(k)
\end{equation}
instead of 
\begin{equation}
 ie\left(g^{\mu\nu}k^\alpha-g^{\mu\alpha}p^\nu-g^{\alpha\nu}q^\mu\right)\varepsilon_\mu(p)\varepsilon_\nu(q)\varepsilon_\alpha(k).
\end{equation}
$q=p-k$ is the momentum of the outgoing vector meson.

%%%%%%%%%%%%%%%%%%%%%%%%%%%%%%%%%%%%%%%%%%%%%%%%%%%%%%%%%%
%
\section{Tensor Mesons}\label{app:tensor}
%
%%%%%%%%%%%%%%%%%%%%%%%%%%%%%%%%%%%%%%%%%%%%%%%%%%%%%%%%%%
In this section we introduce tensor fields $T^{ij}$. The fields are symmetric in the spinor indices and traceless:
\begin{equation}
 T^{ij}=T^{ji}, \qquad T^{ii}=0.
\end{equation}
The corresponding polarization vectors obey
\begin{equation}
 p_i\varepsilon^{ij}_{(\lambda)}(p)=p_j\varepsilon^{ij}_{(\lambda)}(p)=0.
\end{equation}
The polarization sum reads
\begin{equation}
 \sum_\lambda p_i\varepsilon^{ij}_{(\lambda)}(p)p_i\varepsilon^{\dagger kl}_{(\lambda)}(p) =\frac13g^{ij}g^{kl}-\frac12 \left(g^{ik}g^{jl}+g^{il}g^{jk}\right).
\end{equation}
The resulting propagator for a tensor meson with momentum $l$ and mass $M_T$ is
\begin{equation}
 i\frac{\frac13g^{ij}g^{kl}-\frac12 \left(g^{ik}g^{jl}+g^{il}g^{jk}\right)}{[l^0-\vec l^2/2M_T-M_T+i\varepsilon]}.
\end{equation}

%%%%%%%%%%%%%%%%%%%%%%%%%%%%%%%%%%%%%%%%%%%%%%%%%%%%%%%%%%
%
\chapter{Electromagnetic Decay of $\dszero$ and $\dsone$}\label{app:radiativedecays}
%
%%%%%%%%%%%%%%%%%%%%%%%%%%%%%%%%%%%%%%%%%%%%%%%%%%%%%%%%%%

\section{Integrals}\label{app:integrals}
In this section we will calculate the relativistic scalar loop integrals with one, two and three propagators. These are the relevant integrals for the radiative decays of $\dszero$, $\dsone$ and their bottom partners. The first step is to simplify the denominator by introducing so-called Feynman parameters:
\begin{equation}
 \frac{1}{A_1 A_2...A_n}=\int_0^1\!\rm dx_1...\rm dx_n\delta\left(\sum x_i-1\right)\frac{(n-1)!}{(x_1 A_1+x_2A_2+...+x_nA_n)^n}
\end{equation}
The remaining four-momentum integration can further be simplified by shifting the integration parameter. This integration is carried out in dimension $d$ to isolate divergences:
\begin{equation}
 i\int \!\frac{d^4l}{(2\pi)^4}\to \lim_{d\to 4} \mu^{4-d}i\intl
\end{equation}
We can then use the master formula for dimensional regularization, see Ref.~\cite{Peskin:1995ev}:
\begin{eqnarray}
 i \intl\frac{1}{(l^2-\Delta)^n}=\frac{(-1)^{n+1}}{(4\pi)^{d/2}}\frac{\Gamma(n-d/2)}{\Gamma(n)}\left(1/\Delta\right)^{n-d/2}.
\end{eqnarray}
For divergent integrals we need to expand up to $\mathcal O(d-4)$:
\begin{equation}
 \left(1/\Delta\right)^{2-d/2}=1-(4-d)/2\log\Delta+... ,\qquad \mu^{d-4}=1+(d-4)\log\mu+...
\end{equation}
The $\mathcal O(d-4)$ terms can contribute when multiplied by the divergent term
\begin{equation}
 \Gamma\left(\frac{d-4}2\right)=\frac{2}{d-4}-\gamma+\mathscr O(d-4)
\end{equation}
where $\gamma\sim 0.5772$ is the Euler-Mascheroni constant.

We give the results explicitely. The tadpole integral reads
\begin{eqnarray}
  T\left(M^2_1\right){=}i\mu^{4-d}\intl\frac{1}{[l^2-M^2_1+i\varepsilon]}=\frac{M^2_1}{(4\pi)^2}\frac{2}{d-2}\left[\log\frac{M^2_1}{\mu^2}-\bar\lambda\right]+\mathscr O(4-d)
\end{eqnarray}
with the divergence $\bar \lambda=2/(d-4)-\gamma+\log 4\pi$.
The results for two and three propagators are
\begin{eqnarray}
 I^{(0)}(p^2,M_1^2,M^2_2)&{=}&i\mu^{4-d}\intl \frac{1}{[(p-l)^2-M^2_1+i\varepsilon]]l^2-M^2_2+i\varepsilon]}\\
 &{=}&\frac{1}{(4\pi)^2}\intx \left[\log\frac{\Delta_2}{\mu^2}-\bar\lambda\right]+\mathscr O(4-d)
\end{eqnarray}
and
\begin{flalign}
 J^{(0)}(p^2,q^2,(p-q)^2,M^2_1,M^2_2,M^2_3){=}\phantom{xxxxxxxxxxxxxxxxxxxxxxxxxxxxxxxxxxx}\\
 {=}i\mu^{4-d}\intl\frac{1}{[(p-l)^2-M^2_1+i\varepsilon][(q-l)^2-M^2_2+i\varepsilon]l^2-M^2_3+i\varepsilon]}\\
 {=}\frac{1}{(4\pi)^2}\frac12 \intxxx\frac{1}{\Delta_3}+\mathscr O(4-d),
\end{flalign}
respectively, where we have defined
\begin{eqnarray}
 &&\Delta_2\equiv x(x-1)p^2+xM^2_1+(1-x)M^2_2\\
 &&\Delta_3\equiv x_1M^2_1+x_2M^2_2+x_3m^3-x_1x_2p^2-x_2x_3q^2-x_1x_3(p-q)^2.
\end{eqnarray}

\section{Tensor Reduction}
We will use the method of tensor reduction to simplify the occuring integrals in such a way that only the scalar integrals calculated in the previous section remain. The general idea is to make an ansatz for the integral using all orthogonal projectors that can be constructed of the occuring Lorentz structures. In the case of two propagators these are
\begin{align}
  & i\intl\frac{l_\mu}{\propq}=q_\mu I^{(1)}(q^2,M^2_1,M_2^2)  \non\\
  & i\intl\frac{l_\mu l_\nu}{\propq}=  \non\\
  & \phantom{xxxxxxxxxxxxxxxxxxxxx} {=} \left[g_{\mu\nu}-\frac{q_\mu q_\nu}{q^2}\right]I^{(2)}_0(q^2,M^2_1,M_2^2)+\frac{q_\mu q_\nu}{q^2}I^{(2)}_1(q^2,M^2_1,M_2^2). 
\end{align}
The scalar integrals are obtained by simply multiplying by the projectors. Since we constructed them orthogonally we can immediately get the corresponding integrals. By using the identity
\begin{equation}
 q\cdot l=\frac12 \bigg\{\big[l^2-M^2_2\big]-\big[(q-l)^2-M^2_1\big]+q^2-M^2_1+M^2_2\bigg\}
\end{equation}
we can finally reduce the expressions to the integrals $T(M^2)$ and $I^{(0)}(q^2,M^2_1,M^2_2)$. The results are
\begin{eqnarray}
 I^{(1)}(q^2,M^2_1,M_2^2){=}\frac{1}{2q^2}\big\{T(M^2_1)-T(M^2_2)+\left[q^2-M^2_1+M^2_2\right]\Izeroq\big\}
\end{eqnarray}
and
\begin{eqnarray}
 I^{(2)}_0(q^2,M^2_1,M_2^2){=}\frac{1}{d-1}\big\{T(M^2_1)+M^2_2I^{(0)}(q^2,M^2_1,M_2^2)-I^{(2)}_1(q^2,M^2_1,M_2^2)\big\}  \non\\
 I^{(2)}_1(q^2,M^2_1,M_2^2){=}\frac12\big\{T(M^2_1)+\left[q^2-M^2_1+M^2_2\right]I^{(1)}(q^2,M^2_1,M_2^2)\big\}.
\end{eqnarray}
Tensor reduction for three propagators is slighty more complicated since we have two linerly independent momenta $p$ and $q$. We define $q^\mu_\perp=p^\mu-q^\mu(q\cdot p)/q^2$ to contruct orthogonal projectors. For simplicity we drop the argument $\left(p^2,q^2,(p-q)^2,M^2_1,M^2_2,M^2_3\right)$ for all $J_j^{(i)}$.
\begin{eqnarray}
  &&i\intl \frac{l^\mu}{\prop}
  =q^\mu \Joneone+q_{\perp}^\mu \Jonetwo  \non\\
  &&i\intl\frac{l^\mu l^\nu}{\prop}=\\
  &&\phantom{xxxxx}=\left[\gmunu-\frac{q^\mu q^\nu}{q^2}-\frac{q^\mu_{\perp} q^\nu_\perp}{q^2_\perp}\right]J^{(2)}_0
				  +\frac{q^\mu q^\nu}{q^2}\Jtwoone+ 
				  +\frac{q_\perp^{\mu} q^\nu_\perp}{q_\perp^2}\Jtwotwo 
				  +\frac{q_\perp^{\mu} q^\nu +q^\mu q_\perp^\nu}{q^2}\Jtwothree\non
\end{eqnarray}
and further calculation yields to
\begin{eqnarray}
 \Joneone&{=}&\frac{1}{2q^2}\bigg\{\Izerok-\Izerop +\left[q^2-M^2_2+M^2_3\right]\Jzero\bigg\}  \non\\
 \Jonetwo&{=}&\frac{1}{2q_\perp^2}\bigg\{\Izerok-\Izeroq  \non\\
 &&+\left[p^2-M^2_1+M^2_3\right]\Jzero -2q\cdot p\Joneone\bigg\}
\end{eqnarray}
and
\begin{eqnarray}
  \Jtwozero&{=}&\frac{1}{d-2}\bigg\{ \Izerok+M^2_3\Jzero  -\Jtwoone-\Jtwotwo\bigg\}  \non\\
  \Jtwoone&{=}&\frac{1}{2q^2}\bigg\{ k\cdot q\Ionek+q^2\Izerok \non\\
  && -p\cdot q \Ionep   +\left[q^2-M^2_2+M^2_3\right]q^2\Joneone\bigg\}  \non\\
  \Jtwotwo&{=}&\frac{1}{2q^2}\bigg\{ -k\cdot q\Ionek+q\cdot p\Ionep  \non \\
 && +\left[k\cdot q q^2 -q^2 M^2_1+q\cdot p M^2_2-k\cdot q M^2_3 \right]\Jonetwo\bigg\}  \non\\
  \Jtwothree&{=}&\frac{1}{2}\bigg\{ \Ionek-\Ionep  +\left[q^2-M^2_2+M^2_3\right]\Jonetwo\bigg\}
\end{eqnarray}
%%%%%%%%%%%%%%%%%%%%%%%%%%%%%%%%%%%%%%%%%%%%%%%%%%%%%%%%%%%%%%%%%%%%%%%%%%%
%
\section{Amplitudes}\label{app:amplitudes_radiative}
%
%%%%%%%%%%%%%%%%%%%%%%%%%%%%%%%%%%%%%%%%%%%%%%%%%%%%%%%%%%%%%%%%%%%%%%%%%%%
%

%
%%%%%%%%%%%%%%%%%%%%%%%%%%%%%%%%%%%%%%%%%%%%%%%%%%%%%%%%%%%%%%%%%%%%%%%%%
\begin{figure*}%[h]%[htbp]
\centering
\includegraphics[width=.9\linewidth]{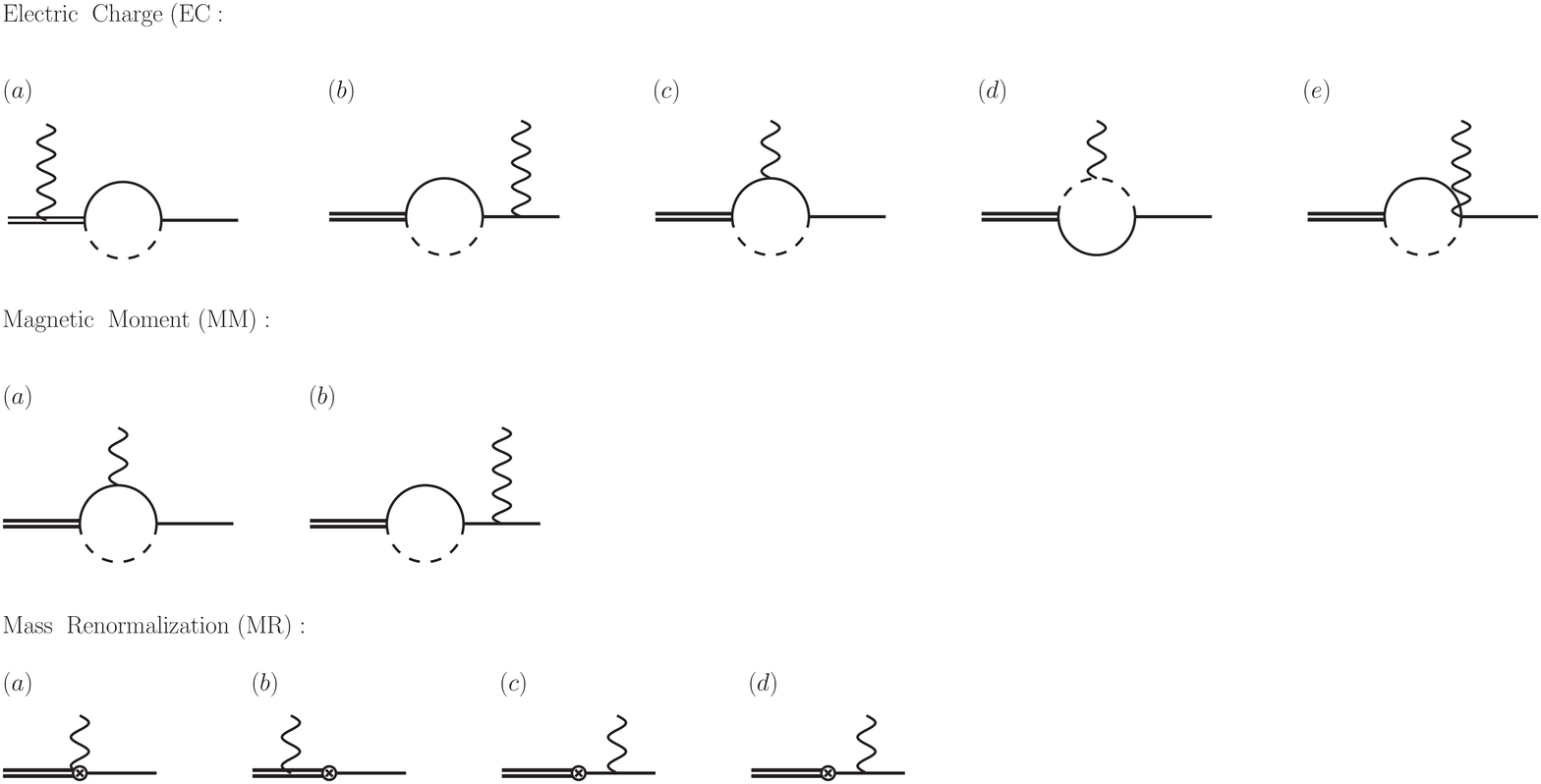}
\caption{\label{fig:diagrams_EC2} All possible diagrams. Double lines denote the molecules, single lines the charmed mesons and dashed lines the kaons. In the last line (c) and (d) denote the diagrams where the photon couples to the electric charge and magnetic moment of the outgoing heavy meson, respectively.}\vspace{6mm}
\end{figure*}

In this section we will present the amplitudes for the radiative decays of the molecular states $\dszero$, $\dsone$ and their open bottom partners. The kinematics are such that the incoming particle has the momentum $p$, the outgoing heavy meson carries the momentum $q$ and the photon $k=p-q$. We can parametrize the amplitudes with the polarization vectors for the photon and heavy mesons:
\begin{eqnarray}
 && \mathcal A^{\dszero\to D_s^*\gamma}=\mathcal A_{\mu\beta}^{\dszero\to D_s^*\gamma}\varepsilon^\mu(\gamma,k)\varepsilon^\beta(D_s^*,q)  \non\\
 && \mathcal A^{\dsone\to D_s\gamma}=\mathcal A_{\mu\alpha}^{\dsone\to D_s\gamma}\varepsilon^\mu(\gamma,k)\varepsilon^\alpha(\dsone,p)  \non\\
 && \mathcal A^{\dsone\to D^*_s\gamma}=\mathcal A_{\mu\alpha\beta}^{\dsone\to D^*_s\gamma} \varepsilon^\mu(\gamma,k)\varepsilon^\alpha(\dsone,p)\varepsilon^\alpha(D_s^*,q)  \non\\
 && \mathcal A^{\dsone\to \dszero\gamma}=\mathcal A_{\mu\alpha}^{\dsone\to\dszero\gamma}\varepsilon^\mu(\gamma,k)\varepsilon^\alpha(\dsone,p)  \non\\
 && \mathcal A^{\bszero\to B_s^*\gamma}=\mathcal A_{\mu\beta}^{\bszero\to B_s^*\gamma}\varepsilon^\mu(\gamma,k)\varepsilon^\beta(B_s^*,q)  \non\\
 && \mathcal A^{\bsone\to B_s\gamma}=\mathcal A_{\mu\alpha}^{\bsone\to B_s\gamma}\varepsilon^\mu(\gamma,k)\varepsilon^\alpha(\bsone,p)  \non\\
 && \mathcal A^{\bsone\to B^*_s\gamma}=\mathcal A_{\mu\alpha\beta}^{\bsone\to B^*_s\gamma}\varepsilon^\mu(\gamma,k)\varepsilon^\alpha(\bsone,p)\varepsilon^\alpha(B_s^*,q)  \non\\
 && \mathcal A^{\bsone\to \bszero\gamma}=\mathcal A_{\mu\alpha}^{\bsone\to\bszero\gamma}  \varepsilon^\mu(\gamma,k)\varepsilon^\alpha(\bsone,p)
\end{eqnarray}
Furthermore we can define overall constants to simplify the amplitudes:
\begin{eqnarray}
 && \mathcal C_{DK}=\frac{e g_{\pi } M_D M_{D^*} g_{DK}}{F_\pi},\quad \mathcal C_{D^*K}=\frac{e g_{\pi } M_D M_{D^*} g_{D^*K}}{F_\pi}, \non\\
 && \mathcal C_{BK}=\frac{e g_{\pi } M_B M_{B^*} g_{BK}}{F_\pi},\quad \mathcal C_{B^*K}=\frac{e g_{\pi } M_B M_{B^*} g_{B^*K}}{F_\pi}, \non\\
 && \mathcal C_{DD^*K}=eM_D M_{D^*}g_{DK} g_{D^*K},\quad \mathcal C_{BB^*K}=eM_B M_{B^*}g_{BK} g_{B^*K}.
\end{eqnarray}
The notation is as follows: 1 means that the intermediate states contain charged kaons and neutral $D-$mesons. $EC$ and $MM$ denotes if the photon couples to the electric charge or the magentic moments, respectively. a-e denote the type of diagram as given in Fig.~(\ref{fig:diagrams_EC2}).

\begin{table}%[htbp]
\centering
\scriptsize
\caption{\label{tab:intermediate} All intermediate loop mesons for the diagrams of the type $EC$, first line in  Fig.~(\ref{fig:diagrams_EC2})}
\renewcommand{\arraystretch}{1.3}
\begin{tabular}{|c|c|c|c|c|c|c|}
\hline\hline

	& $\dszero\to D_s^*\gamma$	& $\dsone\to D_s\gamma$	& $\dsone\to D_s^*\gamma$	& $\bszero\to B_s^*\gamma$	& $\bsone\to B_s\gamma$	& $\bsone\to B^*_s\gamma$ \\ \hline
$(1a)$	& 	$D^0 K^+$		&	$D^{*0} K^+$	&	$D^{*0} K^+$		&	$-$			&	$-$		&	$-$			\\
$(1b)$	& 	$D^0 K^+$		&	$D^{*0} K^+$	&	$D^{*0} K^+$		&	$-$			&	$-$		&	$-$			\\
$(1c)$	& 	$-$			&	$-$		&	$-$			&	$B^-B^-K^+$		&  $B^{*-}B^{*-}K^+$	&	$B^{*-}B^{*-}K^+$	\\
$(1d)$	& 	$K^+K^+D^0$		&	$K^+K^+D^{*0}$	&	$K^+K^+D^{*0}$		&	$K^+K^+B^-$		&	$K^+K^+B^{*-}$	&	$K^+K^+B^{*-}$		\\
$(1e)$	& 	$D^0K^+$		&	$D^{*0}K^+$	&	$D^{*0}K^+$		&	$B^-K^+$		&	$B^{*-}K^+$	&	$B^{*-}K^+$		\\ \hline
$(2a)$	& 	$D^+ K^0$		&	$D^{*+} K^0$	&	$D^{*+} K^0$		&	$-$			&	$-$		&	$-$			\\
$(2b)$	& 	$D^+ K^0$		&	$D^{*+} K^0$	&	$D^{*+} K^0$		&	$-$			&	$-$		&	$-$			\\
$(2c)$	& 	$D^+D^+K^0$		& $D^{*+}D^{*+}K^0$	&	$D^{*+}D^{*+}K^0$	&	$-$			&	$-$		&	$-$			\\
$(2d)$	& 	$-$			&	$-$		&	$-$			&	$-$			&	$-$		&	$-$			\\
$(2e)$	& 	$D^+K^0$		&	$D^{*+}K^0$	&	$D^{*+}K^0$		&	$-$			&	$-$		&	$-$			\\
\hline\hline
\end{tabular}
%%%%%%%%%%%%%%%%%%%%
\caption{\label{tab:intermediate2} All intermediate loop mesons for the diagrams of the type $MM$, second line in  Fig.~(\ref{fig:diagrams_EC2})}
\begin{tabular}{|c|c|c|c|c|c|c|}
\hline\hline

	& $\dszero\to D_s^*\gamma$	& $\dsone\to D_s\gamma$	 & $\dsone\to D_s^*\gamma$	& $\bszero\to B_s^*\gamma$	& $\bsone\to B_s\gamma$	& $\bsone\to B^*_s\gamma$	\\ \hline
(1a)	& $D^0 D^{*0} K^+$ 		&$D^{*0} D^{*0} K^+$ 	 &$D^{*0} D^{*0} K^+$ 		&	$B^0 B^{*0} K^+$  	&$B^{*-} B^{*-} K^+$ 	&$B^{*-} B^{*-} K^+$  		\\
	&  				&	 		 &$D^{*0} D^{0} K^+$ 		&				& 			&$B^{*-} B^{-} K^+$  		\\
(1b)	& $D^0  K^+$ 			&$D^{*0}  K^+$ 		 &$D^{*0} K^+$	 		&	$B^-  K^+$ 	 	&$B^{*-}  K^+$ 		&$B^{*-}  K^+$	  		\\
	&  				& 			 &$D^{*0} K^+$	 		&	 		 	& 			&$B^{*-}  K^+$	  		\\
(2a)	&  $D^+ D^{*+} K^0$		&$D^{*+} D^{*+} K^0$	 &$D^{*+} D^{*+} K^0$		&	$B^0 B^{*0} K^0$	&$B^{*0} B^{*0} K^0$	&$B^{*0} B^{*0} K^0$		\\
	&  				& 			 &$D^{*+} D^{+} K^0$		&	 			& 			&$B^{*0} B^{0} K^0$		\\
(2b)	&  $D^+  K^0$			&$D^{*+} K^0$		 &$D^{*+}  K^0$ 		&	$B^0 K^0$		&$ B^{*0}  K^0$		&$B^{*0} K^0$			\\
	& 				&			 &$D^{*+}  K^0$ 		&				&			&$B^{*0} K^0$			\\
\hline\hline
\end{tabular}
\end{table}
%%%%%%%%%%%%%%%%%%%%%%%%%%%%%%%%%%%%%%%%%%%%%%%%%%%%%%%%%%%%%%%%%%%%%%%%%%%
%
%%%%%%%%%%%%%%%%%%%%%%%%%%%%%%%%%%%%%%%%%%%%%%%%%%%%%%%%%%%%%%%%%%%%%%%%%%%
%
\subsection{$\dszero\to D_s^*\gamma$}
%
%%%%%%%%%%%%%%%%%%%%%%%%%%%%%%%%%%%%%%%%%%%%%%%%%%%%%%%%%%%%%%%%%%%%%%%%%%%

We start with the diagrams with an intermediate charged kaon. The amplitudes where the photon couples to the electric charge are given by
\begin{eqnarray} %Incoming
 \mathcal A_{\mu\beta}^{\dszero\to D_s^*\gamma}\mathrm{(EC,1a)}=\cdk\frac{q_\beta }{q^2}\frac{ \left(p_\mu+q_\mu\right)}{\left(p^2-q^2\right)} \left[q^2 I^{(0)}\left(q^2,M^2_D,M^2_K\right)-I^{(1)}\left(q^2,M^2_K,M^2_D\right)\right]
\end{eqnarray}
and
\begin{eqnarray} %Outgoing
   \mathcal A_{\mu\beta}^{\dszero\to D_s^*\gamma}\mathrm{(EC,1b)}=\cdk\frac{1}{2 \left(p^2-q^2\right)}\times \phantom{xxxxxxxxxxxxxxxxxxxxxxxxxxxxxx}\non \\
  \bigg\{-I^{(0)}\left(p^2,M^2_D,M^2_K\right) \left[g_{\beta \mu } \left(k^2+p^2-q^2\right)+2 q^\mu \left(k_\beta+q_\beta\right) +2 p_\mu q_\beta\right] \non \\
  +2 g_{\beta \mu } k^\xi I_{\xi}^{(1)}\left(p^2,M^2_K,M^2_D\right)-2 k_\beta I_\mu^{(1)}\left(p^2,M^2_K,M^2_D\right) \non \\
  +2 \left(p_\mu+q_\mu\right) I_\beta^{(1)}\left(p^2,M^2_K,M^2_D\right)\bigg\}
\end{eqnarray}
\begin{eqnarray}%Intermediate D
 && \mathcal A_{\mu\beta}^{\dszero\to D_s^*\gamma}\mathrm{(EC,1c)}=0
\end{eqnarray}

\begin{eqnarray}%Intermediate K
   \mathcal A_{\mu\beta}^{\dszero\to D_s^*\gamma}\mathrm{(EC,1d)}=\cdk\times \phantom{xxxxxxxxxxxxxxxxxxxxxxxxxxxxxxxxxxxxxx}\non \\
   \bigg\{q_\beta \left[2 J_\mu^{(1)}\left(p^2,q^2,k^2,M^2_K,M^2_K,M^2_D\right)-\left(p_\mu+q_\mu\right) J^{(0)}\left(p^2,q^2,k^2,M^2_K,M^2_K,M^2_D\right)\right] \non \\
   +\left(p_\mu+q_\mu\right) J_{\beta}^{(1)}\left(p^2,q^2,k^2,M^2_K,M^2_K,M^2_D\right)-2 J_{\beta  \mu }^{(2)}\left(p^2,q^2,k^2,M^2_K,M^2_K,M^2_D\right)\bigg\},
\end{eqnarray}
\begin{eqnarray}%Kroll Rudermann
 \mathcal A_{\mu\beta}^{\dszero\to D_s^*\gamma}\mathrm{(EC,1e)}=\cdk g_{\beta \mu } I^{(0)}\left(p^2,M^2_D,M^2_K\right),
\end{eqnarray}

We can find two diagrams where the photon couples to the magnetic moment of the charmed meson:
\begin{eqnarray} % MM Intermediate D
  \mathcal A_{\mu\beta}^{\dszero\to D_s^*\gamma}\mathrm{(MM,1a)}=\cdk\frac{2 (d-3) \sqrt{M^2_D M^2_{D^*}} \left(\beta  M^2_Q+1\right)}{3 M^2_Q}\times \phantom{xxxxxxxxxxxxxxxxx}\non \\
  \bigg\{\left(v_\beta k\cdot v-v^2 k_\beta\right) J_\mu^{(1)}\left(p^2,q^2,k^2,M^2_D,M^2_{D^*},M^2_K\right) \non \\
  + \left[k^\nu \left(v^2 g_{\beta \mu }-v_\beta v_\mu\right)+v^\nu \left(k_\beta v_\mu-g_{\beta \mu } k\cdot v\right)\right] J_\nu^{(1)}\left(p^2,q^2,k^2,M^2_D,M^2_{D^*},M^2_K\right) \bigg\}
\end{eqnarray}
and
\begin{eqnarray} % MM Outgoing
 \mathcal A_{\mu\beta}^{\dszero\to D_s^*\gamma}\mathrm{(MM,1b)}=-\cdk\frac{\sqrt{M^2_{D^*}} \left(\beta  M^2_Q+2\right)}{6 M^2_Q \left(p^2-q^2\right)}\times \phantom{xxxxx}\\
 \left[k_\beta I_\mu^{(1)}\left(p^2,M^2_D,M^2_K\right)-k^\xi g_{\beta \mu } I_\xi^{(1)}\left(p^2,M^2_D,M^2_K\right)\right].
\end{eqnarray}
The mass renormalization condition gives 
\begin{equation}
 g_\mathrm{MR}^{D^0K^+}=\frac{\cdk}{e} \frac{I^{(1)}\left(p^2,M^2_K,M^2_D\right)-p^2 I^{(0)}\left(p^2,M^2_D,M^2_K\right)}{p^2}
\end{equation}

\begin{equation}
 \mathcal A_{\mu\beta}^{\dszero\to D_s^*\gamma}\mathrm{(EC,MR)}=\cdk\frac{k_\beta q_\mu-k\cdot q g_{\beta \mu } }{p^2\left(p^2-q^2\right)} \left[p^2 I^{(0)}\left(p^2,M^2_D,M^2_K\right)- I^{(1)}\left(p^2,M^2_K,M^2_D\right)\right]
\end{equation}
The additional diagrams from the mass renormalization term read
\begin{equation}
 \mathcal A_{\mu\beta}^{\dszero\to D_s^*\gamma}\mathrm{(MR,EC,1a)}=eg_\mathrm{MR}^{D^0K^+}\frac{  q_\beta \left(p_\mu+q_\mu\right)}{p^2-q^2}
\end{equation}
\begin{equation}
 \mathcal A_{\mu\beta}^{\dszero\to D_s^*\gamma}\mathrm{(MR,EC,1b)}=-eg_\mathrm{MR}^{D^0K^+}\frac{  \left(g_{\beta \mu } k\cdot p+p_\mu q_\beta+p_\beta q_\mu\right)}{p^2-q^2}
\end{equation}
and
\begin{equation}
 \mathcal A_{\mu\beta}^{\dszero\to D_s^*\gamma}\mathrm{(MR,EC,1c)}=eg_\mathrm{MR}^{D^0K^+}  g_{\beta \mu }.
\end{equation}
The ones with the magentic moment couplings are
\begin{equation}
 \mathcal A_{\mu\beta}^{\dszero\to D_s^*\gamma}\mathrm{(MR,MM,1a)}=-\frac{ \sqrt{M^2_{D^*}} eg_\mathrm{MR}^{D^0K^+} \left(\beta  M^2_Q+2\right) \left(k_\beta p_\mu-g_{\beta \mu } k\cdot p\right)}{6 M^2_Q \left(p^2-q^2\right)}
\end{equation}
%%%%%%%%%%%%%%%%%%%%%%%%%%%%%%%%%%%%%%%%%%%%%%%%%%%%%%%%%%%%%%%%%%%%%%%%%%%
%
% \subsection{$\dszero\to D^+K^0\to D_s^*\gamma$}
%
%%%%%%%%%%%%%%%%%%%%%%%%%%%%%%%%%%%%%%%%%%%%%%%%%%%%%%%%%%%%%%%%%%%%%%%%%%%
We now come to the diagrams with neutral kaons. The diagrams of type EC read
\begin{eqnarray} %Incoming 
 \mathcal A_{\mu\beta}^{\dszero\to D_s^*\gamma}\mathrm{(EC,2a)}=\cdk  \frac{q_\beta}{q^2}\;\frac{ p_\mu+q_\mu }{ p^2-q^2}I^{(1)}\left(q^2,M^2_D,M^2_K\right)
\end{eqnarray}
\begin{eqnarray} %Outgoing
 \mathcal A_{\mu\beta}^{\dszero\to D_s^*\gamma}\mathrm{(EC,2b)}=-\cdk \frac{g_{\beta \mu } k\cdot q+p_\mu q_\beta+p_\beta q_\mu}{p^2 \left(p^2-q^2\right)}I^{(1)}\left(p^2,M^2_D,M^2_K\right)
\end{eqnarray}
\begin{eqnarray} %Intermediate D
  \mathcal A_{\mu\beta}^{\dszero\to D_s^*\gamma}\mathrm{(EC,2c)}=\cdk \times \phantom{xxxxxxxxxxxxxxxxxxxxxxxxxxxxxxxxxxxxxx}\non\\
 \left[2 J_{\beta  \mu }^{(2)}\left(p^2,q^2,k^2,M^2_D,M^2_D,M^2_K\right)-\left(p_\mu+q_\mu\right) J_\beta^{(1)}\left(p^2,q^2,k^2,M^2_D,M^2_D,M^2_K\right)\right]
\end{eqnarray}
\begin{eqnarray} %Intermediate K
  \mathcal A_{\mu\beta}^{\dszero\to D_s^*\gamma}\mathrm{(EC,2d)}=0
\end{eqnarray}

\begin{eqnarray}%Kroll Rudermann
 \mathcal A_{\mu\beta}^{\dszero\to D_s^*\gamma}\mathrm{(EC,2e)}=0
\end{eqnarray}

The diagrams of type MM read
\begin{eqnarray} % MM Intermediate
  \mathcal A_{\mu\beta}^{\dszero\to D_s^*\gamma}\mathrm{(MM,2a)}=\cdk \frac{(d-3) \sqrt{M^2_D M^2_{D^*}} \left(\beta  M^2_Q-2\right)}{3 M^2_Q}\times\phantom{xxxxxxxxxxxx}\non\\
  \bigg\{\left(v^2 k_\beta-v_\beta k\cdot v\right) J_\mu^{(1)}\left(p^2,q^2,k^2,M^2_D,M^2_{D^*},M^2_K\right)\\
  +\left[k^\nu \left(v_\beta v_\mu-v^2 g_{\beta\mu}\right)+v^\nu \left(g_{\beta\mu} k\cdot v-k_\beta v_\mu\right)\right] J_\nu^{(1)}\left(p^2,q^2,k^2,M^2_D,M^2_{D^*},M^2_K\right)\bigg\}
\end{eqnarray}

\begin{eqnarray} % MM Outgoing
  \mathcal A_{\mu\beta}^{\dszero\to D_s^*\gamma}\mathrm{(MM,2b)}=-\cdk \frac{\sqrt{M^2_{D^*}} \left(\beta  M^2_Q+2\right)}{6 M^2_Q \left(p^2-q^2\right)} \times \phantom{xxxxxxxxxxxxxxxxx}\non\\
  \left[k_\beta I_\mu^{(1)}\left(p^2,M^2_D,M^2_K\right)-k^\xi g_{\beta\mu} I_\xi^{(1)}\left(p^2,M^2_D,M^2_K\right)\right]
\end{eqnarray}
The mass renormalization condition gives 
\begin{equation}
 g_\mathrm{MR}^{D^+K^0}=-\cdk\frac{I^{(1)}\left(p^2,M^2_D,M^2_K\right)}{e p^2}
\end{equation}
The additional diagrams from the mass renormalization term read
\begin{equation}
 \mathcal A_{\mu\beta}^{\dszero\to D_s^*\gamma}\mathrm{(MR,EC,1a)}=eg_\mathrm{MR}^{D^+K^0}\frac{  q_\beta \left(p_\mu+q_\mu\right)}{p^2-q^2}
\end{equation}
\begin{equation}
 \mathcal A_{\mu\beta}^{\dszero\to D_s^*\gamma}\mathrm{(MR,EC,1b)}=eg_\mathrm{MR}^{D^+K^0}\frac{  \left(g_{\beta\mu} k\cdot p+p_\mu q_\beta+p_\beta q_\mu\right)}{q^2-p^2}
\end{equation}
and
\begin{equation}
 \mathcal A_{\mu\beta}^{\dszero\to D_s^*\gamma}\mathrm{(MR,EC,1c)}=eg_\mathrm{MR}^{D^+K^0} e g_{\beta\mu}.
\end{equation}
The ones with the magentic moment couplings are
\begin{equation}
 \mathcal A_{\mu\beta}^{\dszero\to D_s^*\gamma}\mathrm{(MR,MM,1a)}=-\frac{ \sqrt{M^2_{D^*}} e g_\mathrm{MR}^{D^+K^0} \left(\beta  M^2_Q+2\right) \left(k_\beta p_\mu-g_{\beta\mu} k\cdot p\right)}{6 M^2_Q \left(p^2-q^2\right)}.
\end{equation}

%%%%%%%%%%%%%%%%%%%%%%%%%%%%%%%%%%%%%%%%%%%%%%%%%%%%%%%%%%%%%%%%%%%%%%%%%%%
%
\subsection{$\dsone\to D_s\gamma$}
%
%%%%%%%%%%%%%%%%%%%%%%%%%%%%%%%%%%%%%%%%%%%%%%%%%%%%%%%%%%%%%%%%%%%%%%%%%%%
The EC amplitudes read
\begin{eqnarray} %Incoming
 \mathcal A_{\mu\alpha}^{\dsone\to D_s\gamma}\mathrm{(EC,1a)}=\cdstark\frac{ -g_{\alpha\mu} k\cdot q+p_\mu q_\alpha+p_\alpha q_\mu }{q^2 \left(p^2-q^2\right)} \times\phantom{xxxxxxxxx}\non \\
 \left[I^{(1)}\left(q^2,M^2_K,M^2_{D^*}\right)-q^2 I^{(0)}\left(q^2,M^2_{D^*},M^2_K\right)\right]
\end{eqnarray}

\begin{eqnarray} % Outgoing 
 \mathcal A_{\mu\alpha}^{\dsone\to D_s\gamma}\mathrm{(EC,1b)}=\cdstark\frac{p_\alpha \left(p_\mu+q_\mu\right)}{p^2 \left(p^2-q^2\right)} \left[p^2 I^{(0)}\left(p^2,M^2_{D^*},M^2_K\right)-I^{(1)}\left(p^2,M^2_K,M^2_{D^*}\right)\right]
\end{eqnarray}
\begin{eqnarray}%Intermediate D
 && \mathcal A_{\mu\alpha}^{\dsone\to D_s\gamma}\mathrm{(EC,1c)}=0
\end{eqnarray}

\begin{eqnarray}%Intermediate K
  \mathcal A_{\mu\alpha}^{\dsone\to D_s\gamma}\mathrm{(EC,1d)}=\cdstark\times \phantom{xxxxxxxxxxxxxxxxxxxxxxxxxxxxxxxxxxxxxxx}\non\\
  \times\bigg\{q_\alpha \left[\left(p_\mu+q_\mu\right) J^{(0)}\left(p^2,q^2,k^2,M^2_K,M^2_K,M^2_{D^*}\right)-2 J_\mu^{(1)}\left(p^2,q^2,k^2,M^2_K,M^2_K,M^2_{D^*}\right)\right] \\
  -\left(p_\mu+q_\mu\right) J_\alpha^{(1)}\left(p^2,q^2,k^2,M^2_K,M^2_K,M^2_{D^*}\right)+2 J_{\alpha  \mu }^{(2)}\left(p^2,q^2,k^2,M^2_K,M^2_K,M^2_{D^*}\right)\bigg\}
\end{eqnarray}

\begin{eqnarray}%Kroll Rudermann
 \mathcal A_{\mu\alpha}^{\dsone\to D_s\gamma}\mathrm{(EC,1e)}=-\cdstark g_{\alpha\mu} I^{(0)}\left(p^2,M^2_{D^*},M^2_K\right)
\end{eqnarray}

The MM amplitudes read
\begin{eqnarray}
 \mathcal A_{\mu\alpha}^{\dsone\to D_s\gamma}\mathrm{(MM,1a)}=\cdstark\frac{\sqrt{M^2_{D^*}} \left(\beta  M^2_Q-1\right)}{3 M^2_Q}\times\phantom{xxxxxxxxxxxxxxxxxxxxx}\non\\
 \left[k_\alpha J_\mu^{(1)}\left(p^2,q^2,k^2,M^2_{D^*},M^2_{D^*},M^2_K\right)-k^\nu g_{\alpha\mu} J_\nu^{(1)}\left(p^2,q^2,k^2,M^2_{D^*},M^2_{D^*},M^2_K\right)\right]
\end{eqnarray}
and
\begin{eqnarray}
\mathcal A_{\mu\alpha}^{\dsone\to D_s\gamma}\mathrm{(EC,1a)}= \cdstark\frac{(d-3) \sqrt{M^2_D M^2_{D^*}} \left(\beta  M^2_Q-2\right)}{3 M^2_Q \left(M^2_{D_s^*}-p^2\right)} \times\phantom{xxxxxxxxxxxx}\non\\
\bigg\{\left(v_\alpha k\cdot v-v^2 k_\alpha\right) I_\mu^{(1)}\left(p^2,M^2_{D^*},M^2_K\right)\\
+ I_\xi^{(1)}\left(p^2,M^2_{D^*},M^2_K\right) \left[k^\xi \left(v^2 g_{\alpha\mu}-v_\alpha v_\mu\right) +v^\xi \left(k_\alpha v_\mu-g_{\alpha\mu} k\cdot v\right)\right] \bigg\}.
\end{eqnarray}
The mass renormalization condition gives 
\begin{equation}
 g_\mathrm{MR}^{D^{*0}K^+}=-\cdk\frac{I^{(1)}\left(p^2,M^2_D,M^2_K\right)}{e p^2}
\end{equation}
The additional diagrams from the mass renormalization term read
\begin{equation}
 \mathcal A_{\mu\beta}^{\dsone\to D_s\gamma}\mathrm{(MR,EC,1a)}=eg_\mathrm{MR}^{D^{*0}K^+}\frac{ \left(-g_{\alpha\mu} k\cdot q+p_\mu q^{\alpha }+p^{\alpha } q_\mu\right)}{p^2-q^2}
\end{equation}
\begin{equation}
 \mathcal A_{\mu\beta}^{\dsone\to D_s\gamma}\mathrm{(MR,EC,1b)}=eg_\mathrm{MR}^{D^{*0}K^+}\frac{ p^{\alpha } \left(p_\mu+q_\mu\right)}{q^2-p^2}
\end{equation}
and
\begin{equation}
 \mathcal A_{\mu\beta}^{\dsone\to D_s\gamma}\mathrm{(MR,EC,1c)}=eg_\mathrm{MR}^{D^{*0}K^+}   g_{\alpha\mu}
\end{equation}
The one with the magentic moment couplings is
\begin{eqnarray}
 \mathcal A_{\mu\beta}^{\dsone\to D_s\gamma}\mathrm{(MR,MM,1a)}=e g_\mathrm{MR}^{D^{*0}K^+}\frac{(d-3) \sqrt{M^2_D M^2_{D^*}}  \left(\beta  M^2_Q-2\right)}{3 M^2_Q \left(M^2_{D_s^*}-p^2\right)}\times\phantom{xxxxxxxxx}\non\\
 \times\left(g_{\alpha\mu} \left(k\cdot v p\cdot v-v^2 k\cdot p\right)+p_\mu \left(v^2 k^{\alpha }-v^{\alpha } k\cdot v\right)+v_\mu \left(v^{\alpha
   } k\cdot p-k^{\alpha } p\cdot v\right)\right)
\end{eqnarray}

%%%%%%%%%%%%%%%%%%%%%%%%%%%%%%%%%%%%%%%%%%%%%%%%%%%%%%%%%%%%%%%%%%%%%%%%%%%
%
% \subsection{$\dsone\to D^{*+}K^0\to D_s\gamma$}
%
%%%%%%%%%%%%%%%%%%%%%%%%%%%%%%%%%%%%%%%%%%%%%%%%%%%%%%%%%%%%%%%%%%%%%%%%%%%
The EC amplitudes with charged $D-$meson read

\begin{eqnarray} % Incoming
 \mathcal A_{\mu\alpha}^{\dsone\to D_s\gamma}\mathrm{(EC,2a)}= \cdstark\frac{ g_{\alpha\mu} k\cdot q-p_\mu q_\alpha-p_\alpha q_\mu }{q^2 \left(p^2-q^2\right)}I^{(1)}\left(q^2,M^2_{D^*},M^2_K\right)
\end{eqnarray}

\begin{eqnarray} % Outgoing
 \mathcal A_{\mu\alpha}^{\dsone\to D_s\gamma}\mathrm{(EC,2b)}= \cdstark\frac{p_\mu+q_\mu}{p^2-q^2} I_\alpha^{(1)}\left(p^2,M^2_{D^*},M^2_K\right)
\end{eqnarray}

\begin{eqnarray} % Intermediate D
  \mathcal A_{\mu\alpha}^{\dsone\to D_s\gamma}\mathrm{(EC,2c)}= \cdstark \times\phantom{xxxxxxxxxxxxxxxxxxxxxxxxxxxxxxxxxxxxxxx}\non\\
  \bigg\{-p^\nu g_{\alpha\mu} J_\nu^{(1)}\left(p^2,q^2,k^2,M^2_{D^*},M^2_{D^*},M^2_K\right)+q^\nu g_{\alpha\mu} J_\nu^{(1)}\left(p^2,q^2,k^2,M^2_{D^*},M^2_{D^*},M^2_K\right)\\
  +\left(p_\mu+q_\mu\right) J_\alpha^{(1)}\left(p^2,q^2,k^2,M^2_{D^*},M^2_{D^*},M^2_K\right)+p_\alpha J_\mu^{(1)}\left(p^2,q^2,k^2,M^2_{D^*},M^2_{D^*},M^2_K\right) \\
  -q_\alpha J_\mu^{(1)}\left(p^2,q^2,k^2,M^2_{D^*},M^2_{D^*},M^2_K\right)-2 J_{\alpha  \mu  }^{(2)}\left(p^2,q^2,k^2,M^2_{D^*},M^2_{D^*},M^2_K\right)\bigg\}
\end{eqnarray}

\begin{eqnarray} % Intermediate K
 \mathcal A_{\mu\alpha}^{\dsone\to D_s\gamma}\mathrm{(EC,2d)}= 0
\end{eqnarray}

\begin{eqnarray} %Kroll Rudermann
 \mathcal A_{\mu\alpha}^{\dsone\to D_s\gamma}\mathrm{(EC,2e)}= 0
\end{eqnarray}
The MM amplitudes are
\begin{eqnarray} % MM Intermediate D
  \mathcal A_{\mu\alpha}^{\dsone\to D_s\gamma}\mathrm{(MM,2a)}= \cdstark\frac{\sqrt{M^2_{D^*}} \left(\beta  M^2_Q+2\right)}{6 M^2_Q} \times\phantom{xxxxxxxxxxxxxxxxx}\non\\
  \times\left[k^\nu g_{\alpha\mu} J_\nu^{(1)}\left(p^2,q^2,k^2,M^2_{D^*},M^2_{D^*},M^2_K\right)-k_\alpha J_\mu^{(1)}\left(p^2,q^2,k^2,M^2_{D^*},M^2_{D^*},M^2_K\right)\right]
\end{eqnarray}

\begin{eqnarray} % MM Outgoing
 \mathcal A_{\mu\alpha}^{\dsone\to D_s\gamma}\mathrm{(MM,2b)}= \cdstark\frac{(d-3) \sqrt{M^2_D M^2_{D^*}} \left(\beta  M^2_Q-2\right)}{3 M^2_Q \left(M^2_{D_s^*}-p^2\right)} \times\phantom{xxxxxxxxxx}\non\\
\bigg\{ \left(v_\alpha k\cdot v-v^2 k_\alpha\right) I_\mu^{(1)}\left(p^2,M^2_{D^*},M^2_K\right)  \non\\
+ I_\xi^{(1)}\left(p^2,M^2_{D^*},M^2_K\right) \left[k^\xi \left(v^2 g_{\alpha\mu}-v_\alpha v_\mu\right)+v^\xi \left(k_\alpha v^{\mu  }-g_{\alpha\mu} k\cdot v\right)\right]  \bigg\}
\end{eqnarray}
The mass renormalization condition gives 
\begin{equation}
 g_\mathrm{MR}^{D^{*+}K^0}=\cdk\frac{I^{(1)}\left(p^2,M^2_{D^*},M^2_K\right)}{e p^2}
\end{equation}
The additional diagrams from the mass renormalization term read
\begin{equation}
 \mathcal A_{\mu\beta}^{\dszero\to D_s^*\gamma}\mathrm{(MR,EC,1a)}=eg_\mathrm{MR}^{D^{*+}K^0}\frac{ \left(-g_{\alpha\mu} k\cdot q+p_\mu q^{\alpha }+p^{\alpha } q_\mu\right)}{p^2-q^2}
\end{equation}
\begin{equation}
 \mathcal A_{\mu\beta}^{\dszero\to D_s^*\gamma}\mathrm{(MR,EC,1b)}=-eg_\mathrm{MR}^{D^{*+}K^0}\frac{ p^{\alpha } \left(p_\mu+q_\mu\right)}{p^2-q^2}
\end{equation}
and
\begin{equation}
 \mathcal A_{\mu\beta}^{\dszero\to D_s^*\gamma}\mathrm{(MR,EC,1c)}=eg_\mathrm{MR}^{D^{*+}K^0}   g_{\alpha\mu}
\end{equation}
The ones with the magentic moment couplings are
\begin{eqnarray}
 \mathcal A_{\mu\beta}^{\dszero\to D_s^*\gamma}\mathrm{(MR,MM,1a)}=eg_\mathrm{MR}^{D^{*+}K^0}\frac{(d-3)  \sqrt{M^2_D M^2_{D^*}}  \left(\beta  M^2_Q-2\right) }{3 M^2_Q \left(M^2_{D_s^*}-p^2\right)}\times\phantom{xxxxx}\non\\
 \left[g_{\alpha\mu} \left(k\cdot v p\cdot v-v^2 k\cdot p\right)+p_\mu \left(v^2 k^{\alpha }-v^{\alpha } k\cdot v\right)+v_\mu \left(v^{\alpha
   } k\cdot p-k^{\alpha } p\cdot v\right)\right]
\end{eqnarray}
%%%%%%%%%%%%%%%%%%%%%%%%%%%%%%%%%%%%%%%%%%%%%%%%%%%%%%%%%%%%%%%%%%%%%%%%%%%
%
 \subsection{$\dsone\to D^*_s\gamma$}
%
%%%%%%%%%%%%%%%%%%%%%%%%%%%%%%%%%%%%%%%%%%%%%%%%%%%%%%%%%%%%%%%%%%%%%%%%%%%
The EC amplitudes read

\begin{eqnarray} % Incoming
 \mathcal A_{\mu\alpha\beta}^{\dsone\to D^*_s\gamma}\mathrm{(EC,1a)}= \frac{\cdstark}{q^2 \left(p^2-q^2\right)}\times \phantom{xxxxxxxxxxxxxxxxxxxxxx} \non\\
  \times\bigg\{\left[q^2 I^{(0)}\left(q^2,M^2_{D^*},M^2_K\right)-I^{(1)}\left(q^2,M^2_K,M^2_{D^*}\right)\right]  \non\\
  \left[k^\kappa q^\lambda v^\rho g_{\alpha\mu} \varepsilon _{\beta \kappa \lambda \rho }-k_\alpha q^{\kappa  } v^\lambda \varepsilon _{\beta \kappa \lambda \mu }+q^\kappa v^\lambda \left(p_\mu+q_\mu\right) \varepsilon _{\alpha \beta \kappa \lambda }\right]\bigg\}
\end{eqnarray}

\begin{eqnarray} % Outgoing
  \mathcal A_{\mu\alpha\beta}^{\dsone\to D^*_s\gamma}\mathrm{(EC,1b)}= \frac{\cdstark}{p^2\left(p^2-q^2\right)} \times  \phantom{xxxxxxxxxxxxxxxxxxxxxxxxxxxxxxx}\non\\
  \bigg\{ I^{(1)}\left(p^2,M^2_K,M^2_{D^*}\right) \left[k^\kappa p^\lambda v^\rho g_{\beta\mu} \varepsilon _{\alpha \kappa \lambda \rho }-k_\beta p^\kappa v^\lambda \varepsilon _{\alpha \kappa \lambda \mu }+p^\kappa v^\lambda \left(p_\mu+q_\mu\right) \varepsilon _{\alpha \beta \kappa \lambda }\right] \non\\
  -p^2 p^\kappa I^{(0)}\left(p^2,M^2_{D^*},M^2_K\right) \left(q^\lambda v^\rho g_{\beta\mu}
   \varepsilon _{\alpha \kappa \lambda \rho }-k_\beta v^\lambda \varepsilon _{\alpha \kappa \lambda \mu } + v^\lambda \left(p_\mu+q_\mu\right) \varepsilon _{\alpha \beta \kappa \lambda }\right)\bigg\}
\end{eqnarray}
\begin{eqnarray}%Intermediate D
 \mathcal A_{\mu\alpha\beta}^{\dsone\to D^*_s\gamma}\mathrm{(MM,1c)}= 0
\end{eqnarray}
\begin{eqnarray}%Intermediate K
 \mathcal A_{\mu\alpha\beta}^{\dsone\to D^*_s\gamma}\mathrm{(MM,1d)}= \cdstark  v^\lambda \varepsilon _{\alpha \beta \kappa \lambda } \times \phantom{xxxxxxxxxxxxxxxxxxxxxxxxxxxxxxxx}\non\\
 \bigg\{q^\kappa \left(2 J_\mu^{(1)}\left(p^2,q^2,k^2,M^2_K,M^2_K,M^2_{D^*}\right)-\left(p_\mu+q_\mu\right)
   J^{(0)}\left(p^2,q^2,k^2,M^2_K,M^2_K,M^2_{D^*}\right)\right)   \non\\
   +\left(p_\mu+q_\mu\right) J_\kappa^{(1)}\left(p^2,q^2,k^2,M^2_K,M^2_K,M^2_{D^*}\right)-2 J_{\kappa  \mu }^{(2)}\left(p^2,q^2,k^2,M^2_K,M^2_K,M^2_{D^*}\right)\bigg\}
\end{eqnarray}
\begin{eqnarray}%Kroll Rudermann
 \mathcal A_{\mu\alpha\beta}^{\dsone\to D^*_s\gamma}\mathrm{(EC,1e)}= -\cdstark v^\kappa \varepsilon _{\alpha \beta \kappa \mu } I^{(0)}\left(p^2,M^2_{D^*},M^2_K\right)
\end{eqnarray}
The MM amplitudes read
\begin{eqnarray} % MM Intermediate D
  \mathcal A_{\mu\alpha\beta}^{\dsone\to D^*_s\gamma}\mathrm{(MM,1a)}= \cdstark\frac{\sqrt{M^2_{D^*}} \left(\beta  M^2_Q-1\right)}{3 M^2_Q} \times \phantom{xxxxxxxxxxxxxxxxxxx}\non\\
  \bigg\{k_\alpha v^\lambda \varepsilon _{\beta \kappa \lambda \mu } J_\kappa^{(1)}\left(p^2,q^2,k^2,M^2_{D^*},M^2_{D^*},M^2_K\right)  \non\\
  -k^\kappa v^\rho g_{\alpha\mu}   \varepsilon _{\beta \kappa \lambda \rho } J_\lambda^{(1)}\left(p^2,q^2,k^2,M^2_{D^*},M^2_{D^*},M^2_K\right)\bigg\}
\end{eqnarray}

\begin{eqnarray} % MM Intermediate D-2
  \mathcal A_{\mu\alpha\beta}^{\dsone\to D^*_s\gamma}\mathrm{(MM,1b)}= -\cdstark\frac{2 \sqrt{M^2_D M^2_{D^*}} k^\kappa  \left(\beta  M^2_Q+1\right)}{3 M^2_Q} \times \phantom{xxxxxxxxxx} \non\\
  v^\lambda \varepsilon _{\alpha \kappa \lambda \mu } J_\beta^{(1)}\left(p^2,q^2,k^2,M^2_{D^*},M^2_D,M^2_K\right)
\end{eqnarray}

\begin{eqnarray} % MM Outgoing
 \mathcal A_{\mu\alpha\beta}^{\dsone\to D^*_s\gamma}\mathrm{(MM,1c)}= \cdstark\frac{\sqrt{M^2_{D^*}} \left(\beta  M^2_Q+2\right)}{6 p^2 M^2_Q \left(p^2-q^2\right)}\times \phantom{xxxxxxxxxxxxxxxxxxxxxx}\non\\
   I^{(1)}\left(p^2,M^2_{D^*},M^2_K\right) \left[k^\kappa p^\lambda v^\rho g_{\beta\mu} \varepsilon _{\alpha \kappa \lambda \rho }-k_\beta p^\kappa v^\lambda \varepsilon _{\alpha \kappa \lambda \mu }\right]
\end{eqnarray}

The mass renormalization condition gives 
\begin{equation}
 g_\mathrm{MR}^{D^{*0}K^+}=-\cdk\frac{I^{(1)}\left(p^2,M^2_D,M^2_K\right)}{e p^2}
\end{equation}
The additional diagrams from the mass renormalization term read
\begin{eqnarray}
 \mathcal A_{\mu\beta}^{\dszero\to D_s^*\gamma}\mathrm{(MR,EC,1a)}=\frac{eg_\mathrm{MR}^{D^{*0}K^+}}{p^2-q^2}\times\phantom{xxxxxxxxxxxxxxxxxxxxxxxxxxxx} \non\\
 \left[q^\kappa v^\lambda \left(k^{\alpha } \varepsilon _{\beta \kappa \lambda \mu }-\left(p_\mu+q_\mu\right) \varepsilon _{\alpha \beta \kappa \lambda }\right)-p^\kappa q^\lambda v^\rho
   g_{\alpha\mu} \varepsilon _{\beta \kappa \lambda \rho }\right]
\end{eqnarray}
\begin{eqnarray}
 \mathcal A_{\mu\beta}^{\dszero\to D_s^*\gamma}\mathrm{(MR,EC,1b)}=\frac{eg_\mathrm{MR}^{D^{*0}K^+}}{p^2-q^2} \times\phantom{xxxxxxxxxxxxxxxxxxxxxxxxxxx}\non\\
 p^\kappa \left[q^\lambda v^\rho g_{\beta\mu} \varepsilon _{\alpha \kappa \lambda \rho }+v^\lambda \left(\left(p_\mu+q_\mu\right) \varepsilon _{\alpha \beta \kappa \lambda }-k_\beta
   \varepsilon _{\alpha \kappa \lambda \mu }\right)\right]
\end{eqnarray}
and
\begin{equation}
 \mathcal A_{\mu\beta}^{\dszero\to D_s^*\gamma}\mathrm{(MR,EC,1c)}=eg_\mathrm{MR}^{D^{*0}K^+}  v^\kappa \varepsilon _{\alpha \beta \kappa \mu }
\end{equation}
The ones with the magentic moment couplings are
\begin{eqnarray}
 \mathcal A_{\mu\beta}^{\dszero\to D_s^*\gamma}\mathrm{(MR,MM,1a)}=-eg_\mathrm{MR}^{D^{*0}K^+}\frac{ \sqrt{M^2_{D^*}}  \left(\beta  M^2_Q+2\right)}{6 M^2_Q \left(p^2-q^2\right)} \times\phantom{xxxxxxxxxxxx} \non\\
 \times \left(\varepsilon _{\alpha \kappa\lambda\rho} k^\kappa p^\lambda v^{\rho} g_{\beta\mu}-k_\beta p^\kappa v^\lambda
   \varepsilon _{\alpha \kappa\lambda\mu }\right)
\end{eqnarray}
%%%%%%%%%%%%%%%%%%%%%%%%%%%%%%%%%%%%%%%%%%%%%%%%%%%%%%%%%%%%%%%%%%%%%%%%%%%
%
% \subsection{$\dsone\to D^{*+}K^0\to D^*_s\gamma$}
%
%%%%%%%%%%%%%%%%%%%%%%%%%%%%%%%%%%%%%%%%%%%%%%%%%%%%%%%%%%%%%%%%%%%%%%%%%%%
The diagrams with neutral kaons are given by
\begin{eqnarray} % Incoming
 \mathcal A_{\mu\alpha\beta}^{\dsone\to D^*_s\gamma}\mathrm{(EC,2a)}= \cdstark\frac{I^{(1)}\left(q^2,M^2_{D^*},M^2_K\right)}{q^2 \left(p^2-q^2\right)} \times\phantom{xxxxxxxxxxxxxxxxxxxxxxxx} \non\\
 \left[k^\kappa q^\lambda v^\rho g_{\alpha\mu} \varepsilon _{\beta \kappa \lambda \rho }
 -k_\alpha q^\kappa v^\lambda\varepsilon _{\beta \kappa \lambda \mu  }+q^\kappa v^\lambda \left(p_\mu+q_\mu\right) \varepsilon _{\alpha \beta \kappa \lambda }\right]
\end{eqnarray}

\begin{eqnarray} % Outgoing
 \mathcal A_{\mu\alpha\beta}^{\dsone\to D^*_s\gamma}\mathrm{(EC,2b)}= -\cdstark\frac{I^{(1)}\left(p^2,M^2_{D^*},M^2_K\right)}{p^2 \left(p^2-q^2\right)} \times \phantom{xxxxxxxxxxxxxxxxxxxxxxx}\non\\
 \left[k^\kappa p^\lambda v^\rho g_{\beta\mu} \varepsilon _{\alpha \kappa \lambda \rho }-k_\beta p^\kappa v^\lambda \varepsilon _{\alpha \kappa \lambda \mu }+p^\kappa v^\lambda \left(p_\mu+q_\mu\right) \varepsilon _{\alpha \beta \kappa \lambda }\right]
\end{eqnarray}
\begin{eqnarray} % Intermediate D
 \mathcal A_{\mu\alpha\beta}^{\dsone\to D^*_s\gamma}\mathrm{(EC,2c)}= \cdstark\bigg\{-k^\kappa v^\rho g_{\alpha\mu} \varepsilon _{\beta \kappa \lambda \rho } J_\lambda^{(1)}\left(p^2,q^2,k^2,M^2_{D^*},M^2_{D^*},M^2_K\right)  \phantom{xxxxx} \non\\
  +v^\lambda\left[k_\alpha \varepsilon _{\beta \kappa \lambda \mu }-\left(p_\mu+q_\mu\right) \varepsilon _{\alpha \beta \kappa \lambda }\right]  J_\kappa^{(1)}\left(p^2,q^2,k^2,M^2_{D^*},M^2_{D^*},M^2_K\right)  \non\\
  +2 v^\lambda \varepsilon _{\alpha \beta \kappa \lambda } J_{\kappa  \mu }^{(2)}\left(p^2,q^2,k^2,M^2_{D^*},M^2_{D^*},M^2_K\right)\bigg\}
\end{eqnarray}
\begin{eqnarray} % Intermediate K
 \mathcal A_{\mu\alpha\beta}^{\dsone\to D^*_s\gamma}\mathrm{(EC,2d)}= 0
\end{eqnarray}
\begin{eqnarray}%Kroll Rudermann
 \mathcal A_{\mu\alpha\beta}^{\dsone\to D^*_s\gamma}\mathrm{(EC,2e)}= 0
\end{eqnarray}
The MM diagrams read
\begin{eqnarray} % MM Intermediate D-1
 \mathcal A_{\mu\alpha\beta}^{\dsone\to D^*_s\gamma}\mathrm{(MM,2a)}= \cdstark\frac{\sqrt{M^2_{D^*}} \left(\beta  M^2_Q+2\right)}{6 M^2_Q}\times \phantom{xxxxxxxxxxxxxxxxxxxxxxx}\non\\
 \left[k^\kappa v^\rho g_{\alpha\mu}g^{\delta\lambda} \varepsilon _{\beta \kappa \lambda \rho }  -k_\alpha v^\lambda g^{\delta\kappa} \varepsilon _{\beta \kappa \lambda \mu } \right]
 J_{\delta }^{(1)}\left(p^2,q^2,k^2,M^2_{D^*},M^2_{D^*},M^2_K\right)
\end{eqnarray}

\begin{eqnarray} % MM Intermediate D-2
 \mathcal A_{\mu\alpha\beta}^{\dsone\to D^*_s\gamma}\mathrm{(MM,2b)}= \cdstark\frac{\sqrt{M^2_D M^2_{D^*}}  \left(\beta  M^2_Q-2\right)  }{3 M^2_Q}\times\phantom{xxxxxxxxxxxxxxxxxxxx} \non\\
 k^\kappa v^\lambda \varepsilon _{\alpha \kappa \lambda \mu } J_\beta^{(1)}\left(p^2,q^2,k^2,M^2_{D^*},M^2_D,M^2_K\right)
\end{eqnarray}

\begin{eqnarray} % MM Outgoing
 \mathcal A_{\mu\alpha\beta}^{\dsone\to D^*_s\gamma}\mathrm{(MM,2c)}= \cdstark\frac{\sqrt{M^2_{D^*}} \left(\beta  M^2_Q+2\right) }{6 p^2 M^2_Q \left(p^2-q^2\right)} \times \phantom{xxxxxxxxxxxxxxxxxxxxx}\non\\
 \left(k^\kappa p^\lambda v^\rho g_{\beta\mu} \varepsilon _{\alpha \kappa \lambda \rho }-k_\beta p^\kappa v^\lambda
   \varepsilon _{\alpha \kappa \lambda \mu }\right)I^{(1)}\left(p^2,M^2_{D^*},M^2_K\right)
\end{eqnarray}
The mass renormalization condition gives 
\begin{equation}
 g_\mathrm{MR}^{D^{*+}K^0}=\cdk\frac{I^{(1)}\left(p^2,M^2_{D^*},M^2_K\right)}{e p^2}
\end{equation}
The additional diagrams from the mass renormalization term read
\begin{eqnarray}
 \mathcal A_{\mu\beta}^{\dszero\to D_s^*\gamma}\mathrm{(MR,EC,2a)}=\frac{eg_\mathrm{MR}^{D^{*+}K^0}}{p^2-q^2}\times\phantom{xxxxxxxxxxxxxxxxxxxxxxxxx}\non\\
 \left[q^\kappa v^\lambda \left(k^{\alpha } \varepsilon _{\beta \kappa \lambda \mu }-\left(p_\mu+q_\mu\right) \varepsilon _{\alpha \beta \kappa \lambda }\right)-p^\kappa q^\lambda v^\rho
   g_{\alpha\mu} \varepsilon _{\beta \kappa \lambda \rho }\right]
\end{eqnarray}
\begin{eqnarray}
 \mathcal A_{\mu\beta}^{\dszero\to D_s^*\gamma}\mathrm{(MR,EC,2b)}=\frac{eg_\mathrm{MR}^{D^{*+}K^0}}{p^2-q^2} \times\phantom{xxxxxxxxxxxxxxxxxxxxxxxxx}\non\\
  p^\kappa \left[q^\lambda v^\rho g_{\beta\mu} \varepsilon _{\alpha \kappa \lambda \rho }+v^\lambda \left(\left(p_\mu+q_\mu\right) \varepsilon _{\alpha \beta \kappa \lambda }-k_\beta
   \varepsilon _{\alpha \kappa \lambda \mu }\right)\right]
\end{eqnarray}
and
\begin{equation}
 \mathcal A_{\mu\beta}^{\dszero\to D_s^*\gamma}\mathrm{(MR,EC,2c)}=eg_\mathrm{MR}^{D^{*+}K^0}  v^\kappa \varepsilon _{\alpha \beta \kappa \mu }
\end{equation}
The ones with the magentic moment couplings are
\begin{eqnarray}
 \mathcal A_{\mu\beta}^{\dszero\to D_s^*\gamma}\mathrm{(MR,MM,2a)}=-eg_\mathrm{MR}^{D^{*+}K^0}\frac{ \sqrt{M^2_{D^*}}  \left(\beta  M^2_Q+2\right)}{6 M^2_Q \left(p^2-q^2\right)}\times\phantom{xxxxxxxxxxxxxxx}\non\\
 \times\left[\varepsilon _{\alpha \kappa\lambda\rho} k^\kappa p^\lambda v^{\rho} g_{\beta\mu}- \varepsilon _{\alpha \kappa\lambda\mu }k_\beta p^\kappa v^\lambda\right]
\end{eqnarray}

%%%%%%%%%%%%%%%%%%%%%%%%%%%%%%%%%%%%%%%%%%%%%%%%%%%%%%%%%%%%%%%%%%%%%%%%%%%
%
\subsection{$\dsone\to \dszero\gamma$}
%
%%%%%%%%%%%%%%%%%%%%%%%%%%%%%%%%%%%%%%%%%%%%%%%%%%%%%%%%%%%%%%%%%%%%%%%%%%%
The only possible diagrams have the MM coupling in the loop:
\begin{equation} % MM Intermediate D-1
 \mathcal A_{\mu\alpha}^{\dsone\to D^*_s\gamma}\mathrm{(MM,1a)}=\mathcal C_{DD^*K}\frac{ \left(\beta  M^2_Q+1\right) }{3 M^2_Q}k^\kappa v^\lambda \varepsilon _{\alpha \kappa\lambda\mu }J^{(0)}\left(p^2,q^2,k^2,M^2_{D^*},M^2_D,M^2_K\right)
\end{equation}
and
\begin{equation} % MM Intermediate D-2
 \mathcal A_{\mu\alpha}^{\dsone\to\dszero\gamma}\mathrm{(MM,2a)}=-\mathcal C_{DD^*K}\frac{ \left(\beta  M^2_Q-2\right) }{6 M^2_Q}k^\kappa v^\lambda \varepsilon _{\alpha \kappa\lambda\mu } J^{(0)}\left(p^2,q^2,k^2,M^2_{D^*},M^2_D,M^2_K\right)
\end{equation}

%%%%%%%%%%%%%%%%%%%%%%%%%%%%%%%%%%%%%%%%%%%%%%%%%%%%%%%%%%%%%%%%%%%%%%%%%%%
%
\subsection{$\bszero\to B_s^*\gamma$}
%
%%%%%%%%%%%%%%%%%%%%%%%%%%%%%%%%%%%%%%%%%%%%%%%%%%%%%%%%%%%%%%%%%%%%%%%%%%%
In this section we will give the amplitudes for the process $\bszero\to B_s^*\gamma$. We can parametrize them as
\begin{equation}
 \mathcal A^{\bszero\to B_s^*\gamma}=\varepsilon_\mu(\gamma,k)\varepsilon_\beta(B_s^*,q)\mathcal A_{\mu\beta}^{\bszero\to B_s^*\gamma}
\end{equation}
The notation is as follows: 1 means that the intermediate states contain charged kaons and $B-$mesons, 2 neutral. $EC$ and $MM$ denotes if the photon couples to the electric charge or the magentic moments, respectively. a-e denote the type of diagram as given in Fig.~(\ref{fig:diagrams_EC2}). The overall constant is 
\begin{equation}
 \mathcal C_{BK}=\frac{e g_{\pi } \sqrt{M^2_B M^2_{B^*}} g_{BK}}{F_\pi}
\end{equation}
The EC amplitudes read
\begin{eqnarray}%Intermediate K
 \mathcal A_{\mu\beta}^{\bszero\to B_s^*\gamma}\mathrm{(EC,1a)}=0
\end{eqnarray}

\begin{eqnarray}%Intermediate B
 \mathcal A_{\mu\beta}^{\bszero\to B_s^*\gamma}\mathrm{(EC,1b)}=0
\end{eqnarray}
\begin{eqnarray}%Intermediate B
 \mathcal A_{\mu\beta}^{\bszero\to B_s^*\gamma}\mathrm{(EC,1c)}=\cbk[ \left(p_\mu+q_\mu\right) J_\beta^{(1)}\left(p^2,q^2,k^2,M^2_B,M^2_B,M^2_K\right) \phantom{xxxxxxxxxx}\non\\
 -2 J_{\beta  \mu }^{(2)}\left(p^2,q^2,k^2,M^2_B,M^2_B,M^2_K\right)]
\end{eqnarray}
\begin{eqnarray}%Intermediate K
 &&\mathcal A_{\mu\beta}^{\bszero\to B_s^*\gamma}\mathrm{(EC,1d)}=\cbk\left\{ q_\beta \left[2 g^{\mu\delta}+\left(p_\mu+q_\mu\right)g^{\beta\delta} \right]J_\delta^{(1)}\left(p^2,q^2,k^2,M^2_K,M^2_K,M^2_B\right)\right. \non\\
 &&\left.-q_\beta\left(p_\mu+q_\mu\right) J^{(0)}\left(p^2,q^2,k^2,M^2_K,M^2_K,M^2_B\right)-2 J_{\beta  \mu }^{(2)}\left(p^2,q^2,k^2,M^2_K,M^2_K,M^2_B\right) \right\}
\end{eqnarray}
\begin{eqnarray}%Kroll Rudermann
  \mathcal A_{\mu\beta}^{\bszero\to B_s^*\gamma}\mathrm{(EC,1e)}=\cbk g_{\beta\mu} I^{(0)}\left(p^2,M^2_B,M^2_K\right)\phantom{xxxxxxxxxxxxxxxxxxxxxxxxx}
\end{eqnarray}
For neutral kaons all five diagrams give zero contribution.

The MM amplitudes read

\begin{eqnarray} % MM Intermediate B-1
\mathcal A_{\mu\beta}^{\bszero\to B_s^*\gamma}\mathrm{(MM,1a)}=\cbk  \frac{(d-3) \sqrt{M^2_B M^2_{B^*}} \left(\beta  m^2_b+1\right)}{3 m^2_b} \times\phantom{xxxxxxxxxxxxxxxx} \non\\
\bigg\{+\left(v^2 k_\beta-v_\beta k\cdot v\right) J_\mu^{(1)}\left(p^2,q^2,k^2,M^2_B,M^2_{B^*},M^2_K\right)  \non\\
 + \left[k^\nu \left(v_\beta v_\mu-v^2 g_{\beta\mu}\right)+v^\nu \left(g_{\beta\mu} k\cdot v-k_\beta v_\mu\right)\right] J_\nu^{(1)}\left(p^2,q^2,k^2,M^2_B,M^2_{B^*},M^2_K\right)\bigg\}
\end{eqnarray}

\begin{eqnarray} % MM Outgoing
 \mathcal A_{\mu\beta}^{\bszero\to B_s^*\gamma}\mathrm{(MM,1b)}=-\cbk \frac{\sqrt{M^2_{B^*}} \left(\beta  m^2_b-1\right)  }{6 p^2 m^2_b \left(p^2-q^2\right)}\times \phantom{xxxxxxxxxxxxxxxxxxxxxxx}\non\\
 \left(k_\beta p_\mu-g_{\beta\mu} k\cdot p\right)I^{(1)}\left(p^2,M^2_B,M^2_K\right) 
\end{eqnarray}

\begin{eqnarray} % MM Intermediate B-1
 \mathcal A_{\mu\beta}^{\bszero\to B_s^*\gamma}\mathrm{(MM,2a)}=\cbk\frac{(d-3) \sqrt{M^2_B M^2_{B^*}} \left(2 \beta  m^2_b-1\right)}{3 m^2_b}\times \phantom{xxxxxxxxxxxxxxxx}\non\\
 \times\bigg\{ \left(v_\beta k\cdot v-v^2 k_\beta\right) J_\mu^{(1)}\left(p^2,q^2,k^2,M^2_B,M^2_{B^*},M^2_K\right)  \non\\
+ \left(k^\nu \left(v^2 g_{\beta\mu}-v_\beta v_\mu\right)+v^\nu \left(k_\beta v_\mu-g_{\beta\mu} k\cdot v\right)\right) J_\nu^{(1)}\left(p^2,q^2,k^2,M^2_B,M^2_{B^*},M^2_K\right)\bigg\}
\end{eqnarray}

\begin{eqnarray} % MM Outgoing
 \mathcal A_{\mu\beta}^{\bszero\to B_s^*\gamma}\mathrm{(MM,2b)}=-\cbk\frac{\sqrt{M^2_{B^*}} \left(\beta  m^2_b-1\right) }{6 p^2 m^2_b \left(p^2-q^2\right)} \left(k_\beta p_\mu-g_{\beta\mu} k\cdot p\right)I^{(1)}\left(p^2,M^2_B,M^2_K\right)
\end{eqnarray}

%%%%%%%%%%%%%%%%%%%%%%%%%%%%%%%%%%%%%%%%%%%%%%%%%%%%%%%%%%%%%%%%%%%%%%%%%%%
%
\subsection{$\bsone\to B_s\gamma$}
%
%%%%%%%%%%%%%%%%%%%%%%%%%%%%%%%%%%%%%%%%%%%%%%%%%%%%%%%%%%%%%%%%%%%%%%%%%%%
The EC amplitudes read
\begin{eqnarray} % Incoming
 \mathcal A_{\mu\alpha}^{\bsone\to B_s\gamma}\mathrm{(EC,1a)}=0
\end{eqnarray}

\begin{eqnarray} % Outgoing
 \mathcal A_{\mu\alpha}^{\bsone\to B_s\gamma}\mathrm{(EC,1b)}=0
\end{eqnarray}
\begin{eqnarray}%Intermediate B
 \mathcal A_{\mu\alpha}^{\bsone\to B_s\gamma}\mathrm{(EC,1c)}=\cbstark \bigg\{g_{\alpha\mu} g^{\nu \xi } J_{\nu  \xi }^{(2)}\left(p^2,q^2,k^2,M^2_{B^*},M^2_{B^*},M^2_K\right) \phantom{xxxxxxxxx} \non\\
+ \left[-q^{\nu } g_{\alpha\mu}g^{\delta\nu} +q^{\alpha }g^{\mu\delta}-q_\mu g^{\alpha\delta}\right]J_{\delta}^{(1)}\left(p^2,q^2,k^2,M^2_{B^*},M^2_{B^*},M^2_K\right)
 \bigg\}
\end{eqnarray}
\begin{eqnarray}%Intermediate K
 &&\mathcal A_{\mu\alpha}^{\bsone\to B_s\gamma}\mathrm{(EC,1d)}=\cbstark\bigg\{-\left[2q_\alpha g^{\mu\delta} + (p_\mu+q_\mu) g^{\alpha\delta}\right] J_\delta^{(1)}\left(p^2,q^2,k^2,M^2_K,M^2_K,M^2_{B^*}\right)  \non\\
&& +q_\alpha (p_\mu+q_\mu ) J^{(0)}\left(p^2,q^2,k^2,M^2_K,M^2_K,M^2_{B^*}\right)+2 J_{\alpha  \mu }^{(2)}\left(p^2,q^2,k^2,M^2_K,M^2_K,M^2_{B^*}\right)\bigg\}
\end{eqnarray}
\begin{equation}%Kroll Rudermann
 \mathcal A_{\mu\alpha}^{\bsone\to B_s\gamma}\mathrm{(EC,1e)}=-\cbstark g_{\alpha\mu}I^{(0)}\left(p^2,M^2_{B^*},M^2_K\right)
\end{equation}
For neutral kaons all five diagrams give zero contribution.

The MM amplitudes read

\begin{eqnarray} % MM Intermediate B
 \mathcal A_{\mu\alpha}^{\bsone\to B_s\gamma}\mathrm{(MM,1a)}=\cbstark \frac{\sqrt{M^2_{B^*}} \left(\beta  m^2_b-1\right)}{6m^2_b}\times\phantom{xxxxxxxxxxxxxxxxxxxxxxxx} \non\\
  \left(k^\nu g_{\alpha\mu} J_\nu^{(1)}\left(p^2,q^2,k^2,M^2_{B^*},M^2_{B^*},M^2_K\right)-k_\alpha J_\mu^{(1)}\left(p^2,q^2,k^2,M^2_{B^*},M^2_{B^*},M^2_K\right)\right)
\end{eqnarray}

\begin{eqnarray} % MM Outgoing
 \mathcal A_{\mu\alpha}^{\bsone\to B_s\gamma}\mathrm{(MM,1b)}=\cbstark \frac{(d-3) \sqrt{M^2_B M^2_{B^*}} \left(\beta  m^2_b+1\right)}{3 p^2 m^2_b \left(p^2-M^2_{B_s^*}\right)}\times \phantom{xxxxxxxxxxxxxxxx}  \non\\
  I^{(1)}\left(p^2,M^2_{B^*},M^2_K\right) \big[g_{\alpha\mu} \left(k\cdot v p\cdot v-v^2 k\cdot p\right)  \non\\
 +k_\alpha \left(v^2 p_\mu-v_\mu p\cdot v\right)+v_\alpha \left(v_\mu k\cdot p-p_\mu k\cdot v\right)\big]
\end{eqnarray}

Neutral Intermediates

\begin{eqnarray} % MM Intermediate B
  \mathcal A_{\mu\alpha}^{\bsone\to B_s\gamma}\mathrm{(MM,2a)}=\cbstark \frac{\sqrt{M^2_{B^*}} \left(2 \beta  m^2_b+1\right)}{6 m^2_b}\times\phantom{xxxxxxxxxxxxxxxxxxxxxxxx}  \non\\
   \left[k_\alpha J_\mu^{(1)}\left(p^2,q^2,k^2,M^2_{B^*},M^2_{B^*},M^2_K\right)-k^\nu g_{\alpha\mu} J_{\nu
   }^{(1)}\left(p^2,q^2,k^2,M^2_{B^*},M^2_{B^*},M^2_K\right)\right]
\end{eqnarray}

\begin{eqnarray} % MM Outgoing
 \mathcal A_{\mu\alpha}^{\bsone\to B_s\gamma}\mathrm{(MM,2b)}=\cbstark \frac{(d-3) \sqrt{M^2_B M^2_{B^*}} \left(\beta  m^2_b+1\right)}{3 p^2 m^2_b \left(p^2-M^2_{B_s^*}\right)} I^{(1)}\left(p^2,M^2_{B^*},M^2_K\right) \times\phantom{xx}  \non\\
  \bigg[g_{\alpha\mu} \left(k\cdot v p\cdot v-v^2 k\cdot p\right) 
 +k_\alpha \left(v^2 p_\mu-v_\mu p\cdot v\right)+v_\alpha \left(v_\mu k\cdot p-p_\mu k\cdot v\right)\bigg]
\end{eqnarray}

%%%%%%%%%%%%%%%%%%%%%%%%%%%%%%%%%%%%%%%%%%%%%%%%%%%%%%%%%%%%%%%%%%%%%%%%%%%
%
\subsection{$\bsone\to B_s^*\gamma$}
%
%%%%%%%%%%%%%%%%%%%%%%%%%%%%%%%%%%%%%%%%%%%%%%%%%%%%%%%%%%%%%%%%%%%%%%%%%%%
The EC amplitudes read
\begin{eqnarray} %Incoming
 \mathcal A_{\mu\alpha\beta}^{\bsone\to B^*_s\gamma}\mathrm{(EC,1a)}=0
\end{eqnarray}
\begin{eqnarray} % Outgoing
 \mathcal A_{\mu\alpha\beta}^{\bsone\to B^*_s\gamma}\mathrm{(EC,1b)}=0
\end{eqnarray}
\begin{eqnarray} %Intermediate B
 \mathcal A_{\mu\alpha\beta}^{\bsone\to B^*_s\gamma}\mathrm{(EC,1c)}=\cbstark\times\phantom{xxxxxxxxxxxxxxxxxxxxxxxxxxxxxxxxx} \non\\
 \bigg\{ J_\kappa^{(1)}\left(p^2,q^2,k^2,M^2_{B^*},M^2_{B^*},M^2_K\right) \left(q^\lambda v^\rho g_{\alpha\mu} \varepsilon _{\beta \kappa \lambda \rho }+q_\mu v^\lambda \varepsilon _{\alpha \beta \kappa \lambda }+q_\alpha
   v^\lambda \varepsilon _{\beta \kappa \lambda \mu }\right) \non\\
   -v^\lambda \varepsilon _{\alpha \beta \kappa \lambda }J_{\kappa  \mu }^{(2)}\left(p^2,q^2,k^2,M^2_{B^*},M^2_{B^*},M^2_K\right)-v^\lambda \varepsilon
   _{\beta \kappa \lambda \mu } J_{\alpha  \kappa }^{(2)}\left(p^2,q^2,k^2,M^2_{B^*},M^2_{B^*},M^2_K\right)\bigg\}\non\\
\end{eqnarray}
\begin{eqnarray}%Intermediate K
 \mathcal A_{\mu\alpha\beta}^{\bsone\to B^*_s\gamma}\mathrm{(EC,1d)}=\cbstark v^\lambda \varepsilon _{\alpha \beta \kappa \lambda }\times \phantom{xxxxxxxxxxxxxxxxxxxxxxxxxxxxxx} \non\\
 \times\bigg\{2 \left[q^\kappa g^{\mu\delta}+\left(p_\mu+q_\mu\right)g^{\kappa\delta} \right]J_\delta^{(1)}\left(p^2,q^2,k^2,M^2_K,M^2_K,M^2_{B^*}\right)  \non\\
-\left(p_\mu+q_\mu\right)
   q^\kappa J^{(0)}\left(p^2,q^2,k^2,M^2_K,M^2_K,M^2_{B^*}\right)    -2 J_{\kappa  \mu }^{(2)}\left(p^2,q^2,k^2,M^2_K,M^2_K,M^2_{B^*}\right)\bigg\}
\end{eqnarray}
\begin{equation}%Kroll Rudermann
 \mathcal A_{\mu\alpha\beta}^{\bsone\to B^*_s\gamma}\mathrm{(EC,1e)}=-\cbstark v^\kappa \varepsilon _{\alpha \beta \kappa \mu } I^{(0)}\left(p^2,M^2_{B^*},M^2_K\right)
\end{equation}
For neutral kaons all five diagrams give zero contribution.

The MM amplitudes read
\begin{eqnarray} % MM Intermediate B
 \mathcal A_{\mu\alpha\beta}^{\bsone\to B^*_s\gamma}\mathrm{(MM,1a)}=\cbstark \frac{\sqrt{M^2_{B^*}} \left(\beta  m^2_b-1\right)}{6 m^2_b} \times \phantom{xxxxxxxxxxxxxxxxxxxx} \non\\
 \left[k^\kappa v^\rho g_{\alpha\mu} \varepsilon _{\beta \kappa \lambda \rho }g^{\lambda\delta }
 -k_\alpha v^\lambda \varepsilon _{\beta \kappa \lambda \mu } g^{\kappa\delta}\right]J_{\delta }^{(1)}\left(p^2,q^2,k^2,M^2_{B^*},M^2_{B^*},M^2_K\right)
\end{eqnarray}

\begin{eqnarray} % MM Intermediate B
 \mathcal A_{\mu\alpha\beta}^{\bsone\to B^*_s\gamma}\mathrm{(MM,1b)}=\cbstark \frac{\sqrt{M^2_B M^2_{B^*}} \left(\beta  m^2_b+1\right) }{3 m^2_b} \times\phantom{xxxxxxxxxxxxxxxxxxx} \non\\
 \times k^\kappa v^\lambda\varepsilon _{\alpha \kappa \lambda \mu } J_\beta^{(1)}\left(p^2,q^2,k^2,M^2_{B^*},M^2_B,M^2_K\right)
\end{eqnarray}

\begin{eqnarray} % MM Outgoing
 \mathcal A_{\mu\alpha\beta}^{\bsone\to B^*_s\gamma}\mathrm{(MM,1c)}=\cbstark \frac{\sqrt{M^2_{B^*}} \left(\beta  m^2_b-1\right) }{6 p^2 m^2_b \left(p^2-q^2\right)}\times \phantom{xxxxxxxxxxxxxxxxxxxxxx}  \non\\
 \left(k^\kappa p^\lambda v^\rho g_{\beta\mu} \varepsilon _{\alpha \kappa \lambda \rho }-k_\beta p^\kappa v^\lambda
   \varepsilon _{\alpha \kappa \lambda \mu }\right) I^{(1)}\left(p^2,M^2_{B^*},M^2_K\right)
\end{eqnarray}

and 

\begin{eqnarray} % MM Intermediate B-1
 \mathcal A_{\mu\alpha\beta}^{\bsone\to B^*_s\gamma}\mathrm{(MM,2a)}=\cbstark \frac{\sqrt{M^2_{B^*}} \left(2 \beta  m^2_b+1\right)}{6 m^2_b} \times \phantom{xxxxxxxxxxxxxxxxxxx}\non\\
  \left[k_\alpha v^\lambda \varepsilon _{\beta \kappa \lambda \mu } g^{\kappa\delta}
 k^\kappa v^\rho g_{\alpha\mu} \varepsilon _{\beta \kappa \lambda \rho } g{\lambda\delta}\right]J_\delta^{(1)}\left(p^2,q^2,k^2,M^2_{B^*},M^2_{B^*},M^2_K\right)
\end{eqnarray}

\begin{eqnarray} % MM Intermediate B-2
 \mathcal A_{\mu\alpha\beta}^{\bsone\to B^*_s\gamma}\mathrm{(MM,2b)}=-\cbstark \frac{\sqrt{M^2_B M^2_{B^*}}  \left(2 \beta  m^2_b-1\right)}{3 m^2_b} k^\kappa \times \phantom{xxxxxxxxxxxxxxx} \non\\
  v^\lambda\varepsilon _{\alpha \kappa \lambda \mu } J_\beta^{(1)}\left(p^2,q^2,k^2,M^2_{B^*},M^2_B,M^2_K\right)
\end{eqnarray}

\begin{eqnarray} % MM Outgoing
 \mathcal A_{\mu\alpha\beta}^{\bsone\to B^*_s\gamma}\mathrm{(MM,2c)}=\cbstark \frac{\sqrt{M^2_{B^*}} \left(\beta  m^2_b-1\right)}{6 p^2 m^2_b \left(p^2-q^2\right)} I^{(1)}\left(p^2,M^2_{B^*},M^2_K\right) \times \phantom{xxxxxxxx}\non\\
  \left(k^\kappa p^\lambda v^\rho g_{\beta\mu} \varepsilon _{\alpha \kappa \lambda \rho }-k_\beta p^\kappa v^\lambda \varepsilon _{\alpha \kappa \lambda \mu }\right)
\end{eqnarray}

%%%%%%%%%%%%%%%%%%%%%%%%%%%%%%%%%%%%%%%%%%%%%%%%%%%%%%%%%%%%%%%%%%%%%%%%%%%
%
\subsection{$\bsone\to \bszero\gamma$}
%
%%%%%%%%%%%%%%%%%%%%%%%%%%%%%%%%%%%%%%%%%%%%%%%%%%%%%%%%%%%%%%%%%%%%%%%%%%%
Only the diagrams with MM coupling in the loop contribute:
\begin{equation}
 \mathcal A_{\mu\alpha}^{\bsone\to \bszero\gamma}\mathrm{(MM,1a)}=-\mathcal C_{BB^*K}\frac{ \left(\beta  m^2_b+1\right) }{6 m^2_b}k^\kappa v^\lambda \varepsilon _{\alpha \kappa\lambda\mu }J^{(0)}\left(p^2,q^2,k^2,M^2_{B^*},M^2_B,M^2_K\right)
\end{equation}
and
\begin{equation}
 \mathcal A_{\mu\alpha}^{\bsone\to \bszero\gamma}\mathrm{(MM,2a)}=-\mathcal C_{BB^*K}\frac{ \left(1-2 \beta  m^2_b\right) }{6 m^2_b}k^\kappa v^\lambda \varepsilon _{\alpha \kappa\lambda\mu }J^{(0)}\left(p^2,q^2,k^2,M^2_{B^*},M^2_B,M^2_K\right)
\end{equation}

%%%%%%%%%%%%%%%%%%%%%%%%%%%%%%%%%%%%%%%%%%%%%%%%%%%%%%%%%%%%%%%%%%%%%%%%%%
%
\chapter{Nonrelativistic Effective Theory}
%
%%%%%%%%%%%%%%%%%%%%%%%%%%%%%%%%%%%%%%%%%%%%%%%%%%%%%%%%%%%%%%%%%%%%%%%%%%
%
\section{NREFT Fundamental Integrals}\label{app:NREFT_integrals}
%
%%%%%%%%%%%%%%%%%%%%%%%%%%%%%%%%%%%%%%%%%%%%%%%%%%%%%%%%%%%%%%%%%%%%%%%%%%
The relativistic three-point scalar integral reads
\begin{equation}
 i \intl\frac{1}{[l^2-M^2_1+i\varepsilon][(P-l)^2-M^2_2+i\varepsilon][(l-q)^2-M^2_3+i\varepsilon]}
\end{equation}
which becomes 
\begin{equation}
 \frac{i}{8M_1M_2M_3}\intl\frac{1}{\left[l^0{-}\frac{\vec l^2}{2M_1}{-}M^2_1+i\varepsilon\right]\left[M{-}l^0{-}\frac{\vec l^2}{2M_2}{-}M^2_2{+}i\varepsilon\right]\left[l^0{-}q^0{-}\frac{(\vec l{-}\vec q)^2}{2M_3}{-}M^2_3+i\varepsilon\right]}
\end{equation}
Since the nonrelativistic normalization always gives a factor $M_1M_2M_3$ we define the nonrelativistic three-point scalar integral as
\begin{eqnarray}
 I_{\rm NR}^{(0)}(M_1,M_2,M_3,\vec q):=\non\\-\frac i8\intl\frac{1}{[l^0-\frac{\vec l^2}{2M_1}+i\varepsilon][l^0+b_{12}+\frac{\vec l^2}{2M_2}-i\varepsilon][l^0+b_{13}-\frac{(\vec l-\vec q)^2}{2M_3}+i\varepsilon]}
\end{eqnarray}
where $b_{12}:=M^2_1+M^2_2-M$ and $b_{13}:=M^2_1-M_3-q^0$. The contour integration over $l^0$ can be worked out and we find
\begin{equation}
\frac{\mu_{12}\mu_{23}}{2}\intll \frac{1}{[\vec l^2+c_{12}-i\varepsilon][\vec l^2-2\mu_{23}\frac{\vec l\cdot \vec q}{M_3}+c_{23}-i\varepsilon]}
\end{equation}
where $\mu_{ij}:=\frac{M_1M_2}{M_1+M_2}$, $c_{12}:=2\mu_{12}b_{12}$ and $c_{23}:=2\mu_{23}(b_{12}-b_{13}+\frac{\vec q\,^2}{2M_3})$. Now we introduce a Feynman parameter and shift the integration variable:
\begin{equation}
 \frac{\mu_{12}\mu_{23}}{2}\intx\intll \frac{1}{[\vec l^2-x^2a+(1-x)c_{12}-xc_{23}-i\varepsilon]^2}
\end{equation}
with $a:=\left(\frac{\mu_{23}}{M_3}\right)^2\vec q\,^2$. Since there are no poles we can take $d=4$ and find 
\begin{equation}
 -\frac{\mu_{12}\mu_{23}}{16\pi}\intx\left(x^2a-(1-x)c_{12}-xc_{23}+i\varepsilon\right)^{-1/2}
\end{equation}
and finally
\begin{equation}
 I_{\rm NR}^{(0)}(M_1,M_2,M_3,|\vec q|)=\frac{\mu_{12}\mu_{23}}{16\pi\sqrt a}\left[\tan^{-1}\left(\frac{c_{23}-c_{12}}{2\sqrt{ac_{12}}}\right)+\tan^{-1}\left(\frac{2a+c_{12}-c_{23}}{2\sqrt{a(c_{23}-a}}\right)\right].
\end{equation}
We also need 
\begin{eqnarray}
 q^iI_{\rm NR}^{(1)}(M_1,M_2,M_3,|\vec q|):= \non\\
 M_1M_2M_3i\intl\frac{l^i}{[l^2-M^2_1+i\varepsilon][(P-l)^2-M^2_2+i\varepsilon][(l-q)^2-M_3+i\varepsilon]}
\end{eqnarray}
Together with what we have found for the scalar integral this means
\begin{equation}
 I^{(1)}_{\rm NR}(M_1,M_2,M_3,|\vec q|)=\frac{\mu_{12}\mu_{23}}{2\vec q\,^2}\intll\frac{\vec l\cdot q}{[\vec l^2+c_{12}-i\varepsilon][\vec l^2-2\mu_{23}\frac{\vec l\cdot \vec q}{M^2_3}+c_{23}-i\varepsilon]}
\end{equation}
Using the method of tensor reduction this equals
\begin{equation}
 -\frac{1}{2a}\left\{\frac{\mu_{12}\mu_{23}}{2}\left[B(c_{12})-B(c_{23}-a)\right]+(c_{12}-c_{23})I_{\rm NR}^{(0)}(\vec q,M_1,M_2,M_3)\right\}
\end{equation}
where
\begin{eqnarray}
 B(c)&:=&\intll\frac{1}{[\vec l^2+c-i\varepsilon]}=-\frac{\sqrt{c-i\varepsilon}}{4\pi}
\end{eqnarray}

\section{Amplitudes}\label{app:AmplitudesZb}
\subsection{$Z_b^{(\prime)}\to h_b(nP)\pi$}\label{app:AmplitudesZb_1}
In terms of the loop function given above, the amplitudes for $Z_b^{+}$ and
$Z_b^{\prime+}$ decays into $h_b\pi^+$ are
\begin{align}
\Amp_{Z_b^{+}h_b} &= \frac{2\sqrt{2}g g_1 z_1}{F_\pi} \sqrt{M_{h_b}M_{Z_b}} \varepsilon_{ijk}q^i
\varepsilon_{Z_b}^j \varepsilon_{h_b}^k \non\\
& \times \left[I_{\rm NR}^{(0)}(M_B,M_{B^*},M_{B^*},|\vec{q}|)+I_{\rm NR}^{(0)}(M_{B^*},M_B,M_{B^*},|\vec{q}|)\right], \label{Aeq:ahb1}\\
\intertext{and} %
\Amp_{Z_b^{\prime+}h_b} &= \frac{2\sqrt{2}g g_1 z_2}{F_\pi} \sqrt{M_{h_b}M_{Z_b'}}
\varepsilon_{ijk}q^i \varepsilon_{Z_b'}^j \varepsilon_{h_b}^k \non\\
& \times
\left[I_{\rm NR}^{(0)}(M_{B^*},M_{B^*},M_B,|\vec{q}|)+I_{\rm NR}^{(0)}(M_{B^*},M_{B^*},M_{B^*},|\vec{q}|)\right], \label{Aeq:ahb2}
\end{align}
respectively. In all these amplitudes, both the neutral and charged bottom and
anti-bottom mesons have been taken into account. Here,
\begin{equation}
 |\vec q|=\sqrt{\frac{(M_{Z_b}^2-(M_{h_b}-M_\pi)^2)(M_Z^2-(M_{h_b}+M_\pi)^2)}{4M_{Z_b}^2}}
\end{equation}
\subsection{$Z_b^{(\prime)}\to \chi_{bJ}(nP)\gamma$}\label{app:AmplitudesZb_2}
The amplitudes for $Z_b^{(\prime)0}$ into $\chi_{bJ}\gamma$ read
\begin{eqnarray}
\Amp_{Z_b^{0}\chi_{b0}\gamma} &=& -\sqrt{\frac23}\beta e g_1 z\sqrt{M_{\chi_b}M_{Z_b}}\varepsilon_{ijk}q^i\varepsilon_{Z_b}^j \varepsilon_{\gamma}^k\non \\
&&\times\left[I_{\rm NR}^{(0)}(M_B,M_{B^*},M_{B^*},|\vec{q}|)-3\,I_{\rm NR}^{(0)}(M_{B^*},M_B,M_B,|\vec{q}|)\right], \label{Aeq:achib0}\\
\Amp_{Z_b^{0}\chi_{b1}\gamma} &=&2i\beta e g_1 z\sqrt{M_{\chi_b}M_{Z_b}} (q^i g^{jk}-q^k g^{ij})\varepsilon_{Z_b}^i \varepsilon_{\gamma}^j\varepsilon_{\chi_b}^k  I_{\rm NR}^{(0)}(M_{B^*},M_B,M_{B^*},|\vec{q}|), \label{Aeq:achib1}\\
\text{and} \non \\
\Amp_{Z_b^{0}\chi_{b2}\gamma} &=&\sqrt2i\beta e g_1 z\sqrt{M_{\chi_b}M_{Z_b}}q^i(g^{jm}\varepsilon_{ikl}+g^{jl}\varepsilon_{ikm})\varepsilon_{Z_b}^j \varepsilon_{\gamma}^k \varepsilon_{\chi_b}^{lm}\non \\ 
&&\times  I_{\rm NR}^{(0)}(M_{B^*},M_{B^*},M_{B^*},|\vec{q}|) \label{Aeq:achib2},
\end{eqnarray}
respectively. Here,
\begin{equation}
 |\vec q|=\frac{M_{Z_b}^2-M_{\chi_b}^2}{2M_{Z_b}}
\end{equation}

\subsection{$Z_b^{(\prime)}\to \Upsilon(mS)\pi$}\label{app:AmplitudesZb_3}
The amplitudes for $Z_b^{(\prime)+}$ into $\Upsilon(mS)\pi^+$ $(m=1,2,3)$ read
\begin{align}
 \Amp_{Z_b^{+}\Upsilon\pi^+} =&-\frac{2 \sqrt{2} g g_2 z}{F_\pi} \sqrt{M_{\Upsilon } M_Z} \varepsilon_{Z_b}^i \varepsilon_{\Upsilon}^j \non\\
 &\times\left\{g^{ij}\vec q\,^2\left[I_{\rm NR}^{(0)}(M_B,M_{B^*},M_{B^*},\vec q)+I_{\rm NR}^{(0)}(M_{B^*},M_B,M_{B^*},|\vec q|)\right.\right.\non\\
 &\left.-2 I_{\rm NR}^{(1)}(M_B,M_{B^*},M_{B^*},|\vec q|)-2 I_{\rm NR}^{(1)}(M_{B^*},M_B,M_{B^*},|\vec q|)\right]\non\\
 &+q^iq^j\left[I_{\rm NR}^{(0)}(M_{B^*},M_B,M_B,|\vec q|)-I_{\rm NR}^{(0)}(M_{B^*},M_B,M_{B^*},|\vec q|)\right.\non\\
 &\left.\left.2 I_{\rm NR}^{(1)}(M_{B^*},M_B,M_{B^*},|\vec q|)-2 I_{\rm NR}^{(1)}(M_{B^*},M_B,M_B,|\vec q|)\right]\right\}\\
\intertext{and }\\
 \Amp_{Z_b^{\prime+}\Upsilon\pi^+} =&\frac{2 \sqrt{2} g g_2 z}{F_\pi} \sqrt{M_{\Upsilon } M_Z} \varepsilon_{Z_b}^i \varepsilon_{\Upsilon}^j \non\\
 &\times\left\{\left(|\vec q|\,^2 g^{a c}+q^a q^c\right) \left[I_{\rm NR}^{(0)}\left(M_{B^*},M_{B^*},M_{B^*},|\vec q|\right)-2 I_{\rm NR}^{(1)}\left(M_{B^*},M_{B^*},M_{B^*},|\vec q|\right)\right]\right.\\
 &\left.-\left(q^a q^c-|\vec q|\,^2 g^{a c}\right)\left[I_{\rm NR}^{(0)}\left(M_{B^*},M_{B^*},M_B,|\vec q|\right)-2 I_{\rm NR}^{(1)}\left(M_{B^*},M_{B^*},M^2_B,|\vec q|\right)\right]\right\}
\end{align}
Here,
\begin{equation}
 |\vec q|=\sqrt{\frac{(M_{Z_b}^2-(M_{\Upsilon}-M_\pi)^2)(M_Z^2-(M_\Upsilon+M_\pi)^2)}{4M_{Z_b}^2}}
\end{equation}
\end{appendix}